\documentclass[12pt]{report}

\usepackage{tamuconfig}

\usepackage{times}
\usepackage[T1]{fontenc}

\usepackage{graphicx}
\usepackage{epstopdf}

\graphicspath{ {./graphic/} }

\usepackage[hidelinks]{hyperref}

\usepackage{macro/humberto_commands}
\usepackage{macro/humberto_variable_names}
\usepackage{xcolor}
\usepackage{mathtools}
\usepackage{tabularx}
\usepackage{multirow}
\usepackage{amsmath}

\usepackage{siunitx}
\usepackage{color}
\definecolor{mycolor2}{rgb}{0, 0, 1}

\definecolor{mycolor2}{rgb}{1, 0, 1}

\usepackage{tikz}
\usetikzlibrary{positioning,graphs,calc,backgrounds,fit,arrows,patterns,shapes}
\usepackage[ruled,vlined,algo2e]{algorithm2e}

\usepackage{caption}
\usepackage{subcaption}

\newcommand\dynStateWidth{0.65}
\newcommand\dynBetaWidth{0.5}

\DeclareMathOperator{\Tr}{Tr}
\DeclareMathOperator{\diag}{diag}

\begin{document}

\renewcommand{\tamumanuscripttitle}{Extending the practical applicability of the Kalman Filter}

\renewcommand{\tamupapertype}{Dissertation}

\renewcommand{\tamufullname}{Jose Humberto Ramos Zuniga}

\renewcommand{\tamudegree}{DOCTOR OF PHILOSOPHY}
\renewcommand{\tamuchairone}{John Hurtado}

\renewcommand{\tamumemberone}{John Valasek}
\newcommand{\tamumembertwo}{Manoranjan Majji}
\newcommand{\tamumemberthree}{Pilwon Hur}
\renewcommand{\tamudepthead}{Rodney Bowersox}

\renewcommand{\tamugradmonth}{August}
\renewcommand{\tamugradyear}{2020}
\renewcommand{\tamudepartment}{Aerospace Engineering}

%
%
%
%


\providecommand{\tabularnewline}{\\}

\begin{titlepage}
\begin{center}
\MakeUppercase{\tamumanuscripttitle}
\vspace{4em}

A \tamupapertype

by

\MakeUppercase{\tamufullname}

\vspace{4em}

\begin{singlespace}

Submitted to the Office of Graduate and Professional Studies of \\
Texas A\&M University \\

in partial fulfillment of the requirements for the degree of \\
\end{singlespace}

\MakeUppercase{\tamudegree}
\par\end{center}
\vspace{2em}
\begin{singlespace}
\begin{tabular}{ll}
 & \tabularnewline
& \cr
Chair of Committee, & \tamuchairone\tabularnewline
Committee Members, & \tamumemberone\tabularnewline
 & \tamumembertwo\tabularnewline
 & \tamumemberthree\tabularnewline
Head of Department, & \tamudepthead\tabularnewline

\end{tabular}
\end{singlespace}
\vspace{3em}

\begin{center}
\tamugradmonth \hspace{2pt} \tamugradyear

\vspace{3em}

Major Subject: \tamudepartment \par
\vspace{3em}
Copyright \tamugradyear \hspace{.5em}\tamufullname 
\par\end{center}
\end{titlepage}
\pagebreak{}


%
%
%
%

\chapter*{ABSTRACT}
\addcontentsline{toc}{chapter}{ABSTRACT} 

\pagestyle{plain} 
\pagenumbering{roman} 
\setcounter{page}{2}

\indent A Schmidt filter is a modification of the Kalman filter that allows to append system parameters as states and considers their uncertainty effect in the filtering process without attempting to estimate such parameters. The states that are only considered but not estimated, are generally known as \textit{consider} or \textit{considered} states. The main contributions of this research are the formulations of a Schmidt-Kalman filter that incorporates the numerical robustness of the well-known square root and factorized filtering forms plus the capacity of actively attempting to update the \textit{considered} states.

The filters formulations proposed in this research are a fundamental extension of the Kalman filter.  Therefore, the formulations of this work also apply within the Extended Kalman filter framework. More importantly, they are shown to handle nonlinearities, larger initial uncertainties, and poorly conditioned systems better than a typical Extended or Schmidt Kalman filter. Because the new filters directly based on the Schmidt filter, they offer a novel and straight-forward filtering framework, allowing the use of a more simple filter where a more advanced or elaborated technique could have been needed.

The proposed contributions of this research are organized as follows. First, the Partial-update Kalman filter, a generalized Schmidt filter that allows updating the user-selected consider states partially, is introduced. The indirect or multiplicative error version of this new filter is also derived. An error stability analysis (for linear systems) of the partial-update filter and a discussion on its numerical stability and the potential numerical robustness improvements is presented. Second, a square root formulation to improve the numerical stability of the partial-update Schmidt filter is developed. The derivation of sequential and vector measurement processing schemes for the square root formulation are both presented along with a brief computational complexity and Montecarlo analysis. Third, to gain computational efficiency but still retaining a numerically robust formulation, as an alternative, a U-D factorized version of the partial-update filter is also developed. Fourth, to improve estimation consistency and accuracy of the partial-update filter, baseline methods are proposed to attempt the estimation of the considered states. Finally, formulations proposed in this research are validated through hardware implementations to solve aerospace engineering-related problems.

\pagebreak{}

%
%
%
%

\chapter*{DEDICATION}
\addcontentsline{toc}{chapter}{DEDICATION}  

\begin{center}
\vspace*{\fill}
To Maria, Eileen and Rihan, my lovely family. To my friends that always have supported and believed in me. 
\vspace*{\fill}
\end{center}

\pagebreak{}

%
%
%
%

\chapter*{ACKNOWLEDGMENTS}
\addcontentsline{toc}{chapter}{ACKNOWLEDGMENTS}  

\indent The completion of a doctoral degree is not a task I could have completed by myself. I have been fortunate to have such wonderful persons with me during this journey always to listen, support, and encourage me. I would first like to thank my lovely wife, Maria, for being by my side along the way, helping me in all ways she can, and always believe in me, but above all for being so brave and accept to leave all we had in our country to support me. 
I would also like to thank my father, Humberto, and my grandmother, Chabelita, my sisters, Mayra and Yajaira, and my brother, Kevin, for always being so supportive even when they are far away. Also, I want to express my gratitude to my wife's family, the Don Garcias, where I have been received as another son and supported in many ways.  

I have no words to express my gratitude to my advisor, Dr. Hurtado. I could not be more fortunate to meet and work with him. Thank you for letting me learn from you in both professional and a personal way. I truly appreciate that you always found a way to support me and in one way or another, keeping me on track. Thank you for all your patience, and the support when things were fine and also when they were not that favorable. There is no way to tell how grateful I am for the time you have invested in me. Thank you for believing in me. You will always have all my respect and admiration.

I would like to thank my committee members, Dr. Valasek, Dr. Majji, and Dr. Hur, for their time and invaluable advice for the completion of this work. It is always a pleasure to visit their office to share ideas. 

Dr. Kevin Brink, from the Air Force Research Laboratory, was also crucial in the development of this dissertation. I want to thank him for all his advice, support, and the time he invested in me so I could complete this research. Thank you for your patience, guidance, and the opportunity of interning with you. The work at your lab, the Autonomous Vehicles Lab at the Research Education and Engineering Facility (REEF) at the University of Florida, was a course-changer in my academic development. From the REEF, I also want to thank Prashant Ganesh, a fantastic person, engineer, and friend with who I had the pleasure to work during two summer internships.

During my stay at Texas A\&M, I met wonderful people that have positively impacted my life and helped me to make the past few years more bearable. I want to thank Vinicius Guimaraes, Niladri Das, Daniel Whitten, Clark Moody, Neil McHenry, Irving Solis, Oscar Barajas, Tim Woodbury, and Davis Adams for their support, encouragement, and friendship. Thanks for all the countless good moments, dinners, and trips that remembered me that there is always time to relax. 

Finally, I again thank Dr. Hurtado. This work had not been possible without your full commitment.

\pagebreak{}
%
%
%
%

\chapter*{CONTRIBUTORS AND FUNDING SOURCES}
\addcontentsline{toc}{chapter}{CONTRIBUTORS AND FUNDING SOURCES}  

\subsection*{Contributors}
This work was supported by a dissertation committee consisting of Professor John Hurtado, Professors John Valasek, and Manoranjan Majji, of the Aerospace Engineering Department, and Professor Pilwon Hur of the Mechanical Engineering Department. Dr. Brink of the Air Force Research Laboratory also served as a mentor for this work.

The data collection for Section 6.2 and 6.3 was performed in part by Prashant Ganesh of the REEF at the University of Florida. Davis Adams of the LASR Laboratory at Texas A\&M helped in part to organize simulation and running code for 5.3 and 5.4. All other work conducted for the dissertation was completed by the student independently.
\subsection*{Funding Sources}
Graduate study was supported in part by Texas A\&M University, Air Force Research Laboratory, and a doctorate fellowship (with ID 411660) from the Mexican National Council for Science and Technology.
\pagebreak{}

%
%
%
%

\phantomsection
\addcontentsline{toc}{chapter}{TABLE OF CONTENTS}  

\begin{singlespace}
\renewcommand\contentsname{\normalfont} {\centerline{TABLE OF CONTENTS}}

\setcounter{tocdepth}{4} 

\setlength{\cftaftertoctitleskip}{1em}
\renewcommand{\cftaftertoctitle}{%
\hfill{\normalfont {Page}\par}}

\tableofcontents

\end{singlespace}

\pagebreak{}


\phantomsection
\addcontentsline{toc}{chapter}{LIST OF FIGURES}  

\renewcommand{\cftloftitlefont}{\center\normalfont\MakeUppercase}

\setlength{\cftbeforeloftitleskip}{-12pt} 
\renewcommand{\cftafterloftitleskip}{12pt}

\renewcommand{\cftafterloftitle}{%
\\[4em]\mbox{}\hspace{2pt}FIGURE\hfill{\normalfont Page}\vskip\baselineskip}

\begingroup

\begin{center}
\begin{singlespace}
\setlength{\cftbeforechapskip}{0.4cm}
\setlength{\cftbeforesecskip}{0.30cm}
\setlength{\cftbeforesubsecskip}{0.30cm}
\setlength{\cftbeforefigskip}{0.4cm}
\setlength{\cftbeforetabskip}{0.4cm}



\listoffigures

\end{singlespace}
\end{center}

\pagebreak{}

%
\phantomsection
\addcontentsline{toc}{chapter}{LIST OF TABLES}  

\renewcommand{\cftlottitlefont}{\center\normalfont\MakeUppercase}

\setlength{\cftbeforelottitleskip}{-12pt} 

\renewcommand{\cftafterlottitleskip}{1pt}

\renewcommand{\cftafterlottitle}{%
\\[4em]\mbox{}\hspace{2pt}TABLE\hfill{\normalfont Page}\vskip\baselineskip}

\begin{center}
\begin{singlespace}

\setlength{\cftbeforechapskip}{0.4cm}
\setlength{\cftbeforesecskip}{0.30cm}
\setlength{\cftbeforesubsecskip}{0.30cm}
\setlength{\cftbeforefigskip}{0.4cm}
\setlength{\cftbeforetabskip}{0.4cm}

\listoftables 

\end{singlespace}
\end{center}
\endgroup
\pagebreak{}  
%
%
%
%


\pagestyle{plain} 
\pagenumbering{arabic} 
\setcounter{page}{1}

\chapter{INTRODUCTION}
\section{Motivation}
In many state estimation applications where the Kalman filter framework is utilized, estimators are often implemented under the assumption that system model parameters are known without error. Although, for some applications, such consideration may not significantly affect the filter, neglecting parameter uncertainty in many other situations can lead to inaccurate state estimates, overconfident filters, and even filter divergence. The consideration of the uncertainty for model parameters, also called nuisance parameters or static biases, may be handled by several approaches. One possibility is to account for their effect via the typical tuning of the measurement covariance matrix $ \R $ or the process noise covariance $ \Qmat $ and amount to be added unknown. Although this approach can be straightforward, it does not appropriately account for the colored property of these parameters. A second alternative, and as that of interest for this work, is to attempt the estimation of the nuisance parameters by appending them to the state vector. If this latter approach is implemented, it can be that an Extended Kalman filter is now needed where a linear Kalman filter sufficed before due to nuisance parameters could then appear, along with the core states, in a nonlinear fashion. 

The inclusion of the parameters in the state vector increases the filter's complexity, and, as the state incorporates more nuisance parameters, observability questions also start to arise. Furthermore, attempting the estimation of the nuisance parameters via the augmented state and directly treating them as if they were other core states, may cause divergence problems since the nuisance parameters are in many cases weakly or mildly observable.

\subsection{The Schmidt-Kalman and the partial-update Schmidt filter}
In the spirit of alleviating the difficulties mentioned above when augmenting the state vector with nuisance parameters, and still retaining the Kalman filtering framework to solve the estimation problem, the Schmidt-Kalman filter approach was proposed in \cite{schmidt1966application}. A typical Schmidt-Kalman filter or \textit{consider} Kalman filter, is a standard solution for the application of the Kalman filter on systems where it is required to account for uncertainty in the model and measurement parameters for either static or dynamic systems \cite{woodbury2011considering}. In contrast with the typical Kalman filter, the Schmidt modification does not attempt to estimate all of the elements of the augmented state vector, instead, it only estimates the core or main states and just \textit{considers} (uses) the nuisance parameters values and their uncertainty for the filter computations. In this way, the uncertainty of the \textit{considered} states is still reflected in the resulting Kalman filter error distribution, generally producing more consistent estimates.

Recent advances have generalized the consider filter a step further, allowing to update the consider states \cite{BrinkPartialUpdate} partially. This partial-update technique has been shown to be effective (and statistically consistent) for the same class of systems where the Schmidt filter is useful. However, this new approach also attempts to estimate the considered states so that the estimation consistency can be improved. Mathematically the partial-update filter is grounded in linear system theory as the Extended or Unscented filters. However, in comparison with the Schmidt filter, which either considers (does not update) or updates a state (fully applies the update), the partial-update approach allows the user to apply the chosen percentage of the nominal update. 
\subsection{Partial-update Schmidt filter challenges}
Although the partial-update Schmidt-Kalman filter addresses the issues of having a nuisance parameter in an augmented state vector and additionally attempts to estimate the nuisance parameters, this novel approach still presents several challenges. First, it inherits the numerical instability problems a typical Kalman filter suffers from, and as such, its actual implementation in hardware could represent difficulties. Second, the selection of the percentages for partially updating the nuisance states is not well defined, and it has been observed to be system dependent. Additionally, the partial-update approach does not leverage occasions where the nuisance parameters observability, and ignores if some other information is available to attempt improving the parameter estimates, and the overall filter performance. Third, the implementation of the partial-update filter has been limited to simulation, and its real-world applicability scope is mostly unknown yet. 

\section{Literature Review}\label{cha:literature}

The Kalman filter technique developed by Rudolf Kalman in the 1960s \cite{kalman1960new}, one of the estimation techniques most used today, was not immediately accepted among Rudolf Kalman's peers when it was first presented. Kalman encountered such a reluctance from his peers that he even chose a mechanical engineering journal to publish his work, instead of an electrical engineering journal \cite{Grewal2001}. Fortunately for him, his perseverance in presenting his formulations brought them significant popularity and acceptance in alternative fields, and his filter rapidly gained the interests of researchers, including NASA engineers. Shortly after a visit to NASA in the fall of 1960 \cite{Gibbs2011a}, where Kalman presented his formulations at the Ames Research Center, he met Stanley Schmidt, a member of the Dynamics Analysis Branch. Schmidt's team then was working on the midcourse navigation and guidance for the Apollo circumlunar mission, and since they needed a solution to solve the navigation problems given the computational restrictions of the state-of-the-art \cite{Gibbs2011a}, Stanley Schmidt quickly recognized the Kalman filter technique as the potential solution for the Apollo. Motivated by Kalman's ideas, S. Schmidt soon would realize that even though Rudolf Kalman's development was originally for linear process and measurement models, he could use the Kalman filter on nonlinear systems if he performed a linearization about a nominal or reference trajectory. Shortly, the NASA staff would also infer that such linearization could be improved if performed about the current estimate (produced by the filter) rather than about a reference trajectory. Thanks to conversations with his peers, especially with Richard Battin from the MIT Instrumentation Laboratory, Schmidt played a crucial role in making the Kalman filter an essential component of the Apollo on-board guidance. However, before NASA engineers were able to utilize the Kalman filter for manned missions to the moon, several concerns needed to be addressed first.

\section{Numerical issues and factorized formulations}
With the increase in popularity of the Kalman filter mainly due to the Schmidt research team's success for the circumlunar midcourse navigation, problems with its implementation also started to arise. NASA studies on the effect on midcourse guidance, when fusing radar and on-board sensing, exhibited problems of divergence for the first time along with numerical stability. According to McGee and Schmidt \cite{McGee1985}, the issues of numerical stability and divergence of the Kalman filter were not noticed sooner due to presumably the low sensitivity of the testing problem to nonlinearities, round-off errors, unmodelled dynamics, and prior statistics. Although the numerical issues that arose first were attributed to the limited 16-bit fixed computational word length of the computers at the time, it was later confirmed that the Kalman filter formulation (structure of operations) itself also affected its numerical stability.

Researchers would soon realize that the numerical instability was mainly due to the Kalman update step involving the subtraction of two positive definite matrices, which implemented on a finite precision computer can considerably degrade the filter estimates and even fracture the theoretical positive definiteness and symmetry of the covariance matrix, leading to a total failure of the estimator. With the impetus of addressing these issues, several techniques were proposed. The fixes were mainly extra ad-hoc operations after propagation and update steps; the key point was maintaining the symmetry and positive definiteness of the covariance matrix. Specifically, according to \cite{McGee1985}, among the prosed methods to enforce covariance matrix symmetry, researches tried
to enforce symmetry by replacing the lower diagonal terms with the upper diagonal terms (or vice-versa),
a technique consisting of averaging corresponding off-diagonal terms, computing correlation coefficients and then checking for correlation coefficients greater than one (corrective method), and
to add small positive numbers to the diagonal terms.
This latter method being essentially inflation of the covariance matrix. Almost two decades later, these ad-hoc techniques would, in fact, be mathematically proved to be appropriate approaches \cite{Verhaegen1986}. 

Artificially inflating the covariance matrix, was one of the solutions proposed to also``control" the Kalman filter divergence problem that engineers started to face at the time \cite{Fitzgerald1971}, \cite{Quigley1973}. Although the Kalman filter divergence was mainly attributed to modeling errors, round-off errors also were recognized to affect the filter stability \cite{Verhaegen1986}. 


%
\subsection{Square-root filtering}
The lack of more fundamental methods to improve the numerical robustness of the Kalman filter when limited hardware specifications were available, was the motivation for many researchers to develop alternative recursive forms of the Kalman equations to improve the precision. The contributions of J.E. Potter from the MIT laboratory were crucial in this regard. According to the literature, Potter was the first researcher in publishing a technique able to increase the numerical robustness of the Kalman filter \cite{battin1999introduction}. A filter that, in fact, was flown on the Apollo manned lunar exploration program \cite{Grewal2001}. Potter realized that using the Cholesky decomposition, one could factorize the covariance matrix and propagate the Cholesky factor rather than the full covariance matrix. By doing this, he rigorously guaranteed the non-negative definiteness of the covariance matrix after each Kalman filter recursion \cite{BELLANTONI1967}. Although Potter's formulation assimilated the measurement in a sequential way (vector measurements were sequentially processed) and had the limitation of handling filtering systems with no process noise only, it increased the accuracy of the filter.

After Potter's work, many algorithmic advances and developments soon emerged in the 1970s. Articles extending Potter's formulation to include process noise were made available \cite{dyer1969extension}. Sometime later, even surveys were being published on the numerical Kalman filter issues and the set of techniques available at the time on square root filtering to alleviate them, like the one by Kaminski and Schmidt in \cite{Kaminski1971}. In this paper, Kaminski and Schmidt also recall that round-off errors caused numerical problems with the Kalman filter implementation, but issues may also arise if some components of the state vector are far more observable than others.

\subsection{The U-D filter formulations}
Along with the momentum of square root filtering developments, an alternative presented as square-root free formulations became available in the literature for the Kalman filter: the U-D factorized filter. The U-D Kalman filter mechanization uses the U-D factorization to propagate the error covariance matrix. In practice, this filter decomposes or factorizes the covariance matrix into the product three matrices and propagates two of them. The factorization involves a unit upper-triangular (U), a diagonal (D), and a unit lower-triangular ($ U\trans $) matrix. Gerald Bierman introduced the U-D formulation in \cite{Bierman1975}, where he shows for a simplified case that his formulation guarantees non-negative-definiteness of the covariance matrix and there was no need to perform square-root operations. Later on, Bierman and Catherine L. Thornton would also publish an article that makes a numerical comparison among the conventional Kalman filter, Potter's square-root form, and the U-D formulations. This paper drew two important conclusions: 1) the superior numerical robustness of the factorized formulations (square root and U-D) against the conventional Kalman filter non-factorized form, 2) the Potter formulation can be computationally expensive in comparison with the standard formulation, while the U-D filter offers computational efficiency and increased numerical robustness \cite{Bierman1977}. Although the U-D filter's execution time may not be faster than the conventional Kalman filter form, it will be competitive compared with it. Despite this, in general, the square root formulation of the Kalman filter is recommended in the literature as a representation that facilitates analytical developments \cite{Base1959}.

Even though factorized formulations were, in part, initially developed to alleviate difficulties when dealing with "problematic" or bad conditioned systems, it was later shown that numerical issues could still occur in well-conditioned systems if the filter was implemented in single-precision \cite{Bierman1977}. Considering this fact and that is not possible to predict if numerical problems will arise, Bierman recommended using a factorized filter in all applications, especially in embedded systems. 
While some authors imply that factorized filtering was more intended to be used in the early days when computers were more limited \cite{Brown}, and others mentioned that these forms might be obsolete \cite{crassidis2011optimal}, some researches also believe that a factorized filter is essential \cite{Carpenter2018}. In contrast, others believe they are an important tool to cross-validate filter results, leaving out numerical issues \cite{Gibbs2011}. 

Since the strong wave of first developments related to factorized filtering, researches have seemed to follow the advice of Bierman. However, and for a few exceptions, today, more than using a factorized filter as being cautious for potential single-precision implementations, researchers see the factorized formulations as a guarantee on the non-negative definiteness of the covariance matrix, and a means to reduce computation error effects, as mentioned in \cite{Soremen1979}. Thus, current literature focuses more on the use of the factorized formulations along with recent filtering techniques with the objective of \textit{robustify} such filters. In \cite{Holmes2008}, for example, the author introduces a square root unscented Kalman filter for visual monocular simultaneous localization and mapping (SLAM). X.Li, in \cite{Li2017} as well applies the square root EKF formulation for underwater SLAM, and in \cite{Lou}, Tai-shan presents an ensemble Kalman filter that incorporates the square root formulation. For very large systems, a U-D filter approach was reported by JPL in \cite{boggs1995stabilized} for systems with sparse matrices and large-states. In the work of Shovan in \cite{Bhaumik2014} a square root cubature-quadrature Kalman filter is developed and shown to gain numerical robustness. Besides improving numerical stability, in work presented in \cite{Wu2016} for applications in mobile devices, it is also shown that faster implementations are obtained since the square root formulation allows using single-precision computations (even if double-precision is available). Further developments in unscented smoothing and angles-only orbital navigation that use square root form, are also available in \cite{Rutten2013} and Jason Schmidt's masters' thesis \cite{KnudsenSchmidt2010}. I. Arasaratnam even tests in \cite{Arasaratnam2008} that the square root formulation appropriately handles a singular covariance matrix and can keep the filter running even with perfect measurements. More recent contributions on the topic include an iterative cubature \cite{Mu2011}, an unscented Schmidt filter \cite{Geeraert2018}, and complementary studies on potential methods to compute the square root of the covariance matrix in \cite{Rhudy2011}. Studies on the robustness by testing factorized filters under different word lengths have also been reported in \cite{Papez2013}. Factorized filtering, unsurprisingly, has even found applications out of the engineering field in the econometrics literature as a fix for discrepancies in quantities in the state vector \cite{Carraro2016}. Moreover, other highly specific factorized filtering techniques for stiff systems as the work presented in \cite{Kulikov2017} are still being investigated. In the aerospace engineering side, GPS vehicle navigation that uses square root formulation also are available in the literature \cite{Kane2016}, \cite{Rhudy2011}, and even adaptive approaches can be found to use factorized filters in \cite{zhou2013new} and \cite{zhou2015new}.
A good background summary on the square root filtering can be found in \cite{Gibbs2011} and more comprehensive treatment in \cite{maybeck1982stochastic}.

Specifically speaking on the U-D filter, this formulation remains being the favorite filter for NASA engineers, as mentioned in one of their reports where a relative navigation filter for an International State Station hosted payload is developed \cite{Galante2016}. Due to its high functionality, the U-D filter has barely undergone modifications since its introduction in Thornton's memorandum in \cite{thornton1976triangular}. Nevertheless, some work regarding its implementation methods, for example, can be found in \cite{Zanetti2019}. Also, in a paper by C. Souza \cite{DSouza2018}, the information formulation of the U-D has been introduced. 
Overall, the U-D filter's attributes provide such numerical robustness and performance that is, in fact, the filter of NASA's Orion vehicle for absolute navigation, as presented in \cite{Holt2012}. In this development, as in other similar cases in the literature, the U-D formulation attempts to robustify a base filter; specifically, a consider filter; a technique also attributed to Stanley Schmidt.

\section{The Schmidt consider filter}
Among other developments, Schmidt also invented what is known in the literature as Schmidt or consider filter \cite{Woodbury}. As described in McGee and Schmidt \cite{McGee1985}, in a tool created for NASA, Schmidt included a functionality that allowed to \textit{consider} for parameter model errors on filter performance evaluation. The idea was to consider the effect of errors and their uncertainty, and reflect such "knowledge" on the estimated states. 
The consider or Schmidt filter, somewhat strangely, has adopted different focuses in the literature since its creation. For example, in \cite{Novoselov2005}, it is proposed as a means to deal with biases in the dynamic models. However, as presented in \cite{Yang2010} is a filter that allows handling errors in measurement parameters, and in \cite{Woodbury} is, in part, introduced as a tool to obtain reduced order systems. For other researchers, the Schmidt filter is just a solution to account for the uncertainty of nuisance parameters, that is, parameters required to perform the state estimation but are not the main states of interest \cite{BrinkPartialUpdate}. In any case, the principles of the filter are the same: consider states without estimating them.

The Schmidt filter's use from the parameter's uncertainty consideration perspective has been applied widely and is mainly found in aerospace applications. In \cite{Lou2015} for example, the filter is used in a Mars entry navigation filter, in \cite{Yang2016} for GPS-based on-board real-time orbit determination, in \cite{Li2018a} is used for GNSS-based attitude determination, in \cite{Yang2010} for target tracking to account for sensor positioning error, and in computer vision for proximity operations as presented in \cite{Ramos2018}. Again, regardless of the motivation to use a consider filter, the principle is to "delete" states that are not core states and perform the estimation with the remaining states. More general, and in any event, the Schmidt filter intent is to expand the class of problems the typical Kalman filter can handle well. Because of the Schmidt filter capacity to allow stable estimation of systems with low observable states, it is today a companion workhorse for the EKF.


Recently, developments have generalized the consider filter a step further, allowing the \textit{consider} states to be updated. This technique, known in the literature as the partial-update Schmidt filter, has been shown to be useful for a broader class of systems than the Schmidt filter can cover \cite{BrinkPartialUpdate}. This new approach attempts to update consider states partially, and by doing so, the estimation consistency and accuracy can be improved. However, there are still challenges on how and when to allow a consider state to be updated. Also, the question of how much the user can partially update a state is open. Moreover, since the partial-update technique is relatively recent, no hardware validation has been made, and no research towards the increase of numerical robustness has been done (as it also suffers from the numerical problems the conventional Kalman filter faces). The significance of closing these gaps for the partial-update filter is that it would extend the practical applicability of the EKF while retaining the EKF structure, which remains a widely used filter in the world.

Although some other methods that attempt to improve the Schmidt filter behavior for non-well-known prior covariance of the consider states have been published in \cite{Lou2014}, they do not remain straightforward to implement (do not keep the Kalman filter structure) and do not include the capacity to update considered states. In contrast, the partial-update filter retains the conventional Kalman filter structure and is an easy modification.

\section{The dissertation objectives and outline}\label{sec:outline}
The proposed dissertation seeks to extend the applicability of the Schmidt-Kalman filter framework by increasing its numerical robustness and attempting considered state estimation. Towards the goal of increasing robustness against numerical issues, the square root formulation and the U-D factorized form of the partial-update Schmidt Kalman filter are developed. The numerical problems that these formulations attempt to alleviate are mainly due to round-off error, processing of highly accurate measurements, severe discrepancy among states observability, and bad-conditioned problems. Towards estimation of considered states, techniques that leverage occasions when the system nonlinearities are not so severe are proposed.

Since this work builds up from the partial-update filter concept, the resulting filter formulations inherit the increased tolerance to nonlinearities and uncertainty level; at the cost of a minimal additional computational burden that still allows online execution. Moreover, as the proposed formulations are a fundamental restructuring of the Kalman filter equations, they can be directly leveraged in applications where the Kalman and Extended Kalman filter is already used. The filters proposed in this dissertation are also demonstrated in hardware implementations for aerospace-related applications.

\subsection{Outline}
This research is divided into the proposed main contributions: (1) The development of factorized formulations that increase the numerical robustness of a more general Schmidt-Kalman filter, (2) The establishment of baselines to attempt the estimation of considered states and (3) the implementation of the proposed concepts in hardware for aerospace-related applications. 

The dissertation chapters are organized as follows. Chapter 1 provides the literature review on the Schmidt filter and starts introducing the partial-update filter concept. Chapter 2 introduces some notation used throughout the dissertation. It includes a more detailed description of the partial-update Schmidt filter and its stability analysis for linear systems. In this chapter, the straightforward extension to a multiplicative extended Kalman filter is presented, along with a brief discussion on the numerical issues that the partial-update filter can face (given that is grounded in the Kalman filter). Chapter 3 develops a square root formulation of the partial-update Schmidt filter to increase its numerical robustness. Simulated cases for the square root partial-update filter are included. Motivated by the high computational burden that the square root formulation can implicate, Chapter 4 develops an alternative factorized formulation: a U-D factorized based filter. This alternative form's objective is to maintain the partial-update and factorized form benefits while having a less computationally expensive filter. Numerical simulations of this filter are presented. Chapter 5 introduces techniques that allow the user to utilize a partial-update filter, without the need for tuning the partial-update weights, and its functionality is shown via numerical simulations. To show the applicability of the partial-update concept in real systems, Chapter 6 includes the results of its hardware implementation. Finally, Chapter 7 presents the conclusions of this research.

%
%
%
%
%
%
%

\chapter{A GENERALIZATION OF THE SCHMIDT KALMAN FILTER}\label{sec:Background}
%
%
This chapter mainly introduces underlying mathematical concepts to be referred to in posterior sections of the dissertation. First, a brief description of the Kalman filter framework is included to establish the filtering context and include most of the nomenclature common among the developments presented. Second, the partial-update Schmidt-Kalman filter, the backbone of this work,  is introduced. Third, the partial-update formulation, in its indirect form, is derived. Fourth, a Lyapunov stability analysis of the partial-update filter concept for linear systems, is performed, and finally, a discussion on the potential numerical issues of the partial-update filter is included.

\section{Discrete Extended Kalman filter framework and notation}\label{sec:kalman_and_nomenclature}
This section mainly serves to provide nomenclature utilized in this work in the context of Kalman filtering.
The Kalman filter was developed to perform state estimation of linear systems originally. In order to support nonlinear systems, one of the more utilized techniques today is the extended Kalman filter (EKF), which operates on the linearized system equations about the current state estimate. This linear approximation of the system is used with the original Kalman filter equations to propagate and update the state and estimated uncertainty. The linearization process, however, can be a source of significant errors in the resulting estimates of a  filter. Section \ref{sec:Background_on_Partial_Update} gives details on how  linearization errors may be better handled with the partial-update approach.  

For the interests of this research, the propagation and update equations correspond to the discrete Extended Kalman filter, which allows one to implement the algorithm in digital computers that may not have the power to integrate the continuous dynamics every time step. Next, the discrete Extended Kalman filter framework is summarized without derivation \cite{crassidis2011optimal}.

Let a discrete nonlinear dynamic system with state vector $\x_k \in \Real^n$, and measurement vector  $\yktilde \in \Real^m$, be represented by 
\begin{equation}\label{eq:nonlinear_equations}
\x_{k}=\f_{k-1}(\x_{k-1},\u_{k-1})+\w_{k-1}  \ , 
\end{equation}
\begin{equation}
\yktilde=\hkofxk+\vk \ ,
\end{equation}
\begin{equation}
\wk \sim \mathcal{N} (\zerovec,\Qmat_k) \ ,
\end{equation}
\begin{equation}\label{eq:measurement_cov}
\vk \sim \mathcal{N} (\zerovec,\R_k) \ ,
\end{equation}
where $\wk \in \Real^n$ and $\vk \in \Real^n$ are zero-mean Gaussian white-noise processes, with covariances $\Qmat_k=\expect[\wk\trans\wk]$ and $\R_k=\expect[\vk\trans\vk]$, respectively; with $\uk \in \Real^r$ being the input vector sequence to the system. The function $\hkofxk$ is the nonlinear measurement model and the sub-indices $k$ denote time instance.

Performing a first-order Taylor series expansion of Equation (\ref{eq:nonlinear_equations}) about the current estimate, $\x_{k-1}=\xpost_{k-1}$, forming the error dynamics, and computing its expectation, the propagation equation for the $n \times n$ error covariance matrix $\stdvec{P}[][k][]$ results in Equation (\ref{eq:cov_propagation}) with $\Fmat_k$ given by Equation (\ref{eq:ch2_F_matrix}):
\begin{equation}\label{eq:cov_propagation}
\stdvec{P}[][k][-] = \Fmat_{k-1}\stdvec{P}[][k-1][+]\Fmat_{k-1}\trans+\Qmat_{k-1} \ .
\end{equation}
The propagation of the state vector $\hat{\x}_k$, is done with the nonlinear dynamics expected value as
\begin{equation}\label{eq:nonlinear_propagation}
\xprior_{k}=\f_{k-1}(\xpost_{k-1},\u_{k-1}) \ .
\end{equation}

Every time an observation is available, the measurement update step for the state and error covariance is performed through the $n \times m$ Kalman gain, $\Kmat_k$, according to the following set of equations. Note that a linearization of the measurement model has also been performed:
\begin{equation}\label{eq:kalman_gain}
\stdvec{K}[][k]=\stdvec{P}[][k][-]\Hmat_k\trans(\Hmat_k\stdvec{P}[][k][-]\Hmat_k\trans+\Rk)^{-1} \ ,
\end{equation}
\begin{equation}\label{eq:covariance_update}
\stdvec{P}[][k][+]=(\stdvec{I-\stdvec{K}[][k]}\Hmat_k)\stdvec{P}[][k][-] \ ,
\end{equation}
\begin{equation}\label{eq:expected_measurement}
\hat{\y}_k = \h_{k}(\xprior_{k}) \ ,
\end{equation}
\begin{equation}\label{eq:state_update}
\xpost_k=\xprior_{k}+\stdvec{K}[][k](\ym_k-\yhat_k) \ ,
\end{equation}
where
\begin{equation}\label{eq:ch2_F_matrix}
\Fk = \partder{\f_{k}}{\x}\Big\rvert_{\xprior_{k-1}} \ ,
\end{equation}
and
\begin{equation}\label{eq:h_matrix}
\quad\Hk=\partder{\hk}{\x}\Big\rvert_{\xprior_k}\qquad\ .
\end{equation}
Here, the hat notation, i.e. $\hat{[\,\cdot\,]}$, denotes an expected or estimated value. The notations $[\,\cdot\,]^+$ and $[\,\cdot\,]^-$ refer to posterior and prior values, respectively. 

The set of equations (\ref{eq:nonlinear_equations}) to (\ref{eq:h_matrix}), constitute the Extended Kalman filter framework. The following chapters adopt the nomenclature here presented, and additional information will be introduced if needed.

\section{The partial-update filter concept}\label{sec:Background_on_Partial_Update}
\subsection{The Schmidt-Kalman filter}
When a Kalman filter state vector involves system parameters or weakly observable states, estimating them in a traditional way can be problematic. In that scenario, the direct use of an EKF can be negatively impacted further, if the system involves measurement  and process model nonlinearities, high uncertainties, or some combination, to the extent of leading to estimates degradation or even filter divergence.
In such cases, the alternative Schmidt-Kalman filter becomes particularly useful. The Schmidt approach consists of not estimating such the problematic states but maintaining their values and respective covariances fixed; allowing it to behave more linearly and builds more appropriate cross-correlation terms with the core states. In other words, the Schmidt approach enables the filter designer to \textit{consider} the uncertainties of certain states into the Kalman filter solution without attempting to estimate them. By doing so, the class of problems where the conventional Kalman filter framework is useful is broadened. The problematic parameters, or those states that complicate the estimation process if treated as a ``traditional'' state, are often referred as nuisance states or nuisance parameters. These nuisance parameters, although are often not the main states of interest, their refinement is needed to improve the overall filtering solution.

Mathematically, the conventional formulation of the Schmidt filter starts by partitioning the state vector, $ \xprior $, measurement matrix,$ \Hmat $, and Kalman gain $ \textbf{K}$ into states and parameters as
\begin{equation}
\xprior = \begin{bmatrix}
\xprior_x\\\xprior_p 
\end{bmatrix} \ ,
\end{equation}
\begin{equation}
\Hmat=\rowvec{\Hmat_x,\Hmat_p} \ , 
\end{equation}
and
\begin{equation}
\textbf{K}=\left[\begin{array}{l}{{\textbf{K}}_{x}} \\ {{\textbf{K}}_{p}}\end{array}\right] \ .
\end{equation}
Then, the partitioned state vector is substituted into the conventional Kalman filter update equations and the corresponding operations are performed. Finally, and fundamental to the Schmidt filter, the optimal Kalman gain is computed after forcing the parameters' Kalman gain to be zero ($ \textbf{K}_p=\zerovec $).

The resulting update equations after using the optimal gain for the case when $ \textbf{K}_p=\zerovec $, as reported in \cite{woodbury2011considering}, are 

\begin{equation}\label{eq:schmidt_covariance_update}
\stdvec{P}[][][+]=    \begin{bmatrix}
(I-K_xH_x)\stdvec{P}[][x x][-]-K_xH_p\stdvec{P}[][x x][-] && (I-K_xH_x)\stdvec{P}[][x p][-]-K_xH_p\stdvec{P}[][p p][-]\\
[(I-K_xH_x)\stdvec{P}[][x p][-]-K_xH_p\stdvec{P}[][p p][-]]\trans &&
\stdvec{P}[][p p ][-]
\end{bmatrix} \ ,
\end{equation}
and
\begin{equation}\label{eq:schmidt_state_update}
\begin{bmatrix}
\xpost_x\\\xpost_p
\end{bmatrix}  = \begin{bmatrix}
\xprior_x\\\xprior_p
\end{bmatrix}+
\begin{bmatrix}
K_x\\
\zerovec
\end{bmatrix}
\begin{bmatrix}
\tilde\y-H_x\xprior_x-H_p\xprior_p
\end{bmatrix} \ .
\end{equation}
From Equations (\ref{eq:schmidt_covariance_update})-(\ref{eq:schmidt_state_update}) it can be clearly seen that on the Schmidt approach, 1) the cross-correlation terms of the covariance matrix become updated and they account for parameter uncertainty, and 2) both parameters not their respective uncertainties are updated.

The Schmidt filter is popular mainly because it is easy to implement, and it is often sufficient to alleviate issues when dealing with nuisance parameters. Moreover, since the parameters are not estimated, it offers some computational advantage compared to the conventional Kalman filter (KF). 

The \textit{considering} approach that allows the Schmidt filter to work on the described scenarios also limits its capabilities. First, while the Schmidt filter can cope with nuisance parameters and system nonlinearities via \textit{considering} them, the Schmidt approach generally cannot react if the nuisance parameters change or slowly vary. Second, if the considered parameters are tightly related to the overall system performance, the Schmidt approach will intrinsically limit the filter performance as no more information is assimilated to improve the parameter belief, and thus core states' belief. Third, the Schmidt filter by design (not attempting to update parameters) ignores situations where nuisance parameters could have been updated, because their observability increased, or nonlinearities are not so severe, and (negative) parameter impact may not be significant.

With these intrinsic, in part limiting, characteristics of the Schmidt approach in mind, a novel concept was created: the partial-update Schmidt-Kalman filter. The idea of this novel filter is to attempt to gain nuisance's states information while trying to keep Schmidt approach benefits. The next subsection gives a brief background on the partial-update filter.

\subsection{The partial-update Schmidt-Kalman filter}
The partial-update Schmidt-Kalman filter (PSKF or partial-update filter for short) is a recent technique that is useful in accommodating measurement updates in nonlinear systems with mildly observable states, as it is an extension of the Schmidt filter. Examples of successful implementations of the partial-update Kalman filter can be found in \cite{Ramos2018}, \cite{chakraborty2018relative}, and \cite{jurado2018towards}. This concept is, in fact, the backbone of this dissertation.

The partial-update Schmidt-Kalman filter \cite{BrinkPartialUpdate} is a straightforward modification of the Schmidt Kalman filter that effectively increases the range of uncertainties and associated nonlinearities that the filter can tolerate (compared to the EKF or Schmidt filter) while still producing accurate state estimates with appropriate covariance bounds. The approach does so with almost no extra computational cost and maintains the linear system theory's desirable underpinnings, generating unbiased and consistent results, just as the Kalman or Schmidt filter does.  Moreover, the technique is extensible to the UKF \cite{BrinkUnscented} and other minimum mean square error approaches.    

Effectively, in contrast with the Schmidt filter, the partial-update uses a percentage of the nominal Kalman update to correct the nuisance states. In fact, the formulation of the partial-update filter allows to apply partial updates to any state. A commensurate update percentage is also reflected in the error covariance update, as seen in Equation (\ref{eq:partial_update_cov_elem_by_elem}). The partial-update is expressed in an element-wise fashion for the states as
\begin{equation}\label{eq:partial_update}
\xpostplus = \gamma_i\xprior_i + (1- \gamma_i) \xpost_i \ ,
\end{equation}
and for the covariance as
\begin{equation}\label{eq:partial_update_cov_elem_by_elem}
\stdvec{P}[][ij][++] = \gamma_i\gamma_j \stdvec{P}[][ij][-] + (1 - \gamma_i\gamma_j)\stdvec{P}[][ij][+] \ .
\end{equation}
Where the update percentages or weights are represented by $ \beta_i $, and related to $ \gamma_i $ via
\begin{equation}\label{eq:betas}
\gamma_i = 1 - \beta_i \ .
\end{equation}
The notation $ [\,\cdot\,]^{++}$ denotes the partial-update value that will overwrite the full state estimates from the conventional Kalman equations to be used at the next propagation step. 

In words, the partial-update blends the updated (posterior) vector $\xpost_k$ computed with Equation (\ref{eq:state_update}) with the prior state vector obtained with Equation (\ref{eq:nonlinear_propagation}) via an update or percentage weight $ \beta $. The partial-update covariance equation (\ref{eq:partial_update_cov_elem_by_elem}) follows the same idea. The weight $ \beta_i \in [0,1]$ in Equation (\ref{eq:betas}), can be thought as the percentage of the updated state being used. Then if $ \beta_i $ equals zero, the Kalman update is totally dropped and the prior is completely kept (Schmidt filter), whereas $ \beta_i = 1 $ corresponds to a regular EKF (full) update; however, the weight can be set anywhere in between. The notation $ \boldsymbol{\beta} $ referring to a vector containing the elements $ \beta_i $ for $ i = 1,2,...,n$, or alternatively, a diagonal matrix $ \betamat=\diag[\beta_1,\beta_2,\dots,\beta_n] $, is used throughout the dissertation as convenient. In any case, $ n $, is the total number of states in the filter (as any state can be partially updated).

While there are more efficient ways to implement the partial-update, specifically by modifying the update step itself, this implementation is still reasonably efficient. Moreover, although Equations (\ref{eq:partial_update}) and (\ref{eq:partial_update_cov}) can be expressed in terms of $\boldsymbol{\beta}$ only, the equations presented are convenient for both proofs and a more natural discussion (i.e. describing weights in terms of the percentage of the update to be applied). The selection of the percentages $ \beta $ is system dependent, but heuristically fast, or core, or more observable states are updated with values of 1, whereas nuisance or weakly observable states, are partially updated with values that can be anywhere inside the permissible range. Methods for online partial-update weight selection are presented in Chapter \ref{ch:dynamic_methods}.

The partial-update filter was originally developed to be used within a filter that uses an additive correction, however, and especially in the aerospace community, the use of filters involving attitude  multiplicative corrections is common. In the interest of this work, the partial-update formulation is next incorporated into the widely adopted multiplicative extended Kalman filter (MEKF), extending the partial-update application to multiplicative attitude filtering.

%

\section{The partial-update for the indirect Kalman filter}
The quaternion or Euler parameters, often the chosen attitude parametrization, conveniently provides a singularity-free, although non-unique orientation representation \cite{hurtado2012kinematic}. Although, some care has to be taken since it is an attitude over-parametrization, and a unit norm constraint needs to be preserved to ensure a correct attitude representation, well-known analytical and numerical techniques exist to deal appropriately with quaternion-based attitude developments \cite{Trawny2005}.

Quaternions are very popular in navigation systems and are often utilized within a Kalman filter \cite{Scholarsarchive2017b},\cite{Zanetti2016},\cite{NikolasTrawnyAnastasiosI.Mourikis2007},\cite{Li2013b}. Specifically, they are used in an indirect Kalman filter formulation as to preserve the unit-norm constraint and avoid a singular covariance matrix. This indirect Kalman filter formulation is referred to as the Multiplicative Extended Kalman filter (MEKF). This filter operates on the state error, instead of directly using the state dynamics and it updates attitude though a multiplicative correction. Via the MEKF, it is possible to implement a state estimator that benefits from the quaternion representation while satisfying the unit-norm constraint in a built-in fashion.

The partial-update concept can also be applied within the MEKF to fuse both approaches' benefits. Although it has a slightly different interpretation due to the  MEKF's indirect formulation, the underlying meaning of partially using the nominal Kalman update still holds. The derivation of the partial-update MEKF, or PU-MEKF for short, is developed next. This formulation is the one implemented on the simulated and hardware implementation cases included in this dissertation.

\subsection{Indirect filtering} \label{subsec:PU-MEKF_propagation}
Previous to describing the PU-MEKF update step, indirect filtering, and the conventional MEKF update step are first briefly discussed.

\paragraph{Indirect filter formulation.}
Direct incorporation of the attitude quaternion into the filter state comes with two issues. First, the unit quaternion constraint implies state dependence, and thus a singular covariance (theoretically correct; however, it can cause numerical instabilities). Second, the Kalman update operation involves additions; however, the addition operation is not defined for quaternions, and the unit-norm constraint may be violated. In order to overcome these issues, the indirect Kalman filter formulation can be employed \cite{trawny2005indirect}.

The use of an indirect formulation means that the filter does not directly use/produce the estimate of the state vector; rather, it uses/produces the estimated error of the state vector. And then, with an estimate of the error in hand, the estimate of the actual variable can be recovered. In other words, the PU-MEKF (and MEKF) estimates the departure of the states from the true values. To perform the filtering in this \textit{mode}, the PU-MEKF uses the error dynamics model, instead of the dynamics model.

In the MEKF framework, two definitions of error are used: additive and multiplicative error. In general, additive errors are associated with states other than quaternions, whereas multiplicative errors are associated with the quaternion states. The definition of the error for additive states is that from Equation (\ref{eq:error_definition_additive}), and the definition for multiplicative error is that from Equation (\ref{eq:error_definition_multiplicative}).

\begin{equation}\label{eq:error_definition_additive}
\xerror=\x-\xhat    \ .
\end{equation}

\begin{equation}\label{eq:error_definition_multiplicative}
\dcm(\delta\quat)=\dcm(\qtrue)\dcm(\qhat)\trans \ .
\end{equation}

Intuitively, the additive error, $\xerror$, is simply computed as the difference between the true value, $\x$, and estimated value, $\xhat$. However, as the addition operation is not defined for rotations, a rotation error is defined in a multiplicative manner as in Equation (\ref{eq:error_definition_multiplicative}). That is, the multiplicative error is defined as the (small) rotation $\dcm(\delta\quat)$ that represents the \textit{rotational difference} between the true and estimated attitude. In other words, the (small) rotation error $\dcm(\delta\quat)$, is the required rotation that when compounded with the current estimated attitude $\dcm(\qhat)$, results in the true attitude $\dcm(\qtrue)$. This is readily seen if Equation (\ref{eq:error_definition_multiplicative}) is written as,
\begin{equation}\label{eq:error_definition_multiplicative_rearranged}
\dcm(\qtrue)=\dcm(\delta\quat)\dcm(\qhat) \ .
\end{equation} 
Equation (\ref{eq:error_definition_additive}) may alternatively be rearranged in terms of the estimate and error variables as,
\begin{equation}\label{eq:error_definition_additive_rearranged}
\x    = \xerror + \xhat \ .
\end{equation}
Equations (\ref{eq:error_definition_multiplicative_rearranged}) and (\ref{eq:error_definition_additive_rearranged}) are in fact, the equations that are used to recover the posterior estimates after processing a measurement within the PU-MEKF (or MEKF). Equations (\ref{eq:error_definition_multiplicative_rearranged}) and (\ref{eq:error_definition_additive_rearranged}), reveal how the indirect formulation is used: 1) the indirect filter produces the estimates for the errors, $\dcm(\delta\quat)$ and $\xerror$, and 2) these estimated errors are now combined with the current estimates $\dcm(\qhat)$ and $\xhat$, to produce the improved estimate (posterior).

\subsection{The conventional MEKF update}\label{subsec:PU-MEKF_measurement_update}
Before incorporating the partial-update within the MEKF, the standard MEKF update itself is briefly revisited. The development of the PU-MEKF is deferred for the next section.

After a measurement is received, the measurement update can be performed via standard Kalman filter equations. At time $t=k$, the Kalman gain, and posterior covariance, $\Cov$, are obtained with \cite{crassidis2011optimal} 
\begin{equation}\label{eq:kalman_gain_equation}
\stdvec{K}[][k]=\stdvec{P}[][k][-]\Hmat_k\trans(\Hmat_k\stdvec{P}[][k][-]\Hmat_k\trans+\Rk)^{-1} \ ,
\end{equation}
\begin{equation}
\stdvec{P}[][k][+]=(\stdvec{I-\stdvec{K}[][k]}\Hmat_k)\stdvec{P}[][k][-]\ ,
\end{equation}
and the error state update via
\begin{equation}
\xerrorpost_k=\stdvec{K}[][k](\ym_k-\yhat_k)=\stdvec{K}[][k]\stdvec{r}\ .
\end{equation}
Although the error state estimate, $\xerrorpost_k$, is in fact the MEKF output, recall that the interest is to recover the actual state estimate $\xhat$. This is accomplished in two steps. First, the components of the computed Kalman correction, $\xerrorpost_k$, are split into additive ($\delta\x_{additive}^+$) and multiplicative ($\smallthetahat^+$) corrections: 
\begin{equation}
\stdvec{K}_k\stdvec{r} =  \colvec{\smallthetahat^+,\xerrorhat_{additive}^+}_k\ .
\end{equation}
Second, the actual states are recovered by applying each type of error definition separately. In this manner, the posterior of the actual additive states is obtained by 
\begin{equation}\label{eq:additive_error_correction}
\xhat_k^+ = \xprior_k + \xerrorhat_{additive_k}^+ \ ,
\end{equation} 
while the posterior of actual quaternion states through quaternion multiplication as
\begin{equation}\label{eq:multiplicative_error_correction}
\quat_k^+=\delta\quat_k\otimes\quat_k =\colvec{\frac{1}{2}\smalltheta_k^+,1}\otimes\quat_k \ .
\end{equation}
After the quaternion multiplicative updates take place, a brute force re-normalization is performed. 
\subsection{The partial-update within the Multiplicative EKF}
In this section, the partial-update step for the MEKF is derived. The derivation is focused on the multiplicative correction only since the additive correction remains unaltered with respect to the original partial-update form. The derivation shows that when the indirect filter formulation is used, the multiplicative partial-update can be interpreted as a special case of the original partial-update form. Furthermore, it is found that the partial-update for the MEKF requires a slightly different implementation from that of the original partial-update.
\subsubsection{The PU-MEKF}
Direct use of the partial-update formulation in a filter involving state quaternions, would require a partial-update of the form (ignoring time indices for clarity)
\begin{equation}\label{eq:partial_update_mekf_wrong}
\quat_i^{++} = \gamma_i\quat_i^{-} + (1-\gamma_i)\quat_i^{+}\ ,
\end{equation}
however, this represents two main inconveniences. First, it is not a multiplicative update, making it inconsistent with the multiplicative error definition. Second, if the partial-update is performed in this manner, the quaternion unit norm can be violated. For a multiplicative formulation, however, these issues can be addressed as follows. 

Consider the alternative expression of the original partial-update equations in terms of the prior expected state, $\xhat_i^-$ as,
\begin{align}
\xhat_i^{++} &= \gamma_i\xhat_i^- + (1 -\gamma_i) \xhat_i^+ \\
&=\gamma\xhat_i^- + (1 -\gamma_i) (\xhat_i^- + \mathbf{K}_i\stdvec{r})\\
&=  \gamma_i\xhat_i^- + \xhat_i^- + \mathbf{K}_i\stdvec{r} - \gamma_i \xhat_i^- - \gamma_i\mathbf{K}\stdvec{r}\\
&= \gamma_i\xhat_i^- - \gamma_i \xhat_i^- + \xhat_i^- + (1 - \gamma_i) \mathbf{K}_i\stdvec{r})\\
&= \xhat_i^- + (1-\gamma_i)\mathbf{K}_i\stdvec{r}\label{eq:partial_update_alternative_with_gamma}\\
&= \xhat_i^- + \bar{\beta_i}\mathbf{K}_i\stdvec{r}\label{eq:partial_update_alternative} \ ,
\end{align}
with $\stdvec{K}_i$ being the $i^{ith}$ row of the Kalman gain $\stdvec{K}$.
Equation (\ref{eq:partial_update_alternative}) suggests that even before computing the posterior state, the partial update can be applied directly on the correction term, $\mathbf{K}_i\stdvec{r} $, through $\bar{\beta_i}$. This means that the partial-update can take place in the composition of the multiplicative correction for a filter with multiplicative correction (the MEKF in this case). To elaborate on this
let the quaternion correction $\delta\quat$ from Equation (\ref{eq:multiplicative_error_correction}) be decomposed into three single small rotations (small angular corrections), through the $\phi$, $\theta$ and $\psi$ angles. Also, suppose that such angle corrections are partially applied (in a multiplicative fashion) to an estimated quaternion $ \qhat, $ and they appear scaled by $\beta$ factors as,
\begin{align}
\quat &= \delta\quat\otimes \qhat\\
&=[\delta q_1 \otimes \delta q_2 \otimes \delta q_3] \otimes \qhat\\
&=\colvec{\sin\frac{\beta_1\phi}{2},0,0, \cos\frac{\beta_1\phi}{2}}  \otimes \colvec{0, \sin\frac{\beta_2\theta}{2},0, \cos\frac{\beta_2\theta}{2}}  \otimes \colvec{0,0, \sin\frac{\beta_3\psi}{2}, \cos\frac{\beta_3\psi}{2}}\otimes \qhat \ .
\end{align}
Now, since the angle corrections are considered to be very small for the PU-MEKF formulation, then $ \delta\quat$ is small (which is originally the case for the MEKF), which means $\sin\frac{\phi}{2} \approx 0$ and $\cos\frac{\phi}{2} \approx 1$. Under this considerations, the quaternion multiplication (as defined in \cite{trawny2005indirect}) of the three single small rotations gives
\begin{align}
\quat &=\colvec{\frac{\beta_1\phi}{2},0,0, 1}  \otimes \colvec{0, \frac{\beta_2\theta}{2},0, 1}  \otimes \colvec{0,0, \frac{\beta_3\psi}{2}, 1}\otimes \qhat= \colvec{\frac{\beta_1\phi}{2} -  \frac{\beta_2\theta\psi}{4}, \frac{\beta_2\theta}{2} -  \frac{\beta_1\theta\phi}{4}, \frac{\beta_3\psi}{2} -  \frac{\beta_2\phi\theta}{4}, \frac{\beta_1\beta_2\beta_3\phi\psi\theta}{8} +1}\otimes \qhat \ .
\end{align}
By retaining first-order terms only, results in
\begin{align}
\quat &\approx \colvec{\frac{\beta_1\phi}{2}, \frac{\beta_2\theta}{2}, \frac{\beta_3\psi}{2}, 1}\otimes \qhat=\colvec{\frac{1}{2}\boldsymbol{\beta}\smalltheta,1}\otimes \qhat \ ,
\end{align}
where
\begin{equation}
\boldsymbol{\beta} = \diag[\beta_1, \beta_2, \beta_3] \ ,
\end{equation}
and
\begin{equation}
\smalltheta=\colvec{\phi,\theta,\psi} \ .
\end{equation}
This indicates that by scaling the correction term $\smalltheta$ the partial-update concept can be applied to the quaternion state, and that this is valid up to a first-order approximation, which holds under the MEKF assumptions.

Following the alternative partial-update formulation from Equation (\ref{eq:partial_update_alternative}), the multiplicative partial-update can be written as
\begin{equation}\label{eq:partial_update_mekf}
\mathbf{\delta\hat\theta}_i^{++} = \delta\hat{\boldsymbol\theta}_i^- + \bar{\beta_i}\mathbf{K}_i\stdvec{r}=\delta\hat{\boldsymbol\theta}_i^- + \bar{\beta_i}\delta\hat{\boldsymbol\theta}_i^+\ ,
\end{equation}
which in virtue of the expectation of the state error being zero ($\expect{[\delta\hat{\theta}^-]}=0$), and the scalars $\beta$ and $\bar\beta$ playing the same role, Equation (\ref{eq:partial_update_mekf}) simplifies to
\begin{equation}\label{eq:partial_update_mekf_simplified}
\mathbf{\delta\hat\theta}_i^{++} = {\beta_i}\delta\hat{\boldsymbol\theta}_i^+\ ,
\end{equation}
or in vector form to
\begin{equation}
\delta\hat{\boldsymbol\theta}^{++} = {\boldsymbol{\beta}}\delta\hat{\boldsymbol\theta}^+ \ ,
\end{equation}
with 
\begin{equation}
{\boldsymbol{\beta}} = \diag[{\beta_1}, {\beta_2}, {\beta_3}] \ .
\end{equation}
From this development, it can be concluded that to perform a multiplicative partial-update while maintaining the quaternion unit-norm (up to first order), the partial-update needs to happen when constructing the quaternion error correction $\delta\quat$ and not afterwards. Following partial-update notation, a multiplicative partial-update, of a prior attitude, $ \quat^- $, is performed via
\begin{align}\label{eq:partial_update_quaternion}
\quat^{++} =\colvec{\frac{1}{2}{\boldsymbol{\beta}}\delta\hat{\boldsymbol\theta}^+,1}\otimes \quat^- \ .
\end{align}
Since the traditional MEKF produces the error estimate, $\delta\hat{\boldsymbol\theta}^+$, this can easily be substituted into Equation (\ref{eq:partial_update_quaternion}) to produce the partially updated quaternion, and by accordingly \textit{partial-updating} the covariance matrix, one will have a MEKF converted into a PU-MEKF. Similarly, if $\boldsymbol{\beta}$ is chosen to be identity (full update), one recovers the standard MEKF update from the PU-MEKF formulation. 

Although recovering multiplicative and additive states from the error estimates require different operations, the partial additive and multiplicative corrections can be computed in the same step. The user just needs to properly identify the partial-update percentages $\beta$'s to be used on each state, and form the $\betamat $ matrix to perform the partial-update, e.g.
\begin{equation}
\boldsymbol{\beta} = \diag\rowvec{\beta_{\smalltheta_1},\beta_{\smalltheta_2},\beta_{\smalltheta_3},\beta_{\stdvec{\pchar}[W][I]_1},\dots,\beta_{\smallalpha_3} } \ ,
\end{equation}
and perform the partial-update as 
\begin{equation}\label{eq:ch2_partial_update_with_beta}
\xerror^{++}=\boldsymbol{\beta}\stdvec{K}[][](\ym-\yhat)=\boldsymbol{\beta}\stdvec{K}[][]\stdvec{r}\ ,
\end{equation}
or explicitly
\begin{equation}
\xerror^{++}=\boldsymbol{\beta}\stdvec{K}\stdvec{r} =  \colvec{\boldsymbol{\beta_{\smalltheta}}\smallthetahat^+,\boldsymbol{\beta_{additive}}\xerrorhat_{additive}^+,\boldsymbol{\beta_{\smallalpha}}\smallalphahat^+}\ .
\end{equation}

Once the partial-update posterior state error, $\xerror_k^{++}$, is computed,  the actual state estimates can be recovered by following additive and multiplicative error definitions. 

Finally, since $\expect[\xerror]=0$, the covariance expression for the error state is
\begin{equation}
\Cov=\expect[(\xerror-\expect[\xerror])(\xerror-\expect[\xerror])\trans]=\expect[\xerror\xerror\trans]  \ ,  
\end{equation}
and thus the covariance matrix $\Cov$ can be partially updated using the original partial-update expression,

\begin{equation}\label{eq:partial_update_cov}
\stdvec{P}[][ij][++] = \gamma_i\gamma_j \stdvec{P}[][ij][-] + (1 - \gamma_i\gamma_j)\stdvec{P}[][ij][+] \ ,
\end{equation}
or alternatively
\begin{equation}\label{eq:partial_update_cov_vec}
\stdvec{P}[][][++] = \Dmat(\stdvec{P}[][][-]-\stdvec{P}[][][+])\Dmat + \stdvec{P}[][][+] \ ,
\end{equation}
where $\Dmat$ is a diagonal matrix with elements $\gamma_i$ for $i = 1,2\dots n$ (recall that $\gamma_i = 1 - \beta_i$).

\subsection{Filter key equations and algorithm}\label{subsec:PUMEKF_algorithm}
Although the PU-MEKF is a straightforward modification of the MEKF, for the sake of completeness, the following pseudo-code is included to summarize the PU-MEKF generic implementation. Since the computation of the measurement residual matrix is generally application dependent, no specifics about computing this matrix are given in the pseudo-code. However, an application that uses the PU-MEKF and shows its construction is given in Chapter \ref{cha:hardware_applications}.
\begin{algorithm}[H]

	\SetAlgoLined
	\KwResult{Partial-updated posterior estimates $\xhat^{++}$ and $\stdvec{P}[][][++]$}
	Initialize $\xprior$, $\boldsymbol{\beta}$, $ \Qmat $, $ \R $ and $\stdvec{P}[][][-]$\\
	\For{The next time step}{
		Obtain propagated state $\xprior_k$ using system dynamics\\
		Compute jacobian $ 		\Fk = \partder{\f_{k}}{\x}\Big\rvert_{\xprior_{k-1}} $
		\\
		Propagate covariance matrix $\stdvec{P}[][k][-]$ using $ \stdvec{P}[][k][-] = \Fmat_{k-1}\stdvec{P}[][k-1][+]\Fmat_{k-1}\trans+\Qmat_{k-1} $\\
		\If{New measurement is available}{
			Form the residual measurement matrix $\Hmat_k$\\
			Compute the Kalman gain $\stdvec{K}_k$ with $ \stdvec{K}[][k]=\stdvec{P}[][k][-]\Hmat_k\trans(\Hmat_k\stdvec{P}[][k][-]\Hmat_k\trans+\Rk)^{-1} $\\
			Compute the residual $\stdvec{r}_k$ using the incoming measurement, $\ym_k$, and expected measurement, $\yhat_k$, according to $\stdvec{r}_k=\ym_k-\yhat_k$\\
			Compute the correction $  \xerror^{++}_k=\boldsymbol{\beta}\stdvec{K}_k\stdvec{r}_k =  \colvec{\boldsymbol{\beta_{\smalltheta}}\smallthetahat^+_k,\boldsymbol{\beta_{additive}}\xerrorhat_{additive-k}^+} $\\
			Use $\boldsymbol{\beta_{\smalltheta}}\smallthetahat^+_k$ to partial-update multiplicative states using $ \quat^{++}_k =\colvec{\frac{1}{2}{\boldsymbol{\beta_{\smalltheta}}}\smallthetahat^+_k,1}\otimes \quat^- $. This recovers attitude estimate.
			\\
			Use $\boldsymbol{\beta_{additive-k}}\xerrorhat_{additive-k}^+$ to partial-update additive states using $ \xhat_k^+ = \xprior_k + \boldsymbol{\beta_{additive-k}}\xerrorhat_{additive-k}^+$. This recovers the actual estimates for additive states.
			\\
			Via $ \stdvec{P}[][ij][++] = \gamma_i\gamma_j \stdvec{P}[][ij][-] + (1 - \gamma_i\gamma_j)\stdvec{P}[][ij][+]$, apply partial-update to the covariance matrix $\Cov$ and obtain the current estimate $\stdvec{P}[][][++]$}
		{
		}
	}
	\caption{Partial-update Multiplicative Extended Kalman filter (PU-MEKF)}
		\label{alg:PUMEKF}
\end{algorithm}

\section{Stability analysis of the partial-update filter}
The partial-update filter is, in general, an \textit{intermediate} filter lying in between the conventional and the consider filter, and for specific $ \beta $ weight values, it can act as one or another as well. In terms of filter stability, this is relevant, as the partial-update should share conventional and consider filter properties: for linear systems, the estimation error is stable. In this section, the insight of the partial-update being stable is confirmed via the direct Lyapunov method. The stability analysis is based on the developments presented in \cite{crassidis2011optimal} and \cite{Woodbury2011a}. Only the discrete stability analysis is developed, and it is assumed that the true linear system dynamics are described by
\begin{equation}
	\x_{k+1} = \Fk\x+ \stdvec{B}[][k] \stdvec{u}[][k]+ \Gmat_k\wk \ ,
\end{equation}
\begin{equation}
	\ym_{k+1} = \Hmat_{k+1}\x_{k+1}+\v_{k+1} \ .
\end{equation}

Let the following estimation error weighted function be the candidate Lyapunov function
\begin{equation}
	\stdvec{V}[][k] =\ek\trans\Pinvk\ek \ ,
\end{equation}
with $ \ek=\xhat-\x $.
The necessary condition for the estimation error to be stable in the Lyapunov sense, is that the change in the function $ \stdvec{V}[][k] $ remains at least negative definite after every recursion, in other words,
\begin{equation}
	\Delta\stdvec{V}[][](\e) = \epost\trans\Pinvpost\epost - \ek\trans\Pinvk\ek < \zerovec \ .
\end{equation}
To begin, the error $\ek$ is defined in terms of the transition and measurement matrix. Towards this goal, first recall that the partial update step can be written as
\begin{equation}
	\xpost_{k-1}= \xhat^{-}_{k-1}+(\Identity-\Dmat)\Kk(\ymprior-\Hprior\xhat^-_{k-1}) \ ,
\end{equation}
and calling $ \Ktk=(\Identity-\Dmat)\Kk $ gives
\begin{equation}
		\xpost_{k-1}= \xhat^{-}_{k-1}+\Ktk(\ymprior-\Hprior\xhat^-_{k-1}) \ .
\end{equation}
Similarly, the system dynamics can be rearranged as
\begin{equation}
	\xpost_k = \Fprior\xpost_{k-1}+\Bprior\u=\Fprior\xpost_{k-1}+\Fprior\Ktprior(\ymprior-\Hprior\xhat^-_{k-1})+\Bprior\u \ .
\end{equation}
Next, the system error can be formed as
\begin{equation}
	\ek=\xhat_k-\x_k=\Fprior\eprior+\Fprior\Ktprior(\ymprior-\Hprior\xhat^-_{k-1})-\Gprior\w_{k-1} \ ,
\end{equation}
and using the measurement equation model in the previous equation, and rearranging terms results in 
\begin{equation}
	\ek = \Fprior\eprior+\Fprior\Ktprior\Hprior\x_{k-1}+\Fprior\Ktprior\v_{k-1}-\Fprior\Ktprior\Hprior\xprior_{k-1}-\Gprior\w_{k-1} \ ,
\end{equation}
or
\begin{equation}
	\ek=\Fprior(\Identity-\Ktprior\Hprior)\eprior+\Fprior\Ktprior\v_{k-1}-\Gprior\w_{k-1} \ .
\end{equation}
Finally, the error at time $ k $ can be expressed as
\begin{equation}
	\ek=\Fprior(\Identity-\Ktprior\Hprior)\eprior \ .
\end{equation}

Now, this alternative definition of the estimation error,$ \ek $, is used in the Lyapunov candidate function. This leads to
\begin{align}
\Delta\stdvec{V}[][](\e) &= \epost\Pinvpost\epost - \ek\trans\Pinvk\ek \\
&=\ek\trans(\Identity-\Ktk\Hk)\trans\Fk\trans\Pinvpost      \Fk(\Identity-\Kk\Hk)\ek-\ek\trans\Pinvk\ek\\
&=\ek\trans[(\Identity-\Ktk\Hk)\trans\Fk\trans\Pinvpost      \Fk(\Identity-\Ktk\Hk)-\Pinvk]\ek\trans \ ,
\end{align}
which translates the problem to show that the bracketed term is negative definite, or at least negative semi-definite for stability,
\begin{equation}\label{eq:bracketed_term_stability}
[(\Identity-\Ktk\Hk)\trans\Fk\trans\Pinvpost \Fk(\Identity-\Ktk\Hk)-\Pinvk] < \zerovec \ .
\end{equation}
Next, the bracketed term is required to be expressed in terms of elements that involve time index $ k $ only. To accomplish this a few steps are required. 

First, Equation (\ref{eq:bracketed_term_stability}) is pre-multiplied by $ \Fk^{-T}(\Identity-\Ktk\Hk)^{-T} $, and then post-multiplied by  $(\Identity-\Ktk\Hk)^{-1}\Fk^{-1} $. This gives rise to,
\begin{equation}
	\Pinvpost -\Fk^{-T}(\Identity-\Ktk\Hk)^{-T}\Pinvk(\Identity-\Ktk\Hk)^{-1}\Fk^{-1}<\zerovec \ .
\end{equation}
Pre-multiplying the previous expression by $ \Cov_{k+1}^- $ leads to
\begin{equation}\label{eq:bracketed_term_stability2}
	\Identity - \Cov_{k+1}^-\Fk^{-T}(\Identity-\Ktk\Hk)^{-T}\Pinvk(\Identity-\Ktk\Hk)^{-1}\Fk^{-1}<\zerovec \ .
\end{equation}
Second, the posterior covariance matrix $ \Cov_{k+1} $ is written in \textit{Joseph form} with the objective of using it in the previous equation, and obtain an expression in terms of the same time index. The general covariance update (often called Joseph form), which also applies when using the partial-update concept, is
\begin{equation}
	\Cov_k^+=(\Identity-\Ktk\Hk)\Cov_k^-(\Identity-\Ktk\Hk)\trans+\Ktk\R_k\Ktk\trans \ .
\end{equation}
Third, substituting the Joseph covariance,$ \Cov_k^+ $, into the traditional covariance propagation equation, a propagated covariance can be obtained as
\begin{align}\label{eq:joseph_expanded}
	\Cov_{k+1}^-&= \Fk\Cov_k^+\Fk\trans+\Gmat_k\Qmat_k\Gmat_k\trans\\\nonumber
	&=\Fk(\Identity-\Ktk\Hk)\Cov_k^-(\Identity-\Ktk\Hk)\trans\Fk\trans+\Fk\Ktk\R_k\Ktk\trans\Fk\trans+\Gmat_k\Qmat_k\Gmat_k\trans \ .
\end{align}
Finally, substituting Equation (\ref{eq:joseph_expanded}) into Equation (\ref{eq:bracketed_term_stability2}), the equation in terms of index $ k $ only, is obtained,
\begin{equation}
 - [\Fk\Ktk\R_k\Ktk\trans\Fk\trans+\Gmat_k\Qmat_k\Gmat_k\trans][\Fk^{-T}(\Identity-\Ktk\Hk)^{-T}\Cov_k^{-1}(\Identity-\Ktk\Hk)^{-1}\Fk^{-1}]<\zerovec \ .
\end{equation}
Noticing that the bracketed term on the right is a positive definite matrix, the stability analysis further reduces to check for positive definiteness of the left bracketed term
\begin{equation}
	- [\Fk\Ktk\R_k\Ktk\trans\Fk\trans+\Gmat_k\Qmat_k\Gmat_k\trans]<\zerovec \ ,
\end{equation}
or equivalently
\begin{equation}
- [\Fk\Kk(\Identity-\Dmat)\R_k(\Identity-\Dmat)\trans\Kk\trans\Fk\trans+\Gmat_k\Qmat_k\Gmat_k\trans]<\zerovec \ ,
\end{equation}
or
\begin{equation}
-[\Fk\Kk\betamat\R_k\betamat\trans\Kk\trans\Fk\trans+\Gmat_k\Qmat_k\Gmat_k\trans]<\zerovec \ .
\end{equation}
Since the process noise covariance ,$ \Qmat $, is at least positive semi-definite and the measurement noise covariance, $ \R $, is positive definite, the factor that defines the filter stability is the matrix $ \betamat $. In summary, 

\begin{itemize}
	\item If $ \betamat  $ is singular with one, or up to $ (n-1) $ diagonal elements being zero, then the partial-update filter will be stable, but not asymptotically stable.
	\item If $ \betamat $ has all diagonal entries different from zero, the system is asymptotically stable. 
	\item If $ \betamat $ is the zero matrix, no update is performed at all, and if this condition is kept filter divergence may be observed.
\end{itemize}

\section{Partial-update numerical stability issues}
It is very well known and documented in the literature that early after the Kalman filter development, and as its popularity increased, researchers started observing issues as filter performance degradation, divergence, and even negative covariance values for apparently well-posed problems \cite{Grewal2001}, \cite{McGee1985}, \cite{Verhaegen1986}, \cite{Kaminski1971}. The desire to maintain the Kalman filter as a reliable online solution for the estimation problem, drove researchers to invest time to investigate such issues, from which it was found that round-off computation errors and ill-conditioned problems, along with finite computer word-length, were the common triggers for such problems to appear. Several solutions for alleviating these inconveniences, consisting of propagating covariance factors, were proposed, widely developed, and adopted. Square root and UD factorized filters were and are, in fact, the most used today. When these solutions are integrated into the backbone of \textit{conventional} filters, it is possible to have more numerically robust, and stable filters, while keeping the underlying filter properties (at the cost of extra computations inherent to the factorized forms). 

Although factorized filters can benefit conventional filter formulations, they may not be too common to see in the literature, but they are in use. The infrequent publication of the alternative formulations may be due to the lack of significant or unperceived of numerical issues for the designer to opt for a change in implementation (a common symptom among practitioners as mentioned in \cite{Carpenter2018}). Furthermore, conventional formulations often result in well-behaved filters when used in simulation, and no extra precision seems to be needed. Nonetheless, recent papers start to increasingly incorporate these square root implementations to ensure the positive semi-definiteness of the covariance matrix, but also to leverage modern computers capacities that allow to invest in a more expensive filter form.  

The partial-update filter has been observed to work well on simulation, and when appropriately applied, it can outperform the EKF and Schmidt filter. However, the partial-update formulation is an extension of the Kalman filter, and as such, it inherits the numerical problems bound to the Kalman filter formulations.

The research presented in the following chapters (third and fourth), is devoted to integrating factorized formulations and partial-update filter to provide formulations that incorporate the benefits from both concepts, but that either cannot deliver if implemented separately. The result is a set of filters with augmented ability to support high uncertainties and nonlinearities, and more robust against numerical issues. The proposed formulations, although they are not \textit{bullet-proof}, they certainly extend the class of problems the EKF can handle well.

%
%
%
%



%
%
%
%

\chapter[SQUARE ROOT PARTIAL-UPDATE SCHMIDT KALMAN FILTER]{SQUARE ROOT PARTIAL-UPDATE SCHMIDT KALMAN FILTER\footnote{Adapted with permission from ``Square root partial-update kalman filter'', by J. H. Ramos, K. M. Brink, and J. E. Hurtado, presented at the 22nd International Conference on Information Fusion FUSION 2019, \cite{Ramo1907:Square}, Copyright 2019 by the International Society of Information Fusion. }}\label{ch:square_root_pu_filter}

\section{Introduction \& Motivation}
Shortly after the original Kalman filter paper \cite{kalman1960new}, however, two primary concerns emerged.  The first pertained to the  numerical precision of the filter and the second pertained to the filter's robustness for nonlinear systems or measurements.  Additionally, it was noted that the subtraction that occurs in the covariance measurement update equation could produce numerical issues when high precision measurements were assimilated on finite precision computers.  Moreover, it has been noticed that numerical issues may surface when significant mismatches in state observability are present \cite{Kaminski1971} or when there exists considerable discrepancies in the magnitudes among state elements \cite{simon2006optimal}. 

These challenges motivated researchers to develop alternative recursive forms of the Kalman equations that would be useful in improving the precision when limited hardware specifications were available.  These led to the development of an approach that relied on propagating the square root of the error covariance matrix rather than the error covariance matrix itself.  The first occurrence of the square root form is attributed to Potter \cite{battin1999introduction}.  Potter noticed that using the square root form of the error covariance, he could rigorously enforce non-negative definiteness of the covariance matrix after each recursion \cite{BELLANTONI1967}. Researchers later generalized Potter's formulation to process vector measurements and to include process noise. Alternative factorizations that allowed one to perform the filter recursion without the explicit use of the square root of the error covariance matrix \cite{Bierman1977} also became available.  Regardless of the form, the main goal of the square root and other factorized filters is to overcome the previously mentioned numerical problems related to direct propagation of the error covariance matrix. 

The square root Kalman filter is mainly a covariance reformulation of the standard Kalman equations, and thus it is still a linear filter.  Similar to what is done with a traditional Kalman filer, the square root Kalman filter can be applied in nonlinear systems through a linearized model.  That is, the square root formulation does not enhance a filter's ability in addressing nonlinearity, it simply improves numerical conditioning.

The partial-update has been demonstrated to be effective on a wide range of filtering applications and the square root implementation has long been shown to improve the filter's numerical qualities. This chapter presents a development that combines the square root and partial-update formulations within the same filter. The rest of the chapter is organized as follows: Background information is provided in Section \ref{sec:Background} including a brief explanation on how the square root formulation helps to improve the numerical robustness in filtering. It also presents the associated equations for the particular square root filter version utilized in this chapter, and a description of the partial-update Schmidt-Kalman filter. Section \ref{Derivation:Potter} provides the derivation of the square root partial-update Kalman filter. Finally, a nonlinear filtering numerical example that uses the square root partial-update filter is presented in Section \ref{sec:Simulations}. Monte Carlo runs and computational complexity of the filter are also included. Section \ref{sec:Conclusions} providing a conclusion statement.

%

\subsection{Square Root filtering}\label{sec:Background_on_Square_Root_filtering}
In the estimation field, square root filtering refers to utilize a square root factorized representation of the error covariance matrix for purposes of propagation and correction of the estimation error. The goal of reformulating filters using such ``square roots'' or factorizations, is to increase the precision of the filter itself. By operating on the square root of the error covariance, the filter lowers the condition number of the uncertainty matrix (to be discussed shortly), which is then less prone to numerical issues because fewer significant figures are required during the arithmetic operations. A low condition number is always desirable, mainly for the cases where the computer word length is limited (as in embedded systems), or when the filtering problem is poorly conditioned. Although these type of formulations are numerically more robust, it is at the cost of increasing the computational effort.  Nevertheless, the amount of extra computations can still be reasonable which allows a factorized filter to be used in many applications \cite{Kaminski1971}.

The definition for the ``square root'' of a matrix, in contrast from scalar quantities, can vary from one reference to another.  For the purposes of this research, the definition for the square root of a matrix is based on the idea of finding a matrix $\Smat$ that satisfies (\ref{eq:square_root_definition}). $\Smat$, is what will be referred to as the square root of the error covariance matrix $\Cov$.
\begin{equation}\label{eq:square_root_definition}
\Cov=\Smat\Smat\trans \ .
\end{equation}

Specifically, $\Smat$ is a lower triangular matrix, and $\Smat\trans$ its transposed. Importantly, it can be noticed that the product $\Smat\Smat\trans$ is naturally symmetric and positive semidefinite, regardless of the value of the lower triangular matrix $\Smat$. Thus, numerical difficulties that could cause the covariance matrix $\Cov$ to become non-symmetric or singular, cannot affect the product $\Smat\Smat\trans$, thus preserving the theoretical properties of the covariance matrix $\Cov$ (within the machine precision). 
Also, as for scalars, the square root $\Smat$ is not unique. That is, there may  be several solutions for $\Smat$. One very well known method to compute the matrix $\Smat$ is the Cholesky decomposition \cite{golub2012matrix}. The Cholesky method directly outputs the matrix $\Smat$ that satisfies Equation (\ref{eq:square_root_definition}). This method requires that the matrix to be factorized is positive definite and symmetric, which holds for $\Cov$.   

The set of equations that correspond to the square root filter are included here. The assumptions for this approach are taken from  Equations (\ref{eq:nonlinear_equations})-(\ref{eq:measurement_cov}).  The state propagation in the EKF is done with Equation (\ref{eq:nonlinear_propagation}), whereas the square root of the covariance is propagated by solving for $\mathcal{T}$ in  Equation (\ref{eq:sq_propagation}) as an alternative to  Equation (\ref{eq:cov_propagation}). The details of how to find matrix $\mathcal{T}$ are given in Section \ref{subsec:sq_cov_propagation}. The Kalman gain in of Equation (\ref{eq:kalman_gain}) is replaced by the gain from Equation (\ref{eq:sq_kalman_gain}) while the covariance update, from Equation (\ref{eq:covariance_update}), is now accomplished with Equation (\ref{eq:sq_update}). The state measurement update is computed with the standard form of Equation (\ref{eq:state_update}).
\begin{equation}
\begin{bmatrix}\label{eq:sq_propagation}
(\stdvec{S}[][k][-])\trans \\ 
\zerovec
\end{bmatrix}=
\mathcal{T}
\begin{bmatrix}
(\stdvec{S}[][][+]_{k-1})\trans\Fmat\trans_{k-1}\\
\Qmat_{k-1}^{T/2}
\end{bmatrix} \ ,
\end{equation}
\begin{equation}\label{eq:sq_kalman_gain}
\Kmat_k=a_i(\stdvec{S}[][][-]_{k})\phi_i \ ,
\end{equation}
\begin{equation}\label{eq:sq_update}
(\stdvec{S}[][k][+])=(\stdvec{S}[][k][-])(\Imat-a_ib_i\phi_i\phi_i\trans) \ ,
\end{equation}
where
\begin{equation}
a_i =\frac{1}{\phi_i\trans\phi_i+\R_i}  \ ,
\end{equation}
\begin{equation}
\phi_i=(\stdvec{S}[][k][-])\trans\Hmat_i\trans \ ,
\end{equation}
\begin{equation}\label{eq:b}
b_i = \frac{1}{1\pm\sqrt{a_i\R_i}}  \ . 
\end{equation}

Importantly, it should be noted that this version of square root filter processes the available measurements in a sequential fashion \cite{Kaminski1971}. Thus, for a particular time $k$ when a measurement vector is available, the update Equations (\ref{eq:state_update}), (\ref{eq:sq_kalman_gain}) and (\ref{eq:sq_update}) are executed for $i=1,2,\dots,m$  in order to process each measurement in the measurement vector $\yktilde$ $\in \Real^m$. In other words, when a vector measurement is available,  $m$ updates are performed, one update per element in the measurement vector. Very importantly, for nonlinear systems the Jacobians, $ \Fmat $ and $ \Hmat $ may need to be recomputed after each measurement assimilation.

Due to the sequential nature of the updates, the measurement covariance matrix $\R$ is assumed to be in diagonal form such that $\R_i$ denotes the $i^{th}$ diagonal element that corresponds to the measurement element $\ym_i$ and $\Hmat_i$ represents the $i^{th}$ row of $\Hmat$. Finally, positive sign can be chosen in Equation (\ref{eq:b}) to avoid subtraction. Now with the proper context, in the next section the square root partial-update Schmidt-Kalman filter is derived.

\section{The square root partial-update Schmidt-Kalman filter}\label{Derivation:Potter}
The objective here is to combine the benefits of square root filtering and the partial-update approach into one filter that provides an increase in robustness to uncertainties and numerical issues beyond what is provided by either individual formulation. As with many available techniques, this formulation also has its limits, but its significant contribution in added robustness and its simple implementation, makes it very attractive.

The derivation proposed follows Potter's original form \cite{battin1999introduction} as it assimilates the measurements sequentially. The method proposed here, however, handles process noise. For clarity of exposition, the derivation considers that the $i^{th}$ element of the measurement vector is being processed and the indices are omitted for $a$ and $b$, since it will be clear that $\R_i$ generates a corresponding $a_i$ and $b_i$. Similarly the index for $\phi$ is omitted. Recall that the resulting update equations will need to be executed $m$ times in order to process all measurements in the vector $\yktilde \in \Real^m$.
\subsection{Measurement update}
It has been previously shown in \cite{BrinkPartialUpdate}, that the partial-update filter is statistically sound; that is, $\expect[\textbf{e}]=0 , \expect[\textbf{e}^2]=0$ for linear systems.  Thus, if a square root form for the covariance partial-update written in Equation
(\ref{eq:square_root_definition}) can be found, this should maintain the statistical consistency once the error covariance matrix is recovered. Specifically this case, the matrix $\stdvec{S}[][][++]$ is sought such that allows to write Equation (\ref{eq:partial_update_cov}) as $\stdvec{P}[][][++] = \stdvec{S}[][][++]\stdvec{S}[][][++]\trans$.
To begin, the partial-update equations are expressed in matrix form. For clarity sake, the time index $k$ is temporarily dropped. First, the partial-update covariance from Equation (\ref{eq:partial_update_cov}) can be reorganized such that
\begin{equation}\label{eq:partial_update_element_form}
\stdvec{P}[][ij][++] = \gamma_i\gamma_j (\stdvec{P}[][ij][-] -\stdvec{P}[][ij][+]) + \stdvec{P}[][ij][+] \ .
\end{equation}

Then, recognizing that the first term on the right-hand side of Equation (\ref{eq:partial_update_element_form}) can be written as $\Dmat(\stdvec{P}[][][-]-\stdvec{P}[][][+])\Dmat$, where $\Dmat$ is a diagonal matrix with elements $\gamma_i$ for $i = 1,2\dots n$ (recall that $\gamma_i = 1 - \beta_i$), the covariance partial-update can be expressed as
\begin{equation}\label{eq:partial_update_cov_vec}
\stdvec{P}[][][++] = \Dmat(\stdvec{P}[][][-]-\stdvec{P}[][][+])\Dmat + \stdvec{P}[][][+] \ .
\end{equation}

Now, from the standard EKF equations (ignoring the time indices for ease of notation),  $\stdvec{P}[][][+]=(\stdvec{I-\stdvec{K}[][]}\Hmat)\stdvec{P}[][][-]$ is incorporated into Equation (\ref{eq:partial_update_cov_vec}).
\begin{equation}
\stdvec{P}[][][++] = \stdvec{P}[][][+] + \Dmat(\Kmat\Hmat\stdvec{P}[][][-])\Dmat \ ,
\end{equation}
then replacing $\stdvec{K}[][]=\stdvec{P}[][][-]\Hmat\trans(\Hmat\stdvec{P}[][][-]\Hmat\trans+\R)^{-1}$ results in
\begin{equation}
\stdvec{P}[][][++] = \stdvec{P}[][][+] + \Dmat(\stdvec{P}[][][-]\Hmat\trans(\Hmat\stdvec{P}[][][-]\Hmat\trans+\R)^{-1}\Hmat\stdvec{P}[][][-])\Dmat \ .
\end{equation}

Then, it is required that $\stdvec{P}[][][-]= (\stdvec{S}[][][-])(\stdvec{S}[][][-])\trans$ and $\stdvec{P}[][][+]= (\stdvec{S}[][][+])(\stdvec{S}[][][+])\trans$, and considering that the measurements are processed one at a time. Thus, if $\Hmat_i$ is the $i$-th row of $\Hmat$, and $\R_i$ the $i$-th diagonal element of $\R$,  $a =(\Hmat_i\stdvec{P}[][][-]\Hmat_i\trans+\R_i)^{-1}$ is an scalar. These actions lead to 
\begin{equation}\label{eq:cov_full_form}
\stdvec{P}[][][++] = (\stdvec{S}[][][+])(\stdvec{S}[][][+])\trans + \Dmat(\stdvec{S}[][][-])(\stdvec{S}[][][-])\trans\Hmat_i\trans a\Hmat_i(\stdvec{S}[][][-])(\stdvec{S}[][][-])\Dmat \ . 
\end{equation}

The sequential processing assumes that $\R$ is already a diagonal matrix.
Further, with the purpose of using Potter's measurement update equations form directly, the scalar $a$ is written as follows:
\begin{equation}
a = \frac{1}{\Hmat_i(\stdvec{S}[][][-])(\stdvec{S}[][][-])\trans\Hmat_i\trans+\R_i}=\frac{1}{\phi\trans\phi+\R_i} \ ,
\end{equation}
where $\phi=(\stdvec{S}[][][-])\trans\Hmat_i\trans$.
Also, from Potter's formulation, the posterior square root error covariance matrix can be obtained as in Equation (\ref{eq:update_cov_potter}). A summary of Potter's equations can be found in \cite{Kaminski1971}.
\begin{equation}\label{eq:update_cov_potter}
\stdvec{S}[][][+] = (\stdvec{S}[][][-])(\Imat-ab\phi\phi\trans) \ ,
\end{equation}
where $b$ is defined as
\begin{equation}\label{eq:b_definition}
b = \frac{1}{1\pm\sqrt{a\R_i}}    \ .
\end{equation}
Now, using the definition of the posterior $\stdvec{S}[][][+]$ from Equation (\ref{eq:update_cov_potter}) in Equation (\ref{eq:cov_full_form}), the partial-update for the error covariance, which is now in terms of the prior $\stdvec{S}[][][-]$, reads
\begin{equation}\label{eq:patial_update_cov_factorized}
\begin{aligned}
\stdvec{P}[][][++] =
[(\stdvec{S}[][][-])(\Imat-ab\phi\phi\trans)]   [(\stdvec{S}[][][-])(\Imat-ab\phi\phi\trans)]\trans + \\ \Dmat(\stdvec{S}[][][-])(\stdvec{S}[][][-])\trans\Hmat_i\trans \sqrt{a}\sqrt{a}\Hmat_i(\stdvec{S}[][][-])(\stdvec{S}[][][-])\trans\Dmat \ .
\end{aligned}
\end{equation}

Because Equation (\ref{eq:patial_update_cov_factorized}) is the sum of two matrices and each of these matrices are factorized in a square root manner, Equation (\ref{eq:s_factorized}) can be set as a candidate square root of $(\stdvec{P}[][][++])$. 
\begin{equation}\label{eq:s_factorized}
\begin{aligned}
\begin{bmatrix}
(\stdvec{S}[][][++])\trans \\ 
\zerovec
\end{bmatrix} = 
\Tmat
\begin{bmatrix}
(\Imat-ab\phi\phi\trans)\trans(\stdvec{S}[][][-])\trans\\
\sqrt{a}\Hmat_i(\stdvec{S}[][][-])(\stdvec{S}[][][-])\trans\Dmat
\end{bmatrix}
\end{aligned} \ .
\end{equation}
Where  matrix $\Tmat = \rowvec{\Tmat_{1},\Tmat_{2}}$ is a $(n+r)\times(n+r)$ orthogonal matrix, with $\Tmat_{1}$ being $(n+r)\times n$ and $\Tmat_{2}$ being $(n+r)\times r$ matrices, meaning that
\begin{equation}
\begin{aligned}
\Tmat\trans\Tmat =
\begin{bmatrix}
\Tmat_{1}\trans \\ \Tmat_{2}\trans
\end{bmatrix}
\begin{bmatrix}
\Tmat_{1} & \Tmat_{2}
\end{bmatrix} \\= 
\begin{bmatrix}
\Tmat_{1}\trans\Tmat_{1} & \Tmat_{1}\trans\Tmat_{2} \\
\Tmat_{2}\trans\Tmat_{1} & \Tmat_{2}\trans\Tmat_{2}
\end{bmatrix}=
\begin{bmatrix}
\Imat & \zerovec\\
\zerovec & \Imat
\end{bmatrix}
\end{aligned} \ .
\end{equation}
Thus,
\begin{equation}\label{eq:s_final_factor}
\begin{aligned}
\begin{bmatrix}
(\stdvec{S}[][][++])\trans \\ 
\zerovec
\end{bmatrix} = 
\begin{bmatrix}
\Tmat_{1} & \Tmat_{2}
\end{bmatrix}
\begin{bmatrix}
(\Imat-ab\phi\phi\trans)\trans(\stdvec{S}[][][-])\trans\\
\sqrt{a}\Hmat_i(\stdvec{S}[][][-])(\stdvec{S}[][][-])\trans\Dmat
\end{bmatrix}
=\begin{bmatrix}
\stdvec{W}_{n\times n}
\\ \zerovec
\end{bmatrix}
\end{aligned} \ .
\end{equation}

Then, if an orthogonal matrix $\Tmat$ can be found such that Equation (\ref{eq:s_final_factor}) produces an ($n\times n$) upper triangular $\stdvec{W}$ matrix stacked on top of a lower zero ($r\times n$) matrix, then it can be recognized that the upper triangular ($n\times n$) matrix $\stdvec{W}$, is actually equal to $(\stdvec{S}[][][++])\trans$, which is the desired result.  The idea behind the use of the orthogonal matrix $\Tmat$, is to use the factorization shown inside the brackets in Equation (\ref{eq:s_factorized}). The transformation $\Tmat$, allows one to find a square root of dimension ($n\times n$). As it may be noted, $\stdvec{S}[][][++]=\rowvec{(\stdvec{S}[][][-])(\Imat-ab\phi\phi\trans) , \Dmat(\stdvec{S}[][][-])(\stdvec{S}[][][-])\trans\Hmat_i\trans \sqrt{a} }$ can also act as a square root of $ \stdvec{P}[][][++]$, however this selection would increase the dimension of the problem to a ($n\times (n+r)$) square root matrix, which is not desirable. 

In the following development, the product from Equation (\ref{eq:product_factorization}) is performed explicitly to show that the candidate form of Equation (\ref{eq:s_factorized}) is a valid square root for $ \stdvec{P}[][][++]$.
\begin{equation}\label{eq:product_factorization}\nonumber
\rowvec{(\stdvec{S}[][][++]),\zerovec}
\begin{bmatrix}
(\stdvec{S}[][][++])\trans\\\zerovec
\end{bmatrix}  = 
\end{equation}
\begin{flalign}\label{eq:sq_development}
&[\Tmat_{1}(\Imat-ab\phi\phi\trans)\trans(\stdvec{S}[][][-])\trans+\Tmat_{2}\sqrt{a}\Hmat_i(\stdvec{S}[][][-])(\stdvec{S}[][][-])\trans\Dmat]\trans[\dots]&&\\
&=(\stdvec{S}[][][-])(\Imat-ab\phi\phi\trans)\Tmat_{1}\trans\Tmat_{1}(\Imat-ab\phi\phi\trans)(\stdvec{S}[][][-])\trans+&&\\\nonumber
&(\stdvec{S}[][][-])(\Imat-ab\phi\phi\trans)\Tmat_{1}\trans\Tmat_{2}\sqrt{a}\Hmat_i(\stdvec{S}[][][-])(\stdvec{S}[][][-])\trans\Dmat+&&\\\nonumber
&\Dmat\sqrt{a}(\stdvec{S}[][][-])(\stdvec{S}[][][-])\trans\Hmat_i\trans\Tmat_{2}\trans\Tmat_{1}(\Imat-ab\phi\phi\trans)(\stdvec{S}[][][-])\trans+
&&\\\nonumber &\Dmat\sqrt{a}(\stdvec{S}[][][-])(\stdvec{S}[][][-])\trans\Hmat_i\trans\Tmat_{2}\trans\Tmat_{2}\sqrt{a}\Hmat_i(\stdvec{S}[][][-])(\stdvec{S}[][][-])\trans\Dmat&&\\\nonumber\\
&=(\stdvec{S}[][][-])(\Imat-ab\phi\phi\trans)\Tmat_{1}\trans\Tmat_{1}(\Imat-ab\phi\phi\trans)(\stdvec{S}[][][-])\trans+&&\\\nonumber
&\Dmat\sqrt{a}(\stdvec{S}[][][-])(\stdvec{S}[][][-])\trans\Hmat_i\trans\Tmat_{2}\trans\Tmat_{2}\sqrt{a}\Hmat_i(\stdvec{S}[][][-])(\stdvec{S}[][][-])\trans\Dmat&&\\\nonumber\\\label{eq:proof_square_root}
&=(\stdvec{S}[][][-])(\Imat-ab\phi\phi\trans)(\Imat-ab\phi\phi\trans)(\stdvec{S}[][][-])\trans+&&\\\nonumber
&\Dmat\sqrt{a}(\stdvec{S}[][][-])(\stdvec{S}[][][-])\trans\Hmat_i\trans\sqrt{a}\Hmat_i(\stdvec{S}[][][-])(\stdvec{S}[][][-])\trans\Dmat \ . && 
\end{flalign}

In Equation (\ref{eq:proof_square_root}) the expression for the partial-update covariance from Equation (\ref{eq:patial_update_cov_factorized}) is recovered, which shows that the factorization proposed in Equation (\ref{eq:s_factorized}) is actually a square root for $\stdvec{P}[][][++]$. Then, the square root partial-update equations are
\begin{equation}\label{eq:s_final}
\begin{aligned}
\begin{bmatrix}
(\stdvec{S}[][][++])\trans \\ 
\zerovec
\end{bmatrix} = 
\begin{bmatrix}
\Tmat_{1} & \Tmat_{2}
\end{bmatrix}
\begin{bmatrix}
(\Imat-ab\phi\phi\trans)\trans(\stdvec{S}[][][-])\trans\\
\sqrt{a}\Hmat_i(\stdvec{S}[][][-])(\stdvec{S}[][][-])\trans\Dmat
\end{bmatrix}
\end{aligned} \ .
\end{equation}
Or equivalently,
\begin{equation}\label{eq:s_final_equivalent}
\begin{aligned}
\begin{bmatrix}
(\stdvec{S}[][][++])\trans \\ 
\zerovec
\end{bmatrix} = 
\Tmat
\begin{bmatrix}
(\stdvec{S}[][][+])\trans\\
\sqrt{a}\phi\trans(\stdvec{S}[][][-])\trans\Dmat
\end{bmatrix}
=\begin{bmatrix}
\stdvec{W}_{n\times n}
\\ \zerovec
\end{bmatrix}
\end{aligned} \ .
\end{equation}
The problem of finding the orthogonal matrix $\Tmat$, but more importantly (for this application), finding the matrix $\stdvec{W}_{n\times n}$, is well known and several well documented algorithms are available in the literature \cite{Kaminski1971}, \cite{golub2012matrix}, \cite{moon2000mathematical}. This work makes use of the algorithm known as the Modified Gram-Schmidt (MGS), which was outlined in \cite{Kaminski1971} specifically for Kalman filtering.

For the implementation of the square root partial-update Kalman filter, the square root of the covariance is being directly propagated, so $\stdvec{S}[][][]$ is available at every time step. For the purposes of this development, $\stdvec{S}[][][]$ is propagated directly through the concept shown in the following sub-section (\ref{subsec:sq_cov_propagation}). Also recall that once the Modified Gram-Schmidt algorithm has been carried out, the matrix $(\stdvec{S}[][][++])\trans$ has been generated and the matrix $\stdvec{\stdvec{W}_{n\times n}}$ is simply retained, since it is the solution $(\stdvec{S}[][][++])\trans$ for the current measurement partial update step.
Finally, notice that one could think about or suggest the direct use of Equation (\ref{eq:partial_update_element_form}) to perform the time update by transforming the square root $\stdvec{S}[][][]$ into the covariance matrix $\stdvec{P}[][][]$, and then going back to the square root via the Cholesky decomposition. However, two main issues would be encountered: 1) it will be computationally expensive to go back and forth for relative large systems and, 2) Equation (\ref{eq:partial_update_element_form}) involves a subtraction, which would bring back the original concern over numerical issues due to finite precision.  

\subsection{Time update}\label{subsec:sq_cov_propagation}
The propagation of the square root error covariance matrix $\stdvec{S}[][][]$, follows the same factorization idea used to compute $(\stdvec{S}[][][++])\trans$ in Equation (\ref{eq:s_final_equivalent}). Recall that the standard propagation of the error covariance matrix is computed as in Equation (\ref{eq:cov_propagation}), which it is included here for convenience \cite{simon2006optimal}.
\begin{equation}\label{eq:cov_propagation2}
\stdvec{P}[][k][-] = \Fmat_{k-1}\stdvec{P}[][k-1][+]\Fmat_{k-1}\trans+\Qmat_{k-1} \ .
\end{equation}
Again, since $\stdvec{P}[][k-1][+]$ is symmetric and positive definite, it can alternatively be written as
\begin{equation}\label{eq:s_cov_propagation}
\stdvec{S}[][k][-]\stdvec{S}[][k][-]\trans = \Fmat_{k-1}\stdvec{S}[][k-1][+]\stdvec{S}[][k-1][+]\trans\Fmat_{k-1}\trans+\Qmat_{k-1}^{1/2}\Qmat_{k-1}^{T/2} \ .
\end{equation}
From which the following factorization can be proposed \cite{anderson2012optimal}
\begin{equation}\label{eq:s_propagation_equivalent}
\begin{aligned}
\begin{bmatrix}
(\stdvec{S}[][k][-])\trans \\ 
\zerovec
\end{bmatrix} = 
\begin{bmatrix}
\mathcal{T}_1 & \mathcal{T}_2
\end{bmatrix}
\begin{bmatrix}
(\stdvec{S}[][][+]_{k-1})\trans\Fmat\trans_{k-1}\\
\Qmat_{k-1}^{T/2}
\end{bmatrix}
=\begin{bmatrix}
\stdvec{\mathcal{W}}_{n\times n}
\\ \zerovec
\end{bmatrix}
\end{aligned} \ ,
\end{equation}
such that,
\begin{equation}\label{eq:s_product_factorization}
\stdvec{P}[][k][-]=\rowvec{(\stdvec{S}[][k][-]),\zerovec}
\begin{bmatrix}
(\stdvec{S}[][k][-])\trans\\\zerovec
\end{bmatrix} \ . 
\end{equation}

In a similar manner as before, $\mathcal{T}= \rowvec{\mathcal{T}_1 & \mathcal{T}_2}$ is a $2n\times 2n$ orthogonal matrix to be found (in this case via Modified Gram-Schmidt), and $\mathcal{W}$ is the solution for  $(\stdvec{S}[][k][-])\trans$. Notice that $\mathcal{T}$ is different from $\Tmat$ in general. Thus, performing MGS for Equation (\ref{eq:s_propagation_equivalent}) provides the direct propagation step for the square root error covariance matrix. Equation (\ref{eq:cov_propagation2}) can be recovered from the product in Equation (\ref{eq:s_product_factorization}), which verifies Equation (\ref{eq:s_propagation_equivalent}) as a valid square root for $\stdvec{P}[][k][-]$. 
In Table \ref{table:SQPSK filter} the square root partial-update Schmidt-Kalman filter equations are summarized.
\begin{table}[h]
	\caption{Square root partial-update Schmidt-Kalman filter. Reprinted with permission from \cite{Ramo1907:Square}.}\label{table:SQPSK filter}
	\centering
	\setlength{\extrarowheight}{8pt}
	\begin{tabular}{ |c|c| } 
		\hline
		\textbf{Model} & $\begin{aligned} &\x_{k}=\f_{k-1}(\x_{k-1},\u_{k-1})+\w_{k-1}\\ &\yktilde=\hkofxk+\vk\\
		&\wk \sim \mathcal{N} (\zerovec,\Qmat_k)\\
		&\vk \sim \mathcal{N}(\zerovec,\R_k)\end{aligned}$\\
		\hline
		\textbf{Initialize} & $\begin{aligned} 
		&\xpost_0=\x_0 \\ &\stdvec{S}[][0][+]= chol(\stdvec{P}[][0][+])\ \cite{golub2012matrix}\ \\
		&\Qmat_0^{1/2} = eigendec(\Qmat_0)\ \cite{strang1993introduction}\\
		&\Dmat = \diag(1-\beta_1,1-\beta_2,...,1-\beta_n)\
		\end{aligned}$\\
		\hline
		\textbf{Propagation} & $\begin{aligned} &\xprior_{k}=\f_{k-1}(\xpost_{k-1},\u_{k-1})\\
		&Perform\ MGS\ \cite{Kaminski1971}\ for \\
		&\begin{bmatrix}
		(\stdvec{S}[][k][-])\trans \\ 
		\zerovec
		\end{bmatrix}=
		\mathcal{T}
		\begin{bmatrix}
		(\stdvec{S}[][][+]_{k-1})\trans\Fmat\trans_{k-1}\\
		\Qmat_{k-1}^{T/2}
		\end{bmatrix}\end{aligned}$\\
		\hline
		\textbf{Gain} & $\begin{aligned}\Kmat_k=a_i&(\stdvec{S}[][][-]_{k})\phi_i\\
		a_i =\frac{1}{\phi_i\trans\phi_i+\R_i};\ &\phi_i=(\stdvec{S}[][][-])\trans\Hmat_i\trans
		\end{aligned}$\\ 
		\hline
		\textbf{Update} & $\begin{aligned}\xpost_k&=\xprior_{k}+\stdvec{K}[][k](\ym_k-\yhat_k)\\&(\stdvec{S}[][k][+])=(\stdvec{S}[][k][-])(\Imat-a_ib_i\phi_i\phi_i\trans)\\&b_i = \frac{1}{1\pm\sqrt{a_i\R_i}}\end{aligned}$\\
		\hline
		\textbf{Partial-Update} & $\begin{aligned} &\xpostplus = \Dmat\xprior + (\Imat- \Dmat) \xpost\\
		&Perform\ MGS\ \cite{Kaminski1971}\ for\\
		&\begin{bmatrix}
		(\stdvec{S}[][k][++])\trans \\ 
		\zerovec
		\end{bmatrix} = 
		\Tmat
		\begin{bmatrix}
		(\stdvec{S}[][k][+])\trans\\
		\sqrt{a_i}\phi_i\trans(\stdvec{S}[][k][-])\trans\Dmat
		\end{bmatrix}\end{aligned}$\\
		\hline
	\end{tabular}
\end{table}

Before moving to examples of the new filter, a few remarks about the development are needed. First, note that the real symmetric matrix $\Qmat^{T/2}$, is required in the propagation step. This is computed by an eigendecomposition procedure which let express $\Qmat=VD V^{T} $, where the columns of $V$ are the eigenvectors of $\Qmat$, and $D$ is a diagonal matrix which entries are the corresponding eigenvalues \cite{strang1993introduction}. In this way,  $\Qmat^{1/2} = VD^{1/2} $, where $D^{1/2}$ is the square root of $D$.
Second, since the measurements are processed sequentially, the filter updates are performed for $i=1,2,..,m$ in order to assimilate the observation vector $\yktilde$ completely. It is highly important that the measurement matrix is computed with the most current estimate, and thus the corresponding Jacobians are to be updated. That is, every time a sequential measurement is assimilated, the measurement (and transition) matrix needs to be re-evaluated at the most recent state posterior available. Finally, since a decomposition to obtain $\Qmat^{T/2}$ is required, the user will have to assess the convenience of this square root filter when dealing with time-varying process noise. A similar assessment may be required in the case of a time-varying $\R$ matrix.

\section{Numerical examples}\label{sec:Simulations}
In this section simulation results that show the effect of using the square root partial-update filter are presented, and compared with the (conventional) square root EKF. A simulation of standard EKF and square root EKF is included as well.
The simulations, also show the agreement between the square root partial-update and the standard partial-update estimates.
\subsection{Body re-entering Earth atmosphere}
In this numerical example, which is based on the example presented in \cite{Julier2000}, the altitude, $\x_1$ (in meters), vertical velocity, $\x_2$ (in meters per second), and the constant ballistic parameter, $\x_3$ (with units of 1/meter), of a body re-entering Earth atmosphere from high altitude and with high velocity, are estimated. It is assumed that the body is constrained to fall vertically, and that a range-measurement system delivers discrete  measurements. The measurements are considered to be affected by a zero-mean Gaussian white-noise process.

The discretized nonlinear dynamics of the system are
\begin{align}
x_{1(k)}&=x_{1(k-1)}+x_{2(k-1)}\Delta t \ ,\\
x_{2(k)}&=x_{2(k-1)}+(e^{\frac{-x_{1(k-1)}}{k_p}}
x_{2(k-1)}^2x_{3(k-1)}-g)\Delta t \ ,\\
x_{3(k)}&=x_{3(k-1)} \ ,
\end{align}
and the range measurement model is
\begin{equation}
y(x_1)=\sqrt{d^2+(x_1-h_0)^2)} + \vk \ ,
\end{equation}
where $k_p=\SI{6.1e3}{\meter}$ is a constant that relates the air density with the altitude, and $g = \SI[per-mode = symbol]{9.81}{\meter\per\second\squared}$ is the acceleration due to the gravity. 

In the measurement equation, $d=\SI{3e4}{\meter}$ is the horizontal distance from the measuring device to the vertical 
line traced by the falling body, and $h_0=\SI{3e4}{\meter}$  is the altitude of the measuring device from ground level. The initial uncertainties, while large, are based on \cite{Julier2000}, but rounded slightly to accommodate SI units. Their values are $\sigma_{x_1} = 300 $, $\sigma_{x_2} = 600 $, and $\sigma_{x_3} = 0.33 $ and the initial guesses for each state were set with a $\pm1\sigma$ error, while $\R=300$. All quantities with appropriate units.
Although this is not a true random draw it is sufficient to exercise the filter for an example.

\begin{figure}[h!]
	\includegraphics[width=1\textwidth]{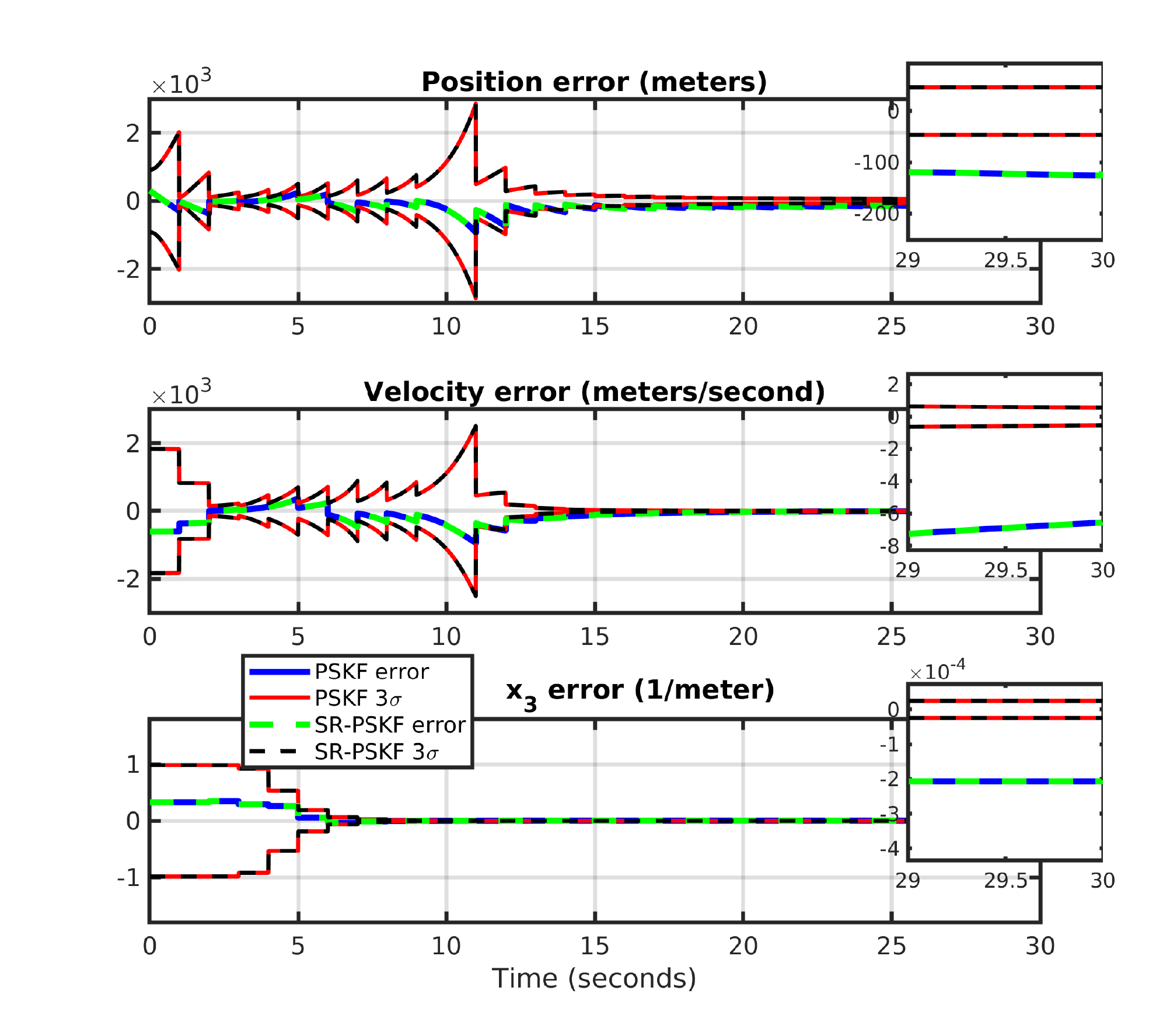}
	\caption{Standard EKF and square root partial-update EKF with full updates. The inset on the right shows a zoom-in for the last second of the simulation, displaying significant filter inconsistency with estimates well outside of $3\sigma$ bounds. Reprinted with permission from \cite{Ramo1907:Square}.}
	\label{fig:full_update}
\end{figure}

Figure \ref{fig:full_update} shows the results for the standard EKF, along with the square root partial-update filter. Both filters are using full measurement updates ($\boldsymbol{\beta} = \rowvec{1.0,1.0,1.0}$) and they begin with estimates within the 3$\sigma$ bounds. However, once the body is affected by increased drag forces the filter makes erroneous updates and the estimates start deviating outside the appropriate bounds. This inconsistency is the result of relatively large uncertainty on the initial guesses and the nonlinear relationship between position, velocity, and the ballistic parameter. Early updates produce slightly inaccurate results, specifically in the ballistic parameter and these errors compound during future propagation and update steps, leading to velocity and then position state and covariance inconsistencies. While this example is from a single run, the results are representative of a typical run for this scenario. It should also be noted that all three plot insets in Figure \ref{fig:full_update} show both the 3$\sigma$ and -3$\sigma$ bounds along with the state error for the last second of the simulation. This is apparent in the position plot, but difficult to see in the other two states. The figure clearly demonstrates the inconsistency between the errors present and associated covariance estimates.

\begin{figure}[h!]
	\includegraphics[width=1\textwidth]{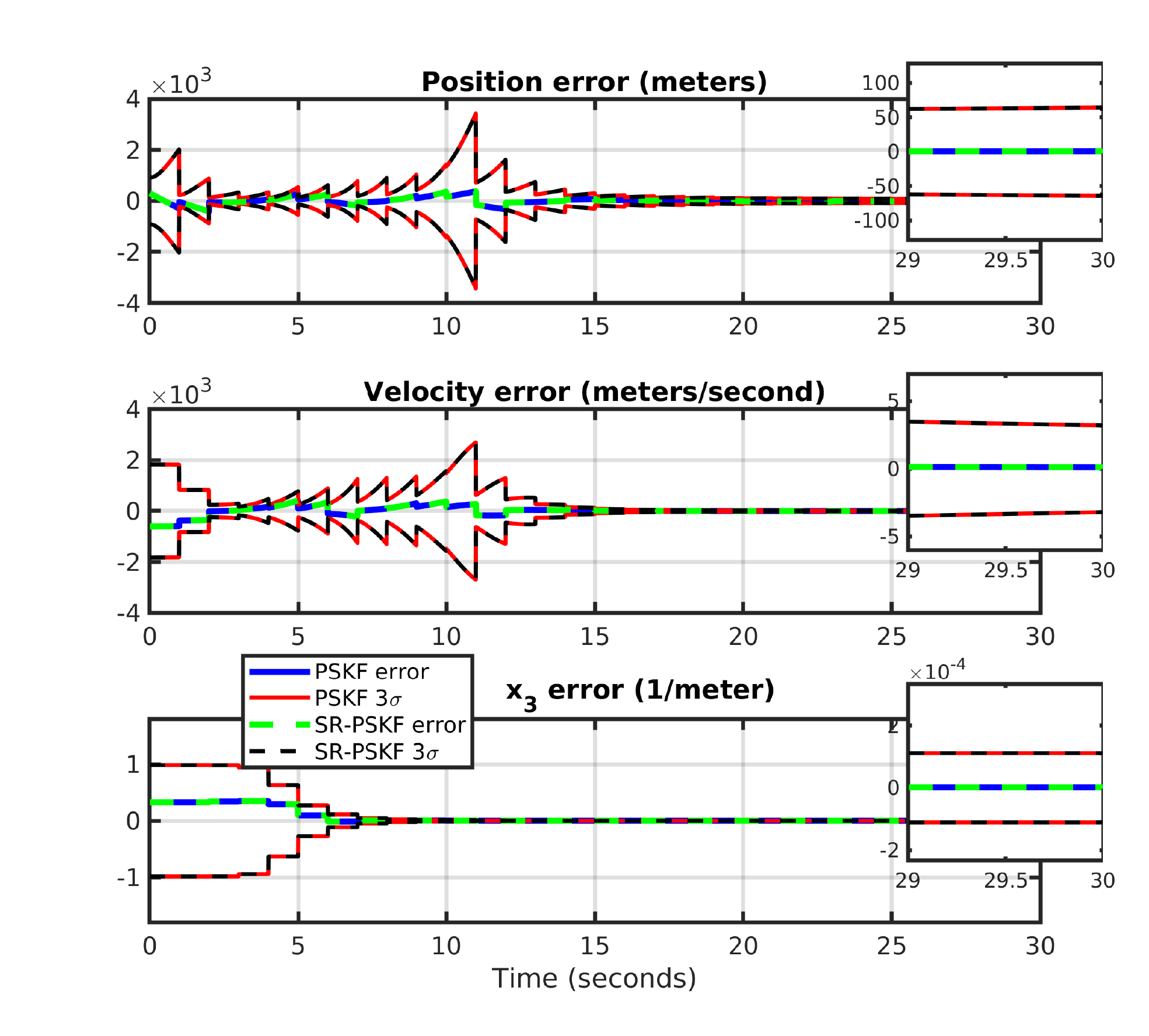}
	\caption{Partial-update EKF and square root Partial-Update EKF with $\boldsymbol{\beta}=\rowvec{0.9, 0.9, 0.75}$ resulting in 90\%, 90\%, and 75\% updates respectively to the position, velocity, and ballistic coefficient states. The inset on the right shows a zoom-in for the last second of the simulation, displaying filter results with estimates within the $3\sigma$ bounds. Reprinted with permission from \cite{Ramo1907:Square}.}
	\label{fig:partial_update}
\end{figure}

If the initial errors from this example were reduced sufficiently for the ballistic parameter, including tighter initial covariance values, the EKF does provide consistent estimates. This fact suggests that poor linearizations are to blame for the filter's poor performance seen in Figure \ref{fig:full_update}. It should also be recalled that the square root implementation provides certain numerical robustness, but not necessarily uncertainty robustness, however the benefit of the square root implementation will be addressed later in this section. 
When the same two filters are used, EKF and square root EKF, but the partial-update is applied, both filters now show consistent estimates for the same scenario. Results are displayed in Figure \ref{fig:partial_update}.

For this example the $\boldsymbol{\beta}$ vector associated with the partial-update was selected to be $\boldsymbol{\beta}\trans=\rowvec{0.9, 0.9, 0.75}\trans$ (or $\Dmat=\diag(0.1 , 0.1, 0.25))$, which means that the position and velocity estimates are updated using 90$\%$ of the original update, whereas the ballistic parameter is updated using only a 75$\%$. These update weight values were selected with the intention of limiting the updates mainly for the ballistic parameter, since it is commonly less observable than the other states.


Generally, the values for $\boldsymbol{\beta}$ are selected based on the idea that static or slowly varying states (with minimal process noise) can receive limited updates, whereas more observable or higher process noise states can receive larger update percentages. Additional ``tuning'' can be utilized to try to achieve the desired filter performance if the necessary data is available. This being a simulation, the data is clearly available for tuning, and the ballistic parameter updates are just sufficiently limited to prevent filter inconsistencies. As shown, the still substantial update of 75\% (along with minor limitations of the other two states) was sufficient to avoid the issues seen in the full update case from Figure \ref{fig:full_update}. One can also note the additional few updates steps taken before $x_3$ covariance values appear to collapse when comparing Figure \ref{fig:partial_update} and Figure \ref{fig:full_update}.  

\begin{figure}[h!]
	\includegraphics[width=1\textwidth]{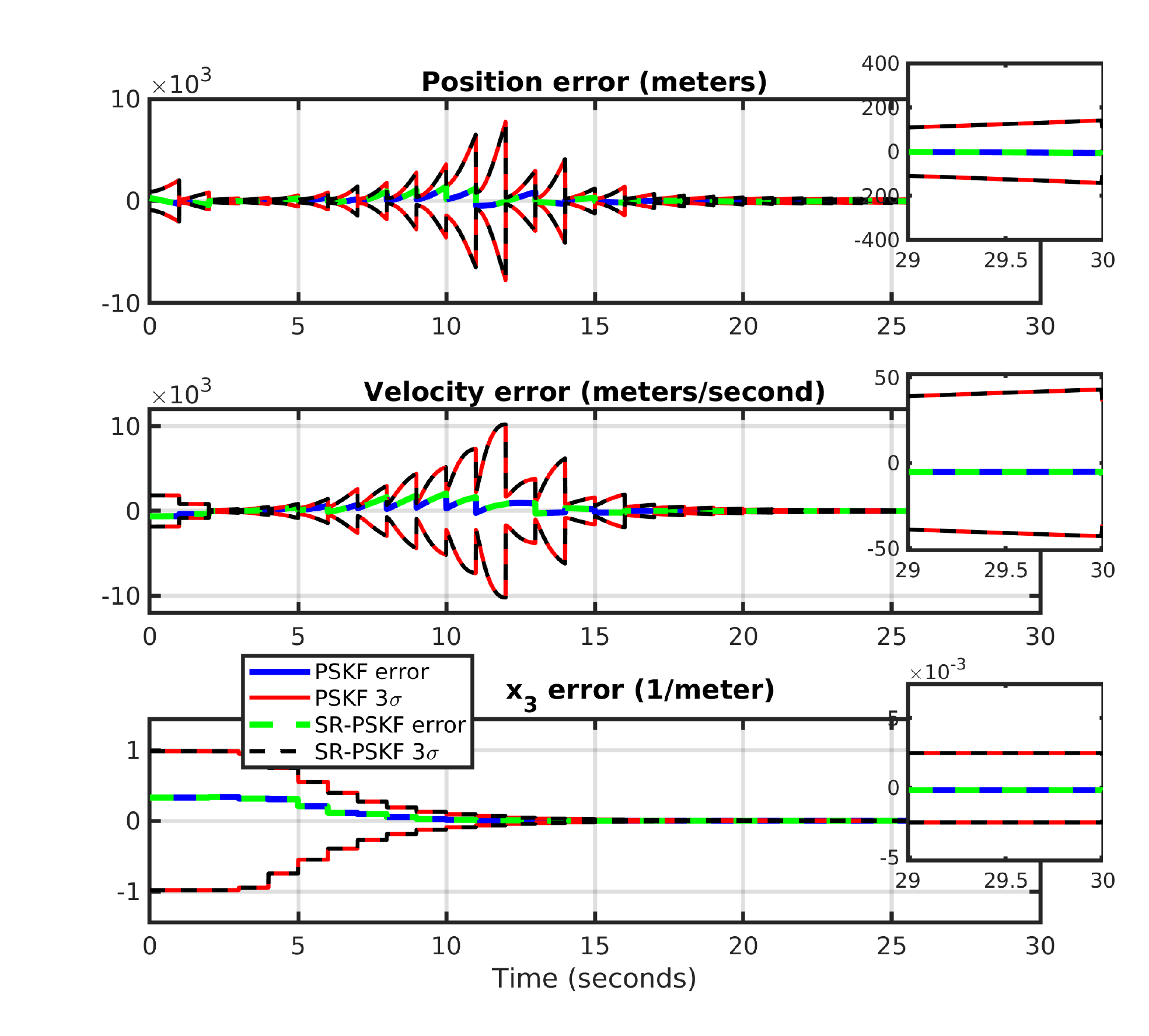}
	\caption{Partial-update EKF and square root Partial-Update EKF with
		$\boldsymbol{\beta}=\rowvec{1.0, 1.0, 0.1}$ resulting in 100\%, 100\%, and 10\% updates respectively to the position, velocity, and ballistic coefficient states. The inset on the right shows a zoom-in for the last second of the simulation, displaying filter results with estimates within the $3\sigma$ bounds. Reprinted with permission from \cite{Ramo1907:Square}.}
	\label{fig:partial_update_NEW}
\end{figure}

The explicit choice of $\boldsymbol{\beta}$ values is not particularly finicky, in fact, similar results are obtained with a weight of $\boldsymbol{\beta}=\rowvec{1.0,1.0,0.1}$ which are shown in Figure \ref{fig:partial_update_NEW}. In observing Figure \ref{fig:partial_update} and Figure \ref{fig:partial_update_NEW}, the reader may note that the estimated covariance of the position and velocity states are significantly larger through the middle of the run. This is due to the more conservative nature of the update weighting. By the end of the simulation, however, the filter is still able to converge even though the ballistic coefficient estimates only received 10\% updates. This example certainly takes more time to converge, yet, like the first partial update example and unlike the full update example, it maintains appropriate estimates and covariance values throughout. 

Also, it may be noted how the velocity uncertainty in both partial-update cases increase faster than those in Figure \ref{fig:full_update}, this is especially notable just after the update at time $t=11 s$. This is because the partial updates did not push the ballistic parameter covariance down as tight, thus allowing uncertainty in velocity and position to grow, permitting the ballistic parameter estimate to settle on the correct value over a few more updates (without over or undershooting due to the linearization errors). Effectively the partial-update prevented the filter from ``sabotaging" itself early on due to some relatively bad linearizations and avoided the associated repercussions. Finally, as before, the plots in Figure \ref{fig:partial_update} and Figure \ref{fig:partial_update_NEW} show the mathematical equivalence between square root and standard versions of the filter for this example.
\subsection{Camera to Inertial Measurement Unit (IMU) calibration}
This example is a multiplicative extended Kalman filter implementation used to calibrate a camera-Inertial Measurement Unit (IMU) system. This is, the filter is used to estimate the rigid body transformation between a camera optical frame and an IMU frame considering that both sensors are fixed to the same rigid platform. For the purposes of demonstrating the functionality of the square root filter, this section is limited to a minimum description of the multiplicative filter and it does not give further details on the process and measurement models. However, a closer examination of the system is presented in Chapter \ref{cha:hardware_applications}, Section \ref{sec:camera_imu_algorithm_description}. This specific problem is selected with the purpose of demonstrating the functionality of the square root partial-update filter when the state involves a larger state vector that includes more nuisance parameters ( in this case the calibration parameters), and vector measurements are to be assimilated.

The filter operation consist of propagating the process model when an IMU measurement is available, and updating when camera images provide landmark features positions of a pre-known map. Although presented later in this dissertation, the process and measurement models are included in this section to facilitate the present discussion. The continuous process model equations are comprised of rotational and translational kinematics as follows:

\begin{equation}\label{eq:SQqdottrue}
\qdeltadot(t)=
\frac{1}{2}
\begin{bmatrix}
-\angratetcross & \angratet \\
-\angratet\trans & 0
\end{bmatrix}
\qdelta(t) \ ,
\end{equation}
\begin{equation}
\pdot=\vtrue \ ,
\end{equation}
\begin{equation}\label{eq:SQvdottrue}
\vdot(t)=\atrue(t) =(\Cdelta \ \Cwizero )\trans \sf(t) + \grav \ ,
\end{equation}
\begin{equation}\label{eq:SQbgtrue}
\bgdott=\nwgt \ ,
\end{equation}
\begin{equation}\label{eq:SQbatrue}
\badott=\nwat \ ,
\end{equation}
\begin{equation}\label{eq:SQcalibration_true_lever}
^I\dot{\mathrmbf{p}}_C = 0 \ ,
\end{equation}
\begin{equation}\label{eq:SQcalibration_true_rotation}
^C_I\dot{\mathrmbf{q}} = 0 \ .
\end{equation}
Such that the state vector is formed as,
\begin{equation}\label{eq:SQstateimu}
\x=\rowvec{\qtrue\trans, \stdvecnb{\pchar}[W][I][\rmT], \stdvecnb{\vchar}[W][I][\rmT], \bg\trans, \ba\trans,  \stdvecnb{\pchar}[I][C][\rmT],\quat[I][C]^T}\trans \ .
\end{equation}
The notation for the filter state variables is as follows: $\qtrue \in \Real^4$, $\stdvecnb{\pchar}[W][I] \in \Real^3$ and $\stdvecnb{\vchar}[W][I] \in \Real^3$ are the attitude quaternion, position and velocity of the IMU with respect to the world frame,. The IMU gyroscope and accelerometer biases are denoted as $\bg \in \Real^3$ and $\ba \in \Real^3$, respectively, and the IMU-camera calibration is denoted as $ ^I\mathrmbf{p}_C$. The sought calibration is considered to be the position of the camera with respect to the IMU frame, $ ^I\mathrmbf{p}_C \in \Real^3$, and the attitude of the camera frame with respect to the IMU frame, $\quat[I][C]^T \in \Real^4$.
The measurement model for the IMU-camera system maps the position vectors of detected features into their corresponding pixels via a pinhole camera model. The position vector of the features is constructed according to
\begin{equation}\label{eq:SQhtrue}
\rowvec{h_x,h_y,h_z}\trans=\pfic=\Cic \big(\Cwi(\pftrue[i]-\ptrue)-\pci\big)\ ,
\end{equation}
where $h_x,h_y$ and $h_z$ represent the components of the $\pfic$ vector (the position vector of the $ i^{th} $ feature coordinatized in the camera frame), and $ \Cic $ is the passive rotation matrix from the IMU to the camera frame. The pinhole camera model is given as, 
\begin{equation}\label{eq:SQpinhole_model}
\yfim=\colvec{\um,\vm}=\colvec{f_x(h_x/h_z)+c_x, f_y(h_y/h_z)+c_y}_i+\vfi \ .
\end{equation}
The term $ \vfi $ in the pinhole camera model of Equation (\ref{eq:SQpinhole_model}), represents a white noise zero-mean Gaussian process with covariance matrix $ \stdvec{R}[][F_i] = E[\stdvec{vv^T}]$ that corrupts the pixel measurements. The simulations shown in this section used the same parameters as those used in Table \ref{tab:initial_conditions_single_run}. The same data is included in Table \ref{tab:SQinitial_conditions_single_run} here with the purpose of making this section more self-contained.

\begin{table}[h!]
	\begin{center}
		\caption{IMU-camera calibration parameters}
		\begin{tabular}{lcr}
			\label{tab:SQinitial_conditions_single_run}
			\textbf{State/parameter} &  Value \\
			\hline\hline
			IMU-camera attitude uncertainty  & 2 deg \\
			Lever arm uncertainty & 5 cm \\
			IMU attitude uncertainty  & 2 deg\\
			IMU position uncertainty & 5 cm\\
			Camera frame rate&20 Hz\\
			IMU rate & 100 Hz\\
			Camera pixel uncertainty & 2 px
		\end{tabular}
	\end{center}
\end{table}

\begin{figure}[h!]
	\centering
	\includegraphics[width=1\linewidth]{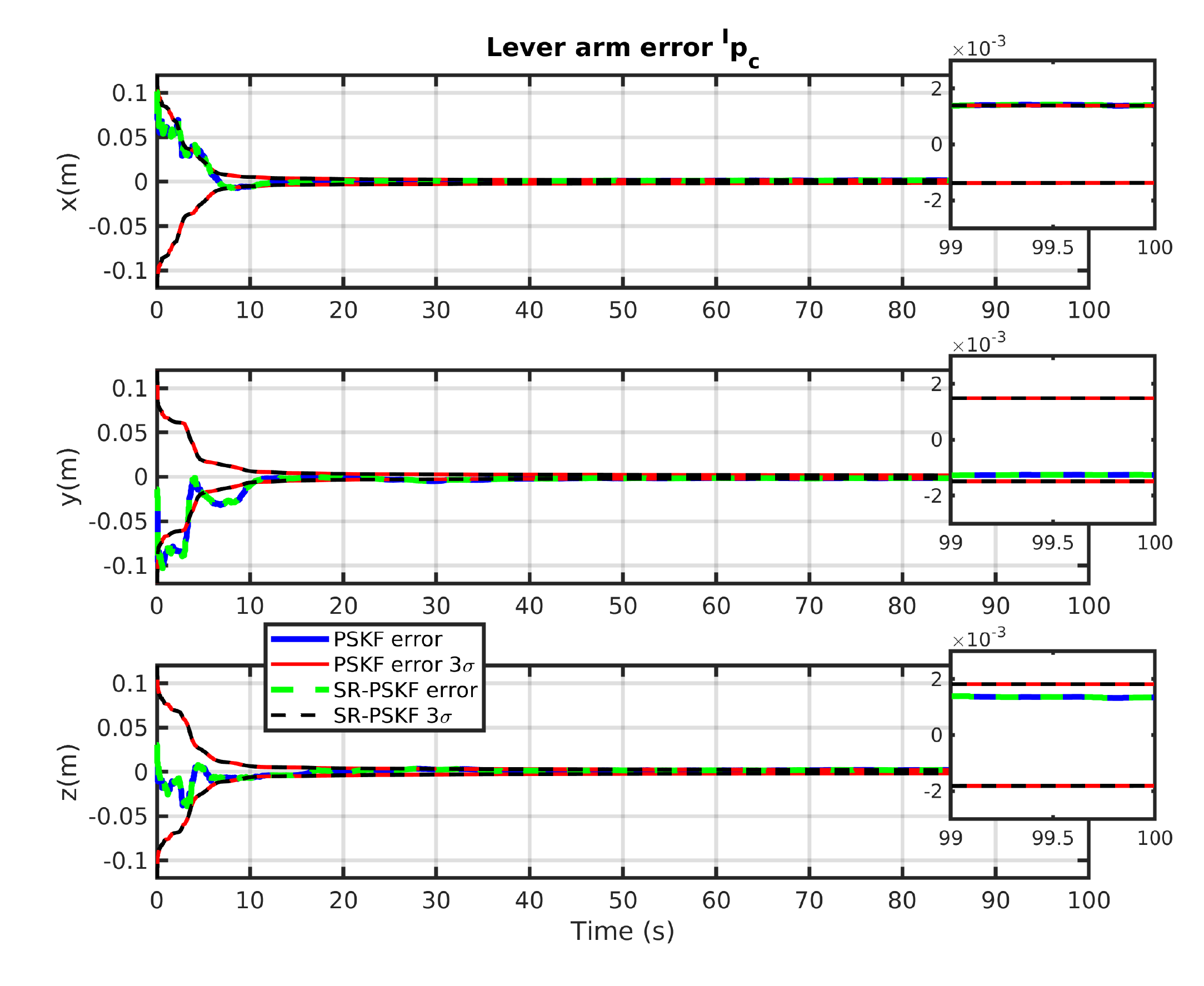}
	\caption{Camera-IMU calibration lever arm estimates when a full update is performed.}
	\label{fig:full_Lever_arm_1_runs}
\end{figure}

\begin{figure}[h!]
	\centering
	\includegraphics[width=1\linewidth]{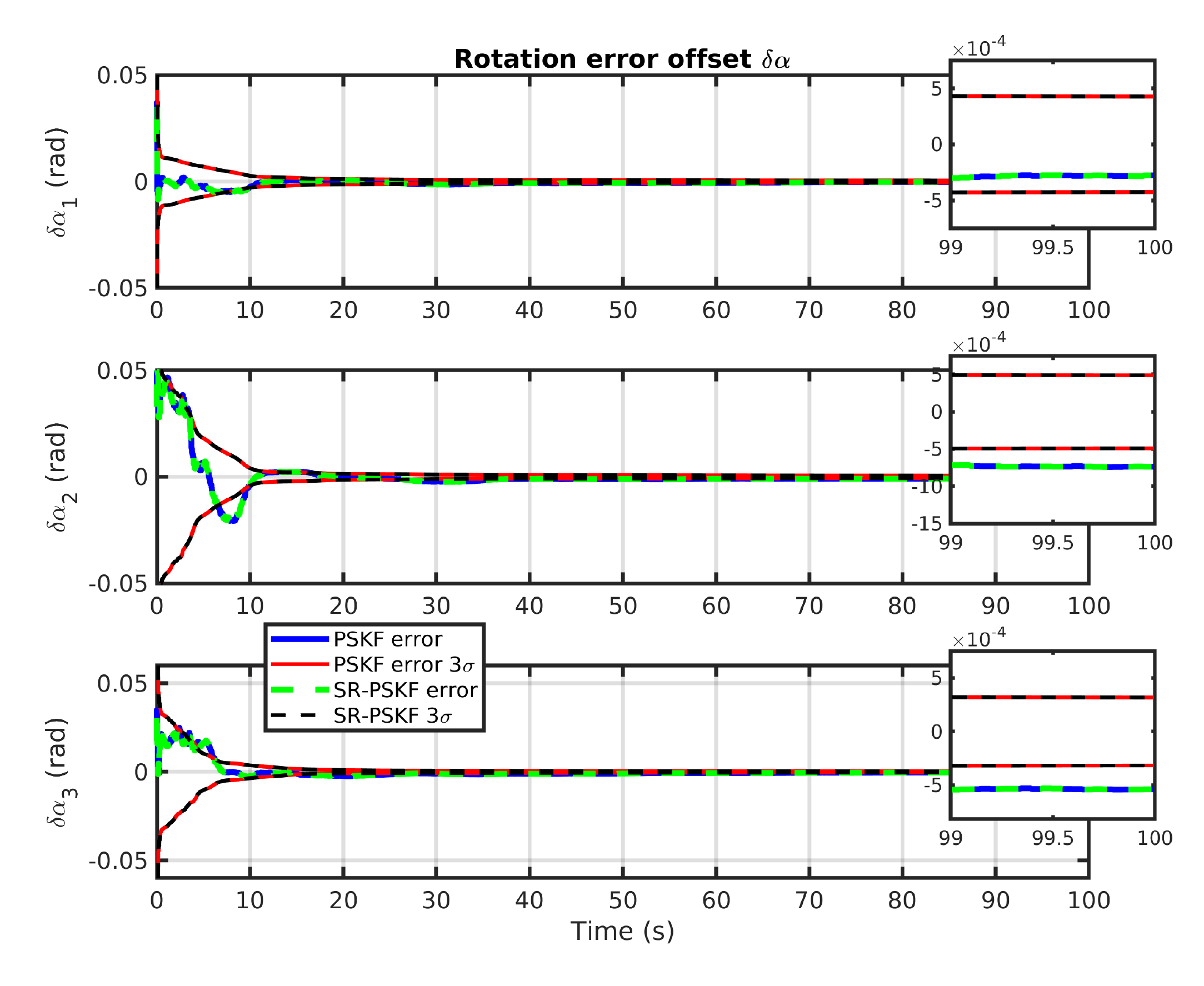}
	\caption{Camera-IMU calibration rotation error estimates when a full update is performed.}
	\label{fig:full_Attitude_offset_1_runs}
\end{figure}

\begin{figure}[h!]
	\centering
	\includegraphics[width=1\linewidth]{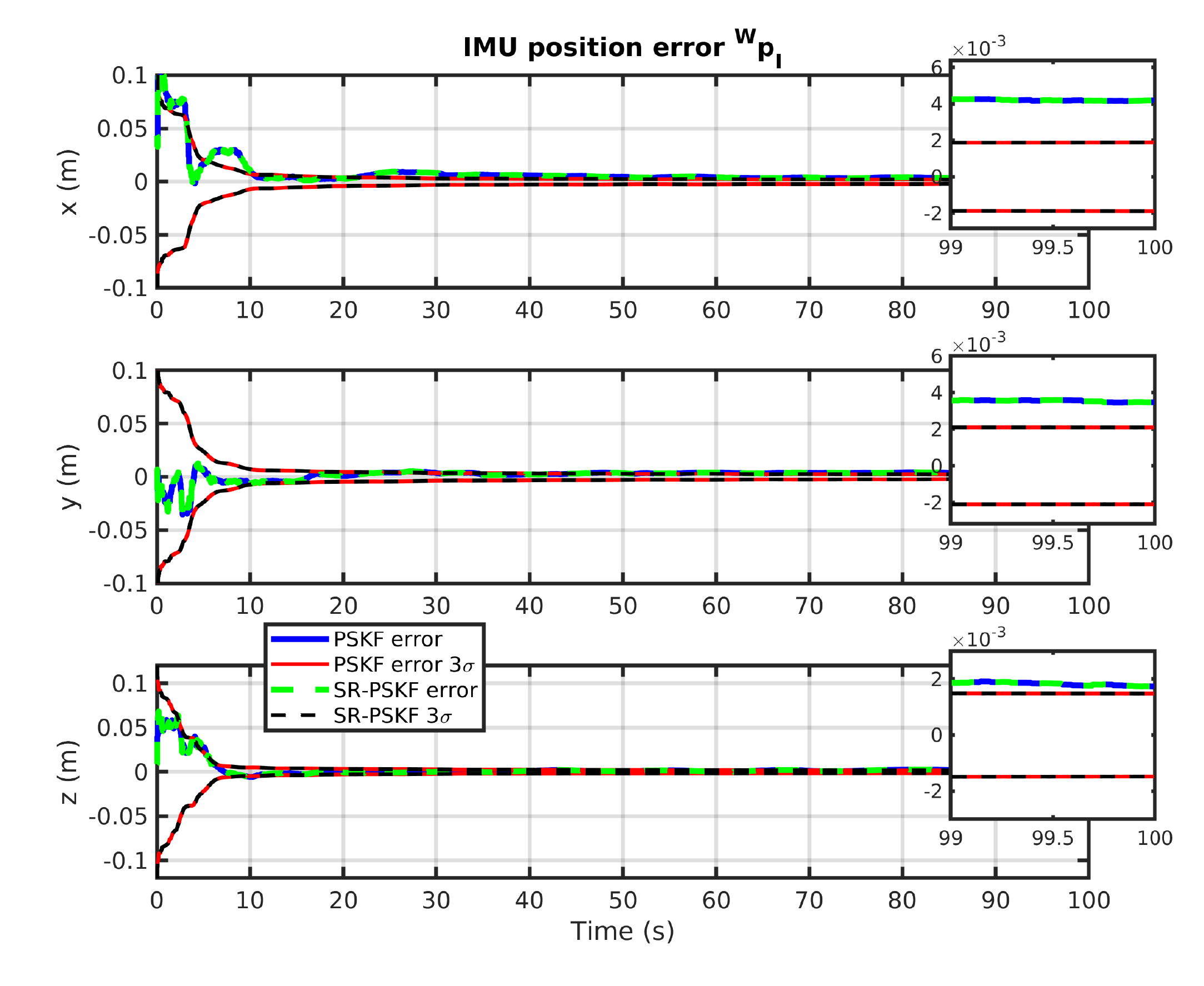}
	\caption{IMU global position when a full update is performed.}
	\label{fig:full_IMU_position_1_runs}
\end{figure}

Figure \ref{fig:full_Lever_arm_1_runs} shows the results for the standard EKF, along with the square root partial-update filter. The partial update filter uses the following update percentages or weights:
\begin{equation}\label{eq:sqrt_cam_imu_weights_theta}
\boldsymbol{\beta}_{^{I}_{W}\mathbf{q}}=\diag\rowvec{0.95, 0.95, 0.95} \ ,
\end{equation}

\begin{equation}
\boldsymbol{\beta}_{^{W}\mathbf{p}_{I}}=\diag\rowvec{0.95, 0.95, 0.95} \ ,
\end{equation}

\begin{equation}
\boldsymbol{\beta}_{^{W}\mathbf{v}_{I}}=\diag\rowvec{1,1,1} \ ,
\end{equation}

\begin{equation}
\boldsymbol{\beta}_{ \bg}=\diag\rowvec{1,1,1} \ ,
\end{equation}
\begin{equation}
\boldsymbol{\beta}_{\ba}=\diag\rowvec{1,1,1} \ ,
\end{equation}

\begin{equation}
\boldsymbol{\beta}_{^{I}\mathbf{p}_{C}}=\diag\rowvec{0.25,0.25,0.25} \ ,
\end{equation}

\begin{equation}
\boldsymbol{\beta}_{^{C}_{I}\mathbf{q}}=\diag\rowvec{0.25,0.25,0.25} \ .
\end{equation}
Such that,
\begin{equation}\label{eq:sqrt_cam_imu_weights_alpha}
\betamat = \diag(\boldsymbol{\beta}_{^{I}_{W}\mathbf{q}},\boldsymbol{\beta}_{^{W}\mathbf{p}_{I}},\boldsymbol{\beta}_{^{W}\mathbf{v}_{I}},\boldsymbol{\beta}_{\bg},\boldsymbol{\beta}_{ \ba},\boldsymbol{\beta}_{^{I}\mathbf{p}_{C}},\boldsymbol{\beta}_{^{C}_{I}\mathbf{q}}) \ . 
\end{equation}
Notice that the quaternion states have only three weights due to the indirect filter's nature of the multiplicative formulation. Due to the indirect form, the notation for the state elements should also change, but it is maintained to facilitate discussion; formally, it should refer to error variables as discussed in Chapter 2. The $ \beta $ weights sub-index indicate correspondence with each different state. The block-diagonal matrix $ \betamat $ contains all of the partial-update weights for the system. This is the matrix that will act on the Kalman gain to perform a partial-update as per Equation \ref{eq:ch2_partial_update_with_beta}.
For the first simulation, both filters (square root and conventional) are using full measurement updates, and they begin with estimates within the 3 $ \sigma $ bounds, but eventually uncertainties and nonlinear effects are not well handled by the filter producing inconsistent estimates. Figures \ref{fig:full_Attitude_offset_1_runs} and \ref{fig:full_IMU_position_1_runs} also show inconsistencies when full update is used.

\begin{figure}[h!]
	\centering
	\includegraphics[width=1\linewidth]{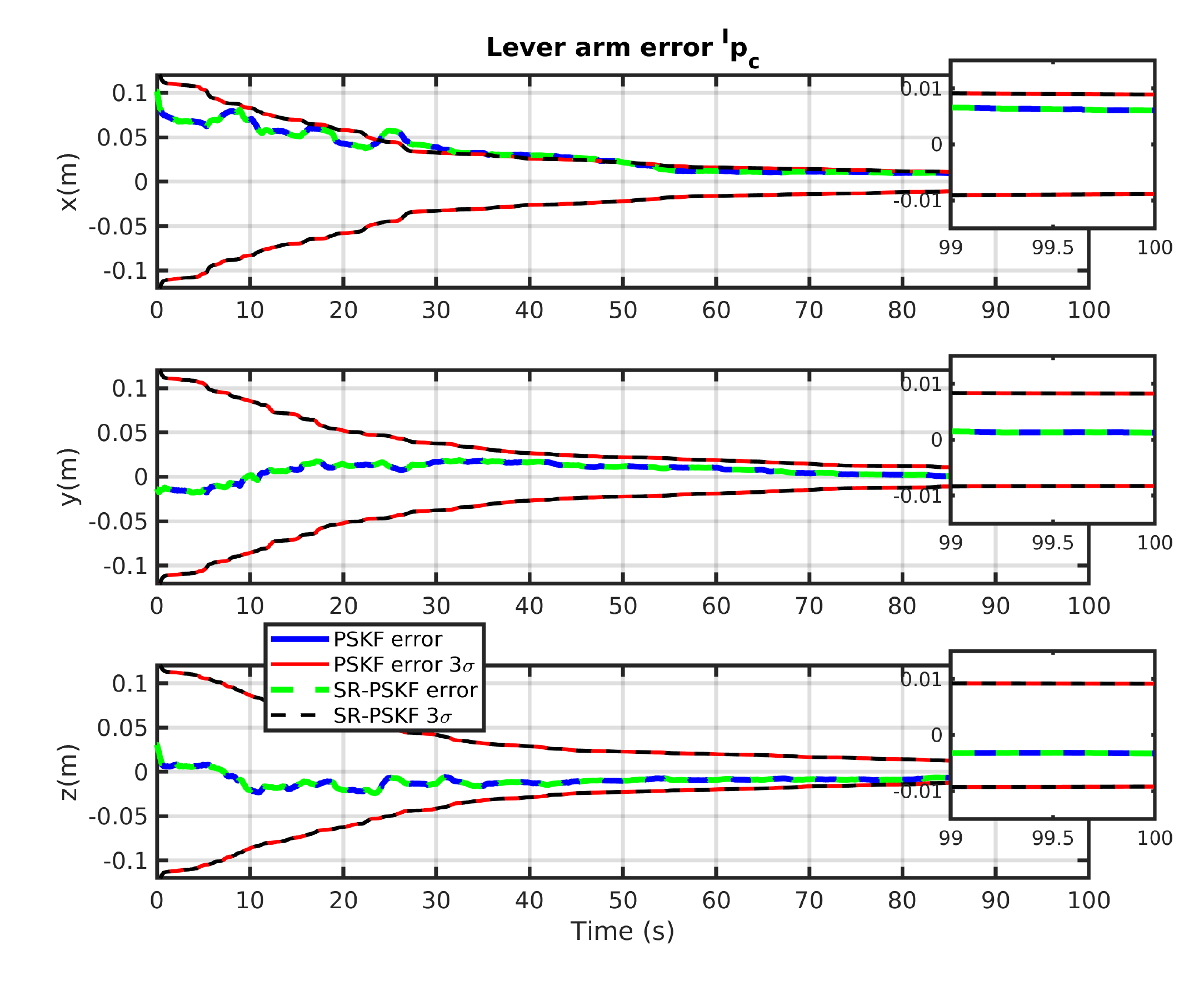}
	\caption{Camera-IMU calibration lever arm estimates when a partial-update is performed.}
	\label{fig:partial_Lever_arm_1_runs_beta95951001001001010}
\end{figure}

\begin{figure}[h!]
	\centering
	\includegraphics[width=1\linewidth]{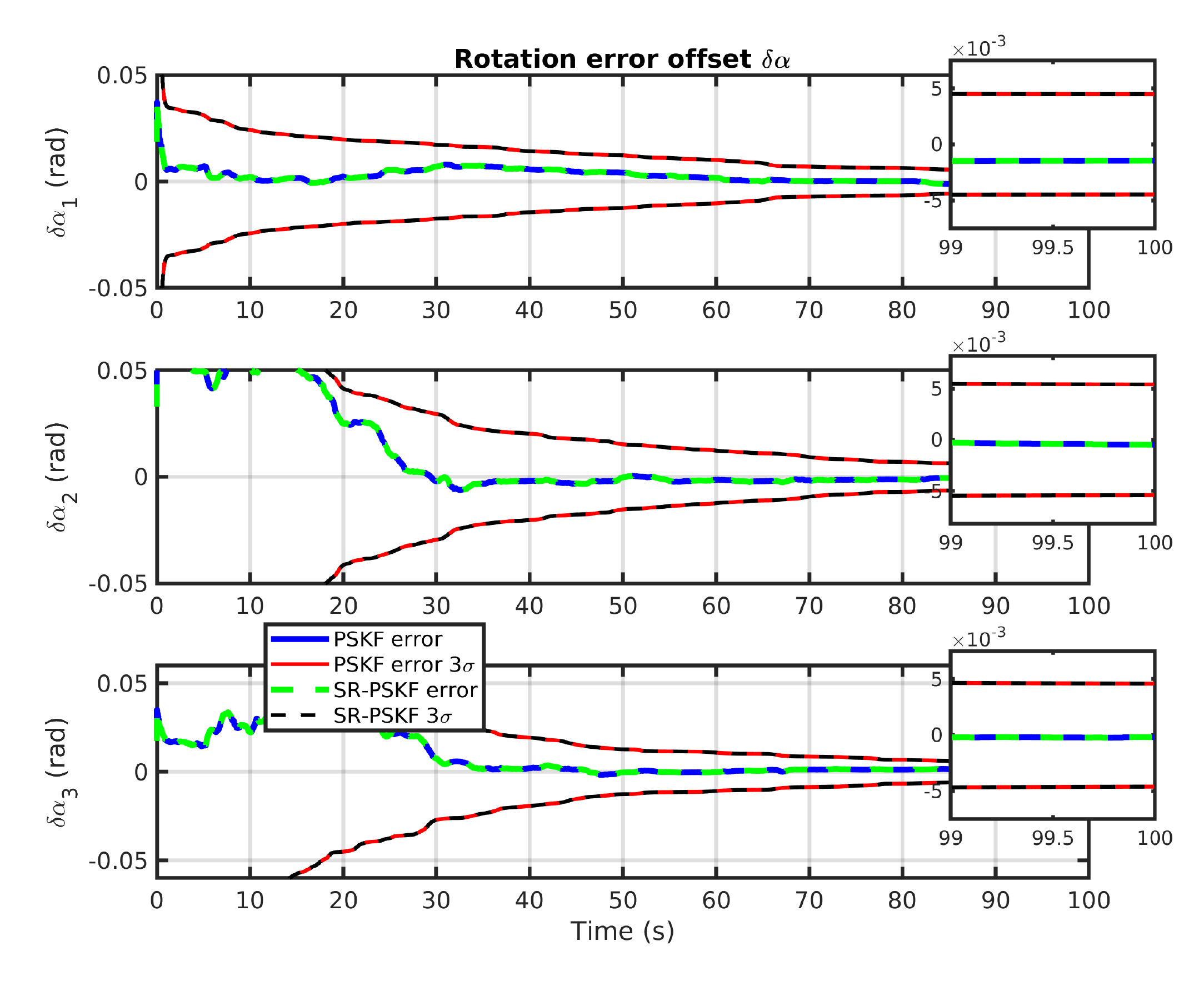}
	\caption{Camera-IMU calibration rotation estimates when a partial-update is performed.}
	\label{fig:partial_Attitude_offset_1_runs_beta95951001001001010}
\end{figure}

\begin{figure}[h!]
	\centering
	\includegraphics[width=1\linewidth]{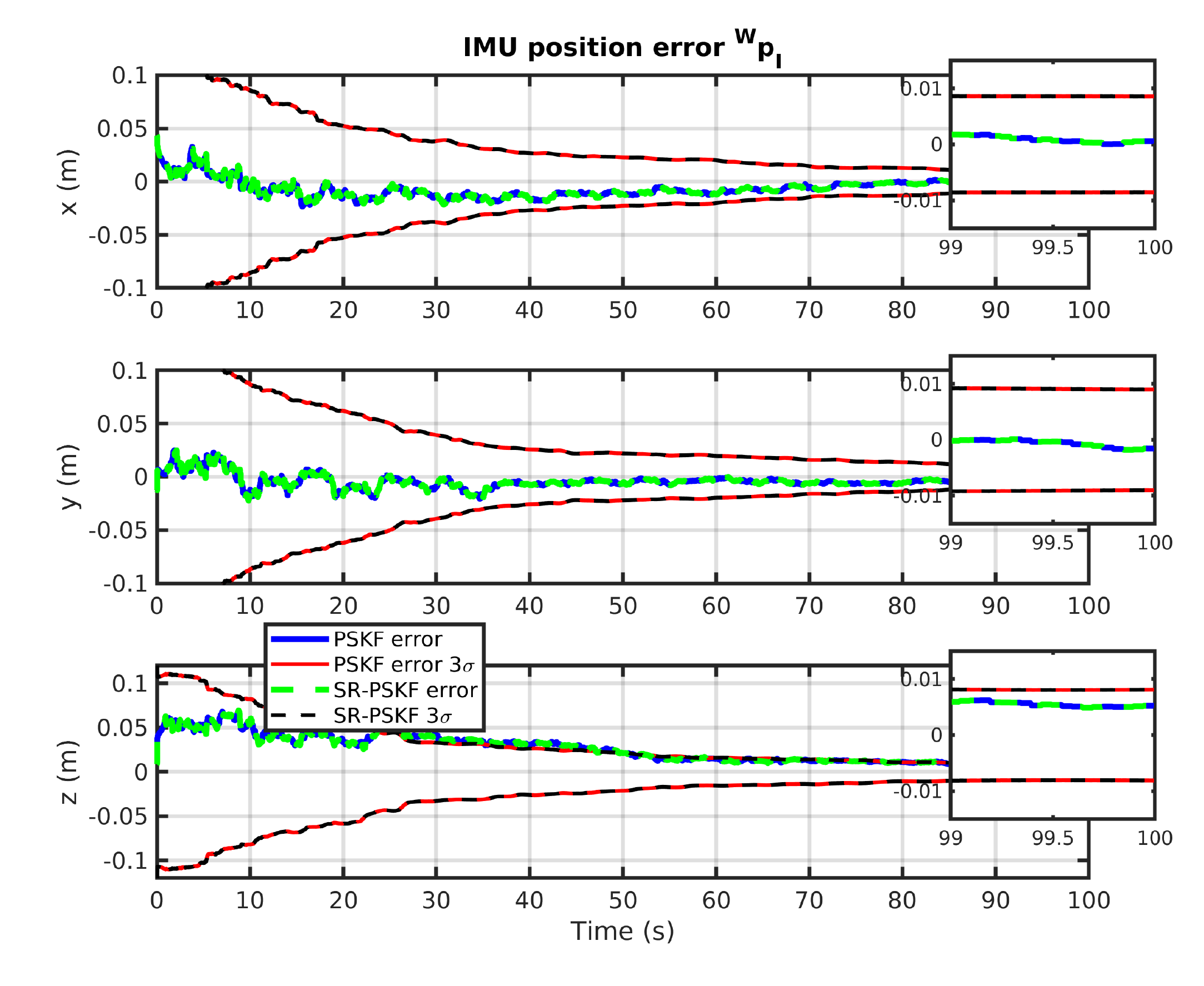}
	\caption{IMU global position when a full update is performed.}
	\label{fig:partial_IMU_position_1_runs_beta95951001001001010}
\end{figure}

On the other hand, as seen in Figures \ref{fig:partial_Lever_arm_1_runs_beta95951001001001010}, \ref{fig:partial_Attitude_offset_1_runs_beta95951001001001010} and  \ref{fig:partial_IMU_position_1_runs_beta95951001001001010} the partial-update filter performs better for both factorized and conventional formulation. Although the filter takes more time to converge, the produced estimates are definitely more consistent than in the full update case. 

\begin{figure}[h!]
	\centering
	\includegraphics[width=1\linewidth]{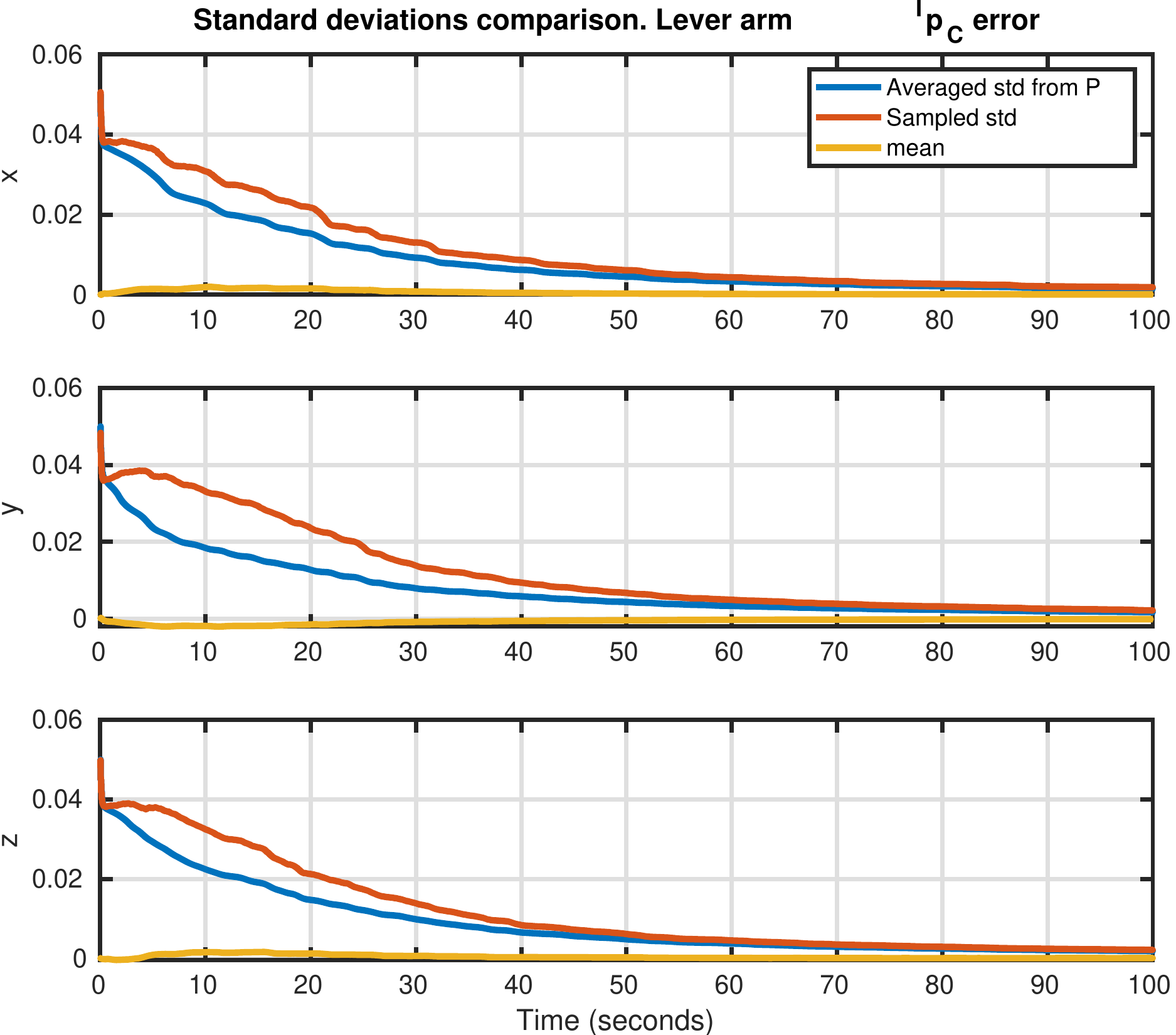}
	\caption{Averaged and sampled standard deviation from 500 Monte Carlo runs for the lever arm components. The mean error across the runs is also plotted. Units are in meters.}
	\label{fig:partial_lever_arm_std_500_runs_beta95951001001002525}
\end{figure}

\begin{figure}[h!]
	\centering
	\includegraphics[width=1\linewidth]{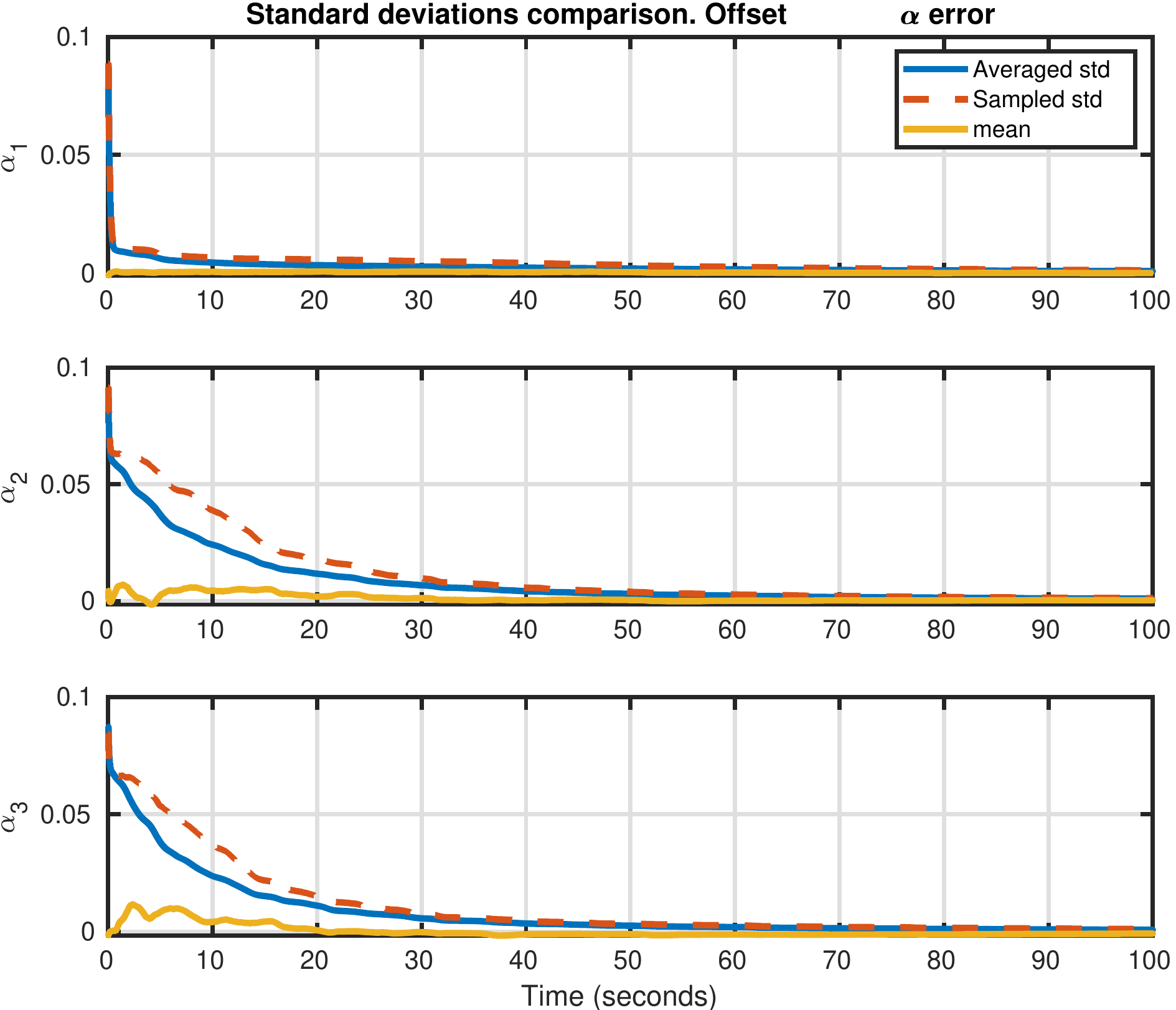}
	\caption{Averaged and sampled standard deviation from 500 Monte Carlo runs for rotation error (IMU to camera rotation). The mean error across the runs is also plotted. Units are in meters.}
	\label{fig:partial_attitude_offset_std_500_runs_beta95951001001002525}
\end{figure}

Monte Carlo simulations were also run for the partial-update filter to show the consistent behavior given this scenario, and are shown in Figure \ref{fig:partial_lever_arm_std_500_runs_beta95951001001002525} and Figure \ref{fig:partial_attitude_offset_std_500_runs_beta95951001001002525}. When comparing the sampled standard deviations against those produced by the partial-update filter for the lever arm in Figure \ref{fig:partial_lever_arm_std_500_runs_beta95951001001002525}, it can be seen that the partial-update filter is slightly overconfident. However, as the uncertainties and nonlinearities impact decays, the partial-update filter is able to improve its consistency. Similarly, for the rotation calibration error, the filter starts slightly overconfident but it gains consistency quickly. Overall, the Monte Carlo runs show consistency improvement over the full update filter for all of the states. Conversely, a full update filter was seen inconsistent, (Monte Carlo runs are not included) always resembling that behavior observed from a single run experiment shown before. 

In general, the partial-update filter was able to improve the behavior of the underlying EKF (MEKF to be more specific) via a well-selected percentage update. Similarly to the previous example, the selection of the weights was not finicky for this problem, and a variety of well behaved filter were also obtained with different $ \beta  $ values. However, the selected weight values were shown to offer better convergence rate and consistency over other weights. Again, the square root formulation certainly provides extra numerical robustness to the filter, not uncertainty or errors robustness.


\section{Monte Carlo runs}
Monte Carlo simulations were ran for both, square root EKF and square root partial-update EKF. A total of 100 runs were executed and the histories of all of the states were recorded. Moreover, the sampled standard deviation and the standard deviation as computed by the filter, were also calculated  and are used to check for filter consistency. For both filters the initial conditions the same as for the single run scenario. For convenience, the parameter values used for the simulations are condensed in Table \ref{table:SQinitial_conditions_SQPU}.

\begin{figure}[h!]
	\includegraphics[width=1\textwidth]{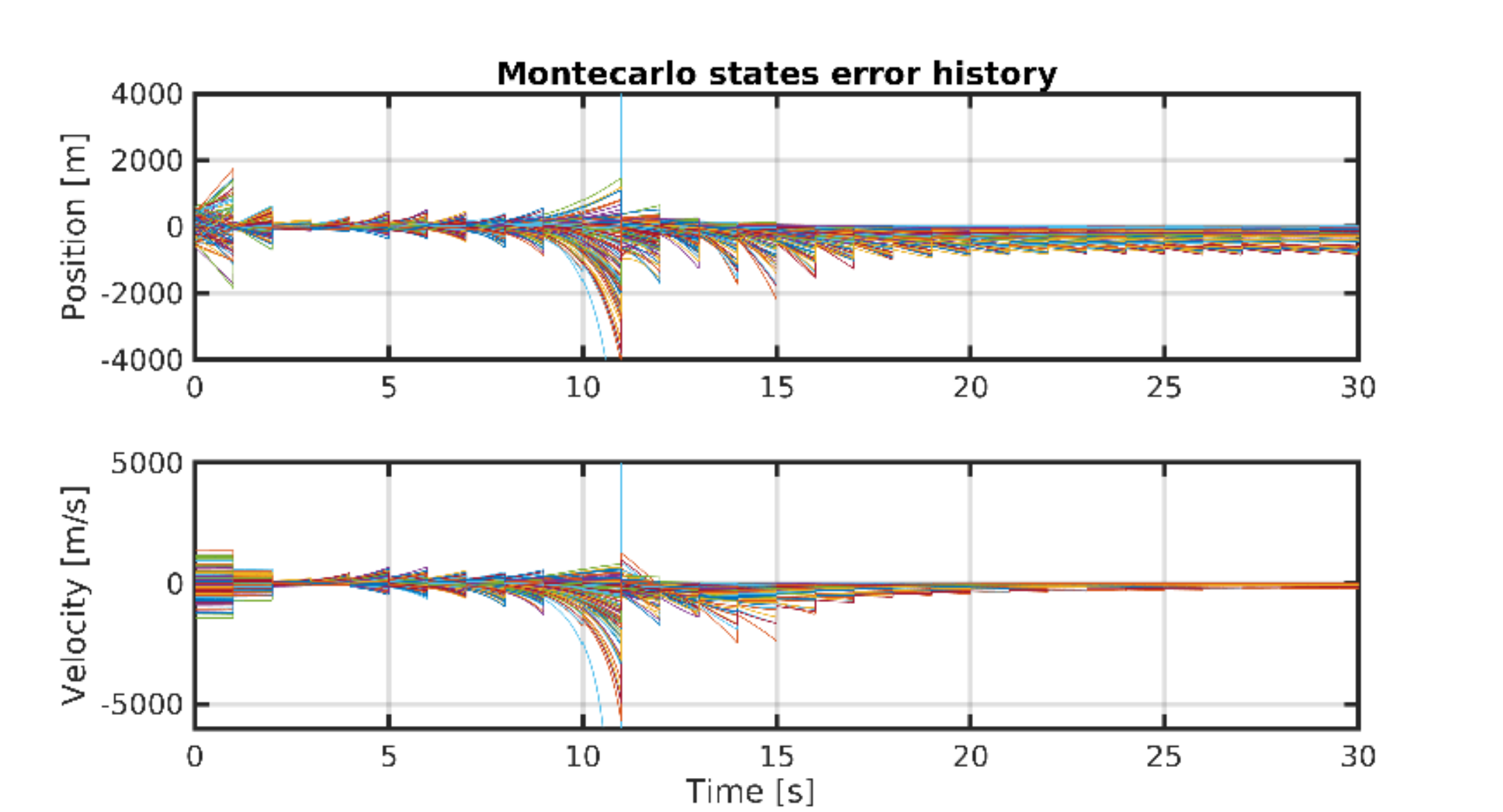}
	\caption{Monte Carlo standard EKF and square root partial-update EKF with full updates.}
	\label{fig:monte_state_ud_full_update}
\end{figure}

\begin{figure}[h!]
	\includegraphics[width=1\textwidth]{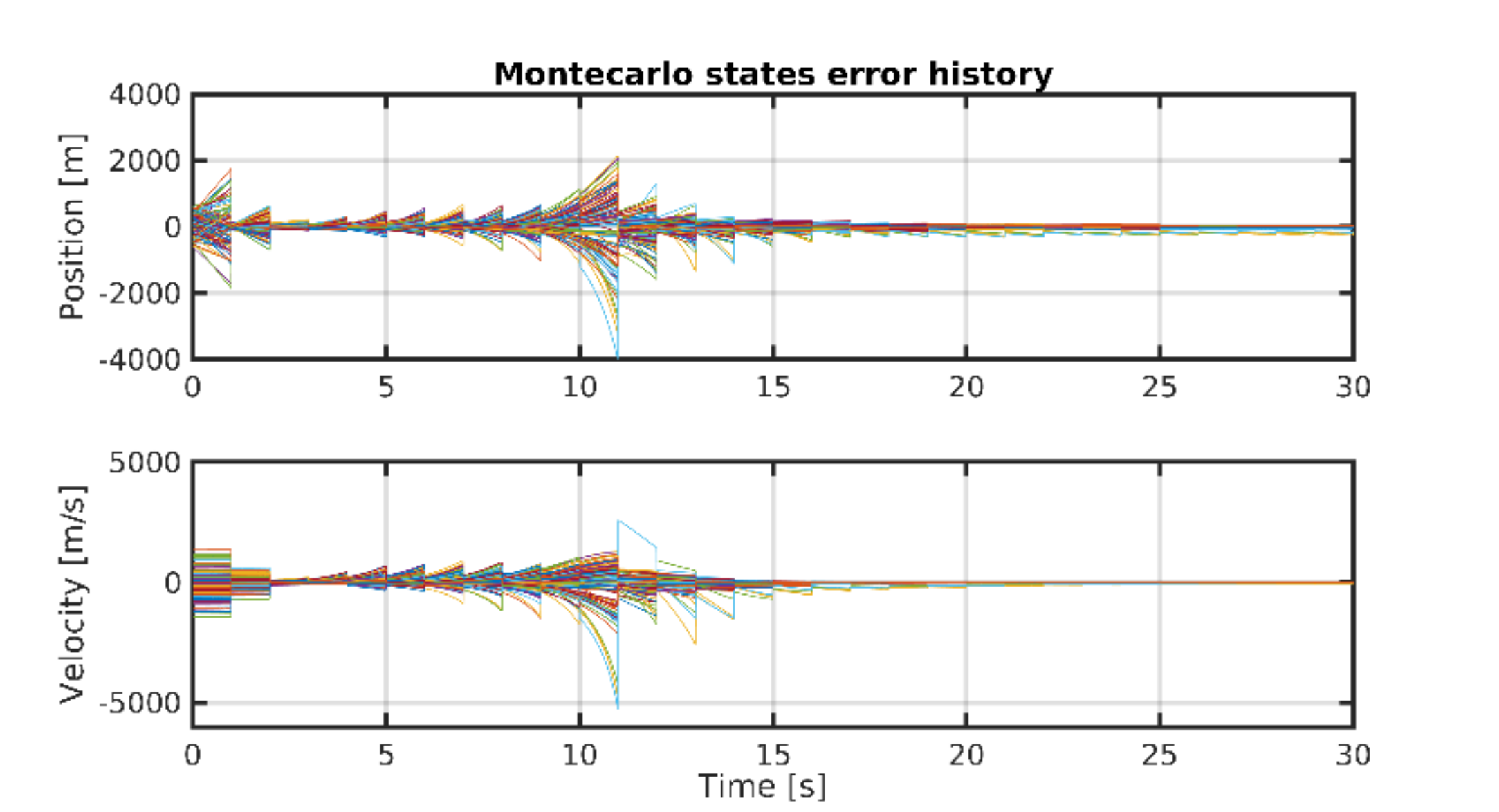}
	\caption{Monte Carlo runs for the partial-Update EKF and square root partial-update EKF with $\boldsymbol{\beta}=\rowvec{0.9, 0.9, 0.75}$ resulting in 90\%, 90\%, and 75\% updates respectively to the position, velocity, and ballistic coefficient states.}
	\label{fig:monte_states_ud_partial_update}
\end{figure}

Figure \ref{fig:monte_state_ud_full_update} shows the 100 EKF runs histories. The EKF shows that in its majority is able to prevent total failure of the filter, except for two cases (blue curves) that are divergent. However, most of the runs show similar behavior as the one showed for the single run: estimates are not within the proper sigma bounds after around $ t = 15 $ seconds and the errors do not converge to zero. The square root partial-update technique on the other hand (as depicted in Figure \ref{fig:monte_states_ud_partial_update}), shows a dramatic improvement over the EKF. First, the filter presents no runs with divergence. Second, the error histories show a significant magnitude reduction, and third, a superior capacity to handle the initial uncertainties by avoiding the overreaction in the update is achieved, showing that the behavior seen in the single run is in fact, the overall filter behavior. 

\begin{figure}[h!]
	\includegraphics[width=1\textwidth]{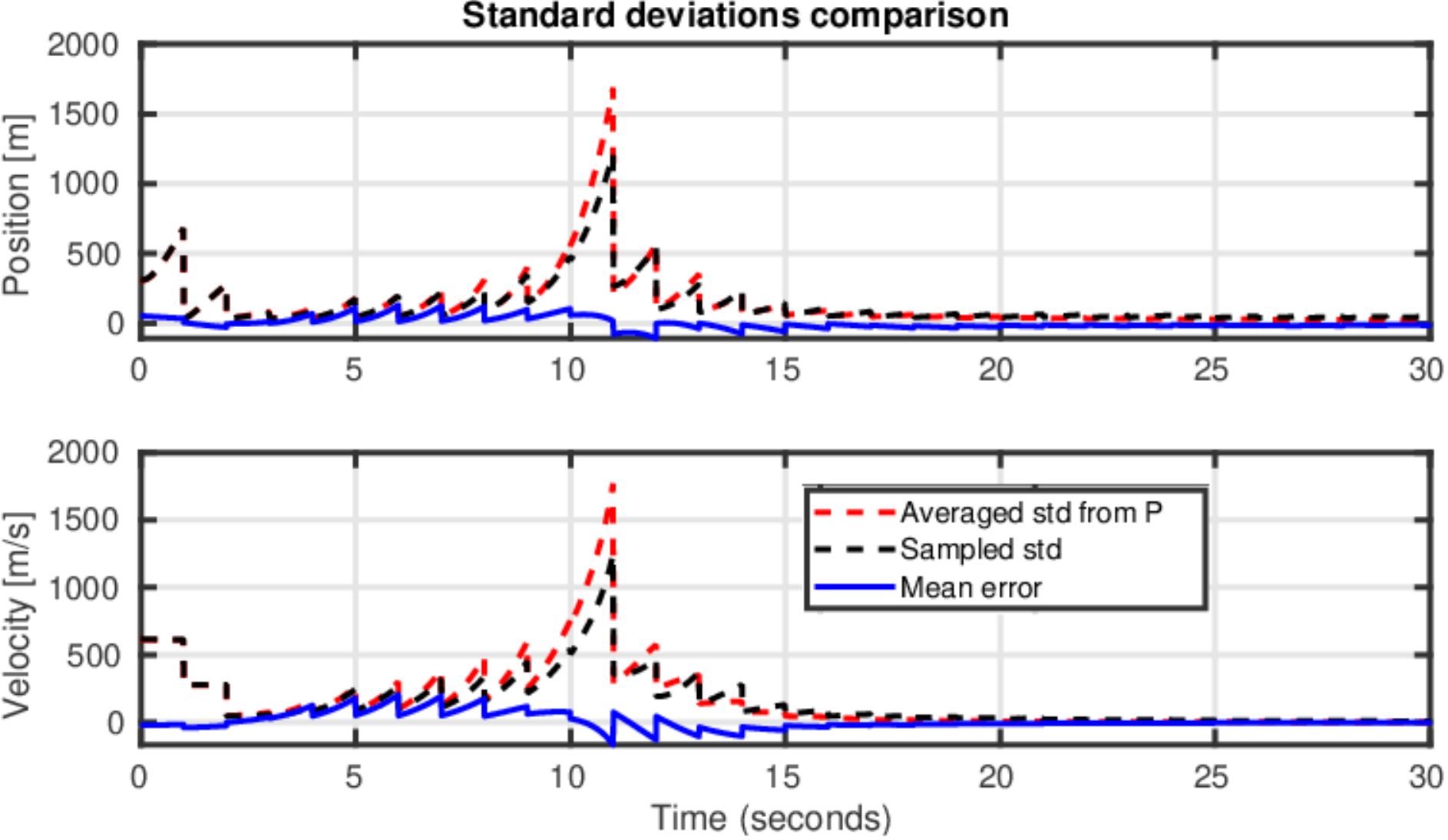}
	\caption{Monte Carlo runs for the partial-Update EKF and square root partial-update EKF with $\boldsymbol{\beta}=\rowvec{0.9, 0.9, 0.75}$ resulting in 90\%, 90\%, and 75\% updates respectively to the position, velocity, and ballistic coefficient states.}
	\label{fig:monte_ud_partial_update}
\end{figure}

The averaged standard deviation from the Monte Carlo runs and the standard deviation estimated by the filter are depicted in Figure \ref{fig:monte_ud_partial_update} and are shown to practically coincide. Since the mean estimation error is around zero, altogether this indicates that the filter is consistent. Since the EKF filter presented divergent cases, only error histories are graphed for it.

\begin{table}[h!]
	\begin{center}
		\caption{Re-entering body parameters}
		\begin{tabular}{lcr}
			\label{table:SQinitial_conditions_SQPU}
			\textbf{State/parameter uncertainty} &  Uncertainty $ 1\sigma $ value \\
			\hline\hline
			Position   & \SI{300}{\meter} \\
			Velocity  & \SI[per-mode = symbol]{600}{\meter\per\second}\\
			Ballistic parameter & \SI{0.33}{\per\meter}  \\
			Measurement & \SI{300}{\meter}\\
		\end{tabular}
	\end{center}
\end{table}

Overall, it was observed that if low initial errors were ensured, the EKF can be functional and quickly converge, but is not robust enough to handle errors at the level of the square root partial-update filter. Conversely, the square root partial-update filter was observed to be consistent, and able to handle higher nonlinearities and uncertainties better than the conventional square root EKF or EKF.

\subsection{Condition number}
Lastly, the numerical stability afforded by the square root implementation \cite{maybeck1982stochastic} is briefly discussed. The chosen example never threatened the numerical integrity of the filter as it was primarily intended to show the benefit of the partial update, and the benefits of the square root form are well known. Nevertheless, a brief analysis of the condition number is shown here simply to assure the reader that the numerical improvements afforded by the square root form were maintained in this development. 

One way to interpret the condition number of a matrix, is to consider it as a measure of how close to be singular the matrix is, with the convention that the matrix is singular when the condition number is infinite. One way to compute the condition number $\kappa$  of a matrix $\Cov$ is through the matrix singular values as:
\begin{equation}
\kappa(\Cov) = {\sigma_{max}(\Cov)}/{\sigma_{min}(\Cov)}
\end{equation}
While the assessment of the condition number is based on heuristics, a condition number is regarded to be a very large (and generally bad) condition number when $\log_{10}(\kappa)>$ (available machine precision) \cite{rao2011fast} and in general it is desirable to keep $\Cov$ condition number low.

\begin{figure}[h!]
	\centering
	\includegraphics[width=1\textwidth]{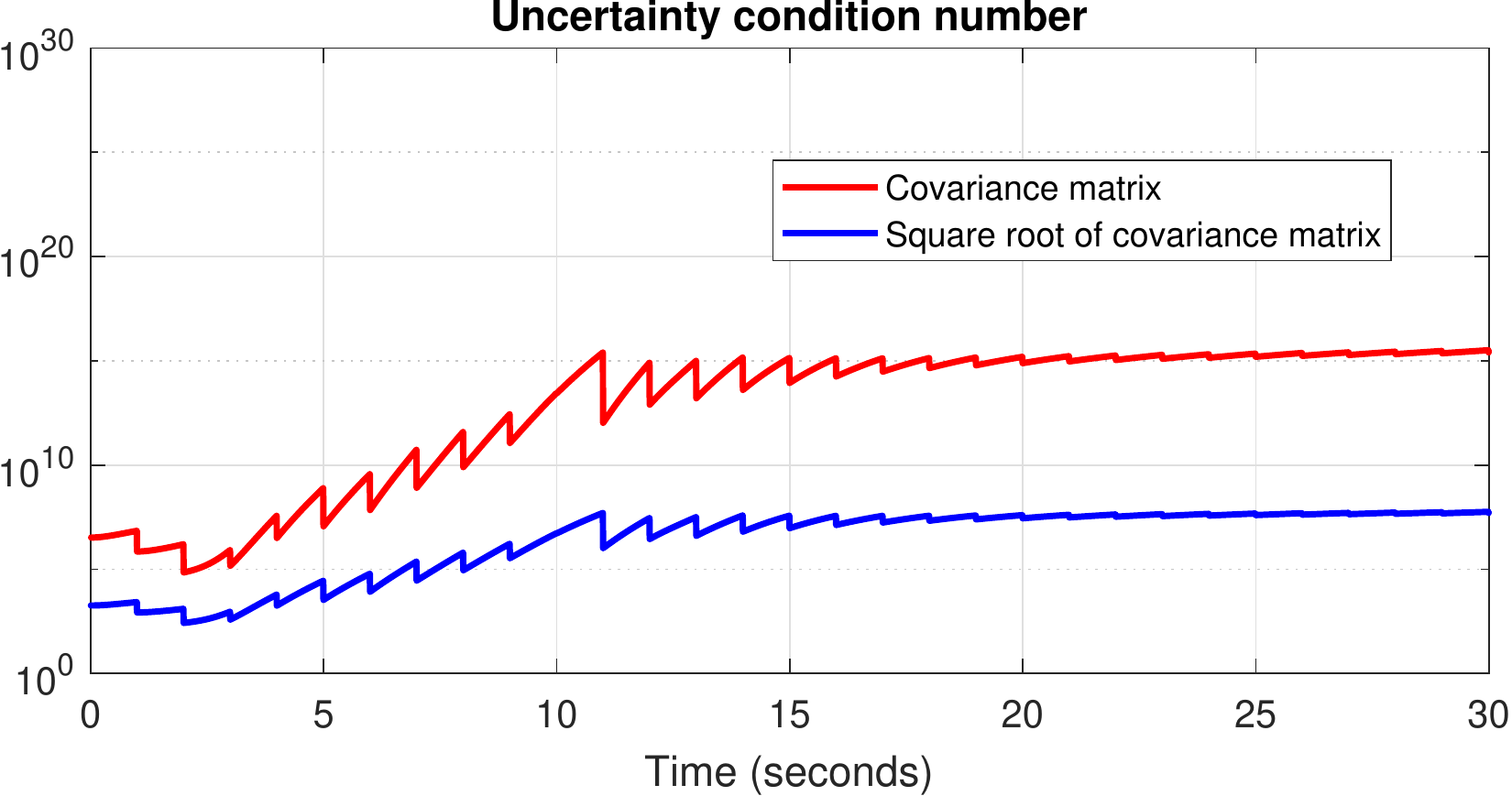}
	\caption{ Uncertainty condition number for the partial-update EKF and square root partial-update EKF filter example for $\boldsymbol{\beta}=\rowvec{0.9,0.9,0.75}$. Reprinted with permission from \cite{Ramo1907:Square}. }
	\label{fig:condition_number}
\end{figure}

The condition numbers for both partial-update examples (standard and square root) are shown in Figure \ref{fig:condition_number}. It is clear that even though both partial-update filters (standard and square root version) produce accurate estimators, the square root partial-update does provide a significantly improved (lower) condition number for the uncertainty matrix, as expected. This indicates that the square root partial-update filter developed in this chapter, maintains the square root filter's numerical precision advantage over the traditional filter approach as desired. 

\section{Processing a vector-valued measurement}
The square root filter can also process vector measurements. Formulations such as those presented in \cite{simon2006optimal} and \cite{bierman2006factorization} are some alternatives. Similarly, the square-root partial-update can process vector measurements, and its derivation is presented in this section. This formulation of the partial-update filter may not be adequate for embedded system implementation due to its high computational complexity, but it can be a useful filter debugging or validation tool. In any case, it is the user's decision to select the filter form based on the design requirements, and if a sequential square root filter is to be used, the engineer also needs to consider the possible extra cost of diagonalizing the measurement noise covariance matrix.

\subsection{Square root partial-update for vector-valued measurements}
The filter version that can process a vector measurement is a slight modification of the sequential version presented in this chapter. However, the complexity of the algorithm increases dramatically with respect to scalar-valued measurement processing. The extension of the square root partial-update filter presented here is mostly based on the factorization presented in \cite{simon2006optimal}. This factorization allows one to perform the conventional update via Gram-modified-Schmidt orthogonalization by constructing an augmented matrix. Regarding the partial-update operation, it is still possible if an extra Cholesky decomposition is performed.

Paralleling the development presented in this chapter for the sequential filter, recall
 
\begin{equation}\label{eq:vec_partial_update_cov_vec}
\stdvec{P}[][][++] = \Dmat(\stdvec{P}[][][-]-\stdvec{P}[][][+])\Dmat + \stdvec{P}[][][+] \ .
\end{equation}

Now, from the standard EKF equations (ignoring the time indices for ease of notation),  $\stdvec{P}[][][+]=(\stdvec{I-\stdvec{K}[][]}\Hmat)\stdvec{P}[][][-]$ is used into Equation (\ref{eq:vec_partial_update_cov_vec}).
\begin{equation}
\stdvec{P}[][][++] = \stdvec{P}[][][+] + \Dmat(\Kmat\Hmat\stdvec{P}[][][-])\Dmat \ ,
\end{equation}
then replacing $\stdvec{K}[][]=\stdvec{P}[][][-]\Hmat\trans(\Hmat\stdvec{P}[][][-]\Hmat\trans+\R)^{-1}$ gives
\begin{equation}
\stdvec{P}[][][++] = \stdvec{P}[][][+] + \Dmat(\stdvec{P}[][][-]\Hmat\trans(\Hmat\stdvec{P}[][][-]\Hmat\trans+\R)^{-1}\Hmat\stdvec{P}[][][-])\Dmat \ .
\end{equation}
Then, requiring that $\stdvec{P}[][][-]= (\stdvec{S}[][][-])(\stdvec{S}[][][-])\trans$ and $\stdvec{P}[][][+]= (\stdvec{S}[][][+])(\stdvec{S}[][][+])\trans$. Additionally it is establish that $\tilde{\R} =(\Hmat\stdvec{P}[][][-]\Hmat\trans+\R)^{-1}$. These actions lead to 
\begin{equation}\label{eq:vec_cov_full_form}
\stdvec{P}[][][++] = (\stdvec{S}[][][+])(\stdvec{S}[][][+])\trans + \Dmat(\stdvec{S}[][][-])(\stdvec{S}[][][-])\trans\Hmat_i\trans \tilde{\R}\Hmat_i(\stdvec{S}[][][-])(\stdvec{S}[][][-])\Dmat \ .
\end{equation}
To achieve the required factorization to partial-update, Cholesky decomposition is used on $\tilde{\R}$ as to obtain,
\begin{equation}\label{eq:chol_R_vector_processing}
		\tilde{\R} = \tilde{\R}^{1/2}\tilde{\R}^{T/2} \ .
\end{equation}
Using Equation (\ref{eq:chol_R_vector_processing}), the expression covariance becomes
\begin{equation}
\stdvec{P}[][][++] = (\stdvec{S}[][][+])(\stdvec{S}[][][+])\trans + \Dmat(\stdvec{S}[][][-])(\stdvec{S}[][][-])\trans\Hmat_i\trans \tilde{\R}^{1/2}\tilde{\R}^{T/2}\Hmat_i(\stdvec{S}[][][-])(\stdvec{S}[][][-])\Dmat \ .
\end{equation}
Similarly to the scalar measurement case the candidate square-root (and in fact, a valid square root) is written as,
\begin{equation}\label{eq:vec_sqrt_partial_update}
\begin{bmatrix}
(\stdvec{S}[][k][++])\trans \\ 
\zerovec
\end{bmatrix} = 
\Tmat
\begin{bmatrix}
(\stdvec{S}[][k][+])\trans\\
\tilde{\R}^{T/2}\Hmat\trans(\stdvec{S}[][k][-])\trans\Dmat
\end{bmatrix} \ .
\end{equation}

\subsubsection{Conventional time update with vector-valued measurement processing}
As the partial-update concept does not affect the propagation step, the method utilized for propagation is identical to the presented for the sequential filter. However, the measurement update (conventional non-partial-update) is modified. In \cite{simon2006optimal} the factorization that allows the conventional update can be used. Paralleling the procedure for the square root filter for scalar measurements, such factorization is set such that an orthogonal matrix needs to be found, but more importantly, that the updated square root of the covariance is computed. Similarly, this process is accomplished via the Modified-Gram-Schmidt. The required factorization for the conventional update is,
\begin{equation}
\begin{bmatrix}
	\tilde{\R}^{T/2} && \Kmat (\R + \Hmat(\stdvec{S}[][][-])(\stdvec{S}[][][-])\trans\Hmat\trans)^{T/2}\\
	\zerovec  && (\stdvec{S}[][][+])\trans
\end{bmatrix} = \tilde{T}
\begin{bmatrix}
\tilde{\R}^{T/2} && \zerovec\\
(\stdvec{S}[][][-])\trans\Hmat\trans && (\stdvec{S}[][][-])\trans
\end{bmatrix} \ .
\end{equation}

Once the MGS is executed with the intent of finding the orthogonal matrix $ (n+r)\times(n+r) $, the bottom-right $( n\times n) $ block, $ (\stdvec{S}[][][+])\trans $ will be generated. With this information in hand, the partial-update procedure (perform MGS on Equation (\ref{eq:vec_sqrt_partial_update})) to process a vector measurement can then be executed.

Table \ref{table:vec_SQPSK filter} summarizes the square root partial-update for vector measurement processing.

\begin{table}[h!]
	\caption{Square root partial-update Schmidt-Kalman filter. Vector measurement processing.}\label{table:vec_SQPSK filter}
	\centering
	\setlength{\extrarowheight}{8pt}
	\begin{tabular}{ |c|c| } 
		\hline
		\textbf{Model} & $\begin{aligned} &\x_{k}=\f_{k-1}(\x_{k-1},\u_{k-1})+\w_{k-1}\\ &\yktilde=\hkofxk+\vk\\
		&\wk \sim \mathcal{N} (\zerovec,\Qmat_k)\\
		&\vk \sim \mathcal{N}(\zerovec,\R_k)\end{aligned}$\\
		\hline
		\textbf{Initialize} & $\begin{aligned} 
		&\xpost_0=\x_0 \\ &\stdvec{S}[][0][+]= chol(\stdvec{P}[][0][+])\ \cite{golub2012matrix}\ \\
		&\Qmat_0^{1/2} = eigendec(\Qmat_0)\ \cite{strang1993introduction}\\
		&\Dmat = \diag(1-\beta_1,1-\beta_2,...,1-\beta_n)\
		\end{aligned}$\\
		\hline
		\textbf{Propagation} & $\begin{aligned} &\xprior_{k}=\f_{k-1}(\xpost_{k-1},\u_{k-1})\\
		&Perform\ MGS\ \cite{Kaminski1971}\ for \\
		&\begin{bmatrix}
		(\stdvec{S}[][k][-])\trans \\ 
		\zerovec
		\end{bmatrix}=
		\mathcal{T}
		\begin{bmatrix}
		(\stdvec{S}[][][+]_{k-1})\trans\Fmat\trans_{k-1}\\
		\Qmat_{k-1}^{T/2}
		\end{bmatrix}\end{aligned}$\\
		\hline
		\textbf{Gain} & $\begin{aligned}\\		&\stdvec{K}[][k]=(\stdvec{S}[][k][-])(\stdvec{S}[][k][-])\trans\Hmat_k\trans\tilde{\R}\\
		&\tilde{\R}_k =(\Hmat_k(\stdvec{S}[][k][-])(\stdvec{S}[][k][-])\Hmat_k\trans+\Rk)^{-1}\\
		&\tilde{\R}^{1/2}_k=chol(\tilde{\R}_k)\\
		\end{aligned}$\\ 
		\hline
		\textbf{Update} & $\begin{aligned} &\xpost_k=\xprior_{k}+\stdvec{K}[][k](\ym_k-\yhat_k)\\
		&\begin{bmatrix}
		\tilde{\R}^{T/2}_k && \Kmat_k (\R_k + \Hmat_k(\stdvec{S}[][k][-])(\stdvec{S}[][k][-])\trans\Hmat_k\trans)^{T/2}\\
		\zerovec  && (\stdvec{S}[][k][+])\trans
		\end{bmatrix} = \tilde{T}
		\begin{bmatrix}
		\tilde{\R}^{T/2}_k && \zerovec\\
		(\stdvec{S}[][k][-])\trans\Hmat_k\trans && (\stdvec{S}[][k][-])\trans
		\end{bmatrix}		
		\end{aligned}$\\
		\hline
		\textbf{Partial-Update} & $\begin{aligned} &\xpostplus = \Dmat\xprior_k + (\Imat- \Dmat) \xpost_k\\
		&Perform\ MGS\ \cite{Kaminski1971}\ for\\
		&\begin{bmatrix}
		(\stdvec{S}[][k][++])\trans \\ 
		\zerovec
		\end{bmatrix} = 
		\Tmat
		\begin{bmatrix}
		(\stdvec{S}[][k][+])\trans\\
		\tilde{\R}^{T/2}_k\Hmat_k\trans(\stdvec{S}[][k][-])\trans\Dmat
		\end{bmatrix}		
		\end{aligned}$\\
		\hline
	\end{tabular}
\end{table}

As previously mentioned, at a first glance the algorithm seems simple to implement as once the MGS and the Cholesky decomposition are available, there should not be any difficulties implementing this version of the square root filter. Nonetheless, the cost incurred makes it non apt for small computers. The extra computations come mainly from the matrix inversion required to compute the gain (also required in the conventional Kalman filter formulation), a Cholesky decomposition to obtain $ \tilde{\R}^{1/2}$ every time a measurement is available, an extra MGS (of a $ (n+m)\times(n+m) $ matrix) to perform the conventional update, and finally, a second MGS for a $ (n+m)\times n  $.

\subsection{Measurement and process noise covariance decorrelation}
If the user desires to process vector measurements in a sequential fashion, measurement covariance matrix must be diagonalized, if needed. Decorrelation can be achieved by different methods. In this section the modified Cholesky decomposition is described, as the factors would be available already (if the square root filter is implemented). Documentation and more details on this diagonalization algorithm is vast and can be found in many linear algebra or filtering books \cite{crassidis2011optimal}, \cite{simon2006optimal},  \cite{strang1993introduction}. The use of the decorrelation concept similarly applies to the measurement noise covariance matrix.

Let the noise measurement covariance be decomposed by Cholesky method into its \textit{square root} 
\begin{equation}
 \R = \sRmat\sRmat\trans \ ,
\end{equation}
and consider the measurement equation $\y = \Hmat\x+\v$, being transformed by $\sRmat^{-1}$ as
\begin{equation}	
\stdvec{z} = \sRmat^{-1}\y = \sRmat^{-1}\Hmat\x+\sRmat^{-1}\v \ . 
\end{equation}
The residual is then
\begin{equation}
    \stdvec{e}=\stdvec{z}-\hat{\stdvec{z}}=\sRmat^{-1}\Hmat(\stdvec{x}-\hat{\stdvec{x}})+\sRmat^{-1}\v  \ , 
\end{equation}
with covariance 
\begin{equation}
\expect[\stdvec{e}\stdvec{e}\trans]=\expect[(\sRmat^{-1}\Hmat(\stdvec{x}-\hat{\stdvec{x}})+\sRmat^{-1}\v)(\sRmat^{-1}\Hmat(\stdvec{x}-\hat{\stdvec{x}})+\sRmat^{-1}\v)\trans ]
\end{equation}
\begin{equation}
\expect[\stdvec{e}\stdvec{e}\trans]=\sRmat^{-1}\Hmat\expect[(\stdvec{x}-\hat{\stdvec{x}})(\stdvec{x}-\hat{\stdvec{x}})]\trans\Hmat\trans\sRmat^{-T} +\sRmat^{-1}\expect[\v\v]\trans\sRmat^{-T}
\end{equation}
\begin{equation}
\expect[\stdvec{e}\stdvec{e}\trans]=\sRmat^{-1}\Hmat\Cov\trans\Hmat\trans\sRmat^{-T} +\sRmat^{-1}\R\sRmat^{-T} \ .
\end{equation}
But since
\begin{equation}
  \sRmat^{-1}\R  \sRmat^{-T}=\stdvec{I} \ ,
\end{equation} 
and naming $\Hmat_z=\sRmat^{-1}\Hmat$, results in
\begin{equation}
	\expect[\stdvec{e}\stdvec{e}\trans] = \Hmat_z\Cov\Hmat_z\trans+\stdvec{I} \ .
\end{equation}
That is, a measurement noise covariance is now available. Thus, in order to properly use the transformed measurement, in the filter $ \sRmat^{-1}\Hmat $ is used instead of $ \Hmat $, and instead of directly computing the residual measurement as $ (\y-\Hmat\x)$ $(\sRmat^{-1}\y-\sRmat^{-1}\Hmat\x)$ is implemented.
\subsection{Computational complexity}
\begin{figure}[h]
	\centering
	\includegraphics[width=1\textwidth]{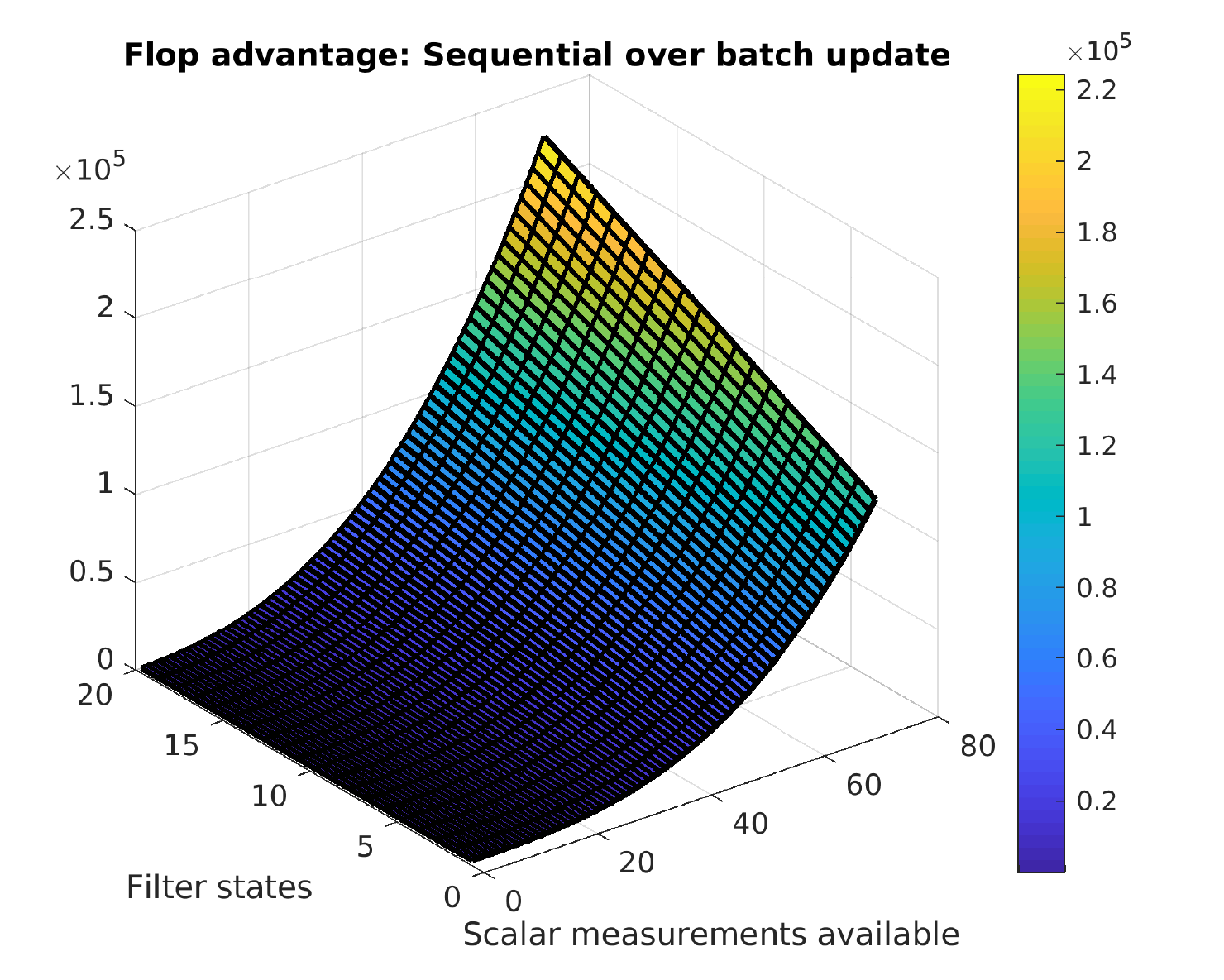}
	\caption{ Rough complexity comparison for sequential and batch measurement processing for the Kalman filter. UDU decorrelation is considered in the cost to use sequential filtering.}
	\label{fig:flops_batch_sequential_advantage}
\end{figure}

As reported in \cite{Grewal2001} and \cite{Kaminski1971}, within the conventional Kalman filter implementation, sequential measurement processing is, in general, more efficient than batch measurement assimilation. Table \ref{table:flops_advantage}, following the computation complexity formulas, as reported in \cite{Grewal2001}, shows a comparison on the number of flops (multiplications and divisions only) needed for batch and sequential measurement assimilation. In the same table, the computational advantage of sequential processing is shown. The graph from Figure \ref{fig:flops_batch_sequential_advantage}  shows the flops advantage of sequential measurement processing versus batch processing for various values of the number of states and available measurements. From this figure, it can be seen that it is always advantageous to process measurements sequentially, but more importantly, it shows that sequential processing becomes more advantageous quickly for filters with a large number of states. The reason for the sequential filter being more efficient is that many operations are saved mainly because the residual covariance inversion is avoided. Table \ref{table:sequential_kalman_operations} and \ref{table:batch_kalman_operations} include a breakdown of the number of flops for both sequential and batch processing for the conventional Kalman filter update step.

\begin{table}[h!]
	\begin{center}
		\caption{Flops required for batch measurement processing. Flop count considers matrices symmetry \cite{Grewal2001}.}
		\begin{tabular}{lcr}
			\label{table:sequential_kalman_operations}
			\textbf{Update stage} $\Hmat_{m\times n}$ and  $\Cov^{-}_{n\times n}$ & Flops (multiply or divide only) \\
			\hline\hline
			$ \Hmat \Cov^-$ & $mn^{2} $ \\
			$ \Hmat(\Hmat \Cov^-)\trans+\R $ & $ n(\frac{1}{2}m^2+\frac{1}{2} m) $ \\
			$ (\Hmat\Cov^{-}\Hmat \trans+\R)^{-1} $ & $m^3 + \frac{1}{2}m^2+\frac{1}{2}m$ \\
			$  \Cov^-\Hmat\trans(\Hmat\Cov^{-}\Hmat \trans+\R)^{-1} $ & $ nm^2 $ \\
			$ \Cov^--\Cov^-\Hmat\trans(\Hmat\Cov^{-}\Hmat \trans+\R)^{-1}\Hmat \Cov^- $ & $  (\frac{1}{2}(n^{2}-n)+n)m $ \\
			 \hline \text { Total } & $ m^{3}+ \frac{3}{2}nm^2+ \frac{1}{2}m^{2}+nm+\frac{1}{2}m+\frac{3}{2}mn^{2} $ \\
			\hline
		\end{tabular}
	\end{center}
\end{table}

\begin{table}[h!]
	\begin{center}
		\caption{Flops required for sequential measurement processing. Flop count considers matrices symmetry  \cite{Grewal2001}.}
		\begin{tabular}{lcr}
			\label{table:batch_kalman_operations}
			\textbf{Update stage} $\Hmat_{1\times n}$ and  $\Cov^{-}_{n\times n}$ & Flops (multiply or divide only) \\
			\hline\hline
			$ \Hmat \Cov^-$ & $n^{2} $ \\
			$ \Hmat(\Hmat \Cov^-)\trans+\R $ & $ n $ \\
			$ (\Hmat\Cov^{-}\Hmat \trans+\R)^{-1} $ & 1 \\
			$  \Cov^-\Hmat\trans(\Hmat\Cov^{-}\Hmat \trans+\R)^{-1} $ & $ n $ \\
			$ \Cov^--\Cov^-\Hmat\trans(\Hmat\Cov^{-}\Hmat \trans+\R)^{-1}\Hmat \Cov^- $ & $ \frac{1}{2} (n^{2}-n) + n $ \\
			\hline \text {Sum} $\times m$ \text { measurements } & $(\frac{3}{2} n^{2}+\frac{5}{2} n+1) m $ \\
			+\text { UDU and decorrelation } & $ \frac{2}{3}m^{3}+m^{2}-\frac{5}{3}m+\frac{1}{2}m^{2} n-\frac{1}{2}m n $ \\
			\hline \text { Total } & $ \frac{2}{3}m^{3}+m^{2}-\frac{2}{3}m+\frac{1}{2}m^{2} n+2mn+\frac{3}{2}m n^{2} $ \\
			\hline
		\end{tabular}
	\end{center}
\end{table}

\begin{table}[h!]
	\begin{center}
		\caption{Flop advantage of sequential over batch processing.}
		\begin{tabular}{lcr}
			\label{table:flops_advantage}
			\textbf{Flops advantage} & Flops (multiply or divide only) \\
			\hline\hline
			 \\
			\text { Batch} & $ m^{3}+ \frac{3}{2}nm^2+ \frac{1}{2}m^{2}+nm+\frac{1}{2}m+\frac{3}{2}mn^{2} $ \\
			\text{Sequential} &$ \frac{2}{3}m^{3}+m^{2}-\frac{2}{3}m+\frac{1}{2}m^{2} n+2mn+\frac{3}{2}m n^{2} $\\ 
			\hline \text { Sequential advantage over batch } & $ \frac{1}{2}m^3 - \frac{1}{2}m^2 + m^2n -mn +\frac{7}{6}m  $ \\
			\hline
		\end{tabular}
	\end{center}
\end{table}

In the case of square root filtering, and specifically for the partial-update version, a counting of flops can also be made.  
The flops required for the conventional square root Kalman filter formulation has been extensively well documented and has been analyzed in several scientific papers and books \cite{thornton1976triangular},\cite{Bierman1975},\cite{Kaminski1971}. Here, the computation complexity data reported for square root is used, and the additional operations to perform the partial update operation are included to obtain an estimated overall cost. Table \ref{table:square_root_plus_partial_update_complexity} shows the approximated cost (multiplications, divides, and square roots) of the square root partial-update filter presented in this chapter, along with the conventional square root filter. Although the use of the partial-update concept requires the extra cost of roughly an extra MGS, now the filter can handle higher uncertainties and nonlinearities at the same time that is more robust numerically. Moreover, it provides the ability to consider on a state-by-state basis at any time and not just a pre-selected set of states, as in the conventional consider filter formulation. Furthermore, the execution of an additional MGS naturally will lead to the triangularization of the square root covariance, which otherwise would be non-triangular with the conventional square root filter. Such a triangularization, in fact, translates into considerable storage savings as the elements above the diagonal are simply zero (for a lower triangular square root matrix), and there is no need to reserve memory space for them. 

In any case, the square root filter formulation can add a considerable amount of computations in general. For that reason, alternative ways of factorizing the covariance matrix were developed for Kalman filtering to increase factorized filter efficiency. The UD or modified Cholesky decomposition is an alternative, and in fact, a very efficient one within the Kalman filter framework. It is a more elaborated algorithm, but its efficiency makes the implementation worth it. The next chapter proposes a UD partial-update filter to increase the efficiency of the partial-update filter presented in this chapter.
\begin{table}[H]
	\begin{center}
		\caption{Conventional and partial-update required flops comparison.}		\begin{tabular}{lcr}
			\label{table:square_root_plus_partial_update_complexity}
			\textbf{Process} & Flops (multiply, divide and square root) \\
			\hline\hline
			\text {Conventional square root Kalman update} & $ 3n^2 + 4n + \sqrt{} + 1
			 $ \\
			\hline
			\text{Incorporation of partial-update extra cost} &$ n^3+3n^2+3n+ (n+1)\sqrt{} \approx 1 \ MGS$\\ 
			\hline
		\end{tabular}
	\end{center}
\end{table} 
\section{Summary}\label{sec:Conclusions}
This chapter presented a square root formulation for the partial-update Schmidt-Kalman filter. This form of the Kalman filter inherits the benefits of the partial-update formulation and combines them with the numerical robustness of the square root form. The result is a filter of higher numerical precision and increased tolerance to nonlinearities and uncertainty level at the cost of almost no additional computational burden. A formulation able to process vector-valued measurements was also presented, and due to its computational burden, it is just seen as a debugging tool and a way to facilitate the implementation of the sequential filter as no re-linearization is needed after measurement assimilation. Lastly, a numerical example, along with Monte Carlo runs, was used to demonstrate the effectiveness of the square root partial-update Schmidt-Kalman filter on a nonlinear system for both numerical stability and robustness to large uncertainties and nonlinearities. 
%
%
%
%


\chapter{U-D PARTIAL-UPDATE KALMAN FILTER}\label{ch:ud_partial_update_filter}
\section{Introduction}
Although the square root formulation was shown to increase the Kalman filter numerical precision and was successfully used in the Apollo missions, the square root  Kalman filter presented the issue of considerably increasing the computational cost of the conventional Kalman filter, especially for large systems. Among the alternative proposed techniques to decrease such computational load, the factorized UD Kalman filter update step, originally developed by Bierman and Thornton \cite{Thornton1976}, was a significant advance. This alternative formulation improved the numerical precision of the filter but in a more efficient way than the square root filter. Further, the UD filter has been shown to be more stable than other factorized Kalman filter implementations, being able to handle state vectors with thousands of variables.  For these reasons,  researchers and even NASA, prefer the UD Kalman filter for hardware implementations \cite{Carpenter2018}. 
In contrast with the original square root filter, the UD filter uses a factorization that involves an upper triangular matrix $ \Umat $ with 1's on its diagonal, and a diagonal matrix $ \D $, such that the covariance matrix is expressed as $ \Cov=\Umat\D\Umat\trans $. The efficiency of the UD Kalman filter mostly lies in a clever algorithm that significantly exploits the structure of the $ \Umat $ and $ \D $ matrices \cite{Gibbs2011}.  

In this chapter, to reduce the computational complexity of the square root partial-update, the UD factorized version of the partial-update filter is developed.

\subsection{The UD filter background}
Compared with the conventional square root filter formulation, the conventional UD filter does not require square root operations. For this reason, the UD filter is sometimes called square-root free filter. Interestingly, the UD factors can be obtained as a corollary of the Cholesky square-root factors. A closer examination of the Cholesky decomposition of a $n\times n$ matrix, reveals that $n$ square root operations are computed and that those same square-roots appear dividing each column, motivating a pair of alternative Cholesky factors: the UD factors. To illustrate this and further see how the UD factors can be obtained, consider a $ 3 \times 3 $ covariance matrix, $ \Cov $, to be analytically factorized via Cholesky decomposition as
\begin{equation}
\Cov = \Smat\Smat\trans \ ,
\end{equation}
with $ \Smat $ written as
%
	
	\begin{equation}
\Smat = \begin{bmatrix}
		\sqrt{\Cov_{11}- \frac{\Cov_{13}^2}{\Cov_{33}}-\frac{(\Cov_{12}-\frac{\Cov_{13}\Cov_{23}}{\Cov_{33}})^2}{\Cov_{22}-\frac{\Cov_{23}^2}{\Cov_{33}}}} &
		\frac{\Cov_{12}-\frac{\Cov_{13}\Cov_{23}}{\Cov_{33}}} {\sqrt{\Cov_{22}-\frac{\Cov_{23}^2}{\Cov_{33}}}}& \frac{\Cov_{13}}{{\sqrt{\Cov_{33}}}}\\
		0 		&	\sqrt{\Cov_{22}-\frac{\Cov_{23}^2}{\Cov_{33}}}	& \frac{\Cov_{23}}{{\sqrt{\Cov_{33}}}}\\
		0 & 0 & \sqrt{\Cov_{33}}
	\end{bmatrix} \ ,
	\end{equation}

which can itself be factorized as

\begin{equation}
\Smat	= \begin{bmatrix}
1 & \frac{\Cov{12}-\frac{\Cov_{13}\Cov_{23}}{\Cov_{33}}} {\Cov_{22}-\frac{\Cov_{23}^2}{\Cov_{33}}} & \frac{\Cov_{13}}{\Cov_{33}} \\
0 & 1 & \frac{\Cov_{23}}{\Cov_{33}} \\
0 & 0 & 1
\end{bmatrix}
\begin{bmatrix}
\sqrt{\Cov_{11}- \frac{\Cov_{13}^2}{\Cov_{33}}-\frac{(\Cov_{12}-\frac{\Cov_{13}\Cov_{23}}{\Cov_{33}})^2}{\Cov_{22}-\frac{\Cov_{23}^2}{\Cov_{33}}}} & 0 & 0 \\
0 & \sqrt{\Cov_{22}-\frac{\Cov_{23}^2}{\Cov_{33}}} & 0 \\
0  & 0 & \sqrt{\Cov_{33}}
\end{bmatrix} \ .
\end{equation}

By calling the matrix with 1's in the diagonal $ \Umat $ and the matrix containing the square roots as $ \sqrt{\D} $, $  \Smat $ is written as
\begin{equation}
	 \Smat = \Umat\sqrt{\D} \ , 
\end{equation}
and thus, the covariance matrix reads
\begin{equation}
\Cov=	 \Smat\Smat\trans = \Umat\sqrt{\D} \sqrt{\D}\Umat\trans = \Umat \D\Umat\trans \ .
\end{equation}


As can be observed from this development, although the $ \Umat \D$ product is equivalent to the Cholesky factor $ \Smat $, the $ \Umat \D$ factors do not involve square root operations. Moreover, via this factorization, the check for singularity or positive definiteness is straightforward as it suffices to revise the sign of the diagonal elements of $ \D $ (when monitoring for numerical problems that can be affecting the covariance matrix). Also, it is important to note that the determinant of $ \Umat $ is equal to one, and thus has an inverse, and inversion of an upper/lower triangular matrix with 1's in the diagonal, is an upper/lower triangular matrix. This fact will become relevant in the following sections.


Specific algorithms to obtain the modified Cholesky factorization are available in the filtering literature \cite{bierman2006factorization}, \cite{simon2006optimal}, \cite{Grewal2001},  or in linear algebra books or matrix operations books \cite{strang1993introduction} \cite{golub2012matrix}. However, it is recommended to use Kalman filter oriented routines as they are specifically structured to compute the required filter quantities in an efficient way, as those presented in \cite{Carpenter2018}. Similarly, to obtain the $ \Umat\D $ factors given the positive semi-definite and symmetric covariance matrix $ \Cov $.

In the following section, and before the development of the UD partial-update filter is presented, a superficial overview of the conventional UD filter is given with the purpose of establishing the appropriate context.

\section{The conventional UD Kalman filter}
Since the partial-update approach modifies the measurement-update step, similar to the previously proposed methods, the conventional UD filter and the UD partial-update form proposed in this chapter share the same temporal update. Although several approaches exist in the literature to execute a time update when using the UD filter, only one form is presented in this work.  More specifically, the time-update used for this work is the same presented in \cite{Carpenter2018} and \cite{simon2006optimal}. The selection of the time propagation method, however, is just a user's preference, and it does not affect the proposed partial-update development. 
\subsection{UD temporal update overview}
Consider the conventional discrete covariance propagation equation to obtain the prior covariance (with no time indices for clarity),
\begin{equation}
	\stdvec{P}^- = \Fmat\Cov^+\Fmat\trans + \Gmat\Qmat\Gmat\trans \ ,
\end{equation}
with $ \Gmat $ being the $ (n \times q) $ matrix mapping process noise to the state.
A direct attempt to obtain a factorization that involves three factors, to start forming $ \Cov^- = \Usub\Dsub\Usub\trans$ (sub-bar indicates a prior quantity), leads to the candidate form of
\begin{equation}
	\stdvec{P}^- = \rowvec{\Fmat\Uplus,\Gmat}\begin{bmatrix}
	\Dplus & \zerovec \\ \zerovec & \Qmat
	\end{bmatrix}\colvec{\Uplus\trans\Fmat\trans,\Gmat\trans}= \Wmat\hat{\D}\Wmat\trans \ ,
\end{equation}
where the plus sign above variables is to indicate a posterior quantity.
The matrix $ \hat\D $ defined above, is diagonal, but is an $ (n+q)\times(n+q) $ matrix . Further, $ \Wmat = \rowvec{\Fmat\Uplus,\Gmat} $ is a $ n\times (n+q)$ matrix and is not upper triangular in general, however, some work can be done to triangularize it. With this in mind, it is sought to satisfy
\begin{equation}\label{eq:udu_equal_wdw}
	\Cov^-=\Usub{\Dsub}\Usub\trans = \Wmat\hat{\D}\Wmat\trans  \ ,
\end{equation}
in a way that the factors $ \Usub $ and $ \Dsub $ can be computed with proper dimensions given $ \Wmat $ and $ \hat{\D} $ . In other words, although at this point a direct matrix-by-matrix relationship (i.e. $ \Usub \neq \Wmat $ nor $ \Dsub \neq \hat{\D} $) cannot be established, it is desired.
To accomplish this, first consider the weighted inner product
\begin{equation}\label{eq:vDv_product}
	\v_k\hat{\D}\v_j^T  = 0 \quad  k\neq j \ ,
\end{equation}
where the $ (n+q) $ $ \v_i$ row vectors can be found via the Weighted Modified Gram-Schmidt (WMGS) orthogonalization procedure (that uses the $ (n+q) $ row vectors, $ \w_i $ , of the matrix $ \Wmat $):
\begin{align}
	\v_n&= \w_n \ , \\
\v_k &= \w_k - \sum_{j=k+1}^{n}u(k,j)\v_j \quad k = n-1,\dots,1 \ ,\\
u(k,j) &= \frac{\w_k\hat{\D}\v_j^T}{\v_j\hat{\D}\v_j}	\quad j,k = 1,\dots, n \ .
\end{align}
Alternatively, 
\begin{equation}
	\w_k=\v_k +  \sum_{j=k+1}^{n}u(k,j)\v_j \quad k=1, \dots,n \ ,
\end{equation}
or
\begin{equation}
\w_k^T=\v_k^T +  \sum_{j=k+1}^{n}u(k,j)\v_j^T \quad k=1, \dots,n \ ,
\end{equation}
which can be expressed in matrix form as
\begin{equation}
	\begin{bmatrix}
	\w_1\\
	\vdots\\
	\w_n
	\end{bmatrix}=
	\begin{bmatrix}
	1 & u(1,2) & \dots & u(1,n) \\
	0 & 1 & \ddots& \vdots\\
	\vdots & \ddots & \ddots & u(n-1,n)\\
	0 & \dots & \dots & 1
	\end{bmatrix}
	\begin{bmatrix}
	\v_1\\
	\vdots\\
	\v_n
	\end{bmatrix} \ ,
\end{equation}
or
\begin{equation}
	\Wmat = \Umat\Vmat \ .
\end{equation}
With this expression in hand, Equation (\ref{eq:udu_equal_wdw}) can now be written as
\begin{equation}
\Cov^-=	\Usub{\Dsub}\Usub\trans = \Wmat\hat{\D}\Wmat\trans = (\Umat\Vmat)\hat{\D}(\Umat\Vmat)\trans = \Umat[][] [\Vmat\hat{\D}\Vmat\trans]\Umat\trans \ .
\end{equation}
Since the product $ \Vmat\hat{\D}\Vmat\trans $ is constructed according to Equation (\ref{eq:vDv_product}), the bracketed term in the previous equation is diagonal. Thus, the propagated factors are given by,
\begin{equation}
	\Usub = \Umat \ ,
\end{equation}
and
\begin{equation}
	\Dsub = \Vmat\hat{\D}\Vmat\trans \ .
	\end{equation}
	
In summary, once $ \Wmat $ and $ \hat{\D} $	are formed, the execution of the WMGS will provide the propagated factors, $ \Usub $ and $ \Dsub $. Regarding the state propagation, this is executed normally via system dynamics.

\section{The UD partial-update derivation}
To begin with the derivation of the UD partial-update filter, recall the  matrix form of partial-update,
\begin{equation}\label{eq:ud_partial_update_cov_vec}
\stdvec{P}[][][++] = \Dmat(\stdvec{P}[][][-]-\stdvec{P}[][][+])\Dmat + \stdvec{P}[][][+] \ ,
\end{equation}
where again, $\Dmat$ is the diagonal matrix with elements $\gamma_i$ for $i = 1,2\dots n$ (recall that $\gamma_i = 1 - \beta_i$).
Now, from the standard EKF equations,  $\stdvec{P}[][][+]=(\stdvec{I-\stdvec{K}[][]}\Hmat)\stdvec{P}[][][-]$ is incorporated into (\ref{eq:ud_partial_update_cov_vec}).
\begin{equation}
\stdvec{P}[][][++] = \stdvec{P}[][][+] + \Dmat(\Kmat\Hmat\stdvec{P}[][][-])\Dmat \ ,
\end{equation}
then replacing $\stdvec{K}[][]=\stdvec{P}[][][-]\Hmat\trans(\Hmat\stdvec{P}[][][-]\Hmat\trans+\R)^{-1}$ gives
\begin{equation}\label{eq:plusplus_P_in_UD}
\stdvec{P}[][][++] = \stdvec{P}[][][+] + \Dmat(\stdvec{P}[][][-]\Hmat\trans(\Hmat\stdvec{P}[][][-]\Hmat\trans+\R)^{-1}\Hmat\stdvec{P}[][][-])\Dmat \ .
\end{equation}
Next, the posterior covariance $ \Cov^+ $ is written in terms of the prior covariance by using  
\begin{align}
	\stdvec{P}[][][+]&=(\stdvec{I-\stdvec{K}[][]}\Hmat)\stdvec{P}[][][-]\\
	&=\Cov^--\stdvec{P}[][][-]\Hmat\trans(\Hmat\stdvec{P}[][][-]\Hmat\trans+\R)^{-1}\Hmat\Cov^- \ ,
\end{align}
and incorporating it into Equation (\ref{eq:plusplus_P_in_UD}) gives
\begin{equation}
	\stdvec{P}[][][++]
	=\Cov^--\stdvec{P}[][][-]\Hmat\trans(\Hmat\stdvec{P}[][][-]\Hmat\trans+\R)^{-1}\Hmat\Cov^-+\Dmat\stdvec{P}[][][-]\Hmat\trans(\Hmat\stdvec{P}[][][-]\Hmat\trans+\R)^{-1}\Hmat\stdvec{P}[][][-]\Dmat \ .
\end{equation}
At this point, a few variables are renamed and expressions rearrangements are done. First, the UD decomposition is indicated for the covariance matrix.
\begin{flalign}
&\stdvec{P}[][][++] = \UDU - \UDU\Hmat_i\trans(\Hmat_i\UDU\Hmat_i\trans + \R_i)^{-1}\Hmat_i\UDU+&&\\ \nonumber
&\Dmat\UDU\Hmat_i\trans(\Hmat_i\UDU\Hmat_i\trans + \R_i)^{-1}\Hmat_i\UDU\Dmat \ .
\end{flalign}
Second, the variables $ \wsub = \Usub\trans\Hmat_i\trans $ and $ A_i=(\Hmat_i\UDU\Hmat_i\trans + \R_i)^{-1} $ are introduced, and used in the previous equation. This results in,
\begin{equation}
	\stdvec{P}[][][++] = \UDU - \UD\wsub A_i\wsub\trans\DU+
	\Dmat\UD\wsub A_i\wsub\trans\DU\Dmat \ ,
\end{equation}
and by factorizing $ \Usub $ and $ \Usub\trans $ , the following expression is obtained 
\begin{align}\label{eq:UD_partial_update_before_udu_operation}
	\stdvec{P}[][][++]&= \Usub[\Dsub-(\Dsub\wsub)A_i(\Dsub\wsub)\trans]\Usub\trans + \Dmat\Usub[(\Dsub\wsub)A_i(\Dsub\wsub)\trans]\Usub\trans\Dmat \\
	&=\Umat\{[\Dsub-(\Dsub\wsub)A_i(\Dsub\wsub)\trans] +	\Usub^{-1}\Dmat\Usub[(\Dsub\wsub)A_i(\Dsub\wsub)\trans]\Usub\trans\Dmat\Usub^{-T}\}\Usub\trans \ .
\end{align}

Now because the curly-bracketed term is positive semi-definite, its UD decomposition can be computed resulting in a decomposition formed by $ \mathcal{U}\mathcal{D}\mathcal{U}\trans $ . The obtained factors, $ \mathcal{U}\mathcal{D}\mathcal{U}\trans $, are then used in Equation (\ref{eq:UD_partial_update_before_udu_operation}) and the result reads
\begin{equation}
	\stdvec{P}[][][++]
	=\Usub\{\mathcal{U}\mathcal{D}\mathcal{U}\trans\}\Usub\trans \ .
\end{equation}

The partial-updated (posterior) covariance matrix can also be expressed in terms of UD factors as
\begin{equation}
	\stdvec{P}[][][++]= \Upp \ \Dpp \ \Upp\trans
	=\Usub\{\mathcal{U}\mathcal{D}\mathcal{U}\trans\}\Usub\trans ,
\end{equation}
and since the product of two upper triangular matrix with 1's in the diagonal is also an upper triangular matrix with 1's in the diagonal, the partial-updated factors can be directly identified as
\begin{equation}
	\Upp = \Usub\mathcal{U} \ ,
\end{equation}
and
\begin{equation}
	\Dpp = \mathcal{D} \ ,
\end{equation}
which gives the expressions to execute the partial-update.

Before moving to the numerical examples that show the functionality of this formulation, a few remarks are given. First, notice that if $ \Dmat $ is zero (full-update), this formulation becomes the conventional UD filter. Second, although it seems that a considerable amount of operations are needed to compute the second term in Equation (\ref{eq:UD_partial_update_before_udu_operation}) (the term involving $ \Dmat $) by taking advantage of the matrix structure (upper triangular and diagonals) the extra burden is barely a fraction of a multiplication of two $ (n\times n) $ matrices. Third, it is also important to mention that even though this extra (curly bracketed) term needs to be formed and involved in the UD decomposition, the size of the matrix that enters the UD decomposition is of the same size as for the conventional filter; and thus no extra computation is incurred here with respect to the conventional UD filter. Finally, note that in the derivation, there is no assumption on the form of $ A_i $ (thus the capital letter notation), allowing that the same formulation can be used to process either vector or scalar measurements. However, the UD factorized filter is generally used in scalar measurement mode as it is more efficient than processing the measurements as a vector. In case of having measurement correlations ($ \R $ not a diagonal matrix), similarly to the square root filter, a decorrelation procedure can be executed. The procedure that decorrelates $ \R $ via the already computed UD factors is presented in the following subsection.

Table \ref{table:UDPU_filter} summarizes the UD partial-update Schmidt-Kalman filter equations.

\begin{table}[h!]
	\caption{U-D partial-update Schmidt-Kalman filter}\label{table:UDPU_filter}
	\centering
	\setlength{\extrarowheight}{8pt}
	\begin{tabular}{ |c|c| } 
		\hline
		\textbf{Model} & $\begin{aligned} &\x_{k}=\f_{k-1}(\x_{k-1},\u_{k-1})+\w_{k-1}\\ &\yktilde=\hkofxk+\vk\\
		&\wk \sim \mathcal{N} (\zerovec,\Qmat_k)\\
		&\vk \sim \mathcal{N}(\zerovec,\R_k)\end{aligned}$\\
		\hline
		\textbf{Initialize} & $\begin{aligned} 
		&\xpost_0=\x_0 \\ &[\Uplus_0,\Dplus_0]= udu(\stdvec{P}[][0][+])\ \cite{Grewal2001}\ \\
		&\Dmat = \diag(1-\beta_1,1-\beta_2,...,1-\beta_n)\
		\end{aligned}$\\
		\hline
		\textbf{Propagation} & $\begin{aligned} &\xprior_{k}=\f_{k-1}(\xpost_{k-1},\u_{k-1})\\
		&Form\ \\
		&\Wmat = \begin{bmatrix}
		\Fmat\Uplus && \Identity
		\end{bmatrix}\ ;
		\Dhat = \begin{bmatrix}
		 \Dplus && \zerovec\\
		 \zerovec && \Qmat
		\end{bmatrix}\\
		&[\Usub,\Dsub]=WMGS(\Wmat,\Dhat) \ \cite{simon2006optimal}\\
		\end{aligned}$\\
		\hline
		\textbf{Gain} & $\begin{aligned}
		&\Kmat_k=\Usub\Dsub\wsub A_i\\
		&A_i =(\wsub\trans\Dsub\wsub+\Rk)^{-1};\ \wsub=\Usub\trans\Hmat\trans
		\end{aligned}$\\ 
		\hline
		\textbf{Partial-Update} & $\begin{aligned}
		&\xpost_k=\xprior_{k}+(\Identity-\Dmat)\stdvec{K}[][k](\ym_k-\yhat_k)\\
		&\Lmat=(\Dsub\wsub)A_i(\Dsub\wsub)\trans\\
		&\mathcal{U}\mathcal{D}\mathcal{U}\trans= udu(\Dsub-\Lmat +\Usub^{-1}\Dmat\Usub\Lmat\Usub\trans\Dmat\Usub^{-T})\\
		&\Upp=\Usub\mathcal{U}\\
		&\Dpp=\mathcal{D}
		\end{aligned}$\\
		\hline
	\end{tabular}
\end{table}

\section{Measurement decorrelation using UD factors}
Let the UD decomposition of a non-diagonal measurement noise covariance, $ \R_c $, be
\begin{equation}
	\R_c = \Ur\Dr\Ur\trans \ ,
\end{equation}
and consider the measurement equation $\y = \Hmat\x+\v$, being transformed by $\Uinv$ as
\begin{equation}	
\stdvec{z} = \Uinv\y = \Uinv\Hmat\x+\Uinv\v \ . 
\end{equation}
The residual is then
\begin{equation}
\stdvec{e}=\stdvec{z}-\hat{\stdvec{z}}=\Uinv\Hmat(\stdvec{x}-\hat{\stdvec{x}})+\Uinv\v \ , 
\end{equation}
with covariance 
\begin{equation}
\expect[\stdvec{e}\stdvec{e}\trans]=\expect[(\Uinv\Hmat(\stdvec{x}-\hat{\stdvec{x}})+\Uinv\v)(\Uinv\Hmat(\stdvec{x}-\hat{\stdvec{x}})+\Uinv\v)\trans ] \ ,
\end{equation}
\begin{equation}
\expect[\stdvec{e}\stdvec{e}\trans]=\Uinv\Hmat\expect[(\stdvec{x}-\hat{\stdvec{x}})(\stdvec{x}-\hat{\stdvec{x}})]\trans\Hmat\trans\Ur^{-T} +\Uinv\expect[\v\v]\trans\Ur^{-T} \ ,
\end{equation}
\begin{equation}
\expect[\stdvec{e}\stdvec{e}\trans]=\Uinv\Hmat\Cov\trans\Hmat\trans\Ur^{-T} +\Uinv\R_c\Ur^{-T} \ .
\end{equation}
But since
\begin{equation}
\Uinv\R_c  \Ur^{-T}=\Dr \ ,
\end{equation} 
and naming $\Hmat_z=\Uinv\Hmat$, results in
\begin{equation}
\expect[\stdvec{e}\stdvec{e}\trans] = \Hmat_z\Cov\Hmat_z\trans+\Dr \ .
\end{equation}
That is, the transformed equation now uses a diagonal measurement noise covariance. In order to properly use the transformed measurement, the filter simply uses $ \Uinv\Hmat $ instead of $ \Hmat $, and instead of directly computing the residual measurement with $ (\y-\Hmat\x)$, $\Uinv(\y-\Hmat\x)$ is used. Rather than using the correlated measurement noise $ \R_c $, the filter now uses $ \Dr $ and if desired, the filter measurement can be processed sequentially.

\section{Numerical example}
The UD partial-update filter is exercised using the re-entering body scenario that was used for the square-root version of the filter. The process and measurement model, as well as the initial conditions, remain the same for these simulations. The partial-update weights, $ \betamat $, were also unaltered. Since the UD partial-update and the square-root partial-update filter are mathematically equivalent, the main objective of this numerical simulations is to show that the numerical properties that were gained with the square root factorization, remained for the UD version; with the only difference being the UD filter version more computationally efficient.
\subsection{Body re-entering Earth atmosphere single run}
\begin{figure}[h!]
\includegraphics[width=1\textwidth]{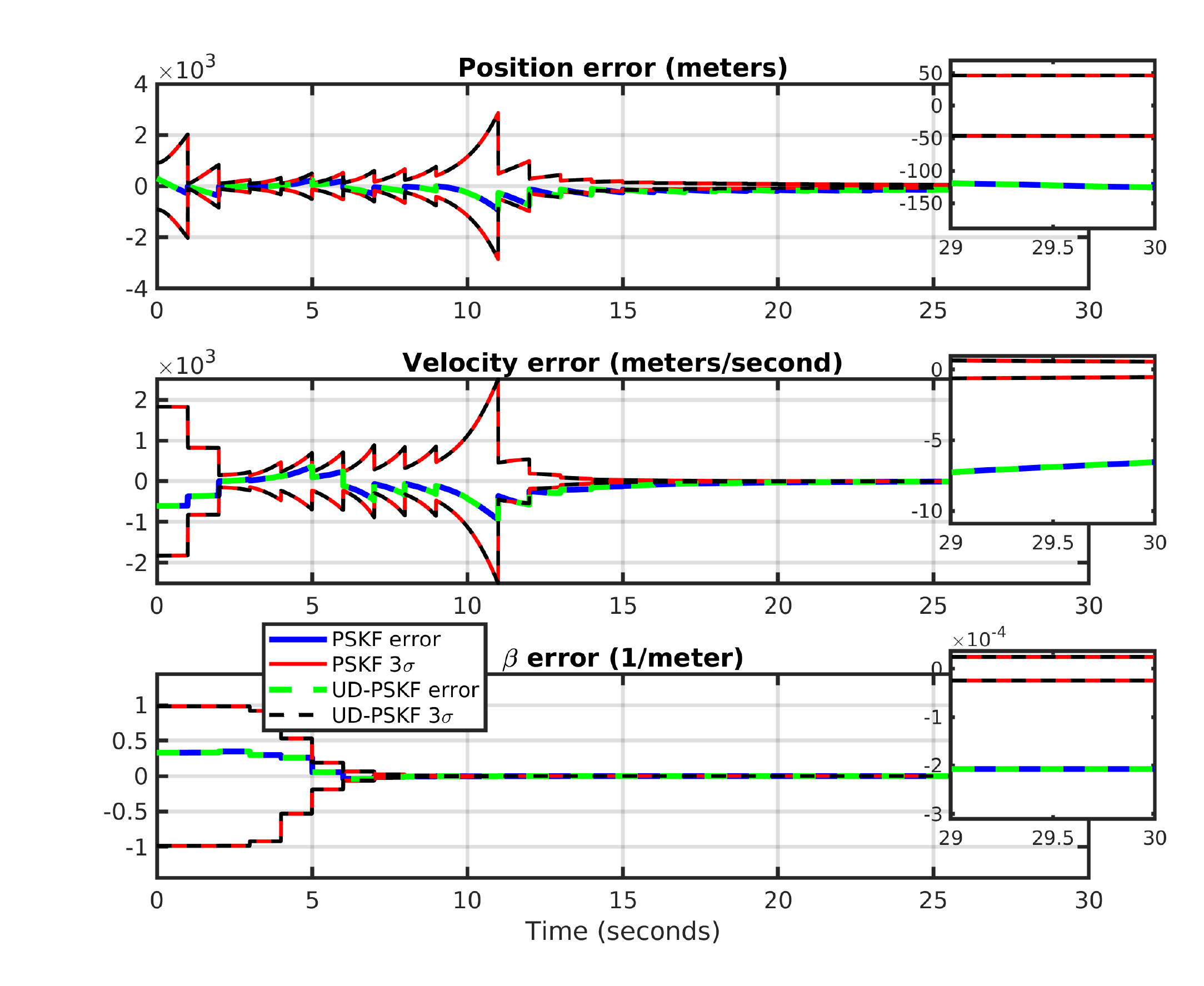}
	\caption{Standard EKF and UD partial-update EKF with full updates. The inset on the right shows a zoom-in for the last second of the simulation, displaying significant filter inconsistency with estimates well outside of $3\sigma$ bounds.}
	\label{fig:ud_full_update}
\end{figure}

Similar to the description of the results from the square root partial-update filter,  three main simulations are presented. First, in Figure \ref{fig:ud_full_update}, the conventional EKF, along with the UD partial-update filter estimates, are depicted for a single run. For this simulation, the partial-update weights are all set to 1 (conventional full update). The estimates for both filters, as expected, show divergence as the EKF cannot handle the simulation's large initial uncertainties. The UD formulation is equivalent to the EKF, and the estimates are not different.

\begin{figure}[h!]
\includegraphics[width=1\textwidth]{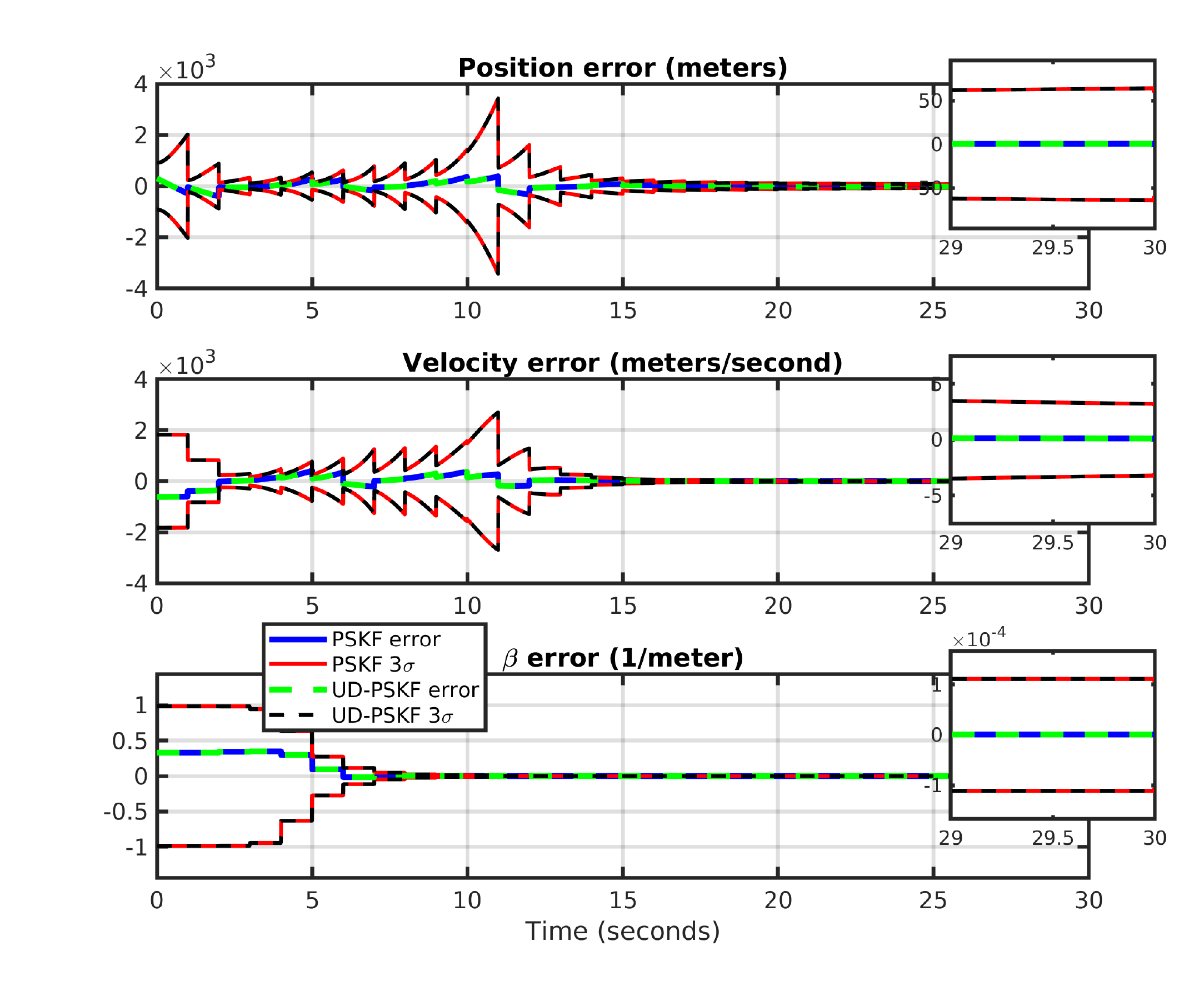}
	\caption{Standard EKF and UD partial-update EKF with  partial updates. The inset on the right shows a zoom-in for the last second of the simulation, displaying significant filter inconsistency with estimates well outside of $3\sigma$ bounds.}
	\label{fig:ud_partial_update}
\end{figure}

\begin{figure}[h!]
\includegraphics[width=1\textwidth]{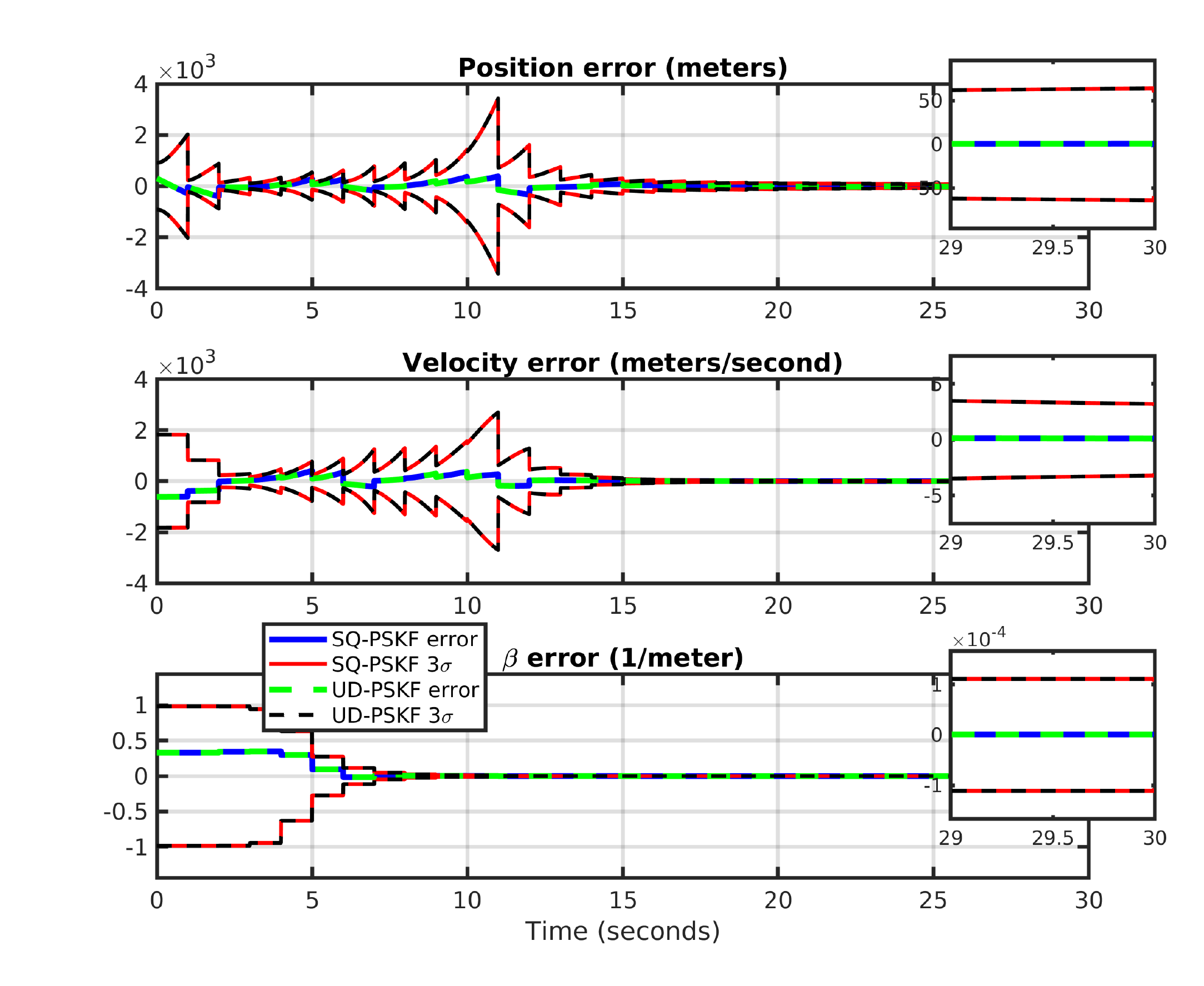}
	\caption{partial-update EKF and UD partial-update EKF with $\boldsymbol{\beta}=\rowvec{0.9, 0.9, 0.75}$ resulting in 90\%, 90\%, and 75\% updates respectively to the position, velocity, and ballistic coefficient states. The inset on the right shows a zoom-in for the last second of the simulation, displaying filter results with estimates within the $3\sigma$ bounds.}
	\label{fig:ud_and_sqrt_partial_update}
\end{figure}

In Figure \ref{fig:ud_partial_update}, the results of applying a partial-update with $ \betamat= \diag \ [0.9,0.96,0.75] $, are shown. Again, these update weight values were selected to limit the updates mainly for the ballistic parameter, since it is commonly less observable than the other states. In this scenario, the filters are seen to be able to handle the given initial conditions. As in the square root filter, the UD formulation now incorporates the partial-update benefits. In the intent to show that the square root and the UD partial-update formulations are mathematically equivalent, they are graphed in Figure \ref{fig:ud_and_sqrt_partial_update} and shown to produce the same state and covariance estimates. Although these results are from a single run, they are representative of a typical run for this scenario. For this filter, the condition number plot is omitted as it is equivalent and conveys the same information than for the graph presented in the square root partial-update chapter: the UD factors have lower condition number than the full covariance matrix.

\section{Monte Carlo runs}
Monte Carlo simulations were run for both, UD EKF and UD partial-update EKF. A total of 100 runs were executed, and the histories of all of the states were recorded. The sampled standard deviation and the standard deviation, as computed by the filter were also calculated and used to check for filter consistency. For both filters, the initial conditions are maintained the same from the single run re-entering body scenario from the previous chapter. For convenience, the parameter values used for the simulations are condensed in Table \ref{table:initial_conditions_SQPU}.

\begin{figure}[h!]
\includegraphics[width=1\textwidth]{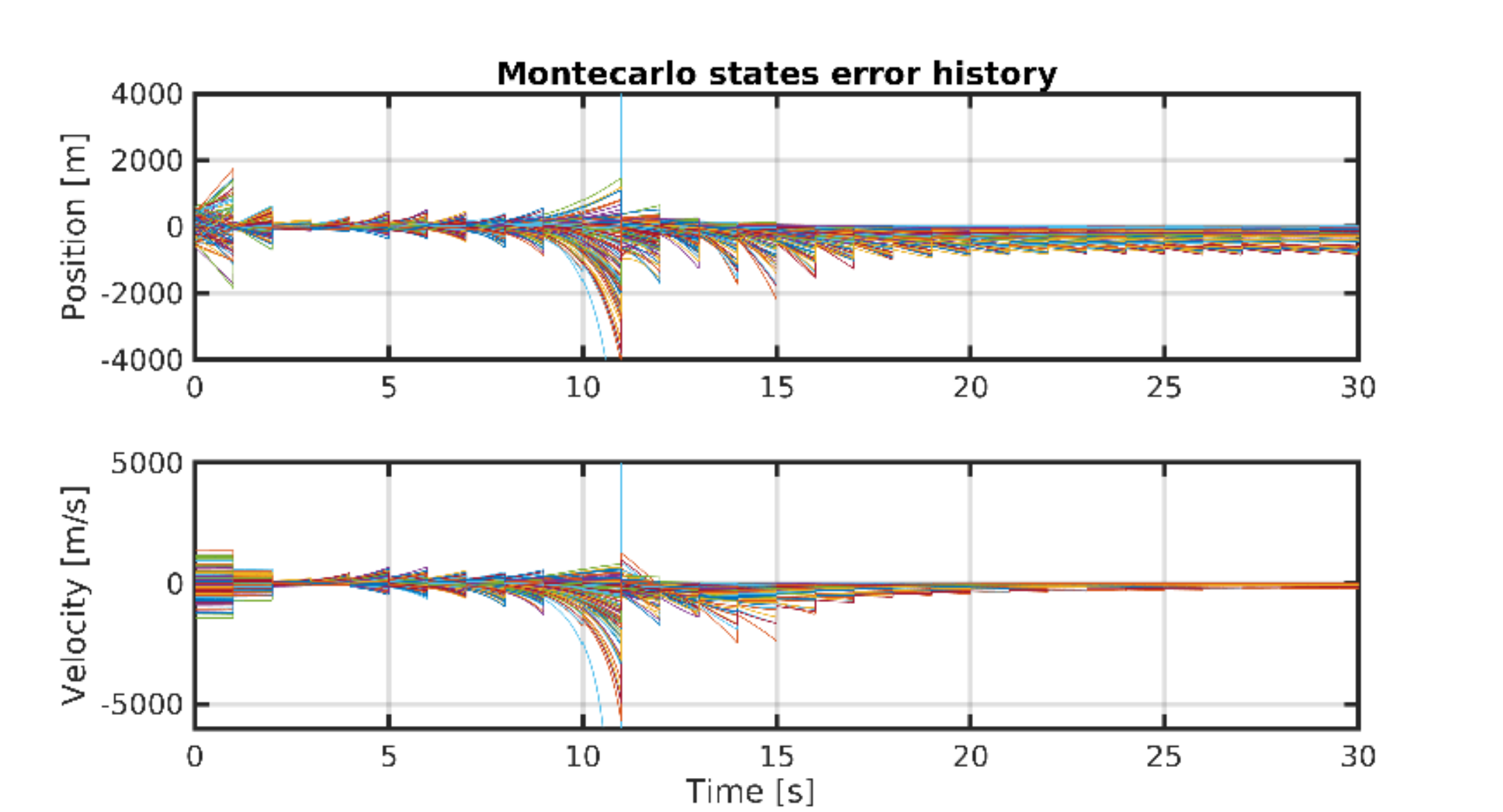}
	\caption{Monte Carlo standard EKF and UD partial-update EKF with full updates.}
	\label{fig:UDmonte_state_ud_full_update}
\end{figure}

\begin{figure}[h!]
\includegraphics[width=1\textwidth]{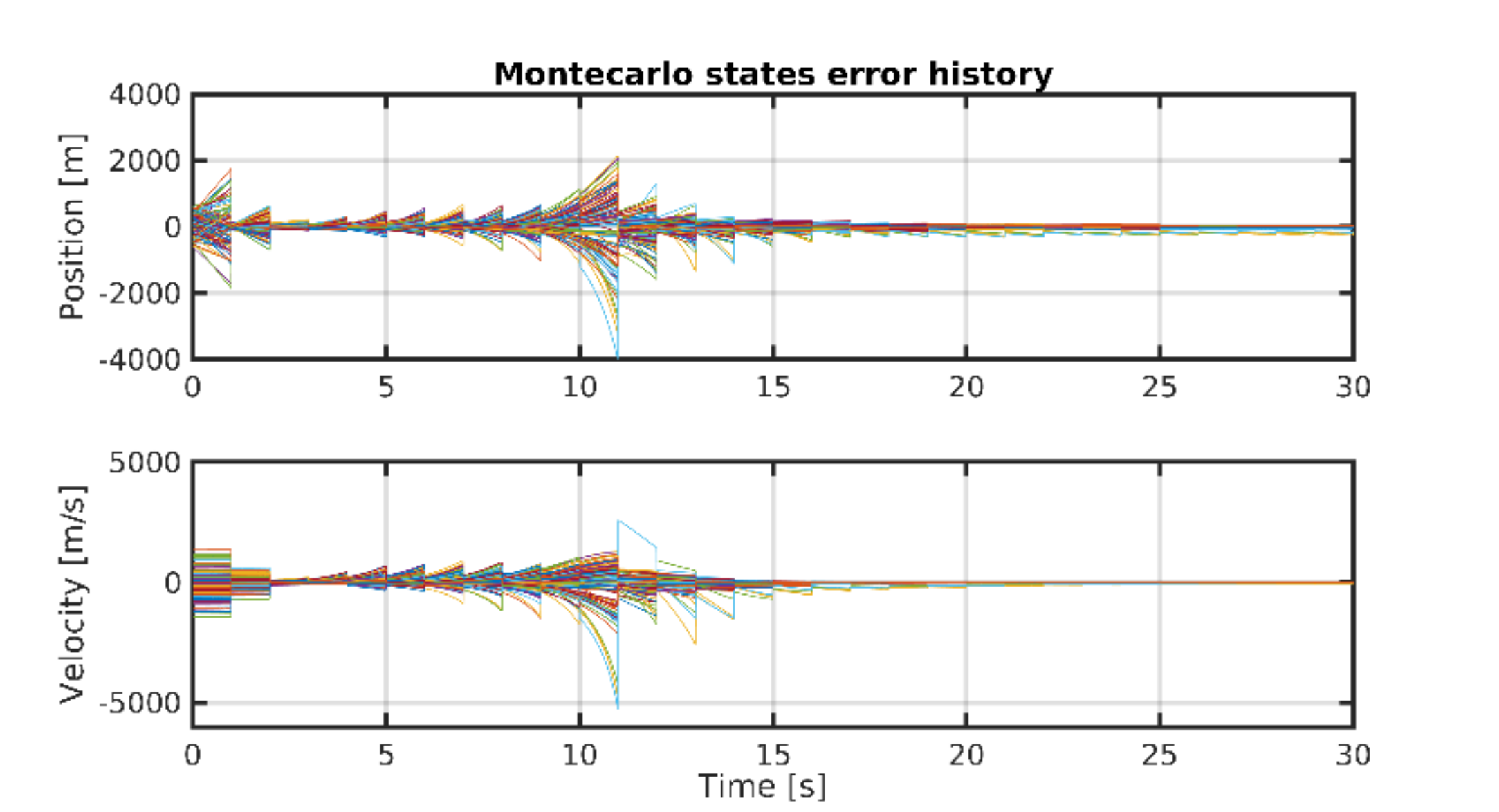}
	\caption{Monte Carlo runs for the partial-Update EKF and UD partial-update EKF with $\boldsymbol{\beta}=\rowvec{0.9, 0.9, 0.75}$ resulting in 90\%, 90\%, and 75\% updates respectively to the position, velocity, and ballistic coefficient states.}
	\label{fig:UDmonte_states_ud_partial_update}
\end{figure}

Figure \ref{fig:UDmonte_state_ud_full_update} shows the 100 EKF runs histories. The EKF shows that, in its majority, it can prevent the filter's total failure, except for two cases (blue curves) that are completely divergent. However, most of the runs show similar behavior as the one shown for the single run: the error estimates are not within the proper sigma bounds after around $ t = 15 $ seconds, and the errors do not converge to zero. The UD partial-update technique, on the other hand (as depicted in Figure \ref{fig:UDmonte_states_ud_partial_update}), and as expected from the results in the previous chapter, shows a dramatic improvement over the EKF. First, for the same sampled initial conditions, the filter presents no runs with divergence. Second, the error histories show a significant magnitude reduction, and third, a superior capacity to handle the initial uncertainties by avoiding the overreaction in the update is achieved. 

\begin{figure}[h!]
\includegraphics[width=1\textwidth]{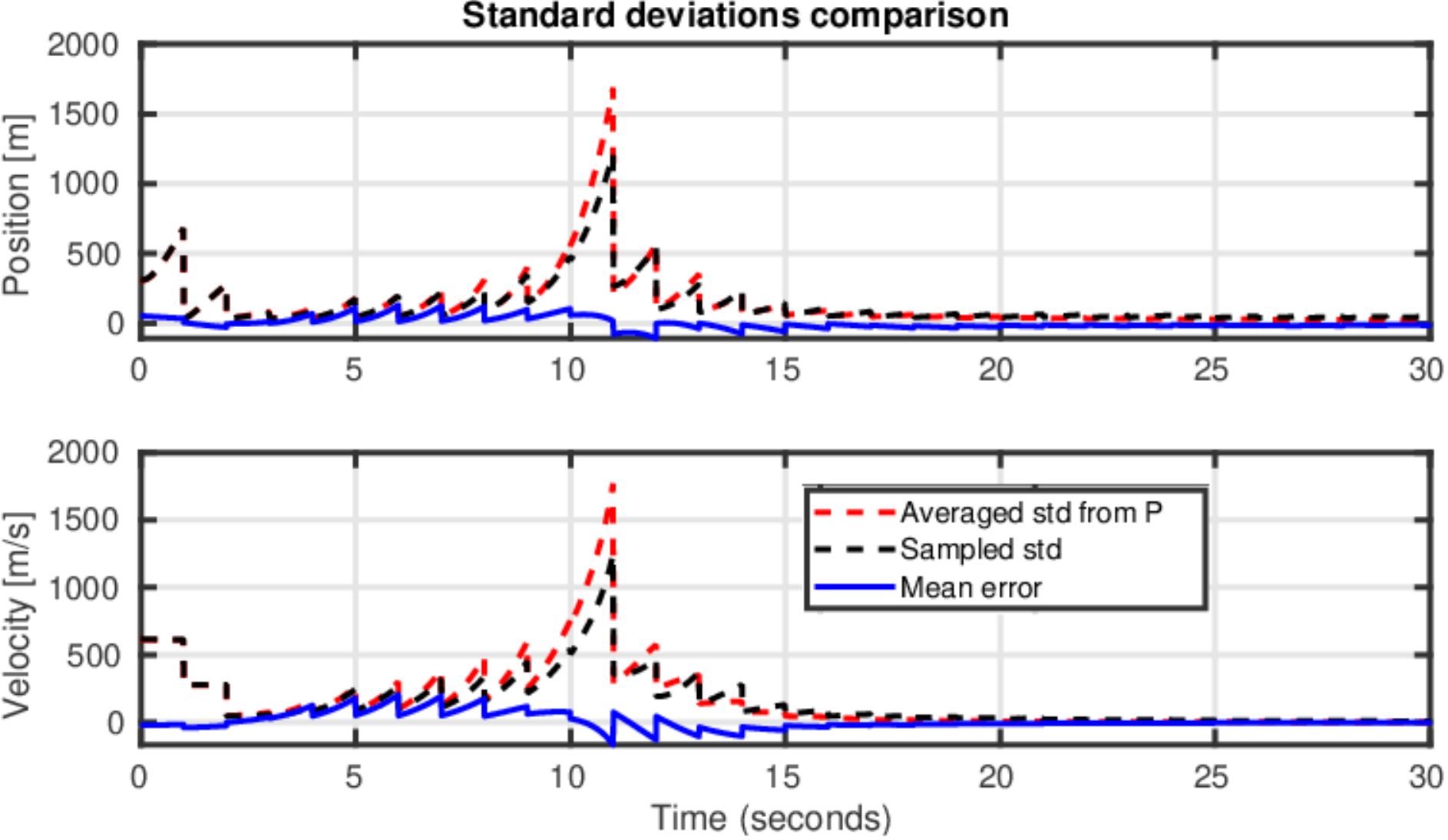}
	\caption{Monte Carlo runs for the partial-Update EKF and UD partial-update EKF with $\boldsymbol{\beta}=\rowvec{0.9, 0.9, 0.75}$ resulting in 90\%, 90\%, and 75\% updates respectively to the position, velocity, and ballistic coefficient states. }
	\label{fig:UDmonte_ud_partial_update}
\end{figure}

The averaged standard deviation from the Monte Carlo runs, and the standard deviation estimated by the filter are depicted in Figure \ref{fig:UDmonte_ud_partial_update}, and are shown to practically coincide, and since the mean estimation error is around zero, this indicates that the filter is consistent. Since the EKF filter presented divergent cases, only standard deviations and mean error were plotted for the partial-update filter (EKF is inconsistent for this scenario).

\begin{table}[h!]
	\begin{center}
		\caption{Re-entering body parameters}
		\begin{tabular}{lcr}
			\label{table:initial_conditions_SQPU}
			\textbf{State/parameter uncertainty} &  Uncertainty $ 1\sigma $ value \\
			\hline\hline
			Position   & \SI{300}{\meter} \\
			Velocity  & \SI[per-mode = symbol]{600}{\meter\per\second}\\
			Ballistic parameter & \SI{0.33}{\per\meter}  \\
			Measurement & \SI{300}{\meter}\\
		\end{tabular}
	\end{center}
\end{table}

Overall, it was observed that if low initial errors were ensured, the EKF can be functional and quickly converge to zero error, but is not robust enough to handle errors at the level of the partial-update UD filter. 
Conversely, the UD partial-update filter was observed to be consistent, and able to handle higher nonlinearities and uncertainties better than the UD EKF or EKF. Again, the UD partial-update filter results, as expected, match those from the square root partial-update filter, since the change in factorization does not provide any additional robustness against uncertainties, and it is just a mechanism to gain computational efficiency.


\section{Numerical complexity}
The UD formulation's main objective is to provide an alternative filter in which the computational burden is lower than for the square root formulation, but retains the numerical properties of a factorized filter. The conventional UD filter is more efficient than the conventional square root formulation, and for the case of the partial-update formulations, this is also the case. Table \ref{table:UD_and_SQ_partial_update_complexity} includes the approximated computational complexity of the square root and UD partial-update formulations. The costs of executing a conventional UD filter and its partial-update version, are shown in Table \ref{table:UD_plus_partial_update_complexity}. The required extra effort to perform the partial update is also included in this table.

\begin{table}[h!]
	\begin{center}
		\caption{Square root and UD partial-update required flops for propagation and update steps combined.}		\begin{tabular}{lcr}
			\label{table:UD_and_SQ_partial_update_complexity}
			\textbf{Filter} & Flops (multiply, divide and square root) \\
			\hline\hline
			\\
			\text {Square root partial-update} & $ 2.5n^3 + (q + 7.5)n^2 + (2\sqrt{} + 6)n + 2\sqrt{} + 1
			
			$ \\
			\hline
			\text{UD partial-update} &$ 2n^3 + (q + 4)n^2 + (q + 1)n + 2
			$\\ 
			\hline
		\end{tabular}
	\end{center}
\end{table}

\begin{table}[h!]
	\begin{center}
		\caption{Conventional UD and UD partial-update required flops for one  scalar measurement update.}		\begin{tabular}{lcr}
			\label{table:UD_plus_partial_update_complexity}
			\textbf{Process} & Flops (multiply, divide and square root) \\
			\hline\hline
			\\
			\text {Conventional UD Kalman update} & $ 1.5n^2+1.5n$\\
			\hline
			\text{UD partial-update Kalman update} & $0.5n^3 + 3.5n^2 + n + 2$
			\\
			\hline
			\text{UD partial-update extra cost} &$ 0.5n^3 + 2n^2 - 0.5n + 2
				$\\ 
			\hline
		\end{tabular}
	\end{center}
\end{table}

Table \ref{table:UD_and_SQ_partial_update_complexity} shows that the UD partial-update formulation is more efficient than the square root formulation, requiring roughly $ 0.5n^3 $ fewer operations. This computational savings, along with the numerical stability, make the UD formulation generally preferred over factorized formulation alternatives. Regarding the comparison between the conventional UD filter and its partial-update formulation, the flop count (multiplication and division) shows that the partial-update version of the filter requires some extra effort which is roughly equivalent to the product of a full $ (n\times n)$ matrix times a $ (n\times n) $ triangular matrix. Although it may not be significant for small to medium-size systems, it may be considerable for large systems. However, that is the price of incorporating the partial-update benefits and gaining robustness against system nonlinearity and uncertainty.

\subsection{IMU-camera example}
The IMU-camera calibration problem introduced in the previous chapter was also implemented in UD form. The simulations, as for the re-entry body problem,  since they were performed for exactly the same scenario as before, show complete agreement with the square root partial-update filter, as expected. The plots for these runs (UD partial-update filter and conventional EKF) are not shown as they match the behavior previously presented.


\subsection{Summary}
The UD partial-update filter was shown to be more efficient than the square root formulation while retaining the numerical robustness properties (since it does not directly operate on the covariance matrix). Similar to the conventional UD filter, the partial-update version heavily relies on the structure of the involved matrices to lower the number of operations to execute a measurement update. However, alternative algorithms may provide even better efficiency for the UD partial-update filter. In terms of consistency and general behavior, since the UD filter is not different from the square root formulation, no changes were expected; the Monte Carlo and single runs corroborated this.

It is important to note that although the UD and Potter formulations are not the only factorized formulations for the Kalman filter, they are taken as the base of the developments presented in this research because they have been successfully and widely applied. Furthermore, since these formulations are among the most fundamental factorized filters, virtually any available extensions that have been applied to square root or UD filters can be incorporated into this work. Although the Carlson algorithm \cite{Carlson1973a} may be very similar in performance to the UD filter and could have been selected as a means to increase efficiency, the UD filter was preferred as it avoids the computation of square roots and is widely used. 

Finally, although the square root filter may not be as attractive as the UD formulation for implementation, it is not all in vain since, in any case, the square root filter can provide a way to corroborate UD filter estimates (alternative factorized filter). Even though the UD formulation can seem complicated at first glance, it only requires the extra coding of the Weighted-Modified Gram-Schmidt routine and the UD decomposition. Once these algorithms are available, the implementation is straightforward, as seen in Table \ref{table:UDPU_filter}. If the filtering engineer is dealing with correlated measurements and the desire is to process measurements sequentially, the UD based decorrelation algorithm can be executed (the corresponding cost needs to be considered in the total computational complexity of the filter).  If the computation complexity is not a concern for any reason, the UD partial-update formulation presented in this chapter can also directly use vector measurements.

\chapter{DYNAMIC PARTIAL-UPDATE KALMAN FILTER}\label{ch:dynamic_methods}
\section{Motivation}
The partial-update formulation has been shown to improve the robustness and overall performance of the underlying filter. To achieve favorable results when using the partial-update method, however, \textit{appropriate} selection of the update percentages, $ \beta $, is needed. Even though the partial-update weights can be selected in a number of different manners, the $ \beta's $ selection could be put into two possibilities: Static partial-update weight and dynamic partial-update weight selection. The numerical examples from the previous chapters, clearly, are cases with static $ \beta's $. 

For a static weight selection, whereas it may not be difficult to adjust the $ \beta $ values to obtain a converging filter in general, it certainly requires some experimentation with the system in question. Overall, when selecting the $ \beta $ values, it is sought to balance the negative impact of certain states by limiting update corrections while using most of the available information, and simultaneously, allowing enough state update to compensate for process noise (if present). An informal technique that has been useful for static weight selection has been based on state dynamics speed, assigning small weights to slowly changing states and larger weights to faster states. In conjunction with this idea, the weights may also be based on how ``close'' to an observation certain states are situated: Directly measured states are almost or fully updated whereas states updated through a long chain of cross-correlation terms are not updated or slightly updated. Nevertheless, even with these ad-hoc rules, the question of how much to slightly or partially update mean, remains. And while static $ \beta $ weights can suffice, filter estimates can certainly be further improved with online selected $ \beta $ weights, as shown with the proposed techniques in this chapter.

\section{Dynamic partial-update weights}

The idea for dynamic weight selection is to provide the partial-update filter (but in general, a Kalman filter) with information that can be used to decide, in a commensurate way, how much of the nominal update should be used.  Specifically, the proposed methods use a system's nonlinearity metric to inform the filter on what weights values can be appropriate at a specific update step, such that it can take full advantage of the incoming measurement when possible but be ``a careful updater'' when required. Following this paradigm, two methods are presented next. Overall, both methods were shown to be of higher capacities than the static partial-update when handling high nonlinearities and uncertainties.

\section{Nonlinearity-aware based method}
For vast low-uncertainty applications, the tolerance of the EKF against slight mismodelling can be enough to prevent the estimator from failing. However, since the conventional EKF retains only the first-order term of the Taylor series expansion of the models in its formulation, if higher-order effects have a significant influence to produce an important mismatch between first order and system models, filter divergence can be originated. After all, the greater the model's mismatch, the more sub-optimal the filter becomes (propagation and update correction would be carried out for a more distinct system). Based on this fact, the idea of trusting (using) the computed update more when the modeling mismatch is small, and trusting the update less for a significant mismatch, is proposed as a selection method of the partial-update percentages. More specifically, the proposed method uses the ratio of second-order to first-order terms of the Kalman update step equations to inform about how \textit{reliable} the EKF equations can be at a given time step so that the filter can decide if it is "safe" to fully update, or partially update, or just \textit{consider} states. Stated differently, in the case of \textit{insignificant}, high, and very high second-order effects, the filter performs a full, partial, or a consider update, respectively. The next subsection provides the mathematical details on this nonlinearity-based $ \beta $ selection method, refereed from now on, as Dynamic nonlinearity-aware partial-update or DNL for short.


\subsection{Nonlinearity-aware partial-update}
To begin, following the same notation for the Kalman filter framework recall the equations for the discrete second-order Kalman filter (EKF2) presented here without derivation \cite{simon2006optimal}.
The dynamics and uncertainty propagation equations are

\begin{align}
	\xprior_k &= f(\xprior_{k-1},\u_{k-1},k-1) + \frac{1}{2}\sum_{i=1}^{n}\phi_i\Tr\rowvec{\frac{\partial^2{f_i}}{\partial{x}^2}\Big\rvert_{\xpost_{k-1}}\Cov_{k-1}^+} \ , \label{eq:ekf2_propagation_state} \\
	\Cov^-_{k}&= \Fmat_{k-1}\Cov_{k-1}^+\Fmat_{k-1}\trans+\Gmat_{k-1}\Qmat_{k-1}\Gmat_{k-1}\trans \ . 
\end{align}
The measurement update equations are
\begin{align}
	\xpost_k &= \xprior_{k} + \Kmat_k\rowvec{\ym_k-h(\xprior_k,k)} - \stdvec{\pi} \ , \label{eq:ekf2_measurement_update}\\
	\stdvec{\pi}  &=\frac{1}{2}\Kmat_k\sum_{i=1}^{m}\phi_i\Tr\rowvec{\D_{k,i}\Cov_k^-} \ ,  \label{eq:ekf2_pi}\\
	\D_{k,i} \ , &= \frac{\partial^2{h_i(\x_k,k)}}{\partial{x}^2}\Big\rvert_{\xprior_k} \ ,\\
	\stdvec{K}[][k]&=\stdvec{P}[][k][-]\Hmat_k\trans(\Hmat_k\stdvec{P}[][k][-]\Hmat_k\trans+\Rk)^{-1} \ ,\\
	\stdvec{P}[][k][+]&=(\stdvec{I-\stdvec{K}[][k]}\Hmat_k)\stdvec{P}[][k][-] \ .
\end{align}
Here, the process and measurement model Jacobian are defined as before, and $ \phi_i $ is the single entry vector (with 1 at the $ i^{th}$ element) given by,
\begin{equation}
	\phi_i\trans=\rowvec{0,0,\dots,0,\dots,1,\dots,0}\trans \  .
\end{equation}
Next, consider the measurement update from Equation (\ref{eq:ekf2_measurement_update}) and the vector $ \pivec $ expressed together as
\begin{equation}
		\xpost_k = \xprior_k + \Kmat_k\rowvec{\ym_k-h(\xprior_{k \ ,}\ k)} - \frac{1}{2}\Kmat_k\sum_{i=1}^{m}\phi_i\Tr\rowvec{\D_{k,i}\Cov_k^-} \ .
\end{equation}
Further, let the prior state, $ \xprior_k $, as defined in Equation (\ref{eq:ekf2_propagation_state}) be substituted into the previous equation to form
\begin{align}
		\xpost_k = f(\xprior_{k-1},\u_{k-1},k-1) + \frac{1}{2}\sum_{i=1}^{n}\phi_i\Tr\rowvec{\frac{\partial^2{f_i}}{\partial{x}^2}\Big\rvert_{\xpost_{k-1}}\Cov_{k-1}^+} &+  \Kmat_k\rowvec{\ym_k-h(\xprior_{k \ ,}\ k)} \\ \nonumber 
		&-\frac{1}{2}\Kmat_k\sum_{i=1}^{m}\phi_i\Tr\rowvec{\D_{k,i}\Cov_k^-} \ .
\end{align}
By reorganizing the terms the posterior estimate can be written as,
\begin{align}
		\xpost_k = f(\xprior_{k-1},\u_{k-1},k-1) +
		\Kmat_k\rowvec{\ym_k-h(\xprior_{k \ ,}\ k)} +\label{eq:ekf2_posterior_state}&
		 \Ymat
		    \ ,
\end{align}
where
\begin{equation}
\Ymat =	\frac{1}{2}
	\Bigg\{\sum_{i=1}^{n}\phi_i\Tr\rowvec{\frac{\partial^2{f_i}}{\partial{x}^2}\Big\rvert_{\xpost_{k-1}}\Cov_{k-1}^+} - \Kmat_k\sum_{i=1}^{m}\phi_i\Tr\rowvec{\D_{k,i}\Cov_k^-}\Bigg\} \ .
\end{equation}
Recalling the partial-update expression for the states,
\begin{equation}
	\xpost_k=\xprior_{k}+(\Identity-\Dmat)\stdvec{K}[][k](\ym_k-\yhat_k) \ ,
\end{equation}
and expressing it in terms of the function dynamics, measurement function (for the same assumed system in the EKF2) and expanding it, leads to
\begin{align}
		\xpost_k&=\xprior_{k}+(\Identity-\Dmat)\stdvec{K}[][k](\ym_k-\yhat_k)\\
		&=\xprior_k+
		\stdvec{K}[][k](\ym_k-\yhat_k)-	
		\Dmat\stdvec{K}[][k](\ym_k-\yhat_k)\\
		&=f(\xprior_{k-1},\u_{k-1},k-1)   + \stdvec{K}[][k](\ym_k-\yhat_k)-	
		\Dmat\stdvec{K}[][k](\ym_k-\yhat_k)\\
		&=f(\xprior_{k-1},\u_{k-1},k-1)   + 	\Kmat_k\rowvec{\ym_k-h(\xprior_{k \ ,}\ k)}-	
		\Dmat	\Kmat_k\rowvec{\ym_k-h(\xprior_{k \ ,}\ k)} \label{eq:long_partial_update__posterior_state} \ .
\end{align}
Interestingly, a direct term-by-term comparison of the partial-update expression from Equation (\ref{eq:long_partial_update__posterior_state}) and Equation (\ref{eq:ekf2_posterior_state}), reveals that the term with the partial-update weights,
\begin{equation}
-\Dmat\Zmat := -\Dmat	\Kmat_k\rowvec{\ym_k-h(\xprior_{k \ ,}\ k)} \ ,
\end{equation}
can be directly related to second-order terms of the EKF2. That is,
\begin{equation} \label{eq:dyn_gamma_2nd_effects_relation}
-	\Dmat	\Kmat_k\rowvec{\ym_k-h(\xprior_{k \ ,}\ k)} \propto \frac{1}{2}
\Bigg\{\sum_{i=1}^{n}\phi_i\Tr\rowvec{\frac{\partial^2{f_i}}{\partial{x}^2}\Big\rvert_{\xpost_{k-1}}\Cov_{k-1}^+} - \Kmat_k\sum_{i=1}^{m}\phi_i\Tr\rowvec{\D_{k,i}\Cov_k^-}\Bigg\} \ ,
\end{equation}
or
\begin{equation}\label{eq:dyn_gamma_2nd_effects_relation_withY}
	-	\Dmat	\Kmat_k\rowvec{\ym_k-h(\xprior_{k \ ,}\ k)} \propto \Ymat \ .
\end{equation}
Further, Equation (\ref{eq:dyn_gamma_2nd_effects_relation_withY}) follows the previously discussed idea of selecting $ \gamma_i \in [0,1]  $ values according to the second-order terms influence, since it suggests that:
\begin{itemize}
	\item $ \Dmat $ should be set with high values if the second-order effects, $ \Ymat $, are large.
	\item If $ \Ymat $ is small, $ \Dmat $ is to be set with small values.
\end{itemize}
Noticing that both left and right term of expression (\ref{eq:dyn_gamma_2nd_effects_relation}) are $ n\times 1 $ vectors, individual relationships between second-order terms and the $ j^{th} $ partial-update weights can be established as,
\begin{equation}
	\Dmat_{jj} \propto \frac{\frac{1}{2}
		\Bigg\{\sum_{i=1}^{n}\phi_i\Tr\rowvec{\frac{\partial^2{f_i}}{\partial{x}^2}\Big\rvert_{\xpost_{k-1}}\Cov_{k-1}^+} - \Kmat_k\sum_{i=1}^{m}\phi_i\Tr\rowvec{\D_{k,i}\Cov_k^-}\Bigg\}_j}{-\Kmat_k\rowvec{\ym_k-h(\xprior_{k \ ,}\ k)}_j} \ ,
\end{equation}
or alternatively,
\begin{equation}
	\Dmat_{jj} \propto \frac{{\Ymat_j}}{\Zmat_j} \ .
\end{equation}
To the end of keeping the $ \gamma $ values within the appropriate bounds, the ratio $ \frac{{\Ymat_j}}{\Zmat_j} $ is saturated to a maximum value of 1. In theory, $ \Zmat_j $ can be equal to zero when the measurement is equal to the expected measurement value, or when the $ i^{th} $ state is known with no uncertainty. But in any case, $ \Dmat $ will take its maximum value of 1. Since only the magnitude of second-order terms is considered, absolute values are taken and an scale factor $ f_r $ is introduced. This gives rise to the equation
\begin{equation}\label{eq:gamma_selection_method1}
		\Dmat_{jj} = f_r \frac{\lvert{\Ymat_j}\rvert}{\lvert\Zmat_j\rvert} \ , 
\end{equation}
which in terms of $ \betamat $ reads
\begin{equation}
	\betamat_{jj} = 1 -  \frac{\lvert{\Ymat_j}\rvert}{\lvert\Zmat_j\rvert} \ .
\end{equation}

From the expressions found through the construction of this nonlinearity-aware method, it is important to highlight the following. First, no assumption on the organization of the states within the filter has been made during the derivation of Equation (\ref{eq:gamma_selection_method1}) such that there is no partition to separate considered states from core states. Thus, Equation (\ref{eq:gamma_selection_method1}) indicates that the partial-update technique could be interpreted as a technique that, to some degree, compensates for second-order effects. Further, the partial-update technique, when using this nonlinearity-aware method for weights selection, can be seen, at least in part, mimicking second-order effects via the term $ \Dmat    \Kmat_k\rowvec{\ym_k-h(\xprior_{k \ ,}\ k)} $ to improve filter behavior. Second, since the motivation behind this weight selection approach is to use the degree of local nonlinearity as an indication of the validity of the linearization, the signs of the quantities are ignored, and the metric is based solely on nonlinear effects magnitudes. Finally, the scale factor $ \f_{r,j} $ is a design variable and is used to make the nonlinearity metric adjustable to the problem in question. As per experiments done with this $ \beta $ selection method, an adaptive scale factor $ f_{r,i} $ (for the $ i^{th} $ partially updated state) that involves the measurement residual and the state uncertainty state covariance was found to be convenient. Specifically, the ratio of the measurement residual covariance trace to the measurement noise covariance trace was used. Mathematically,
\begin{equation}
f_{r,i}=\frac{\sigma_{k,i}}{\sigma_{o,i}}\frac{\Tr(\Hmat_k\Cov_k\Hmat_k\trans+\Rk)}{\Tr(\Rk)} \ .
\end{equation}
The reason for including the ratio of the traces is to weight the second-order terms-effect commensurate to model mismatch. Moreover, the current standard deviation (which appears normalized by the initial value) is included to increase the impact of residual errors when the system uncertainty is high. Additionally, the ratio of the traces varies from infinity (depending on the uncertainty value) to 1, which provides an adequate scale for the filter steady state. In the case of the residual covariance being large, it will make the filter lower the magnitude of the corresponding update.

Although the computation of second-order terms is required to implement this method, they are only used to approximate a nonlinearity measure to assist with weight selection. In a worst-case scenario, an incorrect second-order term can lead to the temporary use of a conventional EKF or a Schmidt filter, which is not critical if linearization errors are within the bounds of what the filter could support for the next time updates. However, this is system dependent, and safety measures may be needed to prevent filter failure. For systems with constant or slowly varying parameters, for example, although conservative, the use of a Schmidt or partial-update could be safer than a full update, but again, this is system dependent.



In the next sections numerical simulations that exercise the nonlinearity-aware partial-update method (DNL) are presented.

\subsection{Numerical example}
\begin{figure}[h!]
	\centering
	\includegraphics[width=\dynStateWidth\textwidth]{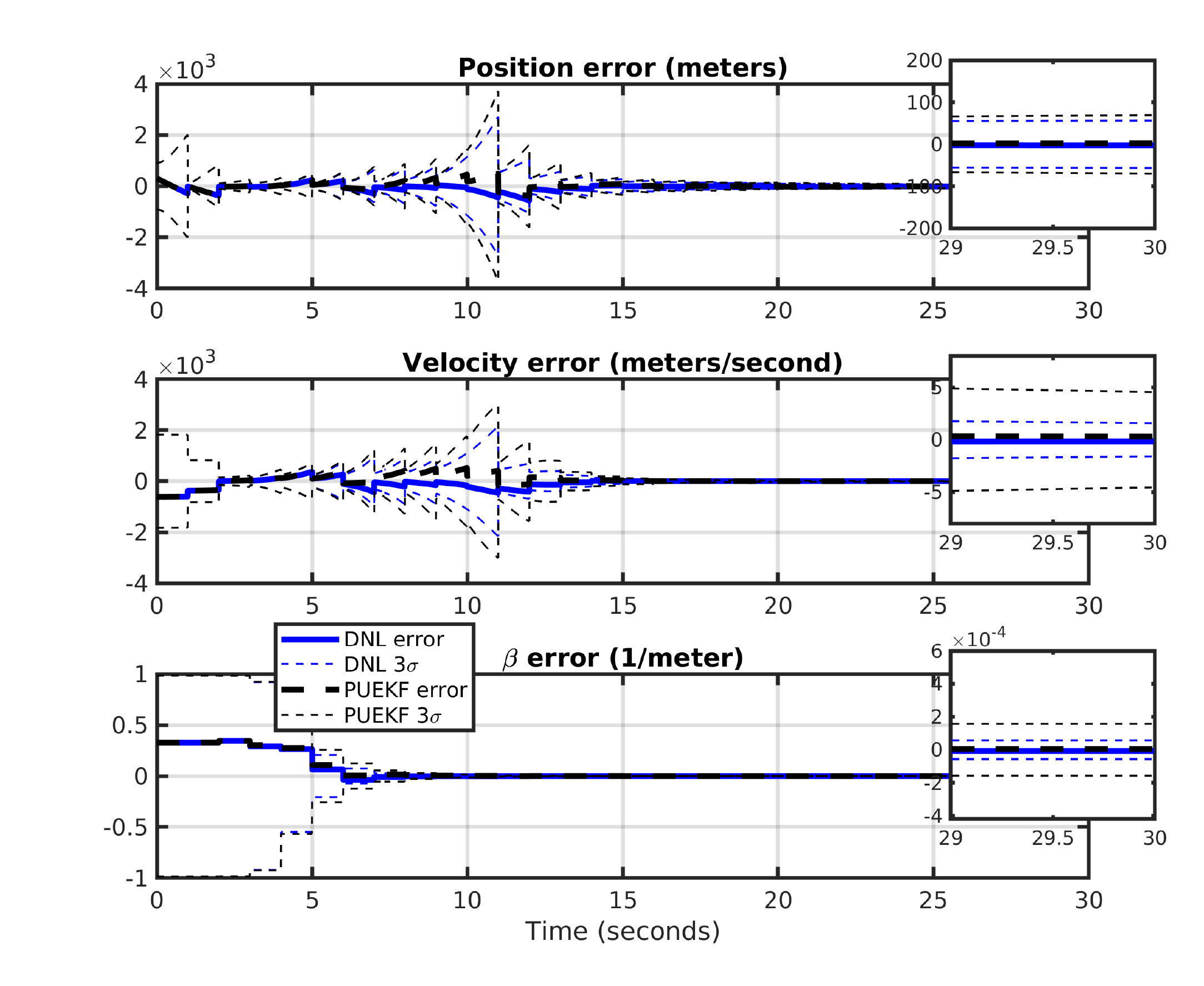}
	\caption{Dynamic nonlinearity-aware method (DNL) without previous tuning, static partial-update with $\betamat=\rowvec{0.9, 0.9, 0.75}$ and conventional EKF single run for body re-entry problem. Initial error within 1$ \sigma $.}
	\label{fig:dyn_DNL_PU_no_baseline}
\end{figure}

\begin{figure}[h!]
	\centering
	\includegraphics[width=\dynBetaWidth\textwidth]{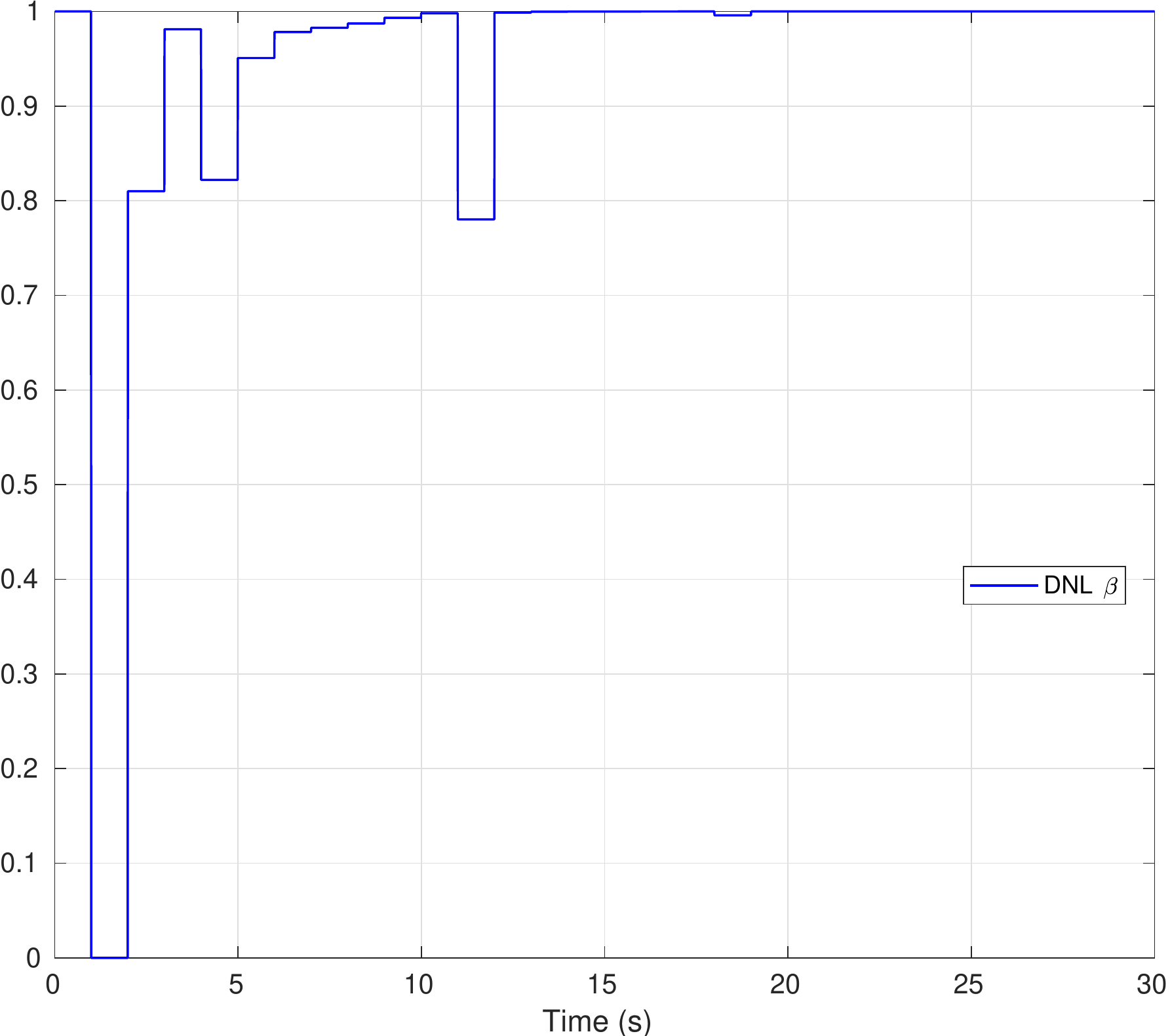}
	\caption{Dynamic nonlinearity-aware method (DNL) without previous tuning, static partial-update with $\betamat=\rowvec{0.9, 0.9, 0.75}$ and conventional EKF single run for body re-entry problem. Initial error within 1$ \sigma $.}
	\label{fig:dyn_DNL_PU_no_baseline_beta_history}
\end{figure}

Again, consider the now familiar falling body that is re-entering Earth. As before, the motion is considered to be constrained to a vertical line, and the filter is used to estimate the body altitude, velocity, and the body's ballistic parameter. If the filter parameters are maintained exactly the same as in previous sections, a typical run is as the one from Figure \ref{fig:dyn_DNL_PU_no_baseline}. The $ \beta $ partial-update weight history is presented in Figure \ref{fig:dyn_DNL_PU_no_baseline_beta_history}. From the states plot for the single run, it can be observed that the dynamic partial-update performs at the level of the static partial-update; however, the DNL method covariance is tighter. The reason for the lower covariance is that overall, the DNL method used more of the nominal update than the static partial-update set with $\betamat=\rowvec{0.9, 0.9, 0.75}$. The DNL method, however, does not require experimentation for tuning or selecting initial $ \beta $. Effectively, the DNL method attempts to make use of most of the information while using state and residual covariances to scale the update percentage commensurate to second-order terms effects. 

In Figure \ref{fig:dyn_DNL_PU_no_baseline_beta_history}, it is important to note that the filter weight adaptation is capturing two key instants and changing $ \beta $ as to maintain consistent estimates. Firstly, at time $ t=1 s$, when the first measurement is assimilated, the filter acts with a considered step since large state uncertainty is present, and the measurement residual is large. Secondly, at time $ t =11 s $ after the filter regains information (after losing observability due to range sensor and body alignment), the filter reduces the update since model mismatch, state uncertainty, and measurement residual increase. Nevertheless, the $\beta $ history allowed estimates to remain within the appropriate bounds and let the filter converge. Although the static partial-update approach achieves comparable results to the dynamic method, a considerable amount of tuning and experimentation was required, whereas the DNL method is run without a trial and error process needed. Also, recall that the conventional EKF is divergent for the run shown in Figure \ref{fig:dyn_DNL_PU_no_baseline}; the plots are not included.

\begin{figure}[h!]
	\centering
	\includegraphics[width=\dynStateWidth\textwidth]{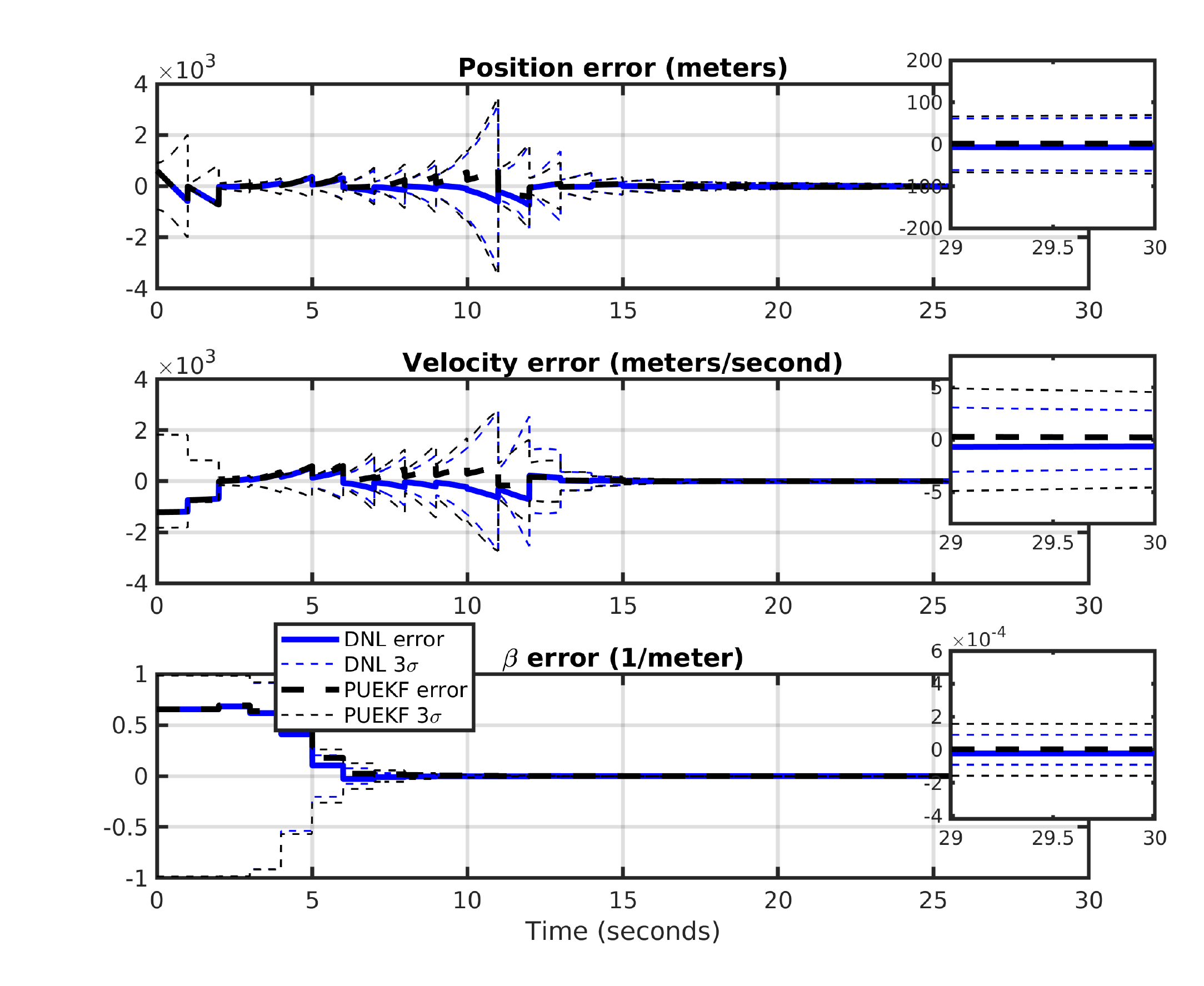}
	\caption{Dynamic nonlinearity-aware method (DNL) without previous tuning, static partial-update with $\betamat=\rowvec{0.9, 0.9, 0.75}$ and conventional EKF single run for body re-entry problem. Initial error within 2$ \sigma $.}
	\label{fig:dyn_DNL_PU_no_baseline2_sigma}
\end{figure}

\begin{figure}[h!]
	\centering
	\includegraphics[width=\dynBetaWidth\textwidth]{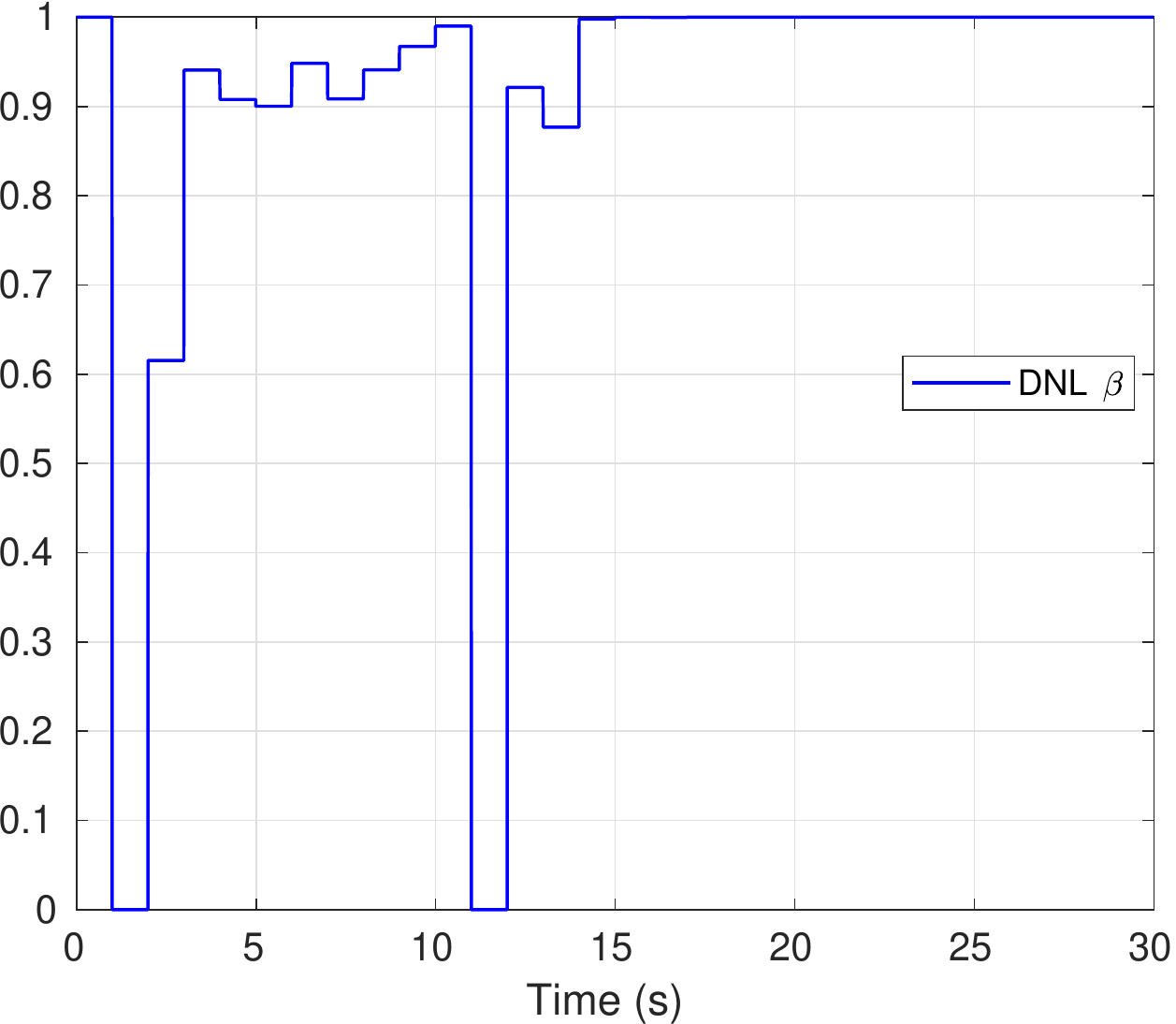}
	\caption{Dynamic nonlinearity-aware method (DNL) without previous tuning, static partial-update with $\betamat=\rowvec{0.9, 0.9, 0.75}$ and conventional EKF single run for body re-entry problem. Initial error within 2$ \sigma $.}
	\label{fig:dyn_DNL_PU_no_baseline_beta_history2_sigma}
\end{figure}

The results for a second run that is initialized with larger state errors are depicted in Figure \ref{fig:dyn_DNL_PU_no_baseline2_sigma}. For this scenario, the filters are also observed to provide consistent estimates for both static and dynamic partial-update filters. The corresponding $ \beta $ history is shown in Figure \ref{fig:dyn_DNL_PU_no_baseline_beta_history2_sigma}. This profile, compared to the $ \beta $ history of Figure \ref{fig:dyn_DNL_PU_no_baseline_beta_history} is seen to undergo more aggressive changes due to the higher initial errors. Nonetheless, the filter can handle such a scenario and maintain the estimates within the 3$ \sigma $ bounds.

\begin{figure}[h!]
\centering
\includegraphics[width=0.8\textwidth]{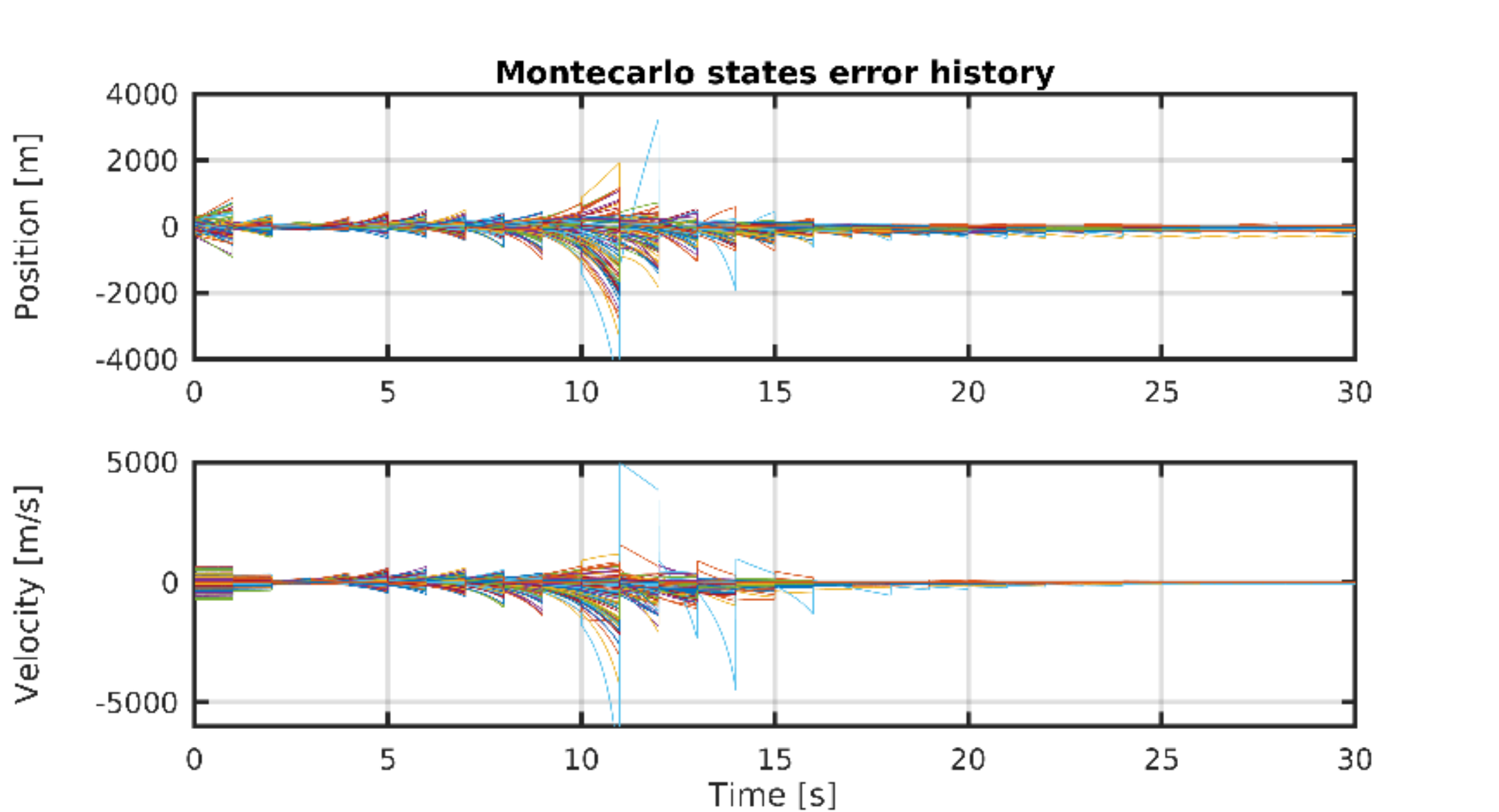}
\caption{Monte Carlo runs state histories for the DNL partial-update. }
\label{fig:Dynchapter_NO_baseline_STATESmonte_falling_body_DNLPUEKF_100_runs-eps-converted-to.pdf}
\end{figure}

\begin{figure}[h!]
\centering
\includegraphics[width=0.7\textwidth]{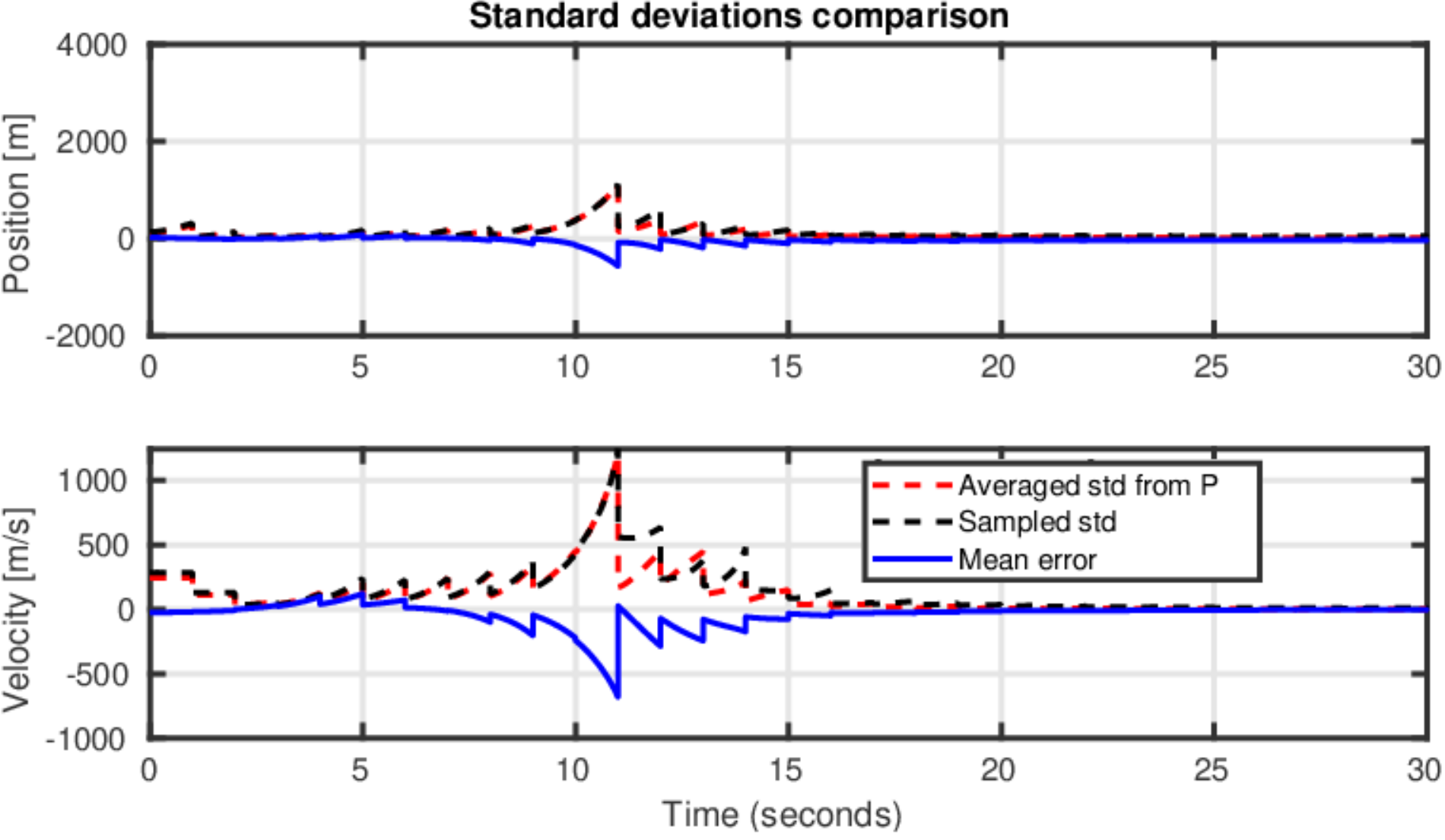}
\caption{Averaged and sampled standard deviation from 100 Monte Carlo runs for the DNL partial-update. The mean error is also shown. Full update is used as nominal value.}
\label{fig:Dynchapter_NO_baseline_monte_falling_body_DNLPUEKF_100_runs}
\end{figure}

The DNL method for $ \beta $ selection, however, as any other filter has its limits, and it cannot provide infinite immunity to initial errors if no additional information is provided to the filter (see subsection \ref{sec:previous_tuning_as_baseline} ). For the re-entry body problem, under the simulation parameters selected, the DNL method has been found to support initial errors up to the equivalent of 2 $ \sigma$. However, the maximum error amount that can be supported is a function of system configuration, including initial uncertainties, initial conditions, measurement frequency, and models nonlinearities. Figure \ref{fig:Dynchapter_NO_baseline_STATESmonte_falling_body_DNLPUEKF_100_runs-eps-converted-to.pdf}  shows the results of 100 runs for this problem for errors lower than 2 $ \sigma $ for position and velocity and for up to 3 $ \sigma $ for the ballistic parameter. Figure \ref{fig:Dynchapter_NO_baseline_monte_falling_body_DNLPUEKF_100_runs} displays the comparison between the estimated and sampled standard deviation. The purpose of this runs is to show that the DNL method performs well under the given initial uncertainties, not just for a single run. From Figure \ref{fig:Dynchapter_NO_baseline_STATESmonte_falling_body_DNLPUEKF_100_runs-eps-converted-to.pdf}, it can be observed that the DNL method is slightly overconfident, which can be due to the filter updates overreaction due to nonlinearities being involved causing the uncertainty to be slightly tighter than it needs to be. Nonetheless, its superior consistency over the EKF is clear since no divergent cases are seen.

\subsection{Pre-tuned partial-update weights as a baseline for DNL method}\label{sec:previous_tuning_as_baseline}
Although, advantageously, the DNL method can provide an operational filter for cases where no prior information on functional $ \beta $ percentages are available (especially for large systems), the DNL method is also able to accommodate known $ \beta $ values. To incorporate known $ \beta $ values, one shifts the nominal update baseline of the DNL method to be the $ \beta $ values, rather than a full update (or $ \beta=1 $). This hybrid method, although effective, it requires tuning and it is over-conservative. In general, dynamic methods alone are recommended and sufficient. The discussion on this hybrid method is presented for the sake of completeness.

\begin{figure}[h!]
	\centering
	\includegraphics[width=\dynStateWidth\textwidth]{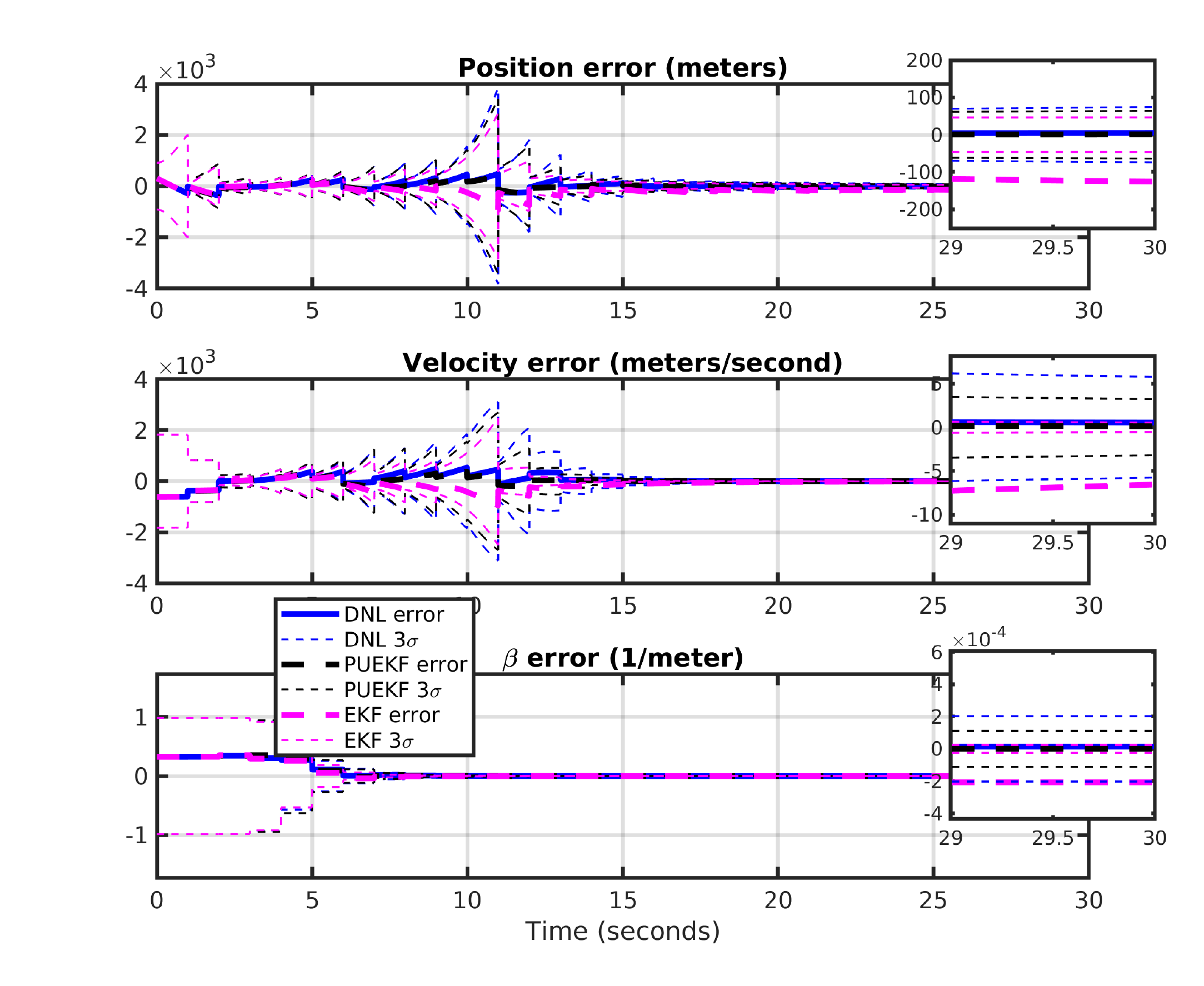}
	\caption{DNL method, static partial-update with $\betamat=[0.9, 0.9, 0.75]$, and conventional EKF single run for body re-entry problem. Initial error within 1$ \sigma $.}
	\label{fig:dyn_DNL_PU_EKF}
\end{figure}

\begin{figure}[h!]
	\centering
	\includegraphics[width=\dynStateWidth\textwidth]{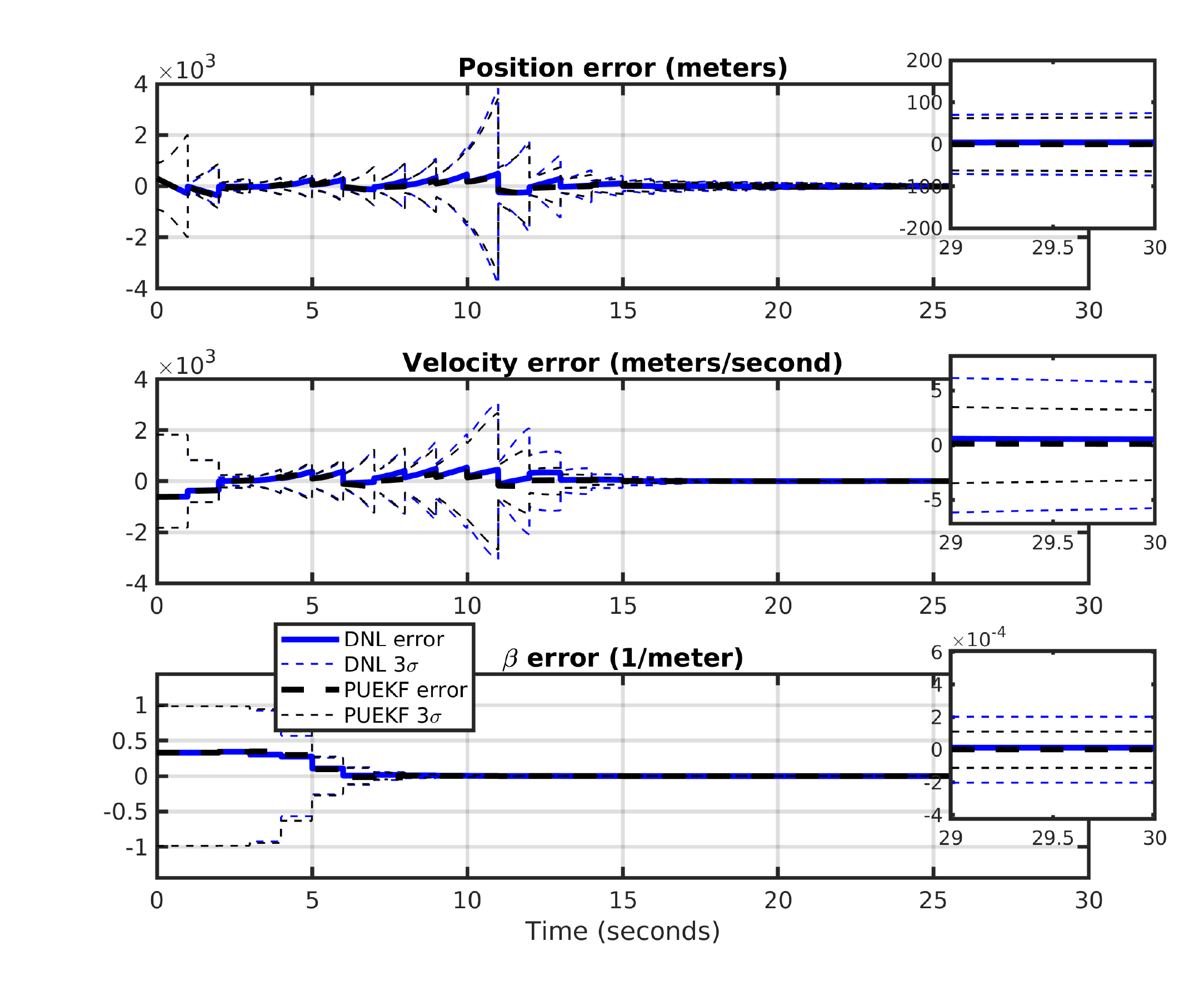}
	\caption{DNL method and static partial-update with $\boldsymbol{\beta}=[0.9, 0.9, 0.75]$ single run for body re-entry problem. Initial error within 1$ \sigma $.}
	\label{fig:dyn_DNL_PU}
\end{figure}

\begin{figure}[h!]
	\centering
	\includegraphics[width=\dynBetaWidth\textwidth]{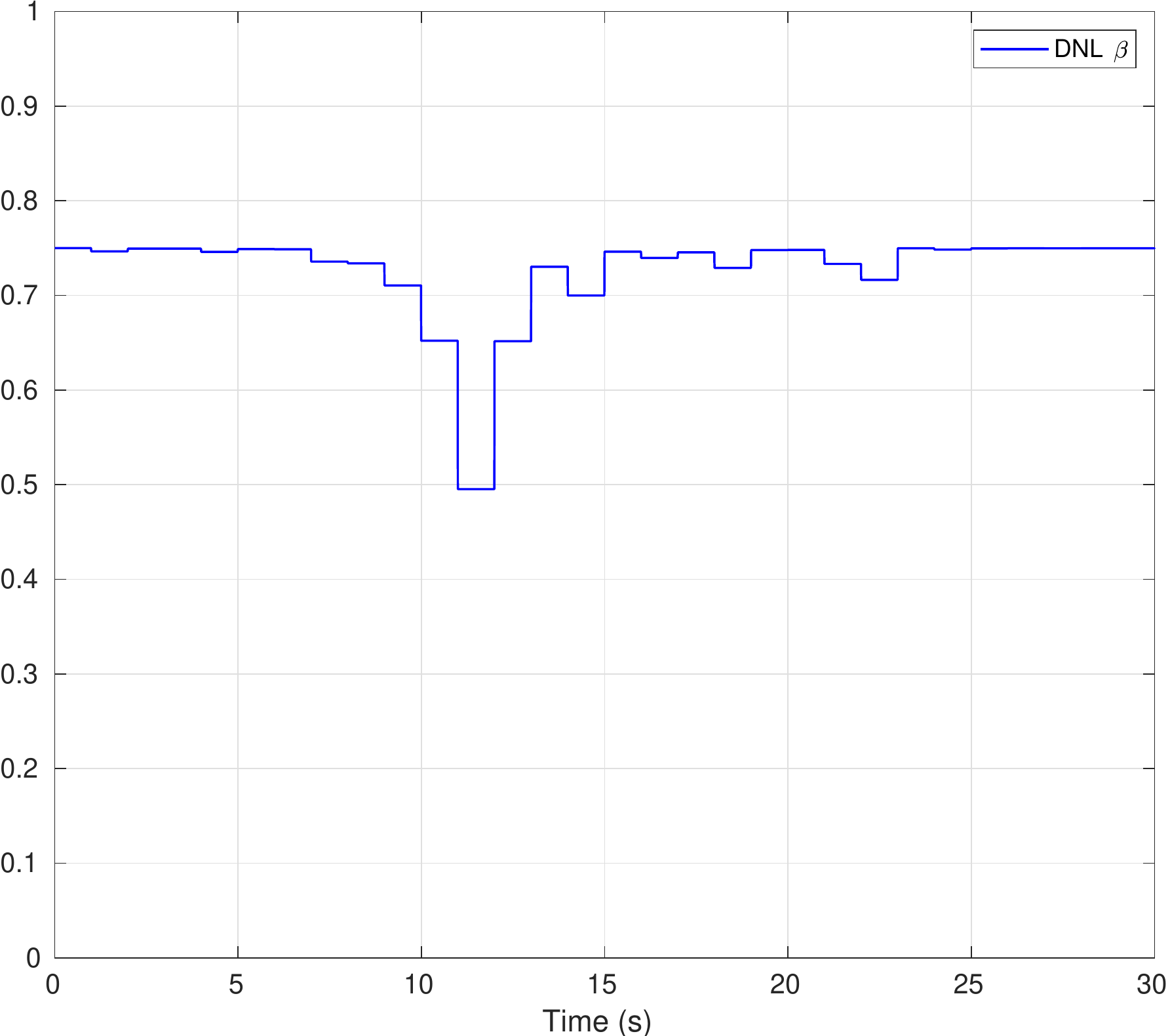}
	\caption{Dynamic nonlinearity-aware method (DNL) $ \beta $ history for the body re-entry problem. Initial error within 1$ \sigma $.}
	\label{fig:dyn_DNL_PU_beta_history}
\end{figure}

To exercise this concept, the same re-entry body problem is employed and the $ \beta $ value from the original example (1 $ \sigma $ initial error and same filter parameters) is taken as the weight baseline ($\betamat=\rowvec{0.9, 0.9, 0.75}$ ). Recall that the dynamic weight selection is only performed for the ballistic parameter. Figure \ref{fig:dyn_DNL_PU},  illustrates the filter estimates that employs the DNL method with tuned betas and the PU method. From such plots, it is apparent that the DNL has not done much to improve the system behavior, yet it has practically accomplished the same results as the static partial-update but with higher uncertainty. The increase of uncertainty is due to, on average, lower updates as observed from the $ \beta $ percentage profile depicted in Figure \ref{fig:dyn_DNL_PU_beta_history}. Nonetheless, this run intends to demonstrate the functionality of the dynamic nonlinearity-aware filter using previously selected (tuned) weights $ \beta $s. 
As a reference only, the conventional Kalman filter was run and plotted in Figure \ref{fig:dyn_DNL_PU_EKF} along with the partial-update methods with the only objective of recalling that it is an inconsistent filter already when the initial condition errors of 1 $ \sigma $. It is important to note from Figure \ref{fig:dyn_DNL_PU_beta_history}, that the $ \beta $ profile has overall changed with respect to the single DNL approach, but more importantly, a decrease in its fluctuation is observed. This is reasonable considering that a partial-update is applied throughout the run. This filter, in contrast with the ``pure'' DNL filter, leaves most of the work to the pre-tuned weight, mostly being nonreactive at the beginning of the run and after the recovery of measurement information after time $ t=11s $. In other words, whereas the DNL method with baseline one seems to start acting early on to handle the initial drag perturbations to maintain consistent estimates, the policy of the DNL base-lined at $\beta = 0.75$ appears more reactive only to instants when the system nonlinearities can have a larger impact due to higher uncertainty.

\begin{figure}[h!]
	\centering
	\includegraphics[width=\dynStateWidth\textwidth]{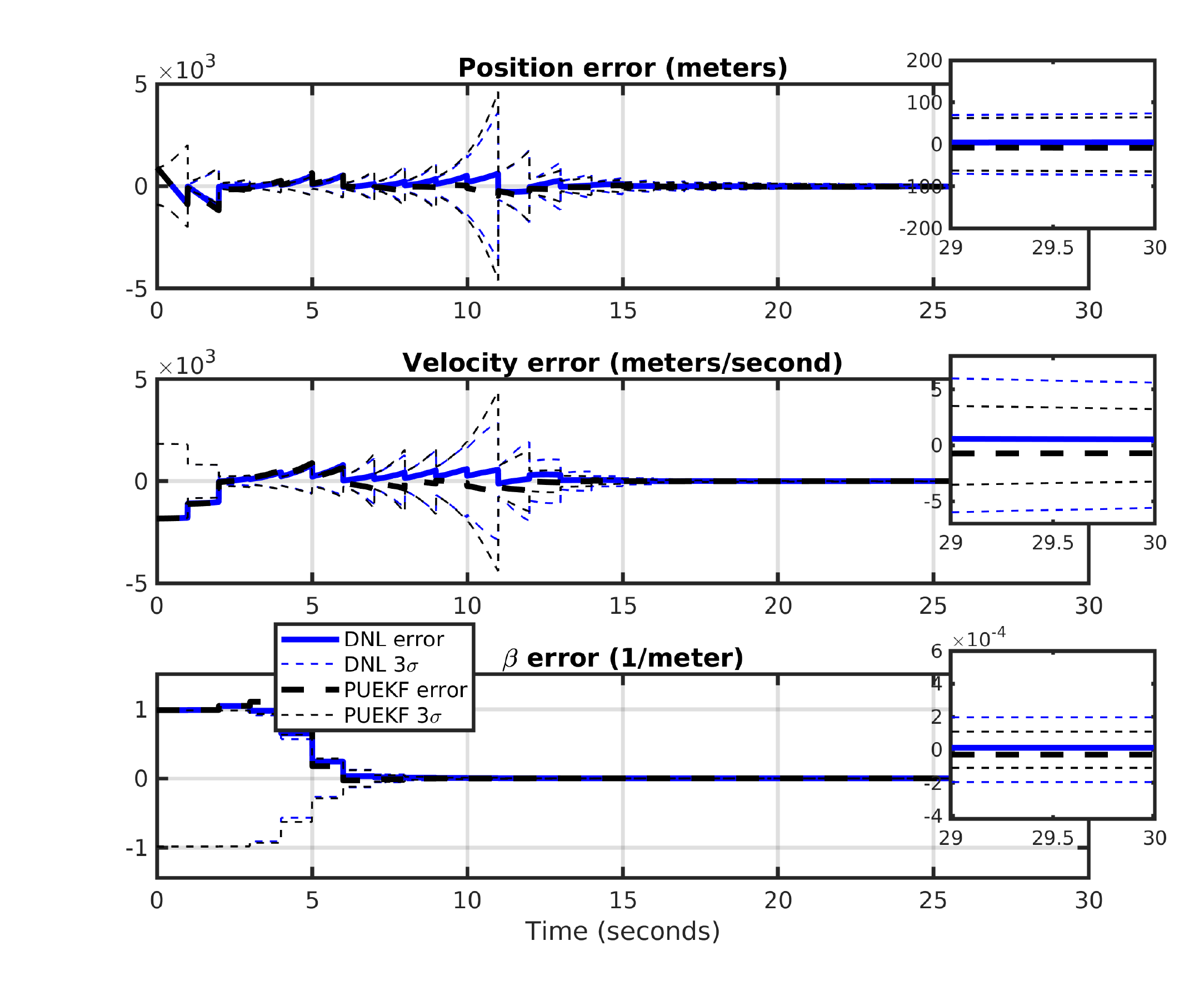}
	\caption{DNL method, static partial-update with $\boldsymbol{\beta}=[0.9, 0.9, 0.75]$ single run for body re-entry problem. 3$ \sigma $ initial errors.}
	\label{fig:dyn_DNL_PU_3sigma}
\end{figure}

\begin{figure}[h!]
	\centering
	\includegraphics[width=\dynBetaWidth\textwidth]{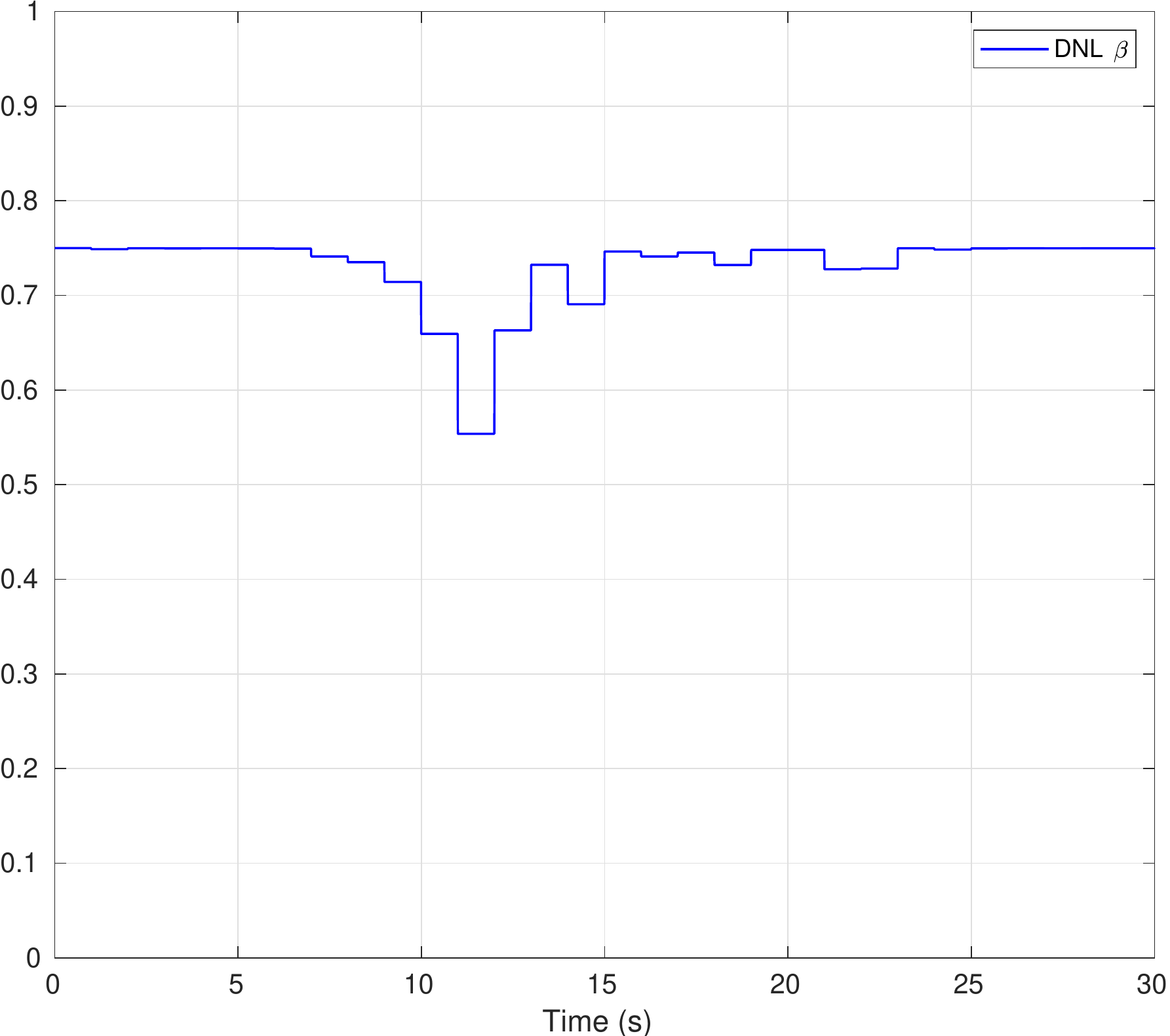}
	\caption{Dynamic nonlinearity-aware method (DNL) $ \beta $ history for the body re-entry problem. 3$ \sigma $ initial errors.}
	\label{fig:dyn_DNL_PU_beta_history_3sigma}
\end{figure}

Now, with the intent of stressing the DNL method further when it incorporates pre-tuned $ \beta $ values, two experiments are performed. First, the initial errors are increased to the equivalent of 3$ \sigma $ for every state. Second, the partial-update filters are required to use a higher percentage to attempt leveraging more measurement information. Recall that a full update causes inconsistent results even for 1 $ \sigma $ errors. Figure \ref{fig:dyn_DNL_PU_3sigma} depicts the results from the single run when a 3$ \sigma $ initial error is used. In addition, Figure \ref{fig:dyn_DNL_PU_beta_history_3sigma} shows the corresponding history for $ \beta $. Under this conditions, as per the estimation error plot from Figure \ref{fig:dyn_DNL_PU_3sigma}, although slight discrepancy is observed between the dynamic and static filter, the results are overall the same as for 1$ \sigma$ errors from Figure \ref{fig:dyn_DNL_PU} with the small difference that the ballistic parameter now appears barely less unbiased in favor to the DNL method. From the $ \beta $ history, it is seen that the dynamic filter required to use larger updates to compensate for the larger errors. The $ \beta $ profile also shows that it is still effectively reacting to high-order effects promptly. 

\begin{figure}[h!]
	\centering
	\includegraphics[width=\dynStateWidth\textwidth]{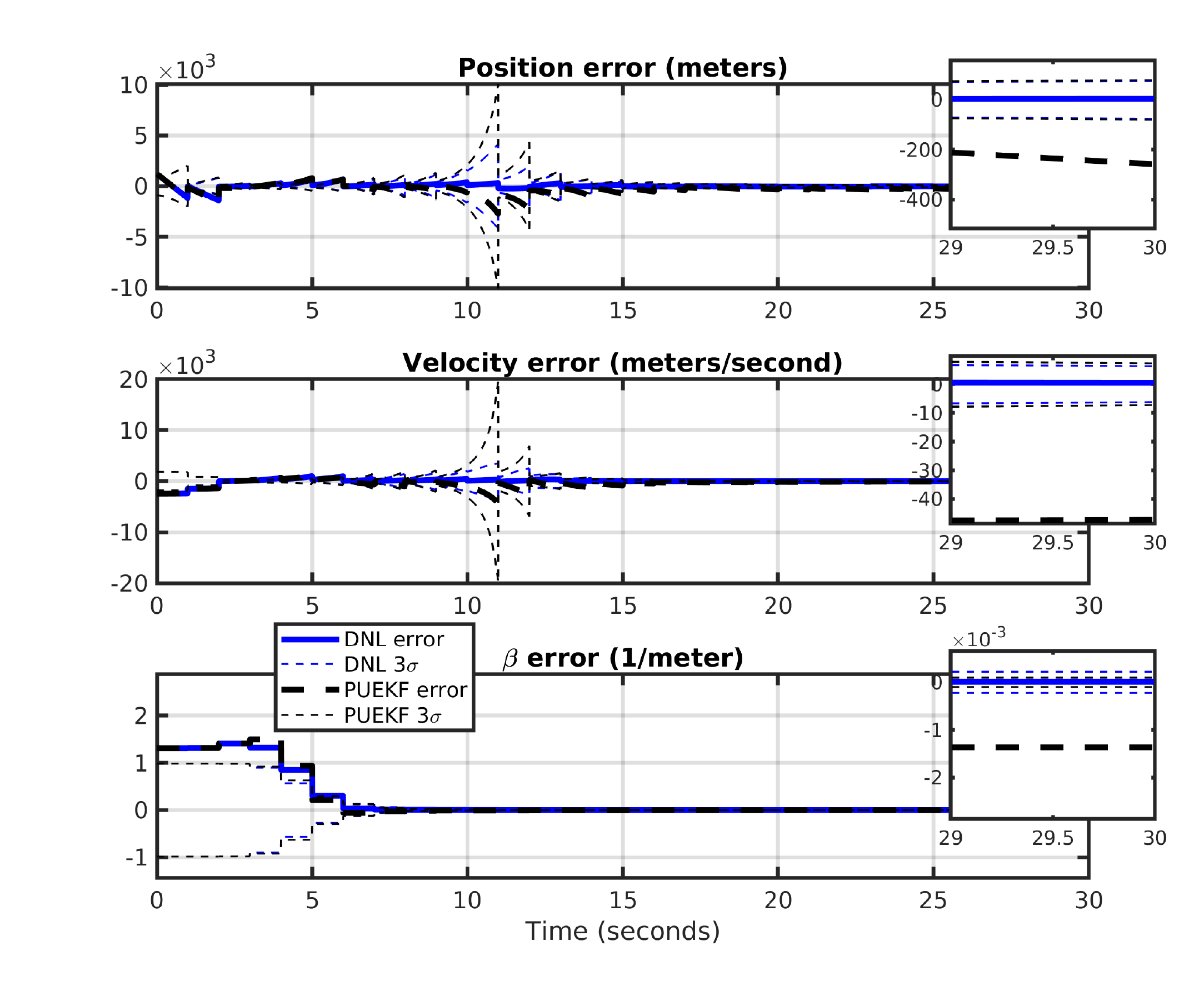}
	\caption{DNL method and static partial-update with $\boldsymbol{\beta}=[0.9, 0.9, 0.75]$ single run for body re-entry problem. 4$ \sigma $ initial errors.}
	\label{fig:dyn_DNL_PU_4sigma}
\end{figure}

\begin{figure}[h!]
	\centering
	\includegraphics[width=\dynBetaWidth\textwidth]{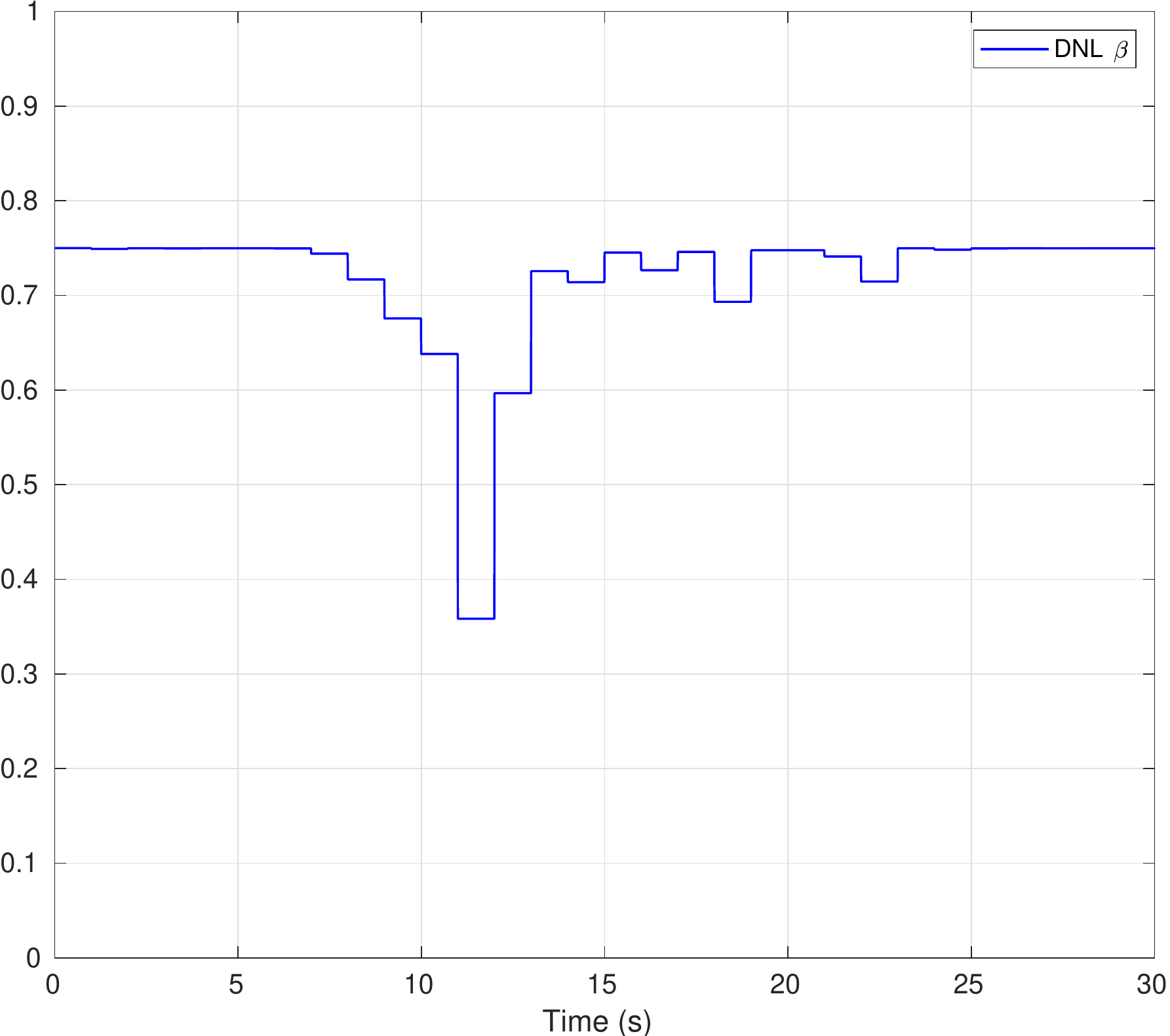}
	\caption{Dynamic nonlinearity-aware method (DNL) $ \beta $ history for the body re-entry problem. 4$ \sigma $ initial errors.}
	\label{fig:dyn_DNL_PU_beta_history_4sigma}
\end{figure}

\begin{figure}[h!]
	\centering
	\includegraphics[width=\dynBetaWidth\textwidth]{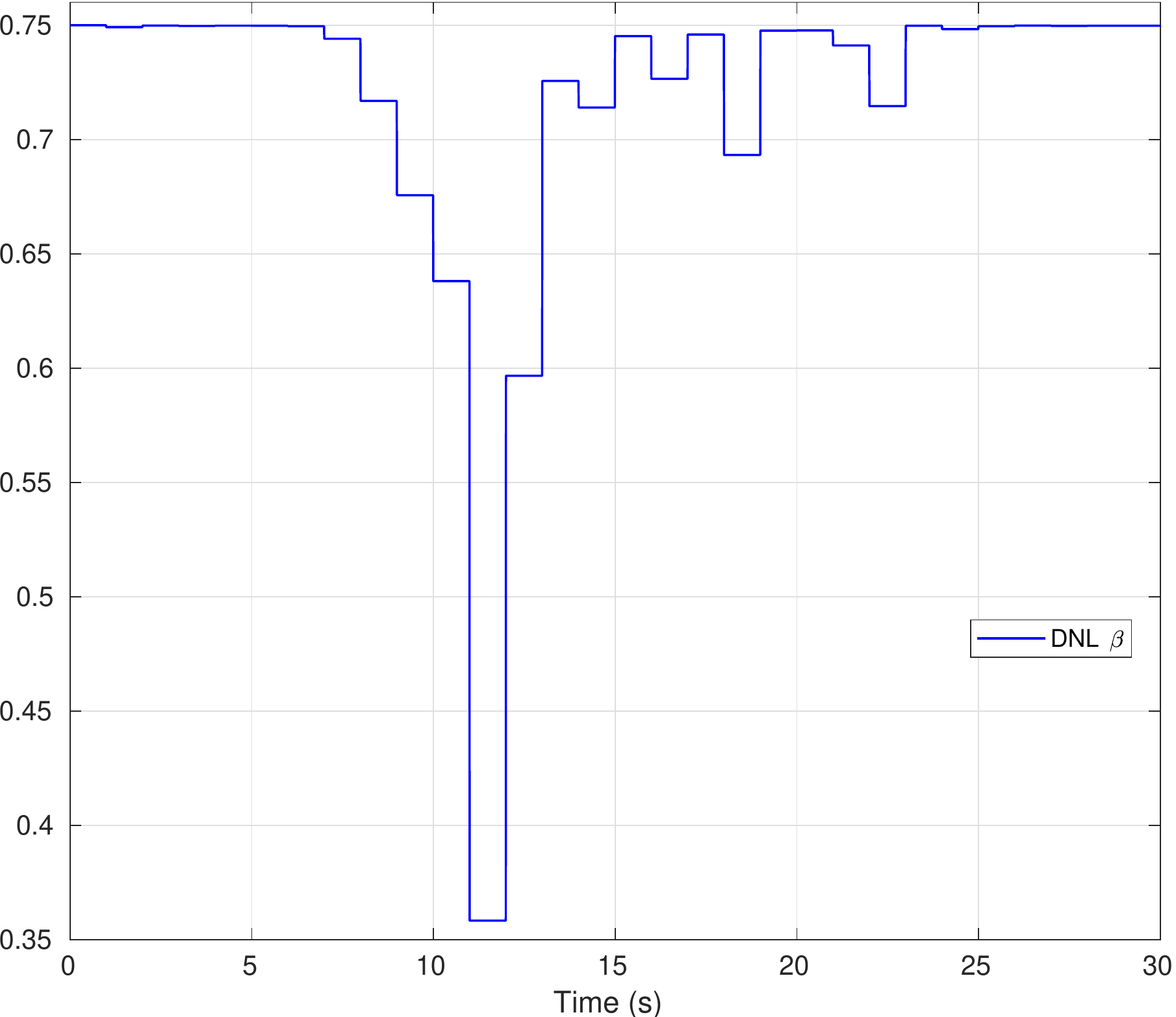}
	\caption{Zoomed-in view of dynamic nonlinearity-aware method (DNL) $ \beta $ history for the body re-entry problem. 4$ \sigma $ initial errors.}
	\label{fig:dyn_DNL_PU_beta_history_4sigma_zoomed}
\end{figure}

An additional experiment that uses initial conditions far in the tail of the assumed Gaussian distribution (4$ \sigma $) is performed. The results illustrated in Figure \ref{fig:dyn_DNL_PU_4sigma} and \ref{fig:dyn_DNL_PU_beta_history_4sigma}, clearly show inconsistency of the static partial-update for the experiment. This run shows the capability of the partial-update filter to deal with very high uncertainties if it is actively adapting the update percentages while having a pre-tuned $\beta$'s as the baseline. For this scenario, the $ \beta $ history, although it displays similar behavior to the two previous experiments, for this run the DNL method is observed to have employed, overall, lower updates as it requires to perform a larger second-order effects compensation. It is important to mention that the DLN not only executed a minimum update precisely when required at time $ t=11s $, but it also prevented filter overreactions due to linearization errors early on during the interval $ t \in [5,10] $. By no overreacting, the filter was able to recover and maintain consistent estimates as seen in Figure \ref{fig:dyn_DNL_PU_beta_history_4sigma_zoomed} (the zoomed-in version (in the interval $ t\in [5,15] \ s $) of Figure \ref{fig:dyn_DNL_PU_4sigma}). On the other hand, the second-order terms contributions are unknown to the static partial-update filter, which overshoots, and it eventually fails due to accumulated error mainly in the drag coefficient. By the time the range sensor becomes aligned (horizontally) with the position of the falling body at time $ t=10 $, the dynamic partial-update filter is far in a better \textit{position} than the static partial-update, and can handle the reacquisition at time $ t=11s $ and on, while the static partial-update errors are large enough such that the filter is unable to recover and eventually diverges.

\begin{figure}[h!]
	\centering
	\includegraphics[width=\dynStateWidth\textwidth]{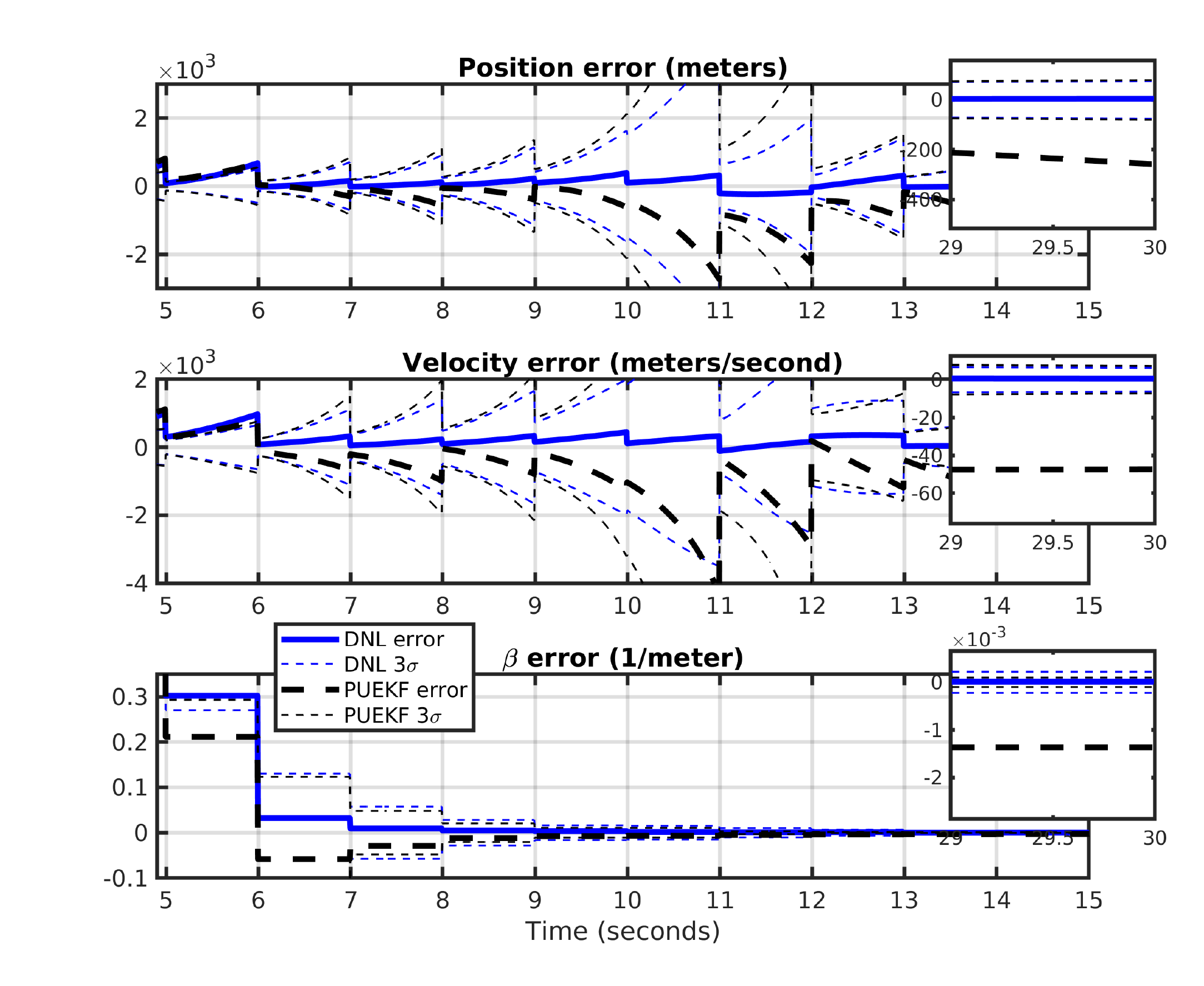}
	\caption{Zoomed-in view of dynamic nonlinearity-aware method (DNL), static partial-update with $\boldsymbol{\beta}=[0.9, 0.9, 0.75]$ single run for body re-entry problem. 4$ \sigma $ initial errors.}
	\label{fig:dyn_DNL_PU_4sigma_zoomed}
\end{figure}

\begin{figure}[h!]
\centering
\includegraphics[width=\dynStateWidth\linewidth]{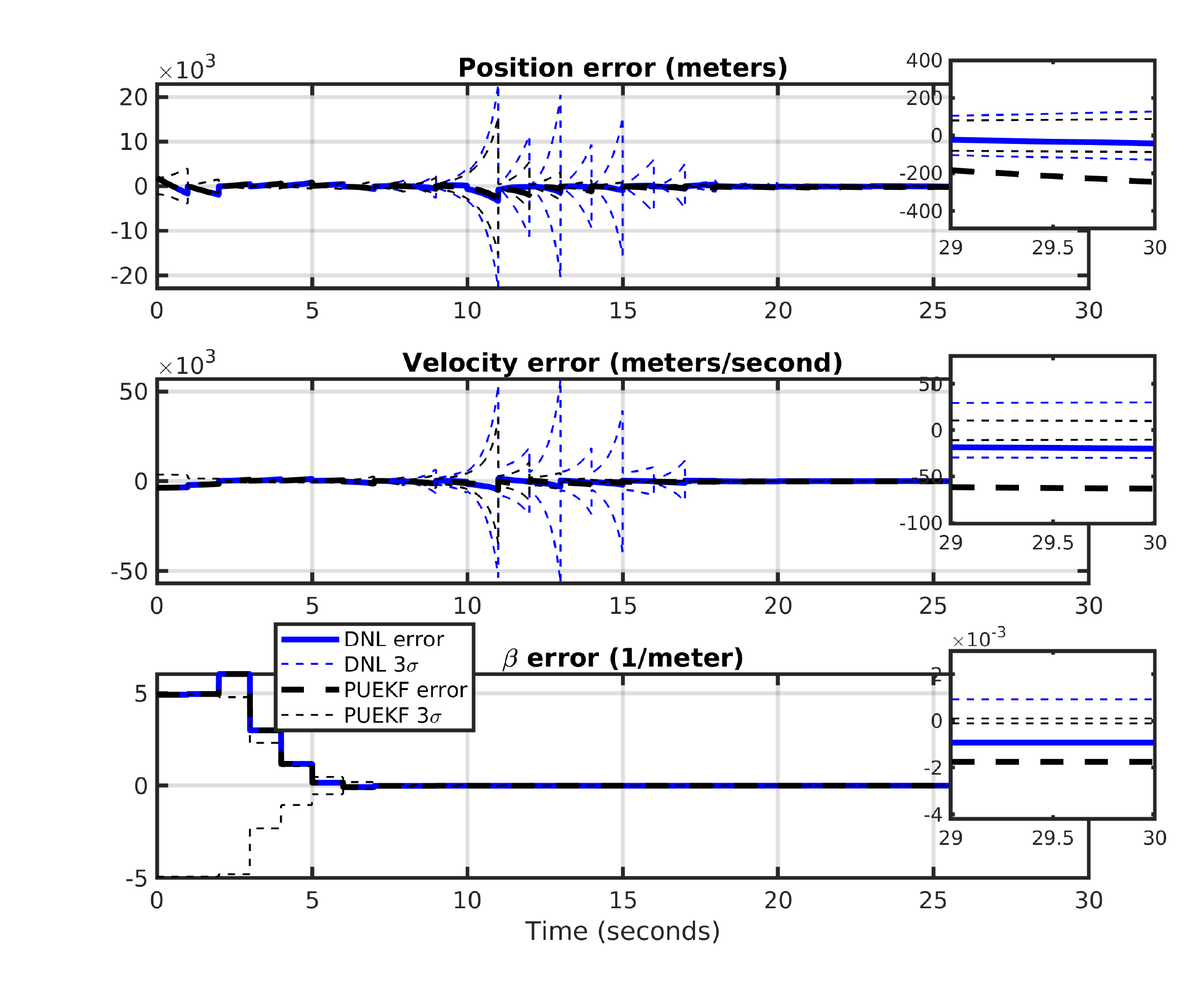}
\caption{Dynamic nonlinearity-aware method (DNL), static partial-update with $\boldsymbol{\beta}=[1, 1, 0.75]$ single run for body re-entry problem subject to higher initial uncertainties and initial errors.}
\label{fig:DNL_and_PU_very_high_uncertainties}
\end{figure}

\begin{figure}[h!]
\centering
\includegraphics[width=\dynBetaWidth\linewidth]{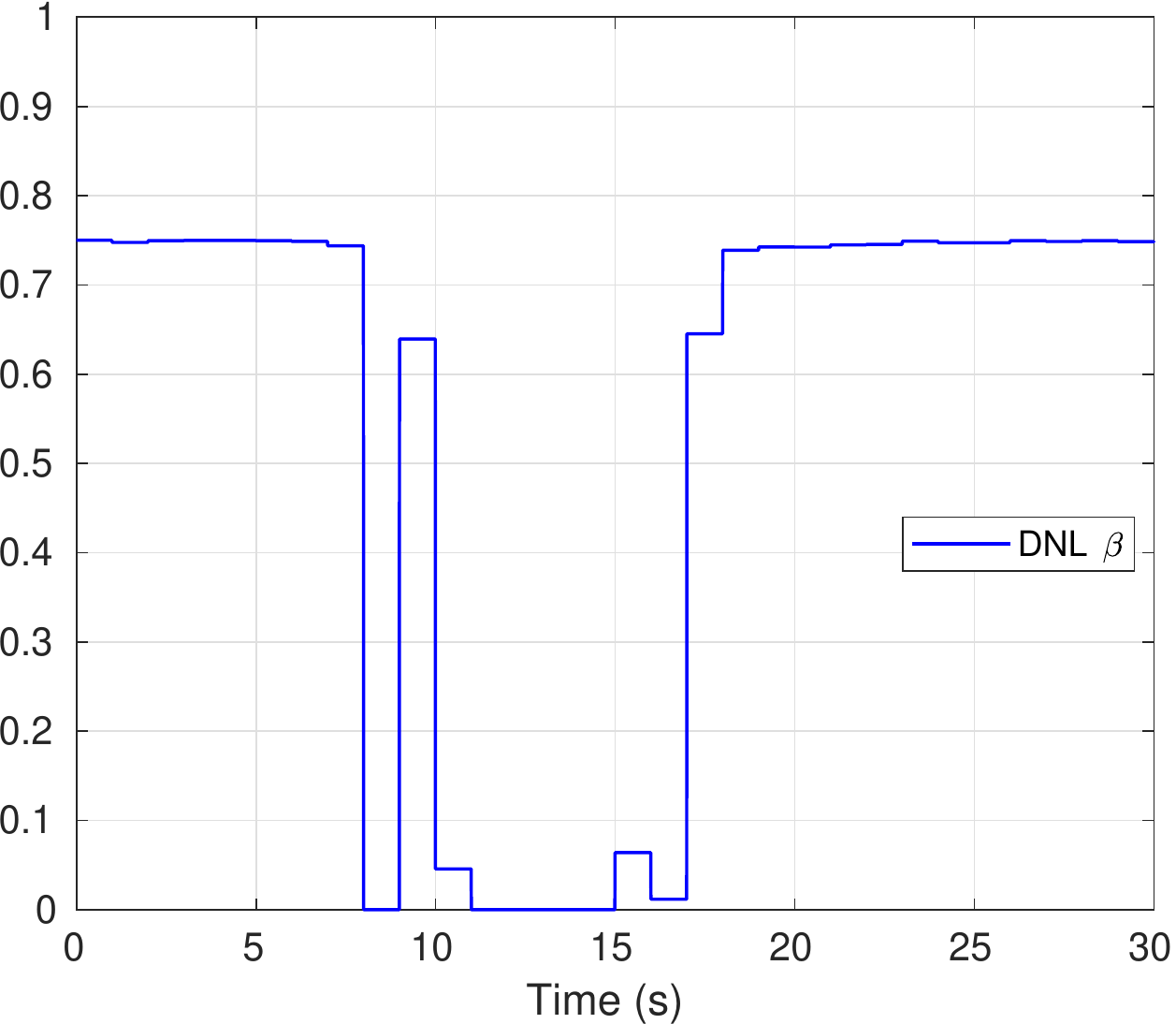}
\caption{Dynamic nonlinearity-aware method (DNL) $ \beta $ history for single run for the body re-entry problem subject to higher initial uncertainties and initial errors.}
\label{fig:DNL_and_PU_very_high_uncertainties_beta_history}
\end{figure}

The value of the partial-update weights were set to $\boldsymbol{\beta}=\rowvec{0.9, 0.9, 0.75}$ since by experimentation, they produced appropriate estimates for a wide variety of initial conditions (within the original 3 $ \sigma $). Although the static partial-update may improve by refining the $ \beta $ values, the DNL method was generally found to be superior in robustness and consistency, especially for scenarios with large uncertainties and initial errors. Figure \ref{fig:DNL_and_PU_very_high_uncertainties}, as an example, shows a single run of the re-entry body problem where the initial errors are maintained at 3 $ \sigma $ levels but in contrast with previous runs, the initial uncertainty has been doubled for position and velocity, and the uncertainty on the ballistic parameter was augmented five times. For this run, the static partial-update which now employs $\boldsymbol{\beta}=\rowvec{1, 1, 0.75}$, is observed to be outside the 3$ \sigma$ bounds for all of the estimation errors, whereas the dynamic $ \beta $ although not fully reaching zero error, the state errors are still within proper bounds for the single run and errors are better than for the static partial-update case. The main observation here is that although the static partial-update filter is enabled to perform larger updates for this experiment to allow it to recover ``faster'' if no adaptation of $ \beta $ is performed, the filter eventually updates considerably when the model mismatch is important, and this leads to estimates inconsistency. Figure \ref{fig:DNL_and_PU_very_high_uncertainties_beta_history} illustrates the \textit{value} of the adaptive method, as for the detected second-order terms the filter practically becomes a consider filter from time $ t=8s $ to $ t=17s $ to maintain an operational filter.

To the end of increasing the capabilities of the nonlinearity-aware method further, a second way for the $ \beta $'s selection is formulated. This alternative method, called the covariance-aware method, is based on monitoring second-order terms of the covariance update equation, but it follows the same idea as the nonlinearity-aware method. The covariance-aware method is presented next.

\section{Covariance-aware based method}
This section presents an alternative way of selecting the partial-update weights in an online fashion called Dynamic Covariance-aware partial-update, or DC for short. Paralleling the previous method that monitors second-order effects on the Kalman update terms, the method proposed in this section monitors second-order covariance terms. As before, the idea is to reduce the update magnitude when the ratio of high-order effects to first-order terms is important, and use more of the nominal update if second-order contributions are small. Although the construction of the expressions for selecting the partial-update weights is based on second-order terms appearing in a second-order hybrid Kalman filter, the dynamic covariance method works for the discrete filter as well. To obtain the expressions for the $ \beta  $ selection using the covariance-aware method, consider the covariance measurement update expression for the second-order Kalman filter (EKF2),

\begin{equation}\label{eq:NC_EKF2_cov}
	\Ppost_k=\Pprior_k-\Pprior_k\Hmat_k\trans(\Hmat_k\Pprior_k\Hmat_k\trans+ \R_k +\Lam_k)^{-1}\Hmat_k\Pprior_k \ ,
\end{equation}
where
\begin{equation}
	 \Lambda_k(i,j)=\frac{1}{2}\Tr(\D_{k,i}\Pprior_k\D_{k,j}\Pprior_k) \ , 
\end{equation}
and 
\begin{equation}
	 \D_{k,i} = \frac{\partial^2{h_i(\x_k,k)}}{\partial{x}^2}\Big\rvert_{\xprior_k} \ .
\end{equation}
Additionally, consider the covariance partial-update expression
\begin{equation}
\stdvec{P}[][k]^{++}=\Ppost_k + \Dmat\Pprior_k\Hmat_k\trans(\Hmat_k\Pprior_k\Hmat_k\trans+ \R_k)^{-1}\Hmat_k\Pprior_k\Dmat \ .
\end{equation}
This equation can be alternatively written in terms of the prior state covariance as,
\begin{equation}
\stdvec{P}[][k]^{++}=\Pprior_k - \Kk\Hmat_k\Pprior_k + \Dmat\Pprior_k\Hmat_k\trans(\Hmat_k\Pprior_k\Hmat_k\trans+ \R_k)^{-1}\Hmat_k\Pprior_k\Dmat \ ,
\end{equation}
and replacing the Kalman gain by $ \Kk=\Pprior_k\Hmat_k(\Hmat_k\Pprior_k\Hmat_k\trans+\R)^{-1} $ leads to
\begin{equation}\label{eq:ekf2_PU_II}
	\stdvec{P}[][k]^{++}=\Pprior_k - \Pprior_k\Hmat_k(\Hmat_k\Pprior_k\Hmat_k\trans+\R)^{-1}\Hmat_k\Pprior_k + \Dmat\Pprior_k\Hmat_k\trans(\Hmat_k\Pprior_k\Hmat_k\trans+ \R_k)^{-1}\Hmat_k\Pprior_k\Dmat \ .
\end{equation}
Next, for the purposes of comparing the partial-update terms against second order uncertainty terms, the $ \Lam  $  term of Equation (\ref{eq:NC_EKF2_cov}) is first extracted from the parenthetical to produce the residual covariance term $ (\Hmat_k\Pprior_k\Hmat_k\trans+ \R_k)^{-1} $ . This is accomplished by applying the matrix inversion lemma. That is,
\begin{align}
	&(\Hmat_k\Pprior_k\Hmat_k\trans+ \R_k +\Lam_k)^{-1}=\\\nonumber&(\Hmat_k\Pprior_k\Hmat_k\trans+ \R_k)^{-1}-\\\nonumber
	&(\Hmat_k\Pprior_k\Hmat_k\trans+ \R_k)^{-1}\Lam_k[(\Hmat_k\Pprior_k\Hmat_k\trans+ \R_k)^{-1}\Lam_k+\Identity]^{-1}(\Hmat_k\Pprior_k\Hmat_k\trans+ \R_k)^{-1} \ .
\end{align}
Substituting this expression into the EKF2 update covariance of Equation (\ref{eq:NC_EKF2_cov}), results in
\begin{align}\label{eq:ekf2_IV}
&\stdvec{P}[][k]^{++}=\\\nonumber 
&\Pprior_k - \Pprior_k\Hmat_k(\Hmat_k\Pprior_k\Hmat_k\trans+\R)^{-1}\Hmat_k\Pprior_k+\\\nonumber
&\Pprior_k\Hmat_k\trans  (\Hmat_k\Pprior_k\Hmat_k\trans+ \R_k)^{-1}\Lam_k[(\Hmat_k\Pprior_k\Hmat_k\trans+ \R_k)^{-1}\Lam_k+\Identity]^{-1}(\Hmat_k\Pprior_k\Hmat_k\trans+ \R_k)^{-1}    \Hmat_k\Pprior \ .
\end{align}

By doing a term-by-term comparison of the partial-update expression of Equation (\ref{eq:ekf2_PU_II}), and the EKF2 update for the error state covariance of Equation (\ref{eq:ekf2_IV}), the following relationship between second-order covariance effects and partial-update terms can be established,
\begin{align}\label{eq:efk2_first_relationship_V}
	&\Dmat\Pprior_k\Hmat_k\trans(\Hmat_k\Pprior_k\Hmat_k\trans+ \R_k)^{-1}\Hmat_k\Pprior_k\Dmat \sim \\ \nonumber
	&\Pprior_k\Hmat_k\trans  (\Hmat_k\Pprior_k\Hmat_k\trans+ \R_k)^{-1}\Lam_k[(\Hmat_k\Pprior_k\Hmat_k\trans+ \R_k)^{-1}\Lam_k+\Identity]^{-1}(\Hmat_k\Pprior_k\Hmat_k\trans+ \R_k)^{-1}    \Hmat_k\Pprior \ ,
\end{align}
where the symbol $ \sim $, is to indicate that terms are related.
Equation (\ref{eq:efk2_first_relationship_V}) can be compactly written as
\begin{equation}\label{eq:ekf2_relationship_V_reduced}
	\Dmat\delta\Pprior_k\Dmat \sim \Kk\Lam_k[(\Hmat_k\Pprior_k\Hmat_k\trans+ \R_k)^{-1}\Lam_k+\Identity]^{-1}\Kk\trans \ ,
\end{equation}
where $ \delta\Pprior_k=\Pprior_k\Hmat_k\trans(\Hmat_k\Pprior_k\Hmat_k\trans+ \R_k)^{-1}\Hmat_k\Pprior_k $. To further simplify relation (\ref{eq:ekf2_relationship_V_reduced}), the matrix on the right is condensed in just one matrix called $ \Nmat $, such that the relationship now reads
\begin{equation}
	\Dmat\delta\Pprior_k\Dmat \sim \Nmat_k \ .
\end{equation}

Based on the previous discussion, the desire is to select $ \Dmat_{jj} $ proportional to second-order effects and with this objective in mind, it is proposed to select the $ \Dmat_{jj} $ values by an straight element-by-element comparison of the diagonal elements of matrix $ \Dmat\delta\Pprior_k\Dmat  $ and $ \Nmat $. This leads to the proportionality relationship
\begin{equation}
	\delta\Pprior_{jj}\gamma_j^2 \propto \Nmat_{jj} \ ,
\end{equation}
or since $ \delta\Pprior $ and $ \Nmat $ are positive semi-definite,
\begin{equation}
\gamma_j \propto \sqrt{\frac{\Nmat_{jj}}{\delta\Pprior_{jj}}} \ .
\end{equation}
Finally, similar to the DNL method, a scale factor $ f_c $ is introduced to account for measurement residual covariance effects as,
\begin{equation}
	\Dmat_{jj}=\gamma_j = f_c \  \sqrt{\frac{\Nmat_{jj}}{\delta\Pprior_{jj}}} \ ,
\end{equation}
or equivalently
\begin{equation}
	\betamat_{jj}=\beta_j=1-f_c \  \sqrt{\frac{\Nmat_{jj}}{\delta\Pprior_{jj}}} \ .
	\end{equation}
The scale factor $ f_c $ used for the covariance-aware method is the same as the one used for the nonlinearity-aware method,
\begin{equation}
f_{c,i}=\frac{\sigma_{k,i}}{\sigma_{o,i}}\frac{\Tr(\Hmat_k\Cov_k\Hmat_k\trans+\Rk)}{\Tr(\Rk)} \ .
\end{equation}
Similar to the DNL method, the value for $ \gamma_i $ is saturated at a maximum magnitude of one. 

In the next section, results from numerical simulations that exercise the dynamic covariance-aware (DC) partial-update filter, are provided.

\subsection{The re-entry falling body}
\begin{figure}[h!]
\centering
\includegraphics[width=\dynStateWidth\linewidth]{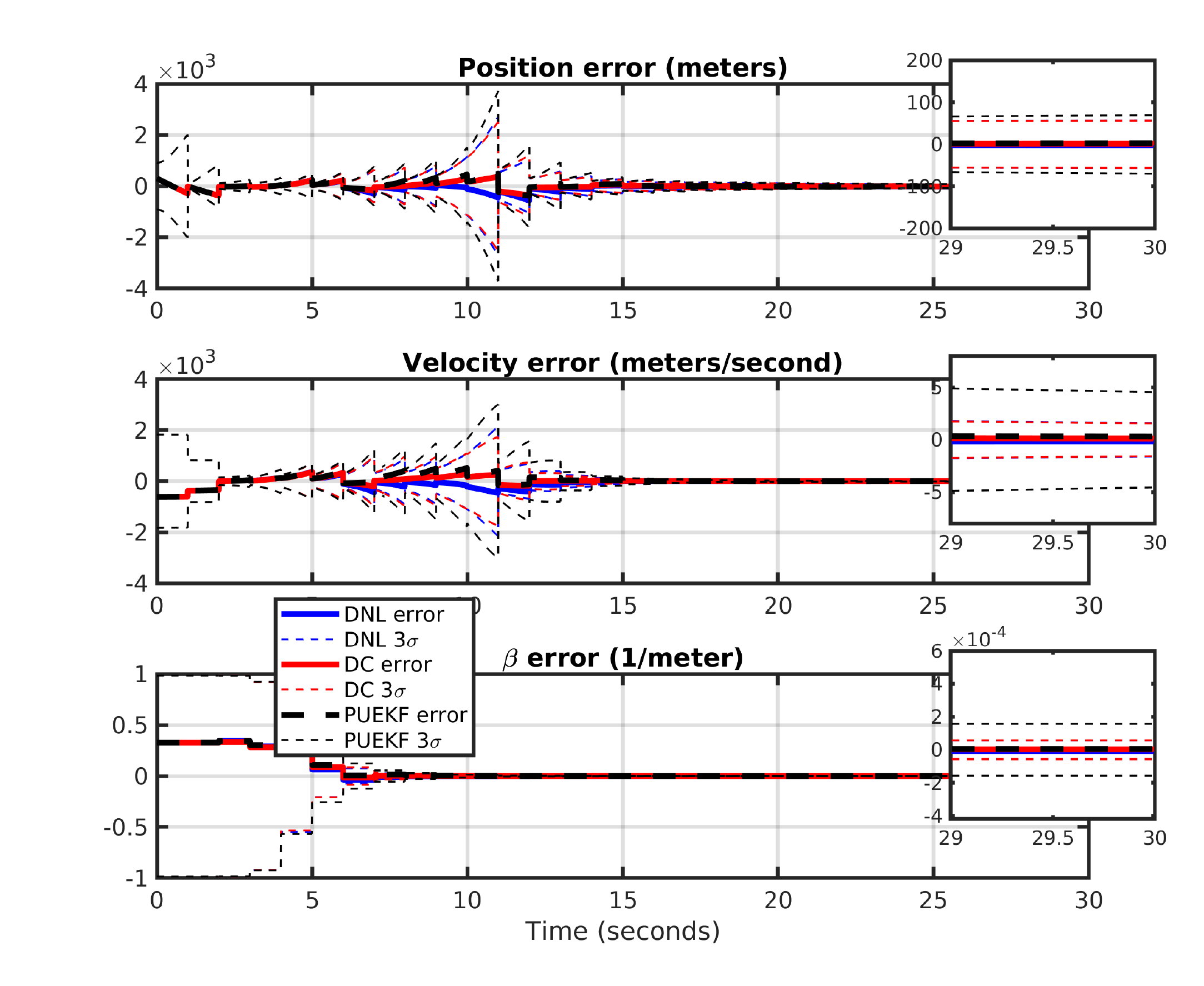}
\caption{Dynamic covariance-aware partial-update method (DC) state errors for a single run of the body re-entry problem subject to 1 $ \sigma $ initial errors. Estimates from static and DNL methods are shown for comparison.}
\label{fig:DC_DNL_and_PU_low_uncertainty_almost_same_result}
\end{figure}

\begin{figure}[h!]
\centering
\includegraphics[width=\dynBetaWidth\linewidth]{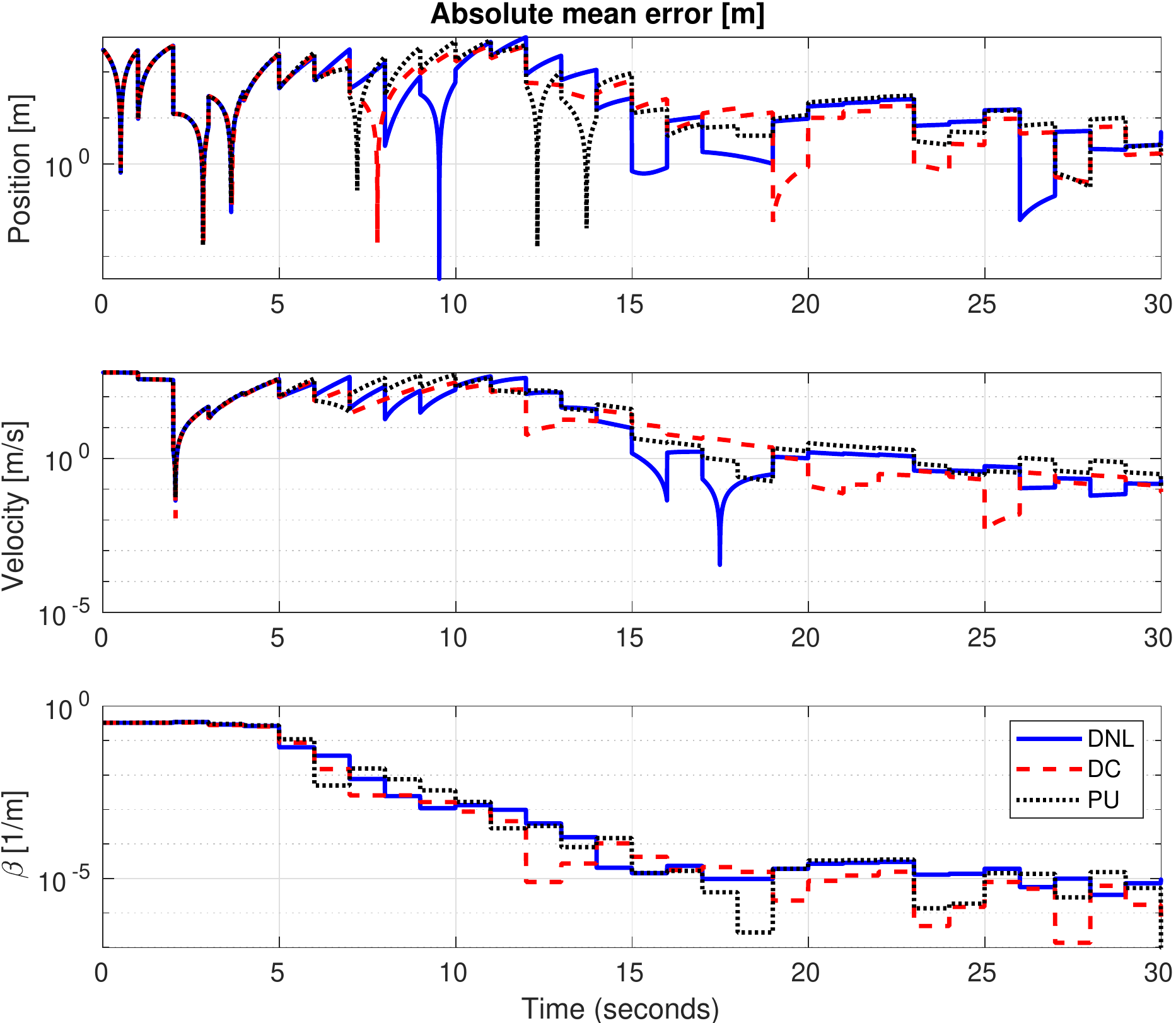}
\caption{Dynamic covariance-aware partial-update method (DC) absolute errors for a single run of the body re-entry problem subject to 1 $ \sigma $ initial errors. Errors from static and DNL methods are shown for comparison.}
\label{fig:log_abs_errors_DC_DNL_and_PU_low_uncertainty_almost_same_result}
\end{figure}

\begin{figure}[h!]
\centering
\includegraphics[width=\dynBetaWidth\linewidth]{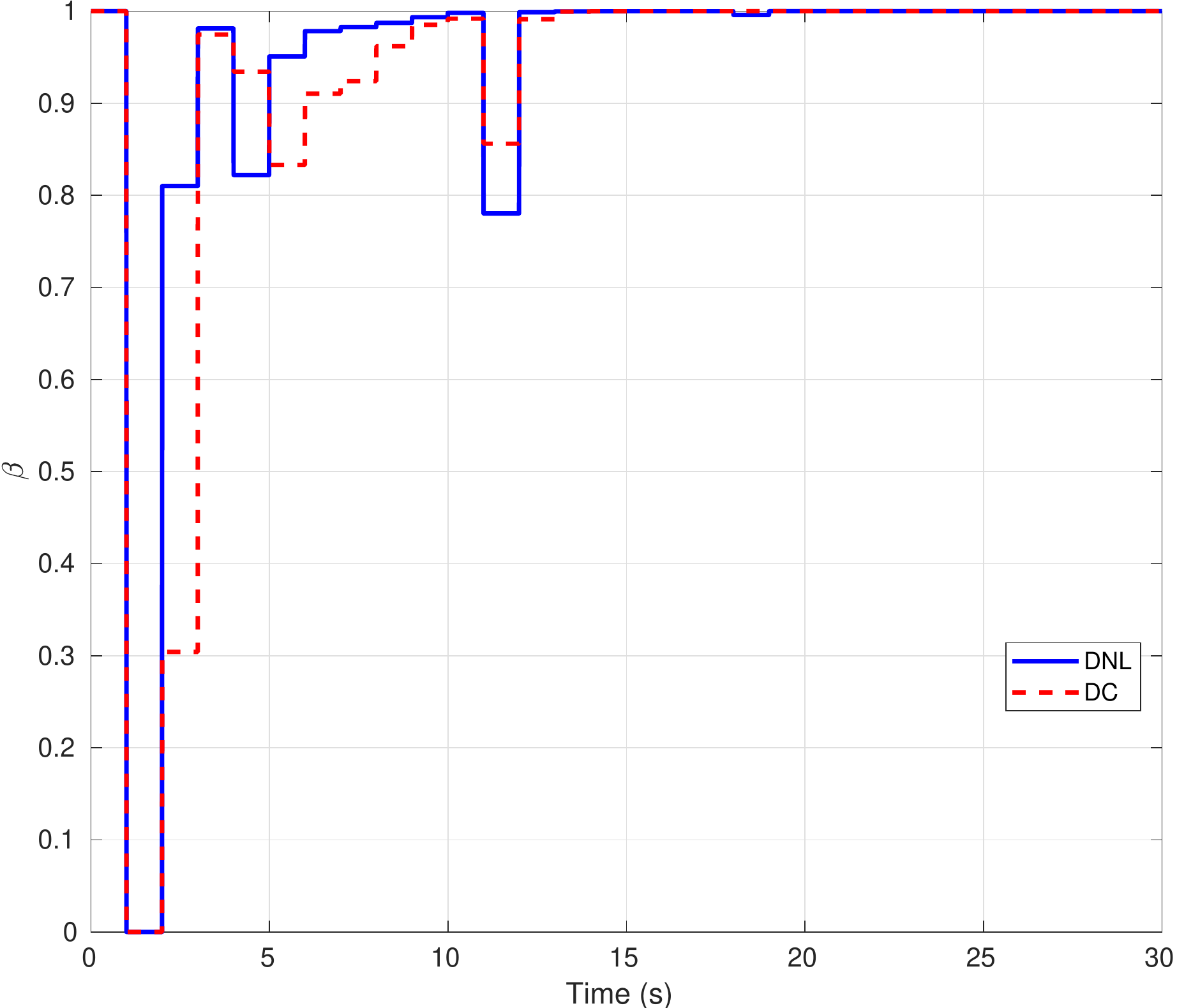}
\caption{Dynamic covariance-aware partial-update method (DC) $ \beta $ history for single run of the body re-entry problem subject to 1 $ \sigma $ initial errors. $ \beta$ produced via DNL method is shown for comparison.}
\label{fig:beta_histories_DC_DNL_and_PU_low_uncertainty_almost_same_result}
\end{figure}

\begin{figure}[h!]
\centering
\includegraphics[width=\dynBetaWidth\linewidth]{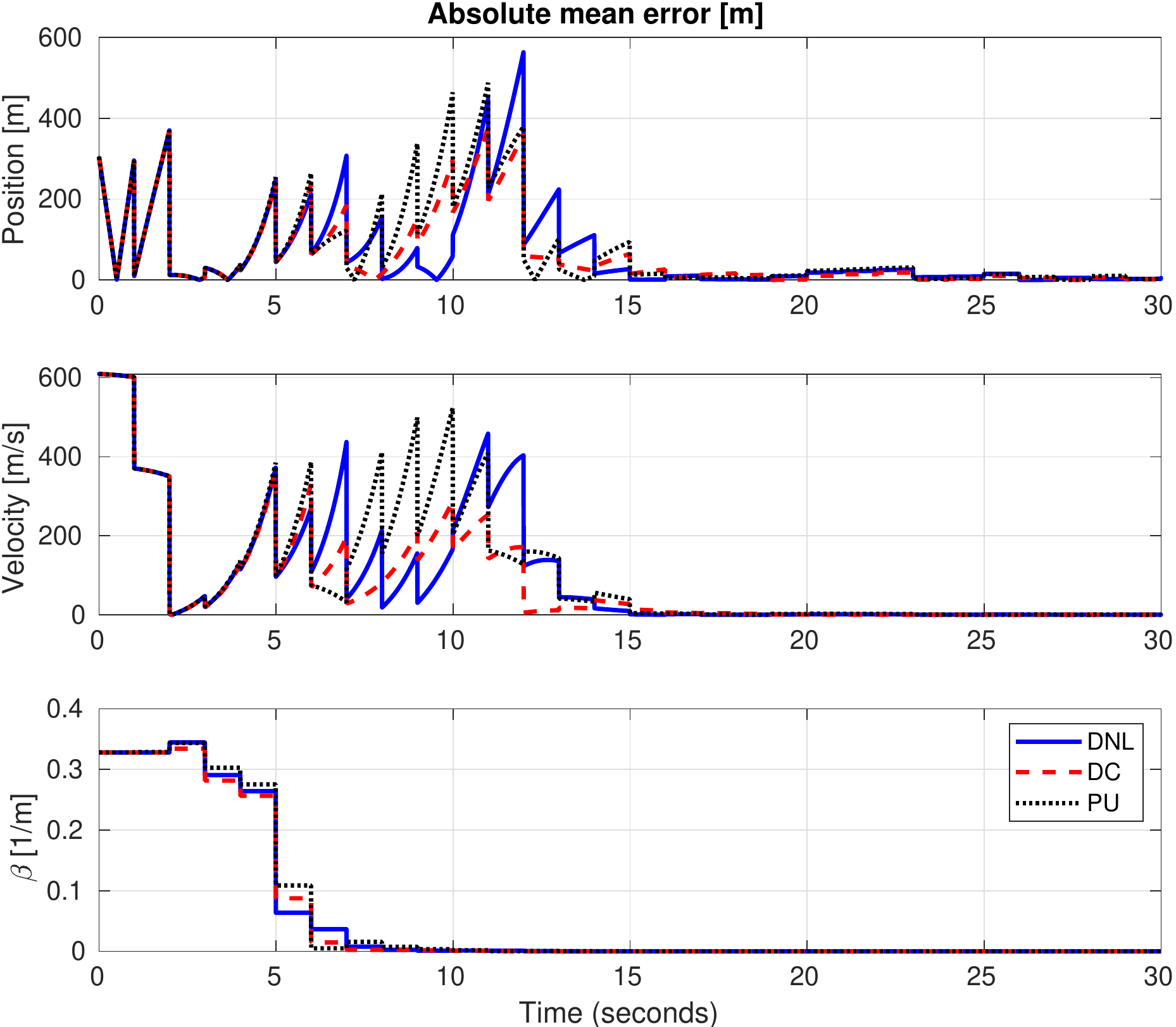}
\caption{Dynamic covariance-aware partial-update method (DC) absolute errors for a single run of the body re-entry problem subject to 1 $ \sigma $ initial errors. Errors from static and DNL methods are shown for comparison.}
\label{fig:abs_errors_DC_DNL_and_PU_low_uncertainty_almost_same_result}
\end{figure}

For convenience, the covariance-aware method was also exercised with the falling body problem. As for the DNL method, two scenarios using the DC method were simulated. First, the filter is used without incorporating any pre-tuned $ \beta $, and second, the filter is tested with a nominal baseline shifted to the already known and well-tuned $ \beta $ value. For the first case scenario, in general, it was observed that for low initial errors in a typical single run, the covariance-aware, the nonlinearity-aware, and static partial-update methods appear to have similar overall performance, as displayed in Figure \ref{fig:DC_DNL_and_PU_low_uncertainty_almost_same_result}. However, a closer examination of the absolute estimation error, shown in Figure \ref{fig:log_abs_errors_DC_DNL_and_PU_low_uncertainty_almost_same_result}, reveals that the DC method performs the best among the three filters (in terms of absolute error amount). In fact, the DC performs the best in general. An important observation for this run is that albeit the similarity of the partial-update weights histories (shown in Figure \ref{fig:beta_histories_DC_DNL_and_PU_low_uncertainty_almost_same_result}) for the dynamic methods, the state errors differences between the methods are significant. This difference in performance highlights the importance of an appropriate dynamic $ \beta $'s selection method, and it demonstrates that a varying $ \beta $ can be advantageous, even for relatively small initial errors. From Figure \ref{fig:abs_errors_DC_DNL_and_PU_low_uncertainty_almost_same_result} is also clear that since the static partial-update filter is unaware of high-order effects, it is less effective if they become significantly large. For this run, this is apparent mostly for the position and velocity states at time $ t=7s$. Finally, it should be noted that due to the DC and DNL method having a nominal update of a hundred percent,  the DC method is less conservative than the static partial-update method, but this also helped obtain a more accurate dynamic partial-update filter.

\begin{figure}[h!]
\centering
\includegraphics[width=\dynStateWidth\linewidth]{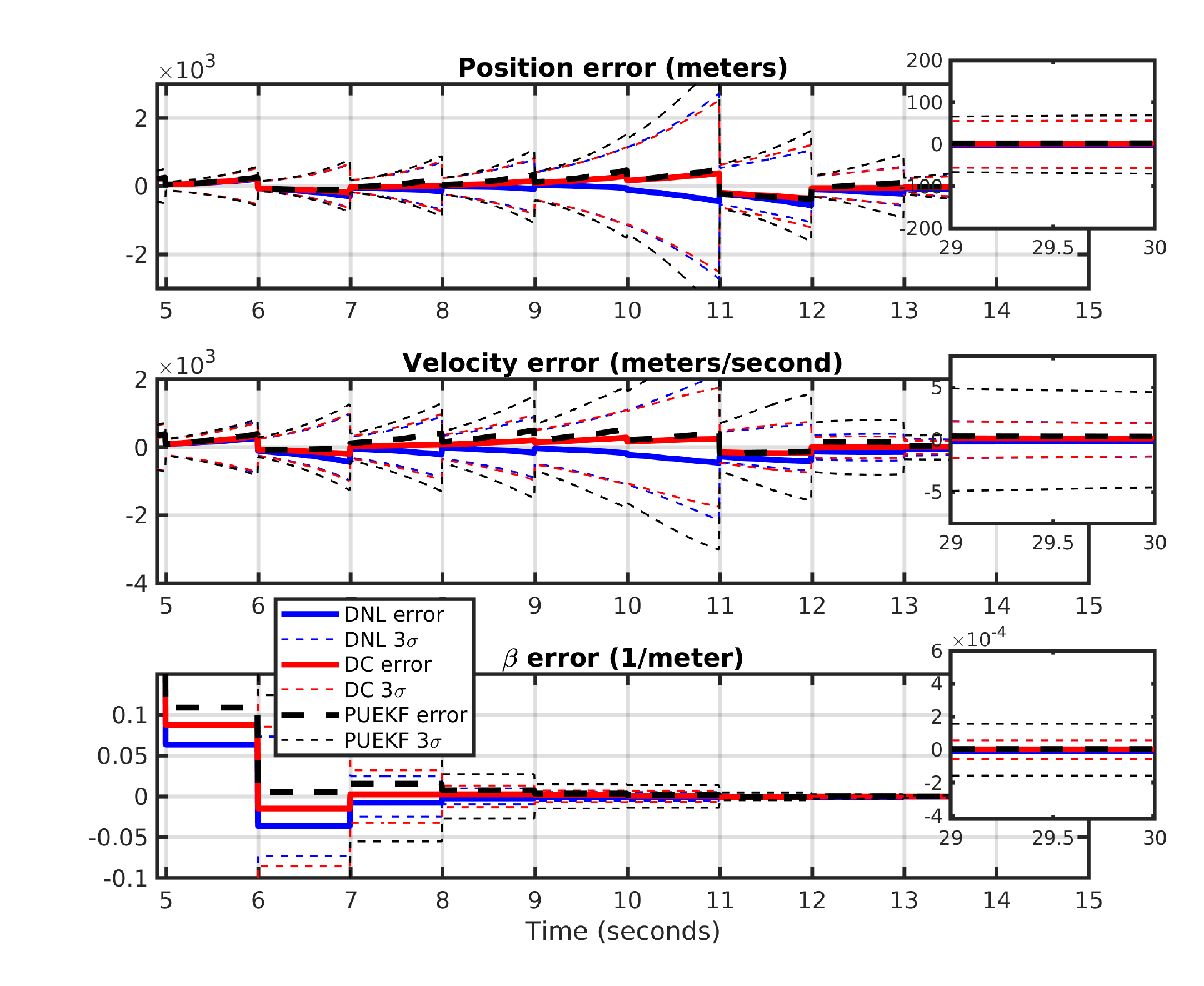}
\caption{Zoomed-in view for dynamic covariance-aware partial-update method (DC) state errors. History of a single run of the body re-entry problem subject to 1 $ \sigma $ initial errors. Estimates from static and DNL methods are shown for comparison.}
\label{fig:zoomed_in_DC_DNL_and_PU_low_uncertainty_almost_same_result}
\end{figure}

Figure \ref{fig:zoomed_in_DC_DNL_and_PU_low_uncertainty_almost_same_result}, a zoomed-in version of Figure \ref{fig:DC_DNL_and_PU_low_uncertainty_almost_same_result}, shows that the main reason for the DC method to have better performance is that although it slightly overshoots on the ballistic parameter early on (at time $ t=6s $) when the drag effects (and thus nonlinearities) influence is larger, it recovers faster than the DNL and static partial-update methods. As a result, even all three filters estimates appear within the 3 $ \sigma $ bounds,  DC partial-update filter is superior. The fast recovery of the DC method in this run, suggests that this method uses better the information provided on second-order effects to select the $\beta$ values.

\begin{figure}[h!]
\centering
\includegraphics[width=\dynStateWidth\linewidth]{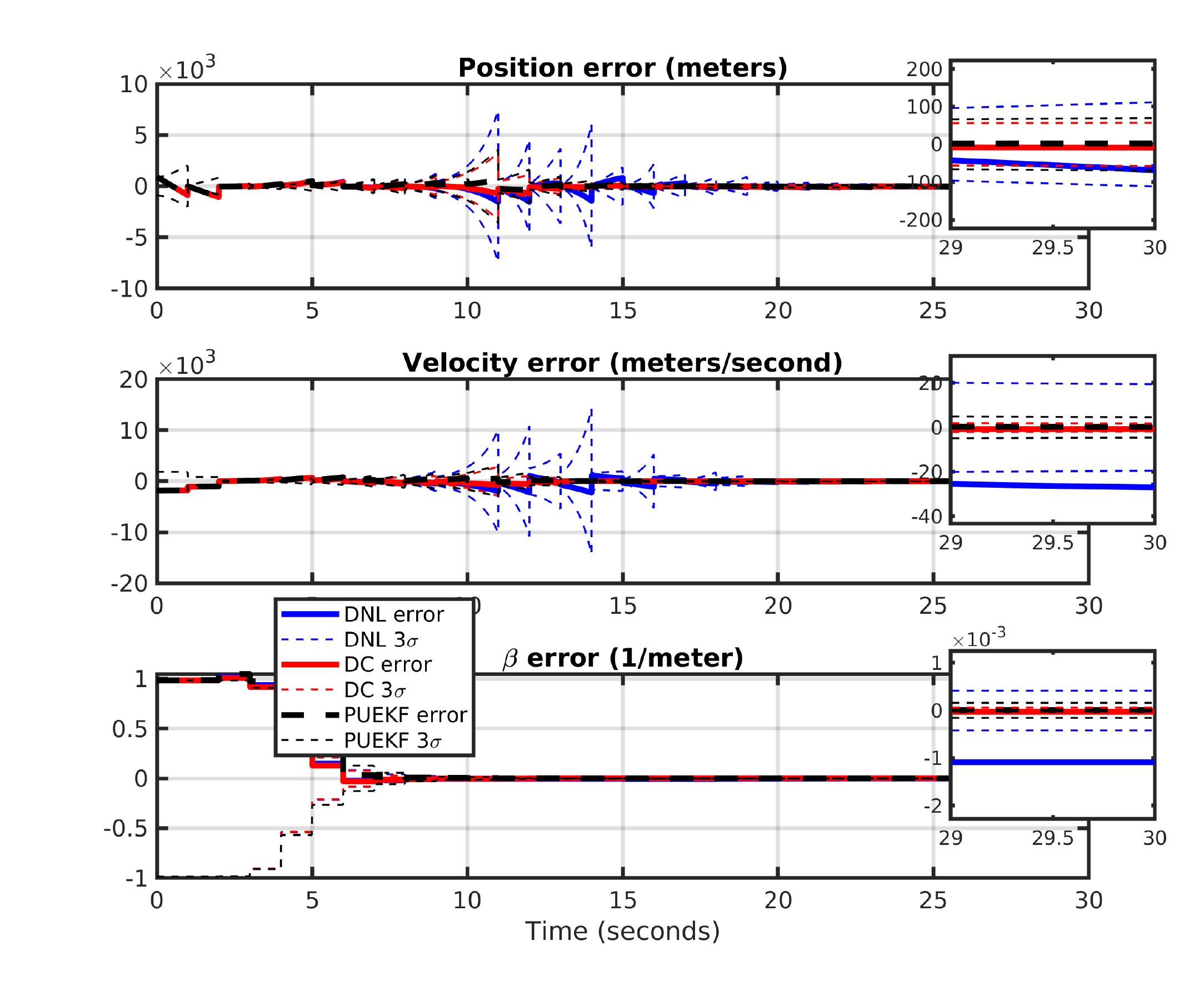}
\caption{Dynamic covariance-aware partial-update method (DC) state errors. History of a single run of the body re-entry problem subject to 3 $ \sigma $ initial errors. Estimates from static and DNL methods are shown for comparison.}
\label{fig:3sigma_in_DC_DNL_and_PU}
\end{figure}

\begin{figure}[h!]
\centering
\includegraphics[width=\dynBetaWidth\linewidth]{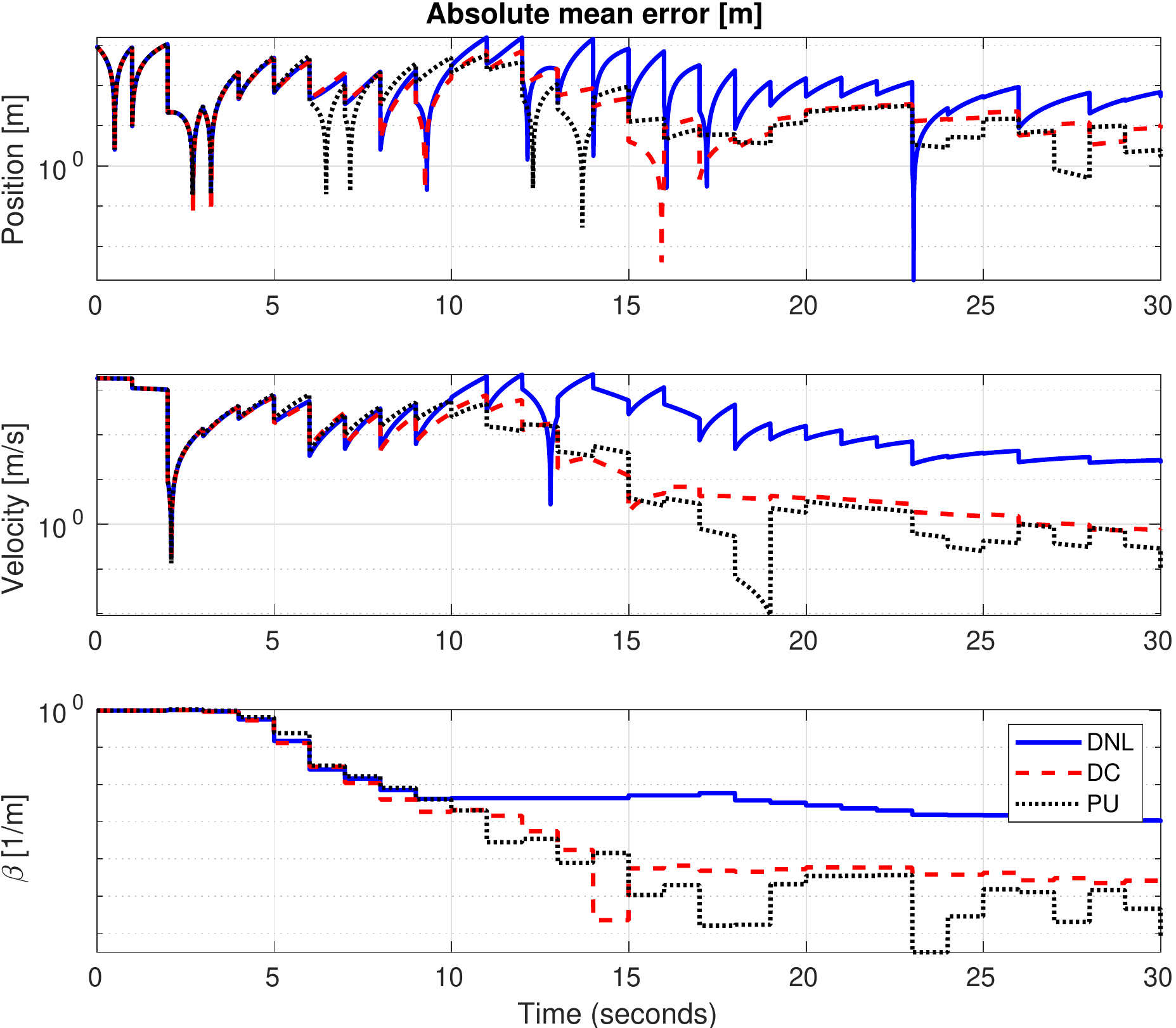}
\caption{Dynamic covariance-aware partial-update method (DC) absolute errors. History of a single run of the body re-entry problem subject to 3 $ \sigma $ initial errors. Errors from static and DNL methods are shown for comparison.}
\label{fig:3sigma_in_DC_DNL_and_PU_log_abs_error}
\end{figure}

Although numerous experiments were conducted to evaluate the robustness of the DC method against initial errors and uncertainties, the results always evidenced that 1) the DC method owns better capabilities than the DNL method and 2) the DC method can perform practically as a carefully and well-tuned static partial-update. Figure \ref{fig:3sigma_in_DC_DNL_and_PU} shows a representative run of the filters when they all undergo initial errors that are 3 $ \sigma $ in magnitude. It is clear that the DC partial-update and the static method performed well, and their estimates are within the 3 $ \sigma  $ bounds, whereas the DNL method does not. For this scenario, the DNL method was found to support maximum initial errors of up to 1 $ \sigma$. Although from Figure \ref{fig:3sigma_in_DC_DNL_and_PU_log_abs_error}, it may seem that the static partial-update method incurred in less error, the computation and comparison of the absolute error histories reveal that the covariance-aware method incurred in the least estimation error among the filters.

\begin{figure}[h!]
\centering
\includegraphics[width=\dynStateWidth\linewidth]{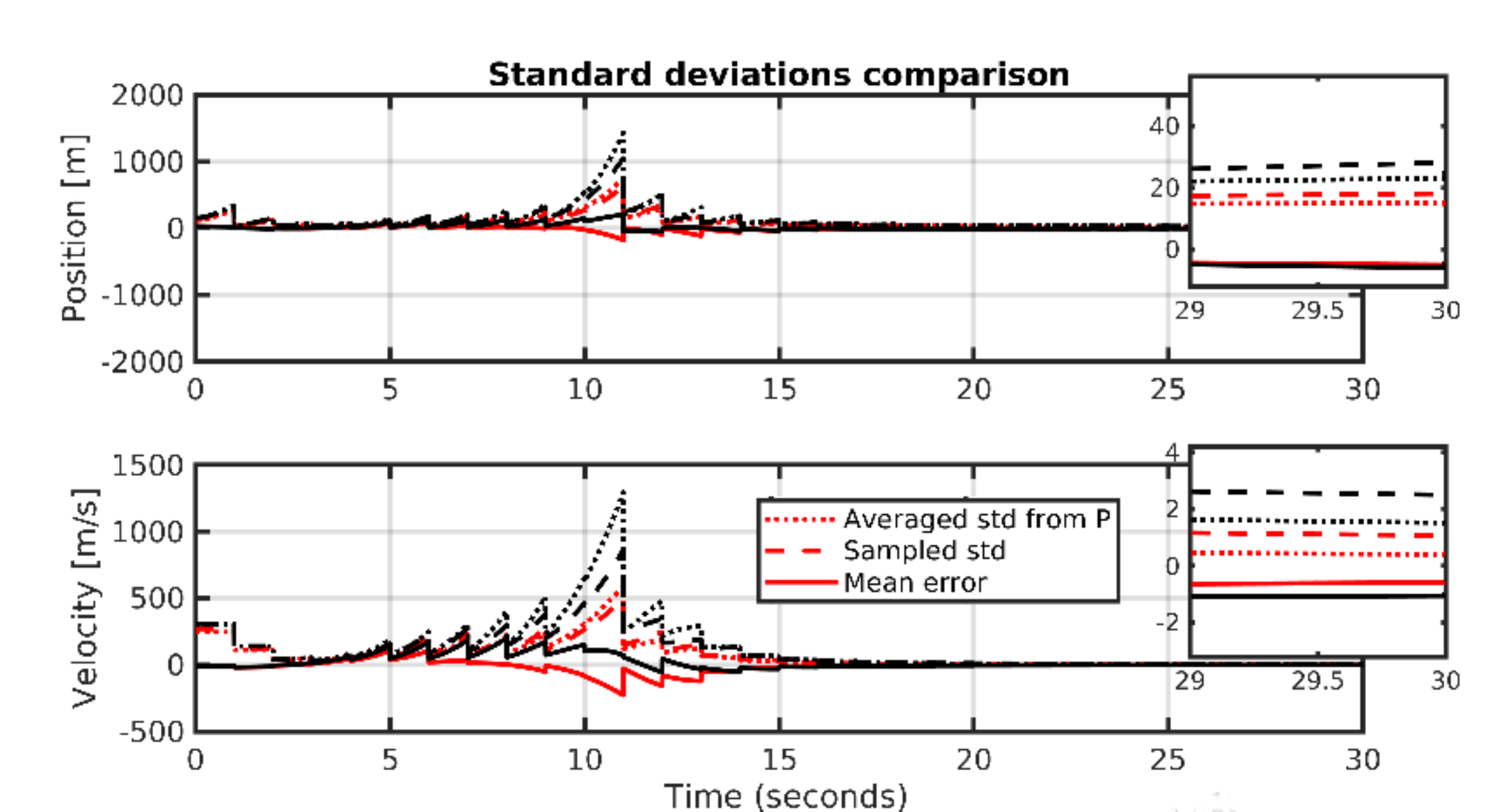}
\caption{Averaged and sampled standard deviation from 100 Monte Carlo runs for dynamic covariance-aware partial-update method (DC) state errors. Estimates from static method are also shown. DNL method is not included as it does not support initial errors higher than 1 $ \sigma $. Full update is used as the baseline for updates.}
\label{fig:Dynchapter_partial_DCvsDNLvsPU_monte_falling_body_DCPUEKFEKF_100_runs_beta10010075}
\end{figure}

Additionally, 100 Monte Carlo runs were executed to show that the robustness observed for the DC method is not specific of the random draw used to initialize the filters for that run. From the resulting Monte Carlo runs are displayed in Figure \ref{fig:Dynchapter_partial_DCvsDNLvsPU_monte_falling_body_DCPUEKFEKF_100_runs_beta10010075}, there are two main observations. First, the filter appears essentially as consistent as the finely tuned static partial-update. Second, the DC method handles the initial uncertainties and higher nonlinearities better than the pre-tuned static partial-update, and it effectively manages to produce less estimation errors overall. It is also important to mention, that even though the consistency of the filter is not perfect, it should be recalled that the filter is still a linear filter and that the conventional EKF was not operational under the scenarios presented in this and the previous sections.
Finally, it must be noted that the static partial-update performs well overall without additional metrics, as in the case for the dynamic partial-update filter. However, recall that the specific $ \beta $ tuning used in this section ($\boldsymbol{\beta}=\rowvec{1,1,0.75}$ ) required extra effort that involved numerous Monte Carlo experiments in covering and successfully running a variety of scenarios whereas the DC method required no tuning at all.

\begin{figure}[h!]
\centering
\includegraphics[width=\dynBetaWidth\linewidth]{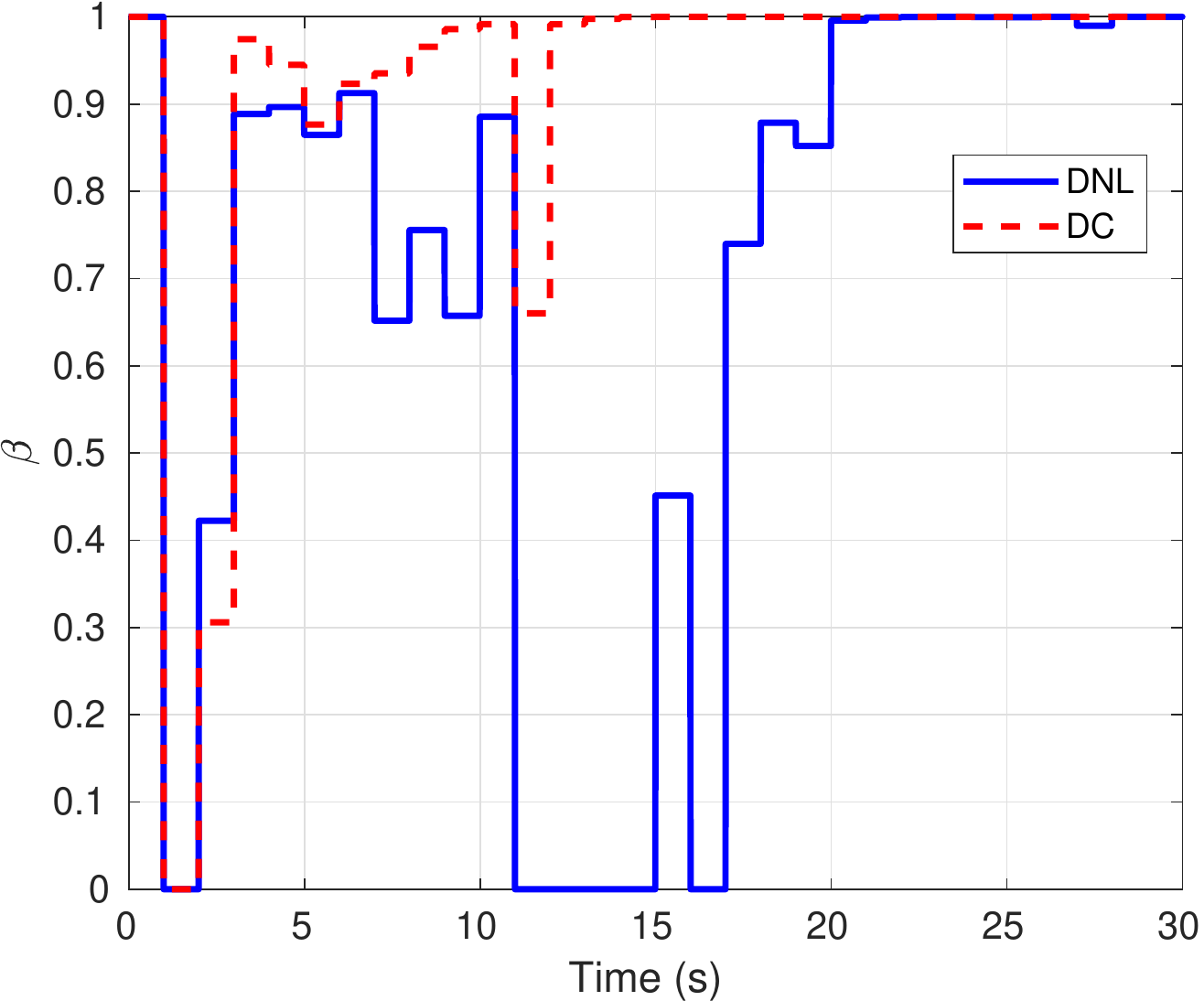}
\caption{Dynamic covariance-aware partial-update method (DC) $ \beta $ history of a single run of the body re-entry problem subject to 3 $ \sigma $ initial errors. The $ \beta $ history produced by the DNL partial-update method is also included for comparison.}
\label{fig:3sigma_in_DC_DNL_and_PU_beta_history}
\end{figure}

An additional comment on the partial-update weight history from Figure \ref{fig:3sigma_in_DC_DNL_and_PU_beta_history}, and in general for the behavior of the $ \beta $ weight when the DNL method is utilized, is that the values, compared to the DC method, tend to fluctuate more aggressively. It appears that this way of quantifying the effect of second-order terms is more \textit{sensitive} than using the second-order covariance terms. Experimentation on scaling the resulting $ \beta $ profile produced by the DNL method was also performed to examine both \textit{metrics}, DNL and DC, when having similar profiles. However, it was found that is not a scale issue, but the experiments suggested that the DNL method does not perform as well as the DC method in general, and the difference is due to the high-order terms quantification. Such sensitivity of the DNL was observed consistently across all experiments involving the dynamic methods. $ \beta $ histories from Figure \ref{fig:dyn_DNL_PU_no_baseline_beta_history2_sigma} and Figure \ref{fig:dyn_DNL_PU_no_baseline_beta_history2_sigma}, for example, also show this behavior.

\subsection{Pre-tuned partial-update weights as a baseline for the DC method}\label{sec:previous_tuning_as_baseline_DC}
\begin{figure}[h!]
	\centering
	\includegraphics[width=\dynStateWidth\linewidth]{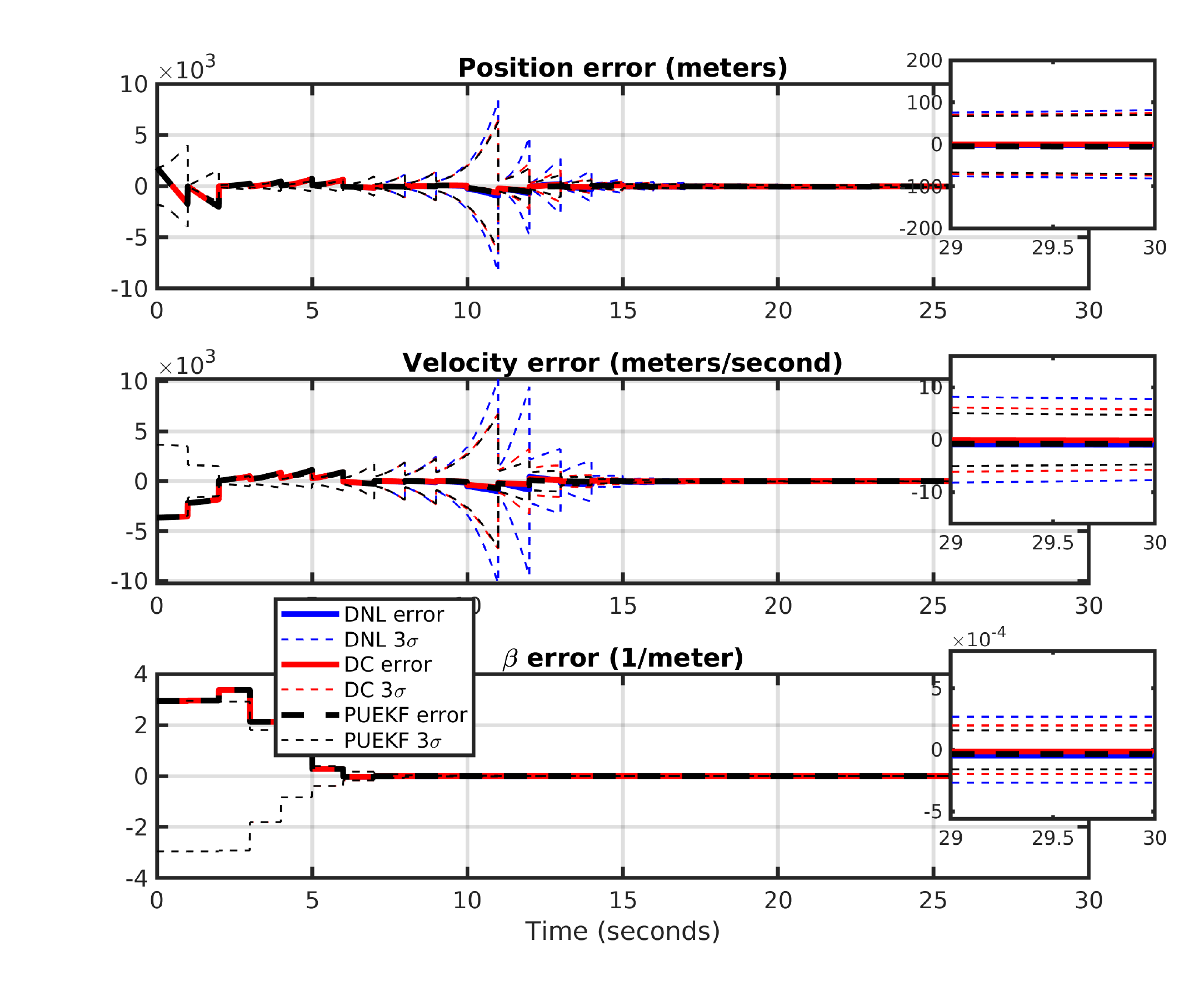}
	\caption{Dynamic covariance-aware partial-update method (DC) state errors. History of a single run of the body re-entry problem subject to moderate uncertainties (around $\rowvec{2\sigma_{p0},2\sigma_{v0},3\sigma_{\beta_0}}$) and 3 $ \sigma $ initial errors. Estimates from static and DNL methods are shown for comparison.}
	\label{fig:moderate_sigma_in_DC_DNL_and_PU}
\end{figure}

\begin{figure}[h!]
	\centering
	\includegraphics[width=\dynBetaWidth\linewidth]{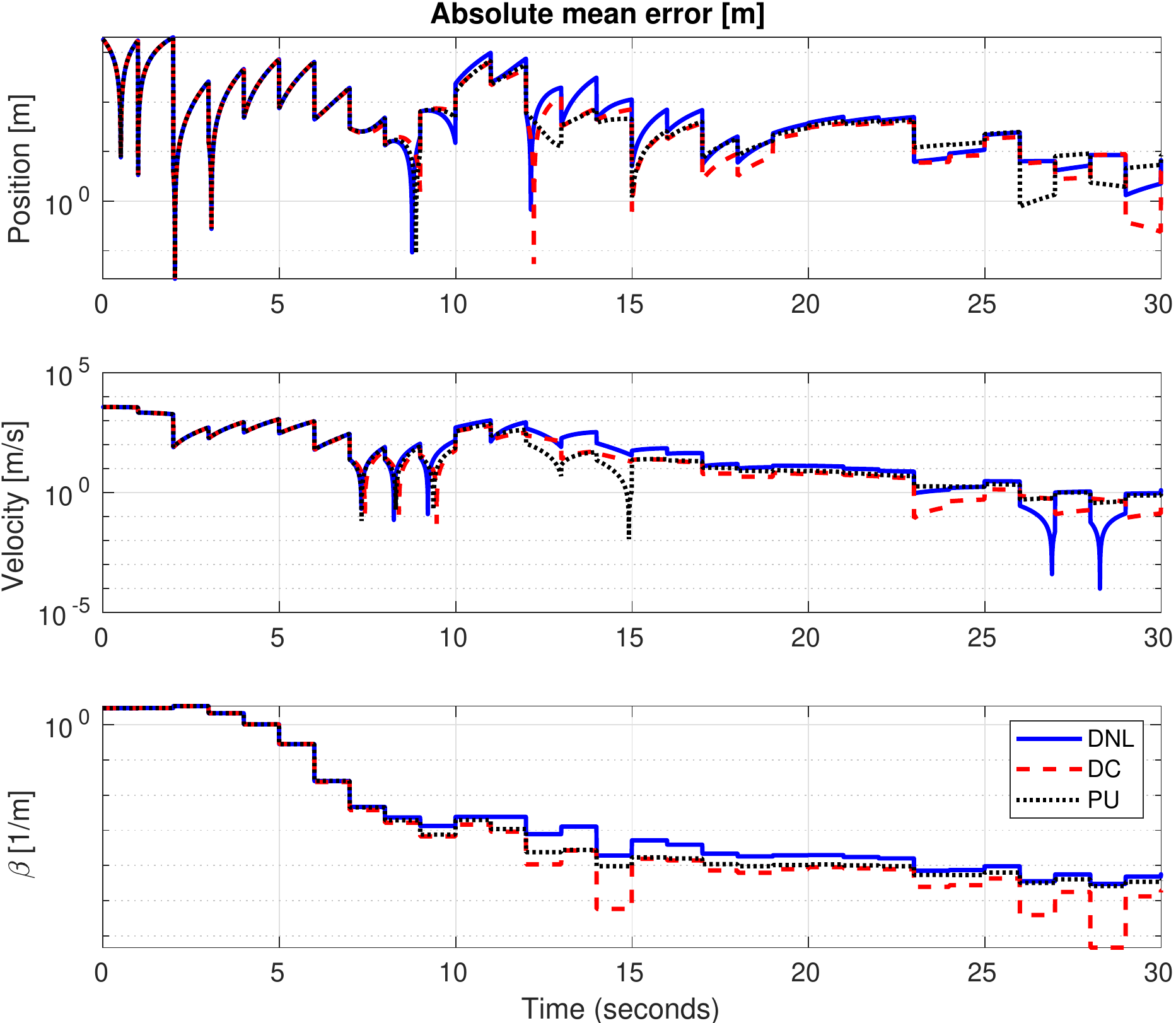}
	\caption{Dynamic covariance-aware partial-update method (DC) absolute errors. History of a single run of the body re-entry problem subject to moderate uncertainties (around $\rowvec{2\sigma_{p0},2\sigma_{v0},3\sigma_{\beta_0}}$) and 3 $ \sigma $ initial errors. Errors from static and DNL methods are shown for comparison.}
	\label{fig:moderate_sigma_in_DC_DNL_and_PU_log_abs_error}
\end{figure}

\begin{figure}[h!]
	\centering
	\includegraphics[width=\dynBetaWidth\linewidth]{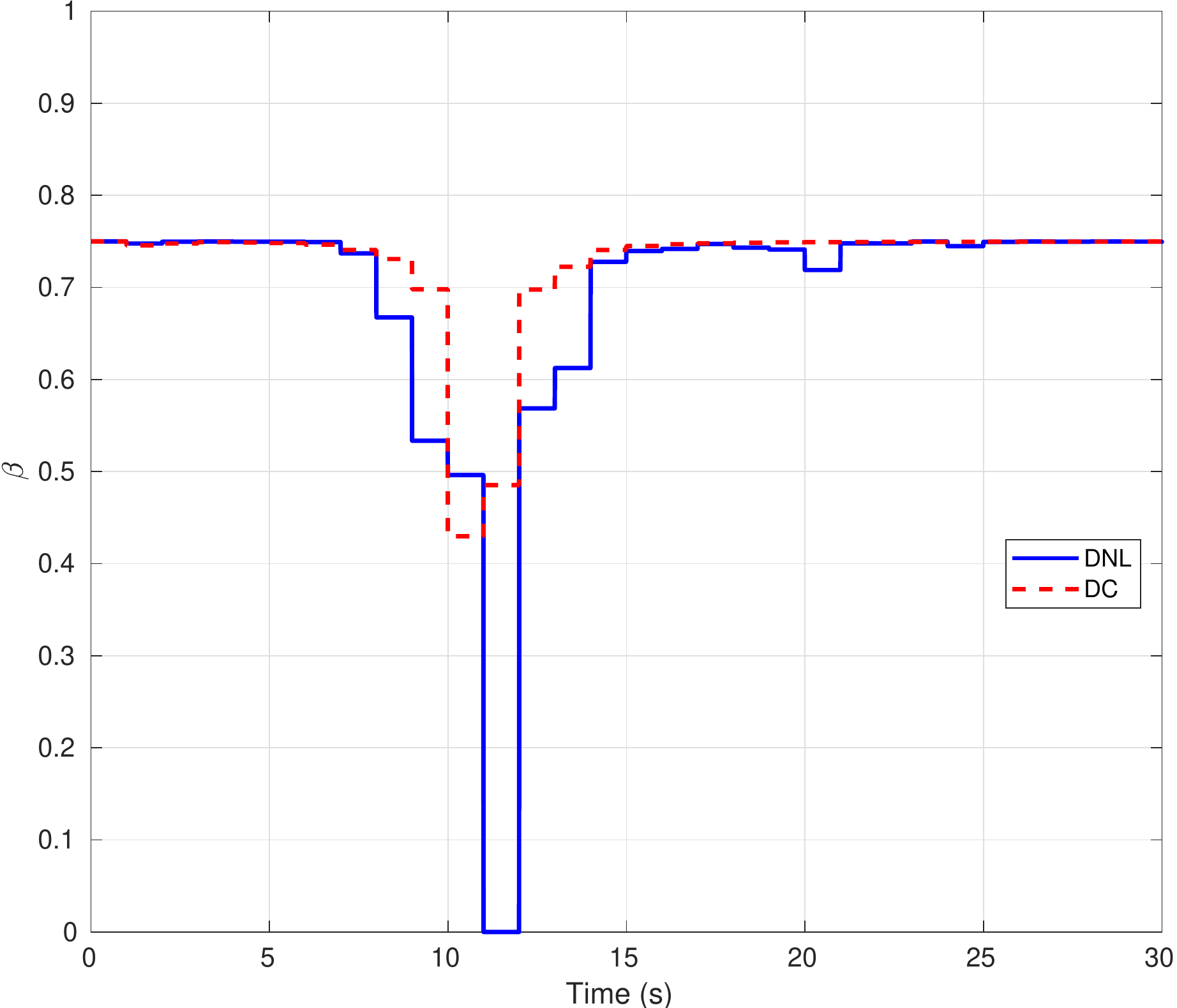}
	\caption{Dynamic covariance-aware partial-update method (DC) $ \beta $ history of a single run of the body re-entry problem subject to uncertainties (around $\rowvec{2\sigma_{p0},2\sigma_{v0},3\sigma_{\beta_0}}$) and 3 $ \sigma $ initial errors. The $ \beta $ history produced by the DNL partial-update method is also included for comparison.}
	\label{fig:moderate_sigma_in_DC_DNL_and_PU_beta_history}
\end{figure}

The DC method using known $ \beta $ values was observed to be almost identical in behavior and consistency as the DNL (with known weights) method when exercised with relatively low initial errors (1 $ \sigma $ or less). For moderate-high initial errors and uncertainties, however, the covariance-aware method was found to perform the best compared with DNL and static methods. A typical run of the DC partial-update filter with high initial errors and uncertainties (errors near 3 $ \sigma $, doubled uncertainty on position and velocity, and tripled uncertainty on ballistic parameter, with respect to base values from Table \ref{table:initial_conditions_SQPU}) are depicted in Figure \ref{fig:moderate_sigma_in_DC_DNL_and_PU}. The corresponding absolute errors are displayed in Figure \ref{fig:moderate_sigma_in_DC_DNL_and_PU_log_abs_error}. It is worth noting that consistently across experiments, the DNL method generally exhibits larger reactions to second-order terms compared to the DC method, as Figure \ref{fig:moderate_sigma_in_DC_DNL_and_PU_beta_history} shows.


Since the methods that use known weights were seen to support high initial errors and uncertainties, several experiments were performed to explore their limits. Overall, it was found that the DC method is more robust, and that to cause divergence, for the re-entry body scenario, extreme cases were needed to make it diverge. Although such bad initial conditions may not be realistic, the objective was to find sufficiently adverse conditions for the dynamic filters to diverge to gain insight into the dynamic methods limits. 

\begin{figure}[h!]
	\centering
	\includegraphics[width=\dynStateWidth\linewidth]{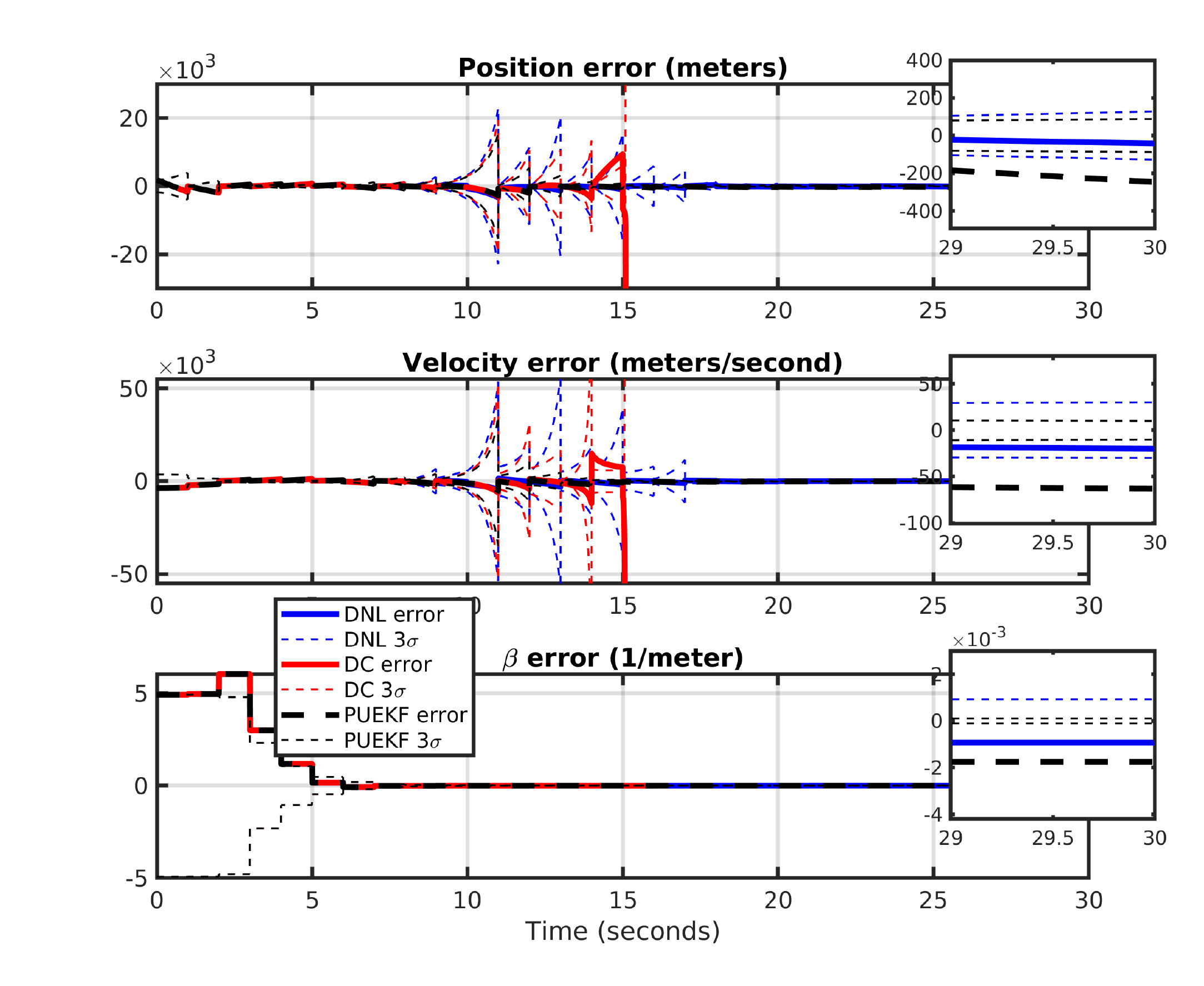}
	\caption{Dynamic covariance-aware partial-update method (DC) state errors. History of a single run of the body re-entry problem subject to high uncertainties (around $\rowvec{2\sigma_{p0},2\sigma_{v0},5\sigma_{\beta_0}}$) and 3 $ \sigma $ initial errors. Estimates from static and DNL methods are shown for comparison.}
	\label{fig:high_sigma_in_DC_DNL_and_PU}
\end{figure}

\begin{figure}[h!]
	\centering
	\includegraphics[width=\dynBetaWidth\linewidth]{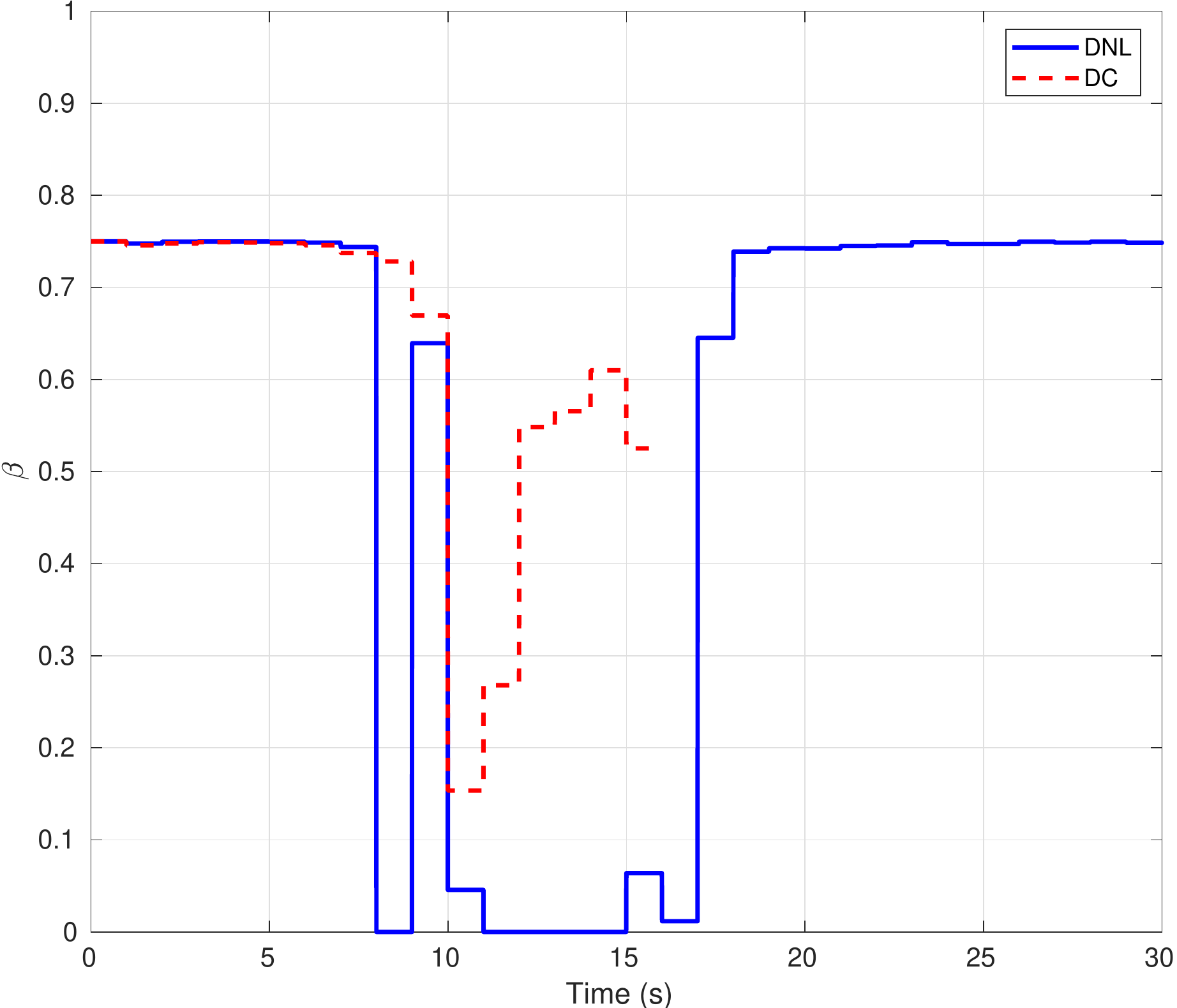}
	\caption{Dynamic covariance-aware partial-update method (DC) $ \beta $ history of a single run of the body re-entry problem subject to high uncertainties (around $\rowvec{2\sigma_{p0},2\sigma_{v0},5\sigma_{\beta_0}}$) and 3 $ \sigma $ initial errors. The $ \beta $ history produced by the DNL partial-update method is also included for comparison.}
	\label{fig:high_sigma_in_DC_DNL_and_PU_beta_history}
\end{figure}

\begin{figure}[h!]
	\centering
	\includegraphics[width=\dynBetaWidth\linewidth]{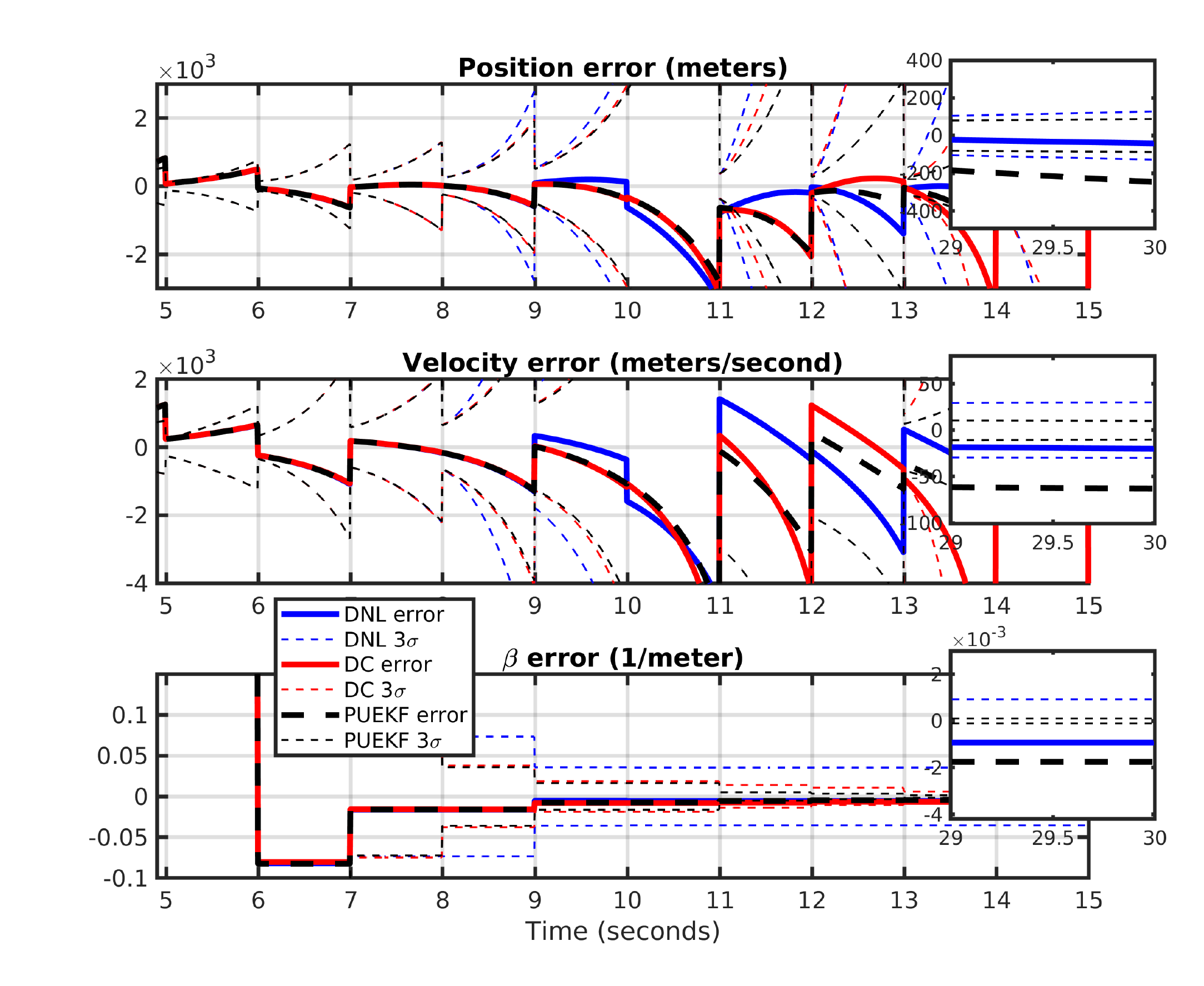}
	\caption{Zoomed-in view for dynamic covariance-aware partial-update method (DC) state errors. History of a single run of the body re-entry problem subject to high uncertainties (around $\rowvec{2\sigma_{p0},2\sigma_{v0},5\sigma_{\beta_0}}$) and 3 $ \sigma $ initial errors. Estimates from static and DNL methods are shown for comparison.}
	\label{fig:zoomed_high_sigma_in_DC_DNL_and_PU}
\end{figure}

One of the experiments showing the covariance-aware method diverging, used initial conditions where the uncertainties on position and velocity states were doubled, and the ballistic parameter uncertainty is five times the original $ \sigma $ (with respect to reference parameters given in Table \ref{table:initial_conditions_SQPU}). 
Figure \ref{fig:high_sigma_in_DC_DNL_and_PU}, shows a filter single run of this scenario with high initial uncertainties. Here, the reason for failure can be attributed to the DC method insufficient reaction to produce large enough weight variations when required as observed from Figure \ref{fig:high_sigma_in_DC_DNL_and_PU_beta_history}, making the filter to eventually fail. The zoomed-in version of the estimates errors, shown in Figure \ref{fig:zoomed_high_sigma_in_DC_DNL_and_PU}, in fact, confirms that insufficiently lowering the $ \beta $ value when high nonlinearities appeared was the root of divergence. Notice that Figure \ref{fig:high_sigma_in_DC_DNL_and_PU} are the results of the same scenario used in the previous section when stressing the DNL method with high uncertainties (see Figure \ref{fig:DNL_and_PU_very_high_uncertainties}), but here, is it clearer how the DNL method managed to handle the large nonlinearities. The DNL method reacted to the extent of even temporarily act as a consider filter, mainly due to its larger sensitivity to second-order effects (Figure \ref{fig:high_sigma_in_DC_DNL_and_PU_beta_history}), being able to retain a consistent filter, which is even better than the finely tuned static partial-update approach for the presented scenario.

Although these are just the results from one run, the overall behavior showed that it takes a relatively bad initialized filter to make the dynamic methods fail, especially for the DC method, which was seen to be more robust than the DNL method overall. Scenarios were the DNL method fails and the DC method works, were also seen and more common for high initial errors, the plots are not included here but they look similar to the divergent DNL of Figure \ref{fig:3sigma_in_DC_DNL_and_PU}.

\subsection{Comments on DNL and DC partial-update methods}
The dynamic beta selection methods showed similar behavior when the initial errors and uncertainties were relatively small. In those cases, it was observed that a finely tuned static partial-update in practical terms would offer similar benefits as a dynamic $ \beta $ selection method, but at the extra effort of tuning weights. However, as seen from the simulated examples, if nonlinearities are significant, the static partial-update will incur in higher estimation errors.
Interestingly, although the dynamic methods appear slightly over or under confident with respect to the static partial-update method, the changes in the 3 $ \sigma $ bounds are relatively small. In other words, the use of a method for dynamically choosing the $ \beta $'s does not sacrifice accuracy importantly; and it makes the filter operational that otherwise would be divergent. As seen from the Monte Carlo runs, the agreement of sample and averaged standard deviation, although it is not perfect (due to mainly to high initial errors), is fairly accurate, and certainly far better than the conventional EKF. Moreover, for scenarios with important nonlinearities and uncertainties, the dynamic methods exhibit lower mean estimation error, especially the covariance-aware method. The virtues of the dynamic beta selection methods, however, come at the cost of the computation of second-order terms. On the other hand, the dynamic methods are still attractive since they effectively allow the use of an EKF filter structure where a more advanced filter may be needed.

Even though different metrics and methods can be proposed to perform online $ \beta $ selection, the methods presented here are functional, do not over-specialize the filter,retain the EKF structure, and use information that is already computed within the KF. Finally, based on the presented results, it is always recommended to use the DC method because it provides higher robustness against high initial errors and uncertainties. If computational cost is a priority, the DNL could be considered, but the user needs to be aware of its lower robustness.

%

\chapter[HARDWARE APPLICATIONS]{HARDWARE APPLICATIONS\footnote{Adapted with permission from ``Reef estimator: A simplified open source
		estimator and controller for multirotors'', by J. H. Ramos, P. Ganesh, W. Warke, K. Volle and K. Brink, presented at the IEEE National Aerospace and
		Electronics Conference 2019 \cite{REEFEstimator}, Copyright 2019 by the  Institute of Electrical and Electronics Engineers; ``Vision-based tracking of non-cooperative
		space bodies to support active attitude control detection'', by J. H. Ramos, T. D. Woodbury, and J. E. Hurtado, presented at the AIAA SPACE and Astronautics Forum and Exposition 2018 \cite{Ramos2018}, Copyright 2018 by J. Humberto Ramos,Timothy Woodbury, and John E. Hurtado.}} \label{cha:hardware_applications}

This chapter describes three aerospace-related hardware applications of the partial-update filter. The specific applications are selected to demonstrate the partial-update filter's capabilities when the system's nonlinearities are significant, and the nuisance states need to be estimated. Furthermore, the implementations show the partial-update filter's capacity to work well where the nuisance states are constant, varying, or are the states of interest. In all three applications, the use of a Schmidt or a traditional EKF failed, or it did not produce acceptable estimates. 

The first application is a filter-based IMU-camera (Inertial Measurement Unit) calibration system. The purpose of this filter is to find the rigid transformation (or calibration) between the camera and IMU to improve the global IMU navigation solution.
The calibration parameters of this application are the constant nuisance states. This hardware application uses a UD partial-update filter and is shown to estimate well the calibration parameters with a sub-centimeter and sub-degree accuracy. 
The second application is a simplified altitude and velocity estimator for Unmanned Aerial Vehicles (UAV). The IMU biases are the nuisance parameters in this case. In contrast to the first application, these nuisance parameters change over time. A UD partial-update filter is also implemented for this application. The results showed a consistent filter for both core and varying nuisance states, providing a highly functional flight platform.
The third application is a vision-based partial-update filter to estimate the angular rates of an uncooperative space body. In contrast with the first and second applications, the nuisance states are the angular rates, the primary states of interest. This last application uses a UD partial-update filter to estimate the angular rates.

\section{Online IMU-camera intersensor parameters calibration}
\subsection{Introduction}
In recent years vision-aided inertial navigation based on cameras and IMU has captured the attention of many researchers. A camera and IMU working together offer a relatively low-weight and complementary team: while the IMU provides high-frequency system motion propagation, the camera provides slower but fixed and more stable (it does not drift in contrast with IMU biases) feedback measurement signal. When the IMU-camera set is used along with a sequential filtering technique, or similar IMU-camera fusion information algorithm \cite{gustafsson2010statistical}, one can obtain an accurate inertial navigation system. Among the factors that contribute to an accurate vision-aided inertial navigation solution, the IMU-camera calibration parameters play a key role.

The knowledge of the relative placement between an inertial measurement unit and a camera, or IMU-camera calibration for short, is crucial \cite{brink2012filter}. Large inaccuracies in this calibration parameters can cause the system estimates to be biased, degraded, and even to diverge \cite{yang2010design}. There exist several methods in the literature for IMU-camera calibration, and are mainly divided into two groups: batch and sequential calibrators. Filter based approaches are commonly more popular than batch approaches because they also provide confidence bounds for the calibration parameters and run online. Filter based approaches essentially estimate ego-states (position, velocity, and attitude) of the IMU, along with the relative pose between the camera and the IMU frame. Systems such as those reported in \cite{mourikis2009vision}, \cite{NikolasTrawnyAnastasiosI.Mourikis2007} and \cite{koch2016multi}, for instance, show the need for accounting and estimating the relative pose between the sensors to improve navigation results, which is the reason to perform the calibration in the first place.  

The IMU-camera calibration of this hardware implementation is a filter-based approach. Generally, due to the high nonlinearities and uncertainties induced by the unknown IMU-camera transformation, and the nuisance parameters included in the state vector, a traditional EKF or Schmidt filter is often not sufficient to obtain acceptable estimates. For this problem, it is common that users alternatively implement an iterative EKF or even a UKF. This hardware application uses the PU-MEKF filter (developed in Chapter \ref{sec:Background}) to accommodate better the calibration parameters, along with system biases (particularly IMU's biases). The PU-MEKF is implemented using the UD factorization presented in Chapter \ref{ch:ud_partial_update_filter}. As commonly done in visual-aided inertial navigation systems, the filter uses the IMU measurements to propagate the rotational kinematics model (of the rigid body carrying the IMU-camera system) while the camera provides image features locations measurements. For this application, the features or landmarks locations, are the extracted corners of a set of arUco markers and are considered to be known in the inertial frame, as emulating navigation using an a priori known map.

\subsection{Related Work} \label{sec:Lit_Review}
One of the most popular batch techniques for IMU-camera calibration was introduced by Furgale et al. in \cite{furgale2013unified}. The work proposes a batch least-squares solution that performs both a temporal and spatial calibration of an IMU and RGB sensor using fiducial markers. The authors provide an open-source implementation of their code along with relevant documentation. Although this technique provides a systematic solution for finding the rigid body transform, it is not suitable for online implementation. If an IMU-camera reconfiguration occurs, as for systems able to tilt and pan the camera, the calibration procedure needs to be repeated offline. Unfortunately, for this type of batch calibrator, the user does not know if the collected data is enough (qualitative and quantitatively) until the algorithm processes the data. Another least-square solution presented in \cite{mair2011spatio}, proposes the use of B-splines, and similarly to \cite{furgale2013unified}, it can also be used to identify temporal IMU-camera alignment. This method is not an online approach neither, but it reports accurate results. 

In regards to filter-based calibration algorithms, there is a variety of flavors and forms. For example, \cite{kelly2011visual} proposes the use of an unscented Kalman filter (UKF)to perform the calibration, and it even includes gravity vector into the state vector to improve results. In \cite{brink2012filter}, the use of a UKF is also introduced and taken further as it generalizes the IMU-camera calibration problem to a multicamera-IMU system. Although these approaches are shown to be functional, they require the user to be familiar with the unscented approach and its implementation. Also, in the Kalman filters line, \cite{mirzaei2008kalman} proposes an extended Kalman filter. This work includes a nonlinear observability analysis (similar to that presented in \cite{kelly2011visual}), showing that the IMU-camera calibration parameters can be estimated from the camera and IMU measurements alone. More specifically, this method uses an iterative multiplicative Extended Kalman filter (IMEFK) to handle the high nonlinearities that appear in the measurement model. The IMEKF is grounded in an extended Kalman filter (EKF), which makes it practical from the implementation perspective, but it may become computationally expensive as each set of camera measurements are re-processed multiple times at each update step.
Other filtering-based approaches include the technique proposed by Li and Mourikis \cite{li2013high}. This technique incorporates the IMU-camera pose into a multi-state constraint Kalman filter, which is shown to achieve acceptable calibration parameters. However, the use of such an algorithm requires more coding effort (with respect to a conventional single-step propagation/update filter) due to the need of book-keeping past system poses.
Yang and Shen in \cite{yang2016monocular} propose a method to initialize velocity, scale, and IMU-camera calibration in real-time without requiring artificial markers. To achieve this, the authors propose a probabilistic optimization-based procedure. The method is shown to be functional, but it may require more experienced users for its implementation. Highly precise techniques, like that presented in \cite{rehder2017camera}, are also available in the literature, but they come at the cost of requiring significantly accurate modeling. Non-filter-based that use closed-form solutions to perform the calibration as the one presented in \cite{lobo2007relative} are also available. However, they may not provide uncertainty quantification on the calibration parameters and are prone to be sensitive to the calibration target pose.

\subsection{Filter-based IMU-camera calibration algorithm }\label{sec:camera_imu_algorithm_description}
In this section, the description of the calibration algorithm is given. As the process is PU-MEKF-based, its description has been divided into propagation and measurement update step. Recall that as conventionally done, the propagation is performed using IMU measurements while the update uses camera measurements. 
Particularly, Section \ref{subsub:pumekf_propagation} details the propagation step and Section \ref{subsec:pumekf} describes the measurement update step. Before going into the specifics of the proposed calibration filter, the following section establishes additional notation used throughout this hardware implementation description.

\subsubsection{Notation}\label{sec:notation}
The notation $\stdvecnb{\pchar}[A][B]$ is used to denote the
vector $\pchar$ of property B \textit{coordinatized} in reference frame A. For example, $\stdvecnb{\pchar}[W][I]$ represents the position vector of the IMU (I) with respect to the world frame (W). The collection of elements in a column vector is written as
$\stdvecnb{\pchar}[A][]$, with no reference frame associated to it.  Rotations are parametrized by quaternions. Quaternions are denoted by $\quat$. A rotation matrix such as $\dcm(_A^B\mathrmbf{\bar{q}})$, for example, is the passive rotation $\dcm$ constructed from the quaternion parametrization $_A^B\mathrmbf{\bar{q}}$. Here $\dcm(_A^B\mathrmbf{\bar{q}})$ is a rotation that takes a vector initially represented in reference frame A, and expresses it in the coordinate frame B. Passive rotations are also written simply as $\dcm[A][B]$, which is considered equivalent to $\dcm(_A^B\mathrmbf{\bar{q}})$. The operator $\skewsym{\vchar}$ constructs a skew-symmetric matrix using the elements of vector $\vchar=\rowvec{v_1,v_2,v_3}\trans$ according to 
\begin{equation}\label{eq:generic_skew_symmetric_matrix}
\skewsym{\vchar} = 
\begin{bmatrix}
0 & -v_3  & v_2 \\
v_3 & 0 & -v_1\\
-v_2 & v_1 & 0\\ 
\end{bmatrix}.
\end{equation}
Variables written with a delta prefix, such as $\xerror$, represent an error quantity. Finally, recall that the error definition is considered as the difference between the true, $ \x $, and expected value, $ \xhat $, as $\xerror=\x-\xhat$. Any variations in notation will be clear from the context or will be clarified as needed.

\subsubsection{The propagation step}\label{subsub:pumekf_propagation}

This section describes the particulars of the propagation step performed in the IMU-camera calibration implementation.


\paragraph{State vector.}
The IMU-camera calibration parameters and the IMU biases are the nuisance states in this filter. Whereas the core or ego-states are the position, velocity, and attitude of the IMU in the inertial frame. The notation for the PU-MEKF states elements as shown in Equation (\ref{eq:stateimu}) is: attitude ($\qtrue \in \Real^4$) and position ($\stdvecnb{\pchar}[W][I] \in \Real^3$) of the IMU with respect to the world frame, the velocity of the IMU with respect to the world frame ($\stdvecnb{\vchar}[W][I] \in \Real^3$), IMU gyroscope ($\bg \in \Real^3$) and accelerometer ($\ba \in \Real^3$) biases, and the IMU-camera calibration ($ ^I\mathrmbf{p}_C$,$\quat[I][C]$). The sought calibration is considered to be the position of the camera with respect to the IMU frame, $ ^I\mathrmbf{p}_C \in \Real^3$, and the attitude of the camera frame with respect to the IMU frame, $\quat[I][C] \in \Real^4$. The state vector denoted by $\x \in \Real^{23}$, encapsulates the ego-states and calibration parameters as indicated in Equation (\ref{eq:stateimu}).

\begin{equation}\label{eq:stateimu}
\x=\rowvec{\qtrue\trans, \stdvecnb{\pchar}[W][I][\rmT], \stdvecnb{\vchar}[W][I][\rmT], \bg\trans, \ba\trans,  \stdvecnb{\pchar}[I][C][\rmT],\quat[I][C]^T} \ .
\end{equation}

\paragraph{True process model.}
In this subsection, the process model is discussed. The measurement model is description is deferred for Section \ref{subsec:camera_measurement_model}.

The PU-MEKF model uses the IMU's gyroscope and accelerometer measurements to propagate rotational and translational rigid body kinematics. Since the raw IMU measurements inevitably deviate from the true values, a measurement model to consider the corrupted measurements is used \cite{crassidis2011optimal} before utilizing the IMU measurements. In this work, the 3$\times$1 gyroscope measurement vector (the angular velocity) $ \angratetildechar $, and the 3$\times$1 accelerometer measurements (specific force)  $ \sftilde$  are related to their respective true values,  $ \angrate $ and $ \sf$, via the following models:

\begin{equation}\label{eq:angratetrue}
\angrate=\angratetildechar-\bg-\ng \ ,
\end{equation}
\begin{equation}\label{eq:bgtrue}
\bgdot=\nwg \ ,
\end{equation}

\begin{equation}\label{eq:sftrue}
\sf=\sftilde-\ba-\na \ ,
\end{equation}
\begin{equation}\label{eq:batrue} 
\badot=\nwa \ .
\end{equation}
Here, $ \ng $ is the gyroscope measurement noise, and $ \nwg $ is the  gyroscope bias drift respectively, whereas $ \na $ and $ \nwa $ are the accelerometer measurement noise and accelerometer bias drift; with $ \bg $ and $ \ba $ being the gyro and accelerometer bias. Both noise and drift (for angular and translational quantities) 3$\times$1 vectors, are modeled as zero-mean white Gaussian noise processes.

This being said,
the true process model for the IMU-camera calibration (rotational and translational kinematics) is now summarized in the following equations \cite{trawny2005indirect}, \cite{mourikis2009vision}, \cite{mirzaei2008kalman}:

\begin{equation}\label{eq:qdottrue}
\qdeltadot(t)=
\frac{1}{2}
\begin{bmatrix}
-\angratetcross & \angratet \\
-\angratet\trans & 0
\end{bmatrix} 
\qdelta(t) \ ,
\end{equation}
\begin{equation}
\pdot=\vtrue \ ,
\end{equation}
\begin{equation}\label{eq:vdottrue}
\vdot(t)=\atrue(t) =(\Cdelta \ \Cwizero )\trans \  \sf(t) + \grav \ ,
\end{equation}
\begin{equation}\label{eq:bgtrue}
\bgdott=\nwgt \ ,
\end{equation}
\begin{equation}\label{eq:batrue}
\badott=\nwat \ ,
\end{equation}
\begin{equation}\label{eq:calibration_true_lever}
^I\dot{\mathrmbf{p}}_C = 0 \ ,
\end{equation}
\begin{equation}\label{eq:calibration_true_rotation}
^C_I\dot{\mathrmbf{q}} = 0 \ .
\end{equation}

The notation $I_k$ is intended to represent the frame from which the attitude evolution starts at time $t_k$ (when a new IMU measurement arrives), and $I$, the frame where the evolution ends (after the integration of the equations, for either continuous or closed-discrete form solution). The vector $ \grav $ is the  3$\times$1 gravity vector in the world frame, and it is assumed to be known without uncertainty. Equation (\ref{eq:vdottrue}) describes the evolution of the IMU velocity. Notice that this equation accounts only for IMU motion-induced measurements, as the gravity vector is being subtracted. Finally, equations (\ref{eq:calibration_true_lever}) and (\ref{eq:calibration_true_rotation}) indicate that the calibration parameters are constant.

\paragraph{Expectation of process model.}
To form the error dynamics for the PU-MEKF (recall it is an indirect filter), that the expected model is required as well. The expected model is obtained by computing the expected value of equations (\ref{eq:qdottrue})-(\ref{eq:calibration_true_rotation}) and the IMU measurement models. That is,
\begin{equation}\label{eq:bghatsolve}
\bghatdott=\mathrmbf0 \rightarrow \bghatt=\bghatzero \ ,
\end{equation}

\begin{equation}\label{}
\angratehatt=\angratetildet-\bghatt=\angratetildet-\bghatzero \ ,
\end{equation}
\begin{equation}\label{eq:qdot}
\qdeltadothat(t)=
\frac{1}{2}
\begin{bmatrix}
-\angratehattcross & \angratehatt \\
-\angratehat\trans(t) & 0
\end{bmatrix}
\qdelta(t) \ ,
\end{equation}

\begin{equation}\label{eq:bahatsolve}
\bahatdott=\mathrmbf0 \rightarrow \bahatt=\bahatzero \ ,
\end{equation}

\begin{equation}\label{}
\sfhatt =\sftildet-\bahatt=\sftildet-\bahatzero \ ,
\end{equation}

\begin{equation}\label{eq:vdot}
\vhatdott=\Cwizerohat\trans \Cdeltahat\trans \sfhatt + \grav \ .\\
\end{equation}
Here, the biases  $\bghatzero$ and $\bahatzero$ indicate that the biases are to remain constant during the propagation step (at time $ t=k $), however, they are allowed to change within the filter (through process noise) as they are modeled as random walks.

\paragraph{Forming the error model.}\label{subsec:error_model_pumekf}
Now the error model for the PU-MEKF can be constructed. To form the error model one needs to accordingly substitute the true and expected models (from the two previous subsections) into the additive and multiplicative error definitions from Equations (\ref{eq:ch6_error_definition_additive}) and (\ref{eq:ch6_error_definition_multiplicative}), and simplify. 
\begin{equation}\label{eq:ch6_error_definition_additive}
\xerror=\x-\xhat    \ .
\end{equation}

\begin{equation}\label{eq:ch6_error_definition_multiplicative}
\dcm(\delta\quat)=\dcm(\qtrue)\dcm(\qhat)\trans \ .
\end{equation}
The expressions reduction results in the following error model equations:

\begin{align}
\delta\dot{\boldsymbol{\theta}} &= \lfloor \Tilde{\boldsymbol{\omega}} - \bghat \rfloor \delta \boldsymbol{\theta} - \delta\bghat - \mathrmbf{n}_g ] \ ,\\
\delta^W\dot{\mathrmbf{p}}_I &= \delta^W\mathrmbf{v}_I \ ,\\
\delta^W\dot{\mathrmbf{v}}_I&= -^I_W\hat{\mathrmbf{C}}^T(\delta \mathrmbf{b}_a + \mathrmbf{n}_a + \lfloor\Tilde{\mathrmbf{s}} - \hat{\mathrmbf{b}}_z\rfloor\delta\boldsymbol{\theta}_I) \ ,\\
\delta\dot{\mathrmbf{b}}_g &= \mathrmbf{n}_{gw} \ ,\\
\delta\dot{\mathrmbf{b}}_a &= \mathrmbf{n}_{aw} \ ,\\
\delta^I\dot{\mathrmbf{p}}_C &= {0} \ ,\\
\delta\dot{\boldsymbol{\alpha}} &= 0 \ .
\end{align}
Here, $\delta$ denotes small departure from true values. Specifically, $\delta\dot{\boldsymbol{\alpha}}$ is the small angle error representation for the attitude offset between the estimated rotation$\quat[I][C][\qhatonly]$ and the true rotation $\quat[I][C]$. Similarly, $\delta\dot{\boldsymbol{\theta}}$ represents the IMU attitude error. The error state vector, $\xerror \in \Real^{21}$ becomes,

\begin{equation}\label{eq:state_error_vector}
\xerror=\rowvec{\delta{\boldsymbol{\theta}}\trans, \delta\stdvecnb{\pchar}[W][I][\rmT], \delta\stdvecnb{\vchar}[W][I][\rmT], \delta\bg\trans, \delta\ba\trans,  \delta\stdvecnb{\pchar}[I][C][\rmT],\delta{\boldsymbol{\alpha}}\trans} \ .
\end{equation}
The reduction in the dimension of the state vector results from the small attitude error representation (small-angle error, which leads to a minimum attitude parametrization). This, allows 1) to approximate quaternion errors with a three elements parametrization instead of four (for both $\delta\boldsymbol{\theta}$ and $\delta{\boldsymbol{\alpha}}$), and 2) remove the explicit quaternion unit-norm constraint, which otherwise would induce a singular covariance matrix.

\paragraph{Covariance propagation.}
Now, with the error dynamics in hand the Jacobians (with respect to the error variables) for uncertainty propagation are obtained. The PU-MEKF covariance propagation equation is identical to the one used for the standard Kalman filter,
\begin{equation}\label{eq:ekfcovprop}
\dot{\Cov}=\Fmat\Cov+\Cov\Fmat\trans+\Gmat\Qmat\Gmat\trans\ ,
\end{equation}
with the Jacobians defined by
\begin{equation} \label{eq:jacobian_F_error}
\Fmat = \partder{\xerrordot_{}}{\xerror}\Big\rvert_{\expect{\xerror}\ ,\ \xhat} \ ,
\end{equation}
and
\begin{equation} \label{eq:jacobian_G_error}
\Gmat = \partder{\xerrordot}{\w}\Big\rvert_{\expect{\xerror}\ ,\ \xhat} \ .
\end{equation}
Calculating the corresponding Jacobians as per Equations (\ref{eq:jacobian_F_error}) and (\ref{eq:jacobian_G_error}), the continuous-time state transition matrix $ \Fmat $, and the input noise matrix $ \Gmat $ are obtained as
\begin{equation}
\Fmat=
\begin{bmatrix}
-\angratehatcross & \zeromatthree & \zeromatthree & -\eye{3} & \zeromatthree & \zeromatthree & \zeromatthree \\
\zeromatthree &  \zeromatthree & \eye{3}& \zeromatthree & \zeromatthree & \zeromatthree & \zeromatthree\\
-\Cwihat\trans \skewsym{(\sftilde-\bahat)} & \zeromatthree & \zeromatthree & \zeromatthree & -\Cwihat\trans & \zeromatthree & \zeromatthree \\
\zeromatthree & \zeromatthree & \zeromatthree & \zeromatthree & \zeromatthree & \zeromatthree & \zeromatthree\\
\zeromatthree & \zeromatthree & \zeromatthree & \zeromatthree & \zeromatthree & \zeromatthree & \zeromatthree\\
\zeromatthree & \zeromatthree & \zeromatthree & \zeromatthree & \zeromatthree & \zeromatthree & \zeromatthree\\
\zeromatthree & \zeromatthree & \zeromatthree & \zeromatthree & \zeromatthree & \zeromatthree & \zeromatthree\\
\end{bmatrix} \ ,
\end{equation}
and
\begin{equation}
\Gmat=
\begin{bmatrix}
-\eye{3} & \zeromatthree & \zeromatthree & \zeromatthree &  \\
\zeromatthree & \zeromatthree & \zeromatthree & \zeromatthree & \\
\zeromatthree & -\Cwihat\trans & \zeromatthree & \zeromatthree & \\
\zeromatthree & \zeromatthree & \eye{3} & \zeromatthree & \\
\zeromatthree & \zeromatthree & \zeromatthree & \eye{3} & \\
\zeromatthree & \zeromatthree & \zeromatthree & \zeromatthree & \\
\zeromatthree & \zeromatthree & \zeromatthree & \zeromatthree & \\
\end{bmatrix} \ ,
\end{equation}
 where
\begin{equation}
\w=\colvectrans{\ng, \na, \nwg, \nwa} \ .
\end{equation}
\subsubsection{PU-MEKF Measurement Update.}\label{subsec:pumekf}
The PU-MEKF uses the linear update from the standard Kalman filter and as such, it assumes a linear measurement model that is corrupted by zero-mean white Gaussian noise $\vk$ of the form
\begin{equation}\label{eq:deltayapprox}
\stdvec{y}[][k] =\Hmat_k(\xhat_k)\x_k+\vk \ .
\end{equation}
However, since this is an indirect filter, the linear measurement model operates in a measurement residual space. 
\begin{equation}\label{eq:deltayapprox}
\stdvec{r}[][k] =\Hmat_k(\xhat_k)\xerror_k+\vk \ .
\end{equation}
This is, the measurement matrix now is a map from state errors to measurement residuals.
To obtain the measurement matrix $\Hmat$ required in the measurement update, the measurement residual $\stdvec{r}$ is formed and then linearized. 

The measurement residual is defined as the difference between the true and the expected measurement model, this is
\begin{equation}
\stdvec{r}[][k]=\stdvec{r}[][k](\xerror_k,\vk,\xhat_k)=  \ym(\xerror_k,\xhat_k) +\vk - \expect[\ym(\xerror_k,\xhat_k)] \ . 
\end{equation}
Consequently, its first-order Taylor series expansion about the nominal state [$\expect[{\xerror_k}]$, $\expect[{\vk}]$, $\xhat_k$], can be carried out as
\begin{equation}
\stdvec{r}[][k]\approx\stdvec{r}[][k](\expect[{\xerror_k}],\expect[{\vk}],\xhat_k)+\partder{\stdvec{r}[][k]}{\xerror_k}(\xerror_k-\expect[{\xerror_k}])+\partder{\stdvec{r}[][k]}{\vk}(\vk-\expect[{\vk}]) \ .
\end{equation}
Note that $ \xhat $ is not considered a variable when computing the Taylor expansion. Or simply expressed as
\begin{equation}\label{eq:linearized_residual_pu_mekf}
\stdvec{r}[][k]\approx\Hmat\xerror_k+\vk \ .
\end{equation}
Equation (\ref{eq:linearized_residual_pu_mekf}) is obtained in virtue of $\partder{\stdvec{r}[][k]}{\xerror_k}=\Hmat_k  $ , $\partder{\stdvec{r}[][k]}{\vk}=\stdvec{I}$, $\expect[{\xerror_k}]=0$, $\expect[{\vk}]=0$ and the expectation of the residual being equal to zero,
\begin{equation}
\expect[{\stdvec{r}[][k]}]=\stdvec{r}[][k](\expect[{\xerror_k}],\expect[{\vk}],\xhat_k)=0 \ .    
\end{equation}
Again, note that measurement matrix $\Hmat$ maps from state errors to residuals; thus, the differentiation of the residual $\stdvec{r}$ is with respect to state errors $\xerror$. In the following section, the specifics to derive the measurement matrix $\Hmat$ for the IMU-camera calibration algorithm are given. \pagebreak
\paragraph{Measurement model.}\label{subsec:camera_measurement_model}
As mentioned at the beginning of Section \ref{sec:camera_imu_algorithm_description}, the IMU-camera calibration uses features positions detected in the camera field of view to perform the filter update step. The measurement model thus relates the state to features positions. More specifically, per the linear measurement model from Equation (\ref{eq:deltayapprox}), the measurement model maps from error state to pixels residuals. 

\begin{figure}[h]
	\centering
	\includegraphics[width=1\linewidth]{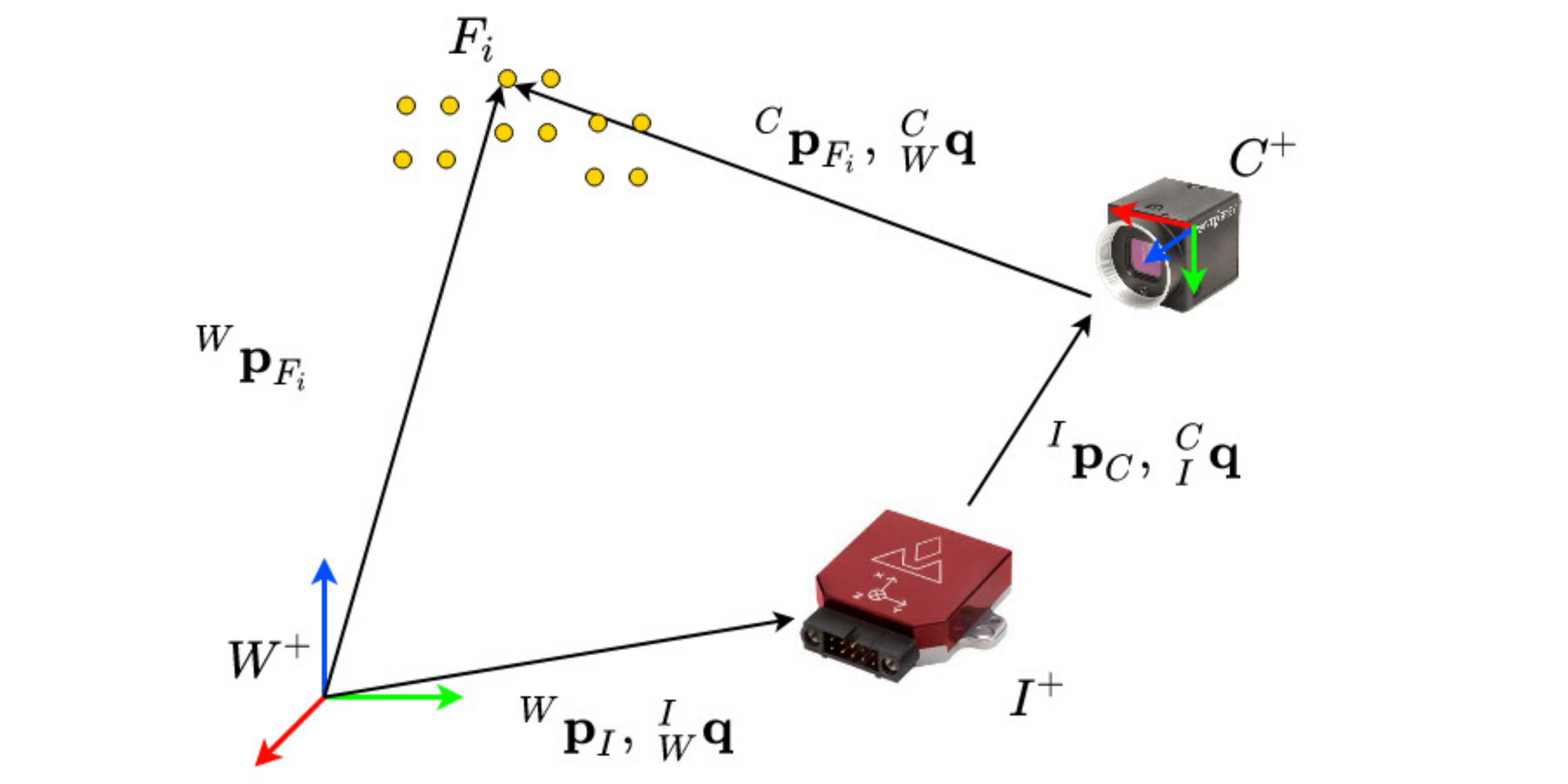}
	\caption{Relationship between World ($ W^+ $), IMU ($ I^+ $), camera ($ C^+ $) and position of the $i^{th}$ feature found in the arUco target. The arUco reference frame is considered to be the world frame $ W^+ $.}
	\label{fig:cam_imu_feat_poses}
\end{figure}

For the IMU-camera system, the measurement model is established by relating the position of a feature (feature position vector) in the camera frame to its position in the image space (pixels) via a camera model. This implementation uses the conventional pinhole camera model from Equation (\ref{eq:pinhole_model}) to associate the pixel position (denoted $\um$ and $\vm$) of the $i^{th}$ feature to its position vector in the camera frame, $\pfic$. From Figure \ref{fig:cam_imu_feat_poses} it is straightforward to show that the position vector of an $F_i$ feature coordinatized in the camera reference frame is,
\begin{equation}\label{eq:htrue}
\rowvec{h_x,h_y,h_z}\trans=\pfic=\Cic \big(\Cwi(\pftrue[i]-\ptrue)-\pci\big)\ ,
\end{equation}
where $h_x,h_y$ and $h_z$ represent the components of the $\pfic$ vector. Once the position vector for a feature is constructed, this can be translated into a projected pixel via the pinhole camera model:
\begin{equation}\label{eq:pinhole_model}
\yfim=\colvec{\um,\vm}=\colvec{f_x(h_x/h_z)+c_x, f_y(h_y/h_z)+c_y}_i+\vfi \ .
\end{equation}
The term $ \vfi $ in the pinhole camera model of Equation (\ref{eq:pinhole_model}), represents a white noise zero-mean Gaussian process with covariance matrix $ \stdvec{R}[][F_i] = E[\stdvec{vv^T}]$ that corrupts the pixel measurements. The expectation of the pinhole camera model is
\begin{equation}
\yhat_{F_i} = \expect[{\ym(\xerror,\xhat)}] = \colvec{\hat{u}_i,\hat{v}_i}=\colvec{f_x(\hxhat/\hzhat)+c_x, f_y(\hyhat/\hzhat)+c_y}_i \ .
\end{equation}

To obtain the measurement matrix $\Hmat$, as described at the beginning of Section \ref{subsec:pumekf}, the measurement residual is formed and then the first-order Taylor series approximation is performed. Since the residual will be a function of $\delta\pfic$, and in turn, $\delta\pfic$ is a function of $\xerror$, the linearized model is obtained via chain rule as
\begin{equation}\label{key}
\Hfi =
\partder{\stdvec{r}_{F_i}}{\delta\pfic}\rowvec{
	\partder{\delta\pfic}{\xerror}}\ .
\end{equation}
Calculation of the partial derivatives for the $ F_i $ feature gives
\begin{equation}
\partder{\stdvec{r}_{F_i}}{\delta\pfic}=
\frac{1}{\hzhat}
\begin{bmatrix}
f_x & 0 & -f_x\hxhat/\hzhat \\
0 & f_y & -f_y\hyhat/\hzhat 
\end{bmatrix}_i\ ,
\end{equation}
and
\begin{align}
\partder{\delta\pfic}{\xerror}&=[
\Cichat\skewsym{\big(\Cwihat(\pfiwhat-\phat)\big)}, -\Cichat\Cwihat, \zero[3][3], \zero[3][3], \zero[3][3], \\ &-\Cichat, \skewsym{\Cichat(\Cwihat(\pfiwhat-\phat)-\pcihat)}] \ .
\end{align} 

As more than one feature may be available at the update step, the measurement matrices for $m$ individual features can be vertically stacked to process all features in a batch manner. However, the measurements also can be processed sequentially (Jacobians need to be re-evaluated after each measurement assimilation). For batch measurement processing, the measurement matrix is constructed as
\begin{equation}\label{eq:H_vertically_stacked}
\Hmat_F=\colvec{\Hmat_{F_1},\vdots, \Hfi,\vdots,\Hmat_{F_m}} \ ,
\end{equation}
and the residual vector as
\begin{equation}\label{eq:residuals_stacked}
\stdvec{r}_F = \colvec{\ym_{F_1}-\yhat_{F_1},\y_{F_2}-\yhat_{F_2},\vdots,\ym_{F_m}-\yhat_{F_m}}   \ . 
\end{equation}
For the same purposes, the measurement noise covariance can be arranged in a block diagonal matrix as
\begin{equation}
\stdvec{R}_F = \diag[\stdvec{R}_{F_1},\dots,\stdvec{R}_{F_{m}}] \ .
\end{equation}

\subsubsection{The partial-update within the Multiplicative EKF}
Once a set of features is received and the measurement matrix, $ \Hmat $, is constructed, the partial-update can be executed following the equations from Algorithm 1 presented in Section \ref{subsec:PUMEKF_algorithm}. To be more specific to the IMU-camera calibration case, the partial-update is done in the following way. First, the $ \betamat $ matrix is formed
\begin{equation}
\boldsymbol{\beta} = \diag\rowvec{\beta_{\smalltheta_1},\beta_{\smalltheta_2},\beta_{\smalltheta_3},\beta_{\stdvec{\pchar}[W][I]_1},\dots,\beta_{\smallalpha_3} } \ .
\end{equation}
Then partial-update correction is computed as
\begin{equation}
\xerror^{++}=\boldsymbol{\beta}\stdvec{K}[][](\ym-\yhat)=\boldsymbol{\beta}\stdvec{K}[][]\stdvec{r}\ ,
\end{equation}
and additive and multiplicative elements are identified as
\begin{equation}
\xerror^{++}=\boldsymbol{\beta}\stdvec{K}\stdvec{r} =  \colvec{\boldsymbol{\beta_{\smalltheta}}\smallthetahat^+,\boldsymbol{\beta_{additive}}\xerrorhat_{additive}^+,\boldsymbol{\beta_{\smallalpha}}\smallalphahat^+}\ .
\end{equation}
After the partial-update posterior state error, $\xerror^{++}$, is calculated, the actual state estimates are recovered by following additive and multiplicative error definitions:
\begin{align}
\phat^+&=\phat^-+\betachar_{\stdvec{p}_I}\delta\phat^+\label{eq:position_partial_update_pumek} \ ,\\
\stdvecnb{\hat\vchar}[W][I]^+&=\stdvecnb{\hat\vchar}[W][I]^-+\betachar_{\stdvec{v}_I}\delta\stdvecnb{\hat\vchar}[W][I]^+ \ ,\\    \bghat^+&=\bghat^-+\betachar_{\bg}\delta\bghat^+ \ ,\\
\bahat^+&=\bahat^-+\betachar_{\ba}\delta\bahat^+ \ ,\\
\stdvecnb{\hat\pchar}[I][C]^+&=\stdvecnb{\hat\pchar}[I][C]^-+\betachar_{\stdvec{p}_c}\delta\stdvecnb{\hat\pchar}[I][C]^+\label{eq:lever_arm_partial_update_pumek} \ ,\\
\end{align}
The IMU attitude and the IMU-camera attitude offset states become updated according to
\begin{equation}\label{eq:attitude_partial_update_pu_mekf}
\quat[W][I]^+=\colvec{\frac{1}{2}\betachar\smalltheta^+,1}\otimes\quat[W][I]^- \ ,
\end{equation}
and
\begin{equation}\label{eq:attitude_offset_partial_update_pu_mekf}
\quat[I][C]^+=\colvec{\frac{1}{2}\betachar\delta\boldsymbol{\alpha}^+,1}\otimes\quat[I][C]^- \ .
\end{equation}
Finally, the covariance matrix is partially updated via the UD covariance update equations that appear in Table \ref{table:UDPU_filter}.

%

\subsection{Simulations}\label{sec:simulations}
In this section, numerical simulations show that the PU-MEKF can perform the IMU-camera calibration achieving sub-centimeter and sub-degree accuracy. The PU-MEKF higher robustness and better consistency are highlighted and compared to those of the conventional MEKF. 

\begin{figure}[ht]
	\centering
	\includegraphics[width=0.7\columnwidth]{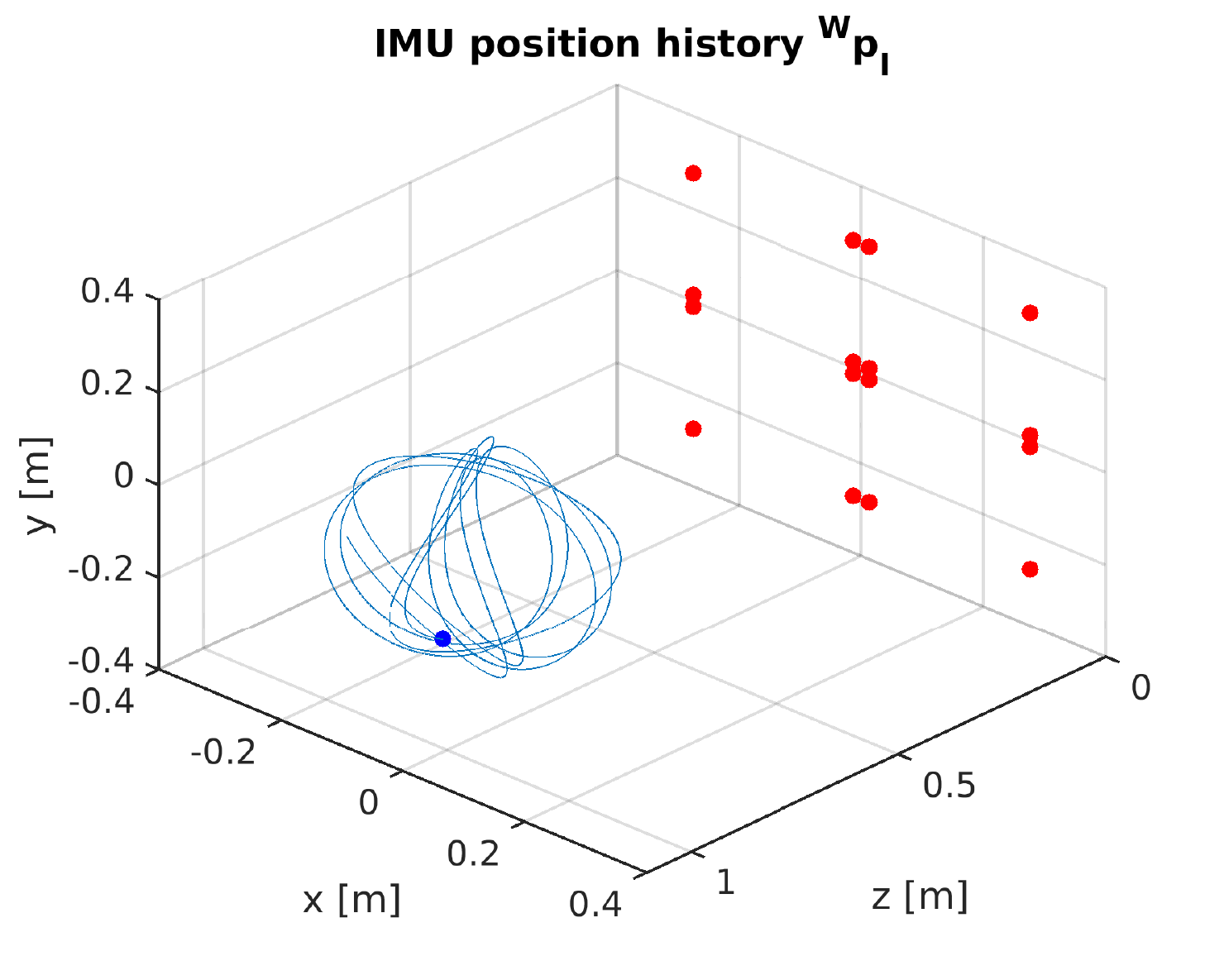}
	\caption{Simulated trajectory for a typical IMU-camera calibration run (in blue) and image features (in red).}
	\label{fig:camera_imu_motion_for_calibration}
\end{figure}

The simulation is set such that the IMU-camera system undergoes motion to stimulate all six degrees of freedom. However, the simulated system motion is constrained to ensure that the pinhole camera model predominately has the simulated markers on sight (based on its field of view-FOV). A set of four simulated arUco markers is used to provide a set of 16 available features. The pinhole camera is simulated with a FOV of 58 degrees horizontal and 45 degrees vertically, and a resolution of 640x480 pixels. The camera measurement error standard deviation is considered to be two pixels. The IMU rate is set at 100 Hz, and it is assumed that the image frame rate is 20 Hz. The process noise (IMU noises and bias drift) is taken directly from the VectorNav IMU VN-100 datasheet (this IMU is used in the hardware implementation). The motion history for a single run typically looks like the trajectory shown in Figure
\ref{fig:camera_imu_motion_for_calibration}. In this figure, the starting IMU position is indicated with the blue dot.

Again, the simulations intend to highlight the PU-MEKF gain in robustness over the (conventional) MEKF capabilities. In this vein, the filters are stressed by using relatively high initial uncertainty on the global IMU attitude, $\delta{\boldsymbol{\theta}}$, and IMU-camera attitude error parameters, $\delta{\boldsymbol{\alpha}}$.

First, the results comparing the MEKF and PU-MEKF outcome for a typical single run under the same motion and conditions, are shown. The initial conditions for the single runs are indicated in Table \ref{tab:initial_conditions_single_run}, along with relevant IMU-camera calibration parameters. The uncertainties indicated in Table \ref{tab:initial_conditions_single_run} represent a 1-$\sigma$ value. The IMU-camera attitude uncertainty was selected as two times the maximum uncertainty the regular MEKF can handle this scenario without diverging. Although the capabilities of the MEKF for this application are also dependent on simulated motion and simulation parameters, the objective is to show the PU-MEKF robustness gain under the same scenario. It is important to mention that although the initial condition is a random draw from the initial error distribution, both filters use that same random draw as the initial condition.

\begin{table}[h!]
	\begin{center}
		\caption{IMU-camera calibration parameters}
		\begin{tabular}{lcr}
			\label{tab:initial_conditions_single_run}
			\textbf{State/parameter} &  Value \\
			\hline\hline
			IMU-camera attitude uncertainty  & 2 deg \\
			Lever arm uncertainty & 5 cm \\
			IMU attitude uncertainty  & 2 deg\\
			IMU position uncertainty & 5 cm\\
			Camera frame rate&20 Hz\\
			IMU rate & 100 Hz\\
			Camera pixel uncertainty & 2 px
		\end{tabular}
	\end{center}
\end{table}

\begin{figure}[!tbp]
	\begin{subfigure}[b]{0.5\textwidth}
		\includegraphics[width=\textwidth]{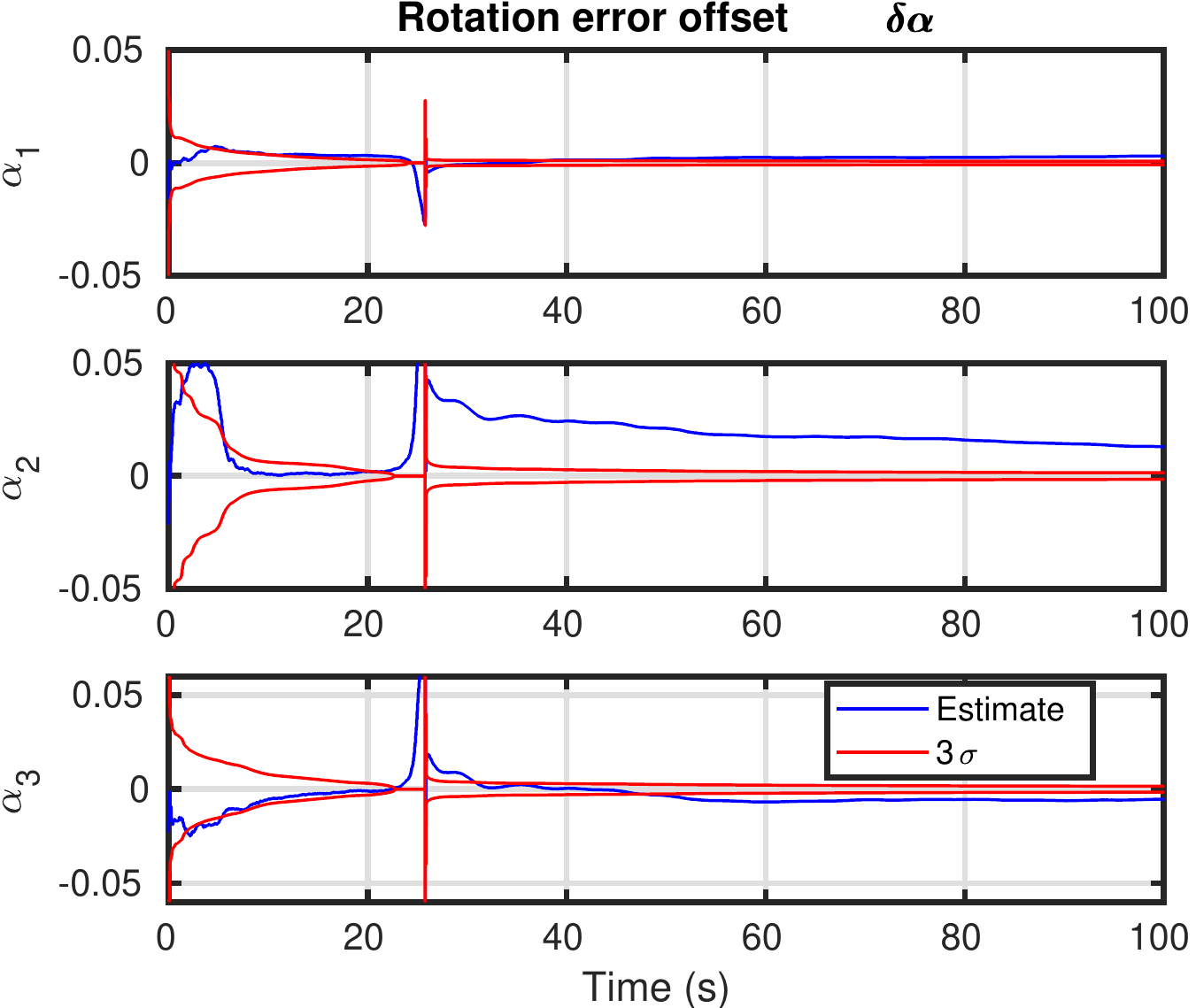}
		\caption{MEKF. Units are in radians.}
		\label{fig:f1}
	\end{subfigure}
	\hfill
	\begin{subfigure}[b]{0.5\textwidth}
		\includegraphics[width=\textwidth]{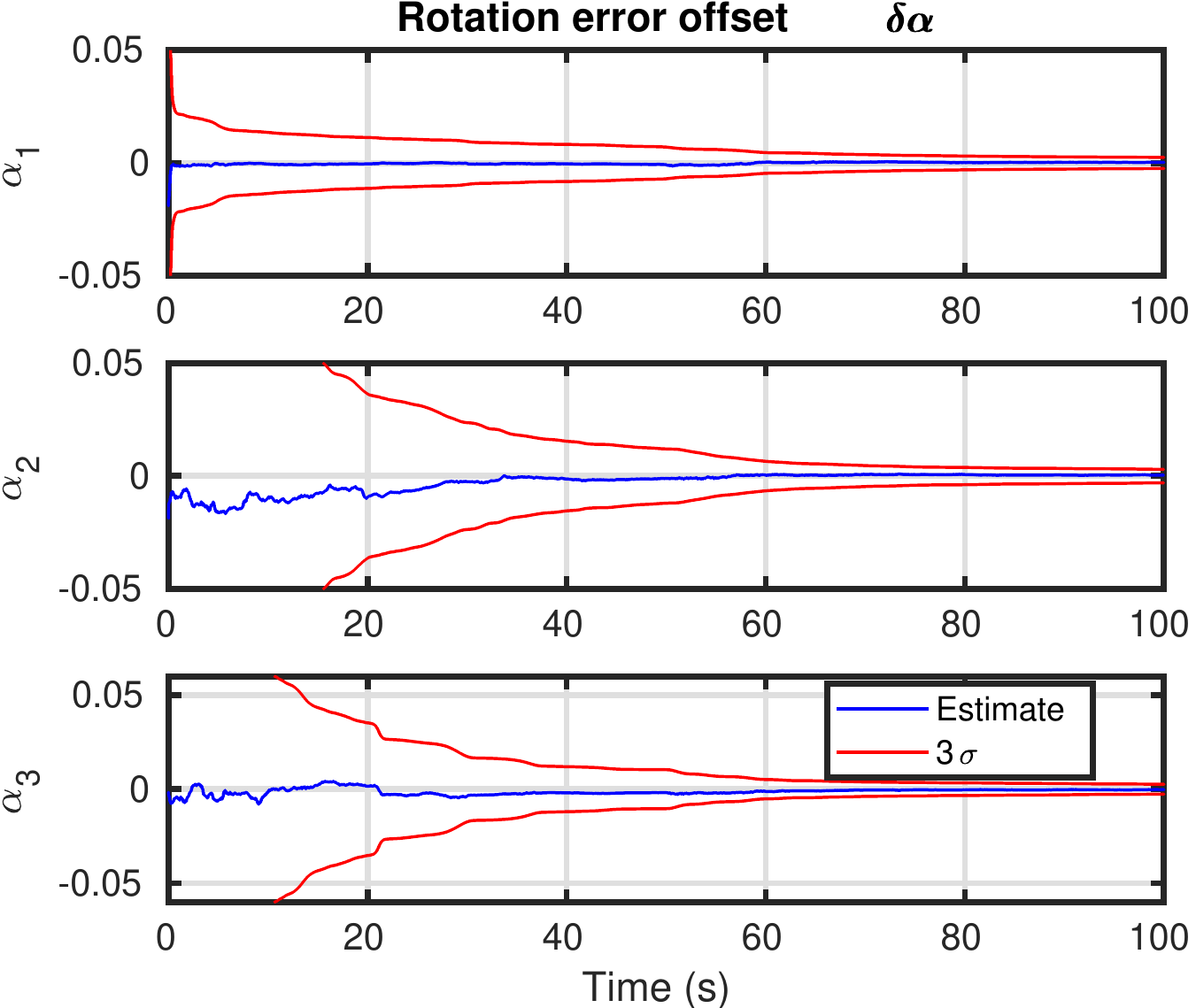}
		\caption{PU-MEKF. Units are in radians.}
		\label{fig:f2}
	\end{subfigure}
	\caption{IMU-camera attitude error for a typical simulation run. Initial condition is a random draw. MEKF and PU-MEKF use the same initial condition.}
	\label{fig:rotation_error_single_run_comparison}
\end{figure}
The partial-update weights for this system were set to the following values:
\begin{equation}\label{eq:cam_imu_weights_theta}
\boldsymbol{\beta}_{\mathbf{\delta\theta}}=\diag\rowvec{0.95, 0.95, 0.95} \ ,
\end{equation}

\begin{equation}
\boldsymbol{\beta}_{\delta ^{W}\mathbf{p}_{I}}=\diag\rowvec{0.95, 0.95, 0.95} \ ,
\end{equation}

\begin{equation}
\boldsymbol{\beta}_{\delta ^{W}\mathbf{v}_{I}}=\diag\rowvec{1,1,1} \ ,
\end{equation}

\begin{equation}
\boldsymbol{\beta}_{\delta \bg}=\diag\rowvec{1,1,1} \ ,
\end{equation}
\begin{equation}
\boldsymbol{\beta}_{\delta \ba}=\diag\rowvec{1,1,1} \ ,
\end{equation}

\begin{equation}
\boldsymbol{\beta}_{\delta^{I}\mathbf{p}_{C}}=\diag\rowvec{0.25,0.25,0.25} \ ,
\end{equation}

\begin{equation}
\boldsymbol{\beta}_{\delta \mathbf{\alpha}}=\diag\rowvec{0.25,0.25,0.25} \ .
\end{equation}
Such that,
\begin{equation}\label{eq:cam_imu_weights_alpha}
	\betamat = \diag(\boldsymbol{\beta}_{\mathbf{\delta\theta}},\boldsymbol{\beta}_{\delta ^{W}\mathbf{p}_{I}},\boldsymbol{\beta}_{\delta ^{W}\mathbf{v}_{I}},\boldsymbol{\beta}_{\delta \bg},\boldsymbol{\beta}_{\delta \ba},\boldsymbol{\beta}_{\delta^{I}\mathbf{p}_{C}},\boldsymbol{\beta}_{\delta \mathbf{\alpha}}) \ . 
\end{equation}

\begin{figure}[!tbp]
	\begin{subfigure}[b]{0.5\textwidth}
		\includegraphics[width=\textwidth,height=8cm]{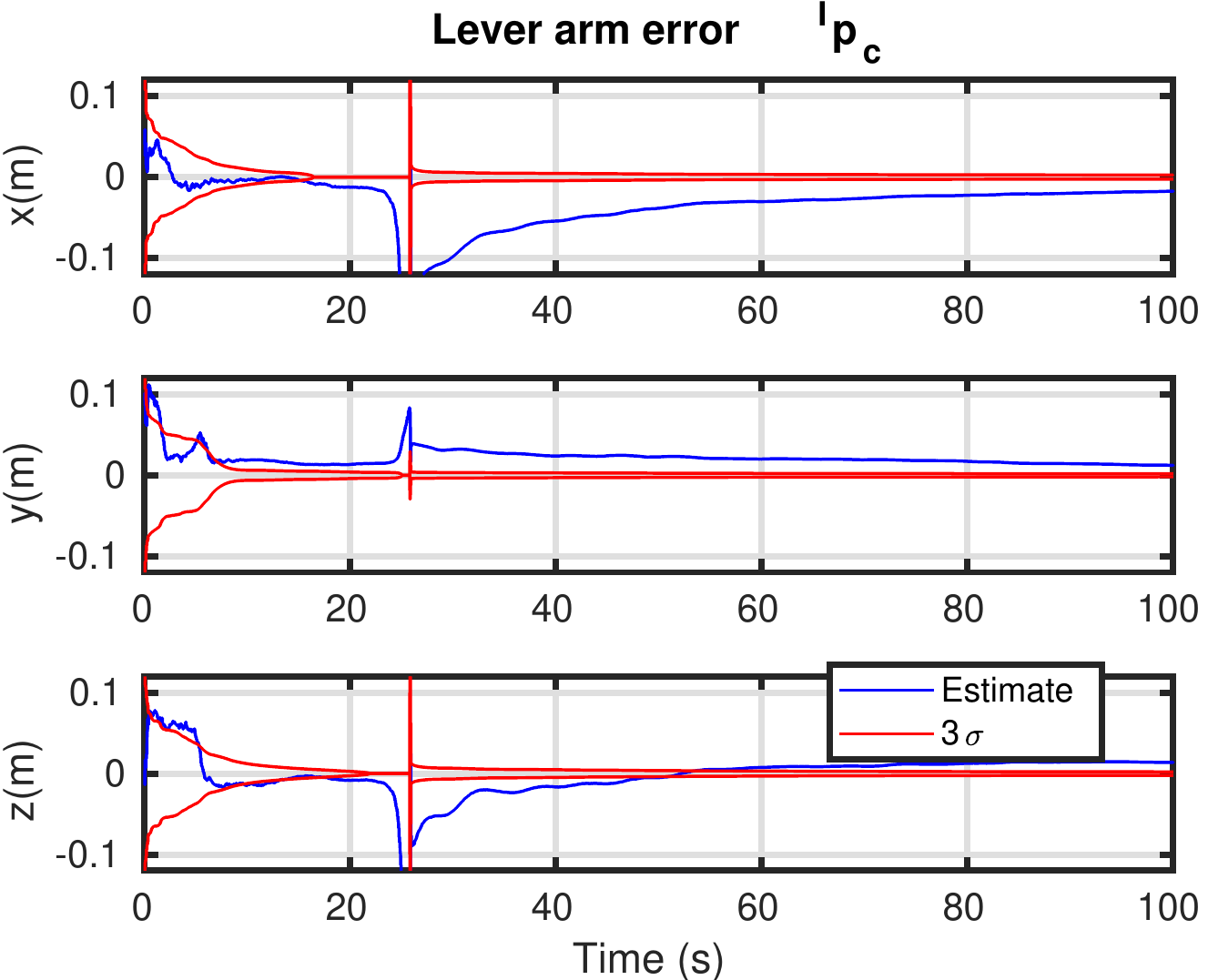}
		\caption{MEKF}
		\label{fig:f1}
	\end{subfigure}
	\hfill
	\begin{subfigure}[b]{0.5\textwidth}
		\includegraphics[width=\textwidth,height=8cm]{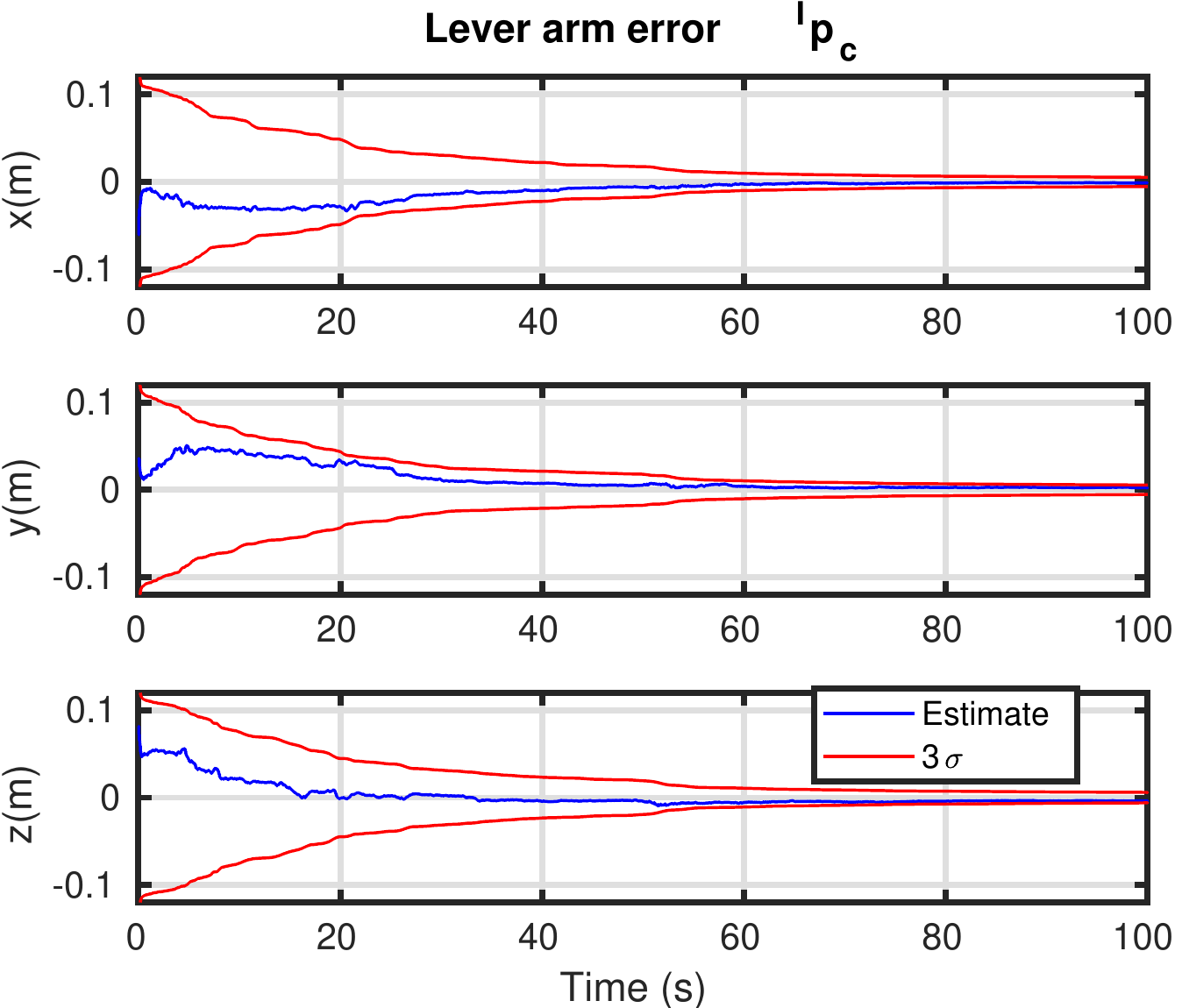}
		\caption{PU-MEKF}
		\label{fig:f2}
	\end{subfigure}
	\caption{IMU-camera lever arm errors for a typical simulation run. Initial condition is a random draw. MEKF and PU-MEKF use the same initial condition.}
	\label{fig:lever_arm_single_run_comparison}
\end{figure}

\begin{figure}[!tbp]
	\begin{subfigure}[b]{0.5\textwidth}
		\includegraphics[width=\textwidth,height=8cm]{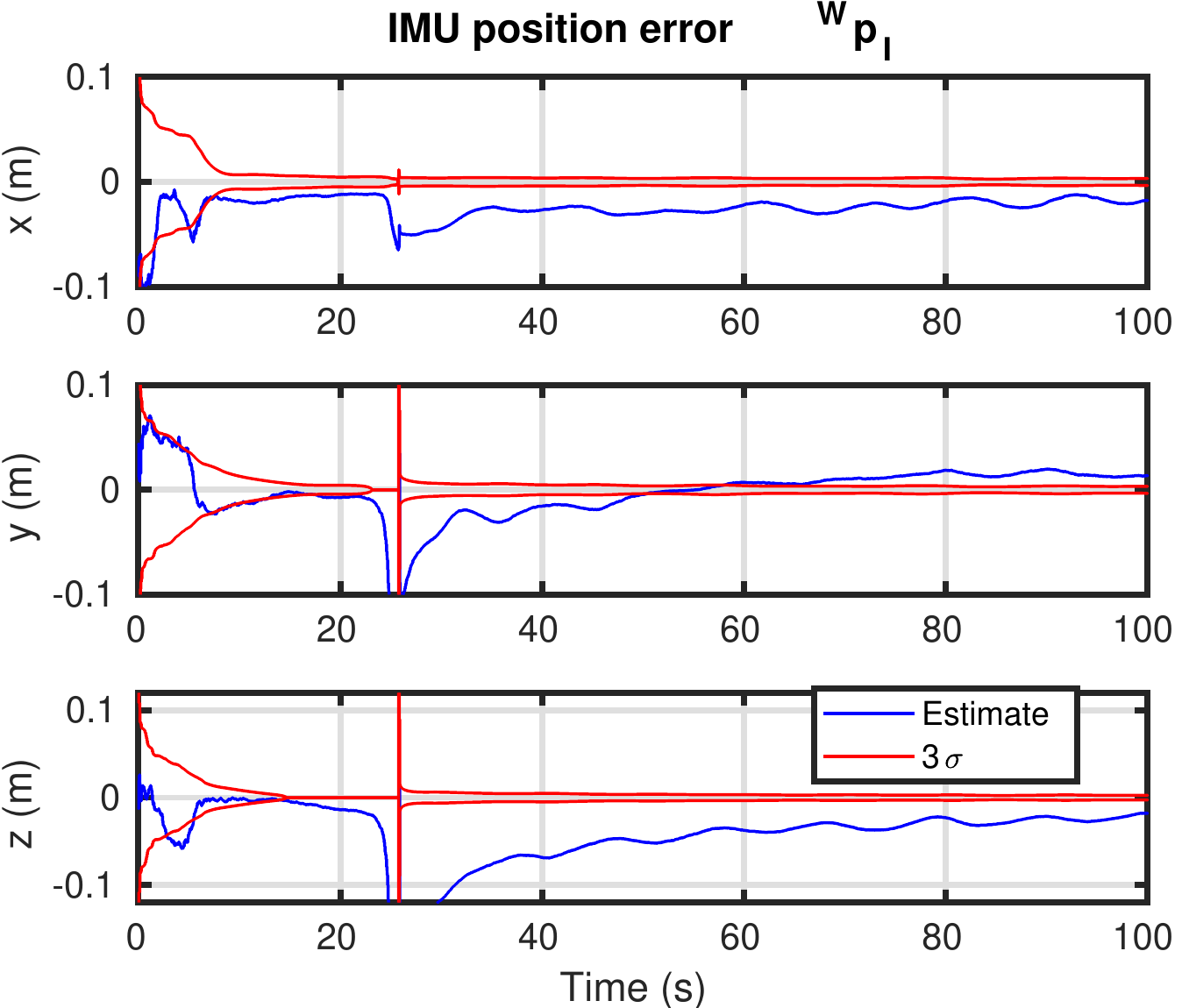}
		\caption{MEKF}
		\label{fig:f1}
	\end{subfigure}
	\hfill
	\begin{subfigure}[b]{0.5\textwidth}
		\includegraphics[width=\textwidth,height=8cm]{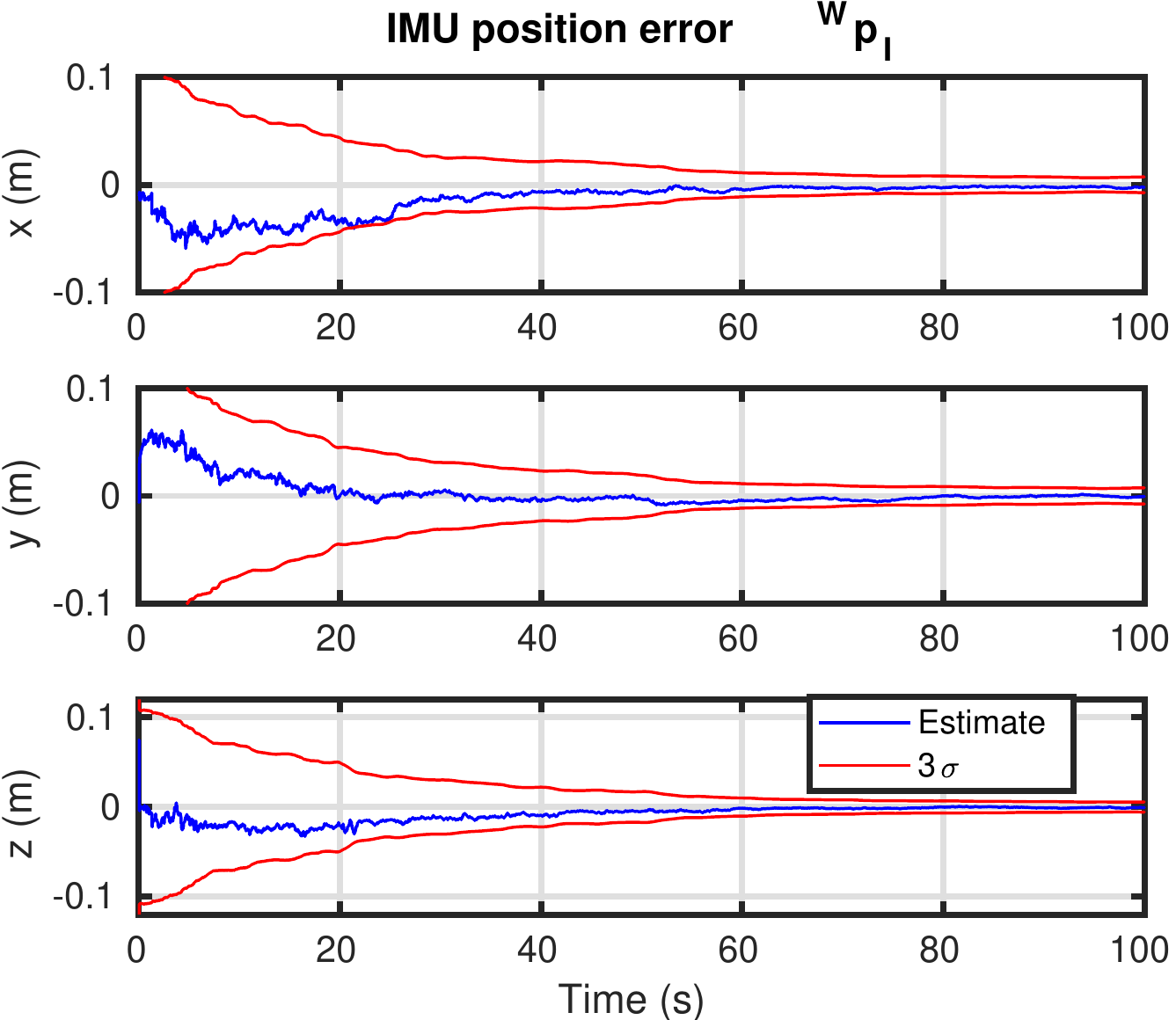}
		\caption{PU-MEKF}
		\label{fig:f2}
	\end{subfigure}
	\caption{IMU global position error for a typical simulation run. Initial condition is a random draw. MEKF and PU-MEKF use the same initial condition.}
	\label{fig:imu_global_position_comparison}
\end{figure}

\begin{figure}[!h]
	\begin{subfigure}[b]{0.52\textwidth}
		\includegraphics[width=1\textwidth,height=8.2cm]{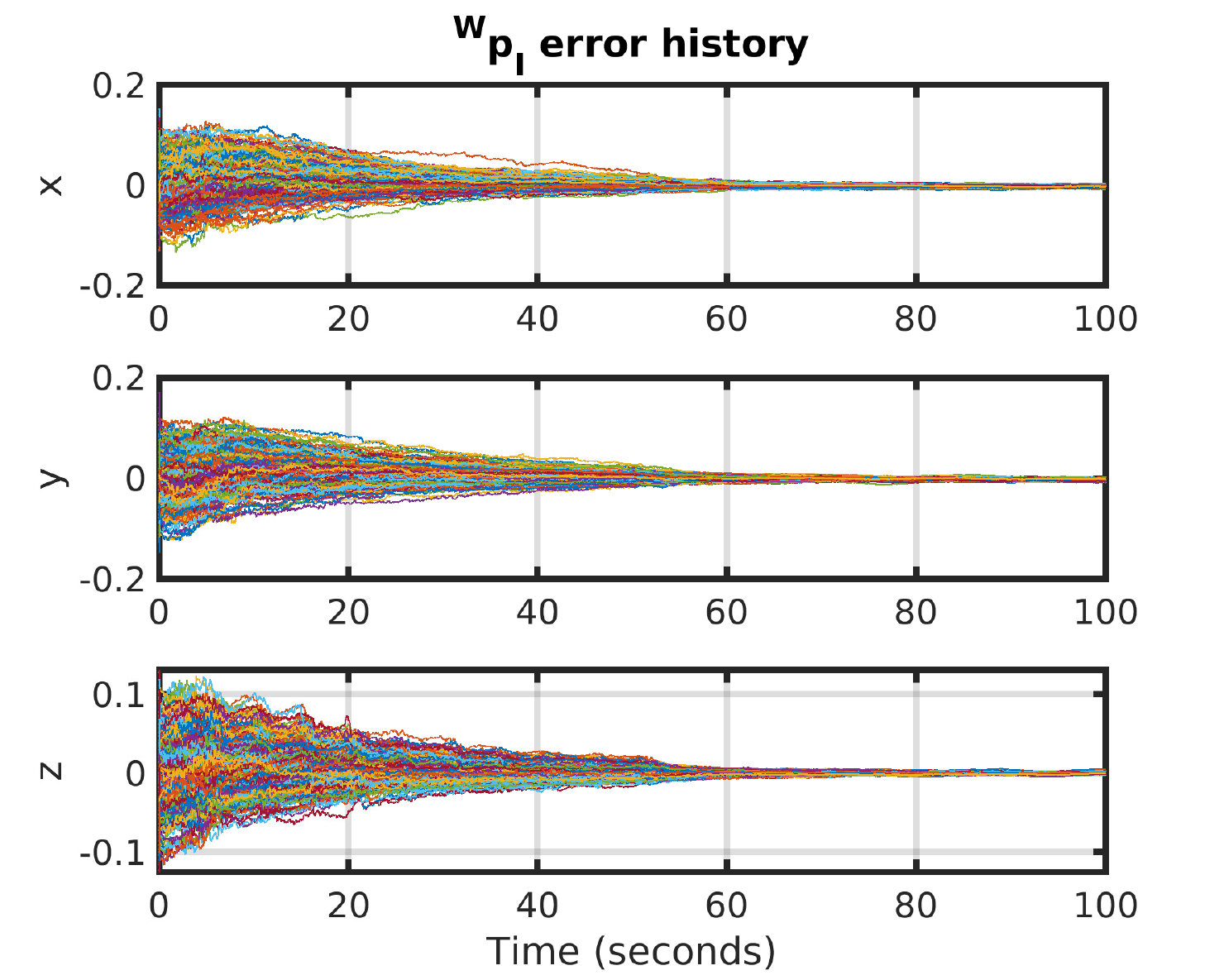}
		\caption{500 Monte Carlo runs for IMU global position \\ error (m).}
		\label{fig:f1}
	\end{subfigure}
	\hfill
	\begin{subfigure}[b]{0.47\textwidth}
		\includegraphics[width=1\textwidth,height=7.4cm]{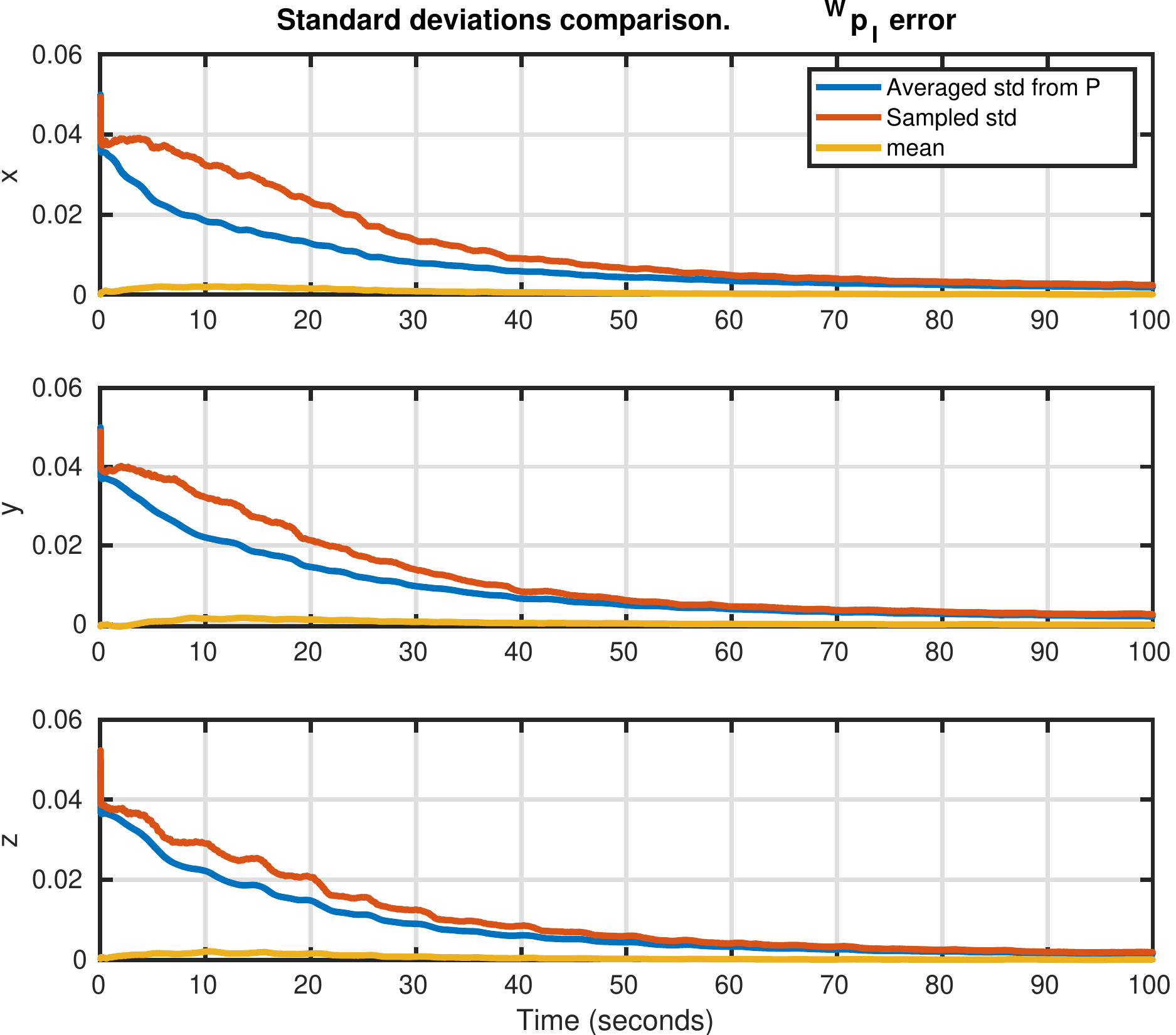}
		\caption{Estimated and actual standard deviation ($ \SI{}{\meter\squared} $) for 500 Monte Carlo runs, and filter mean errors ($ \SI{}{\meter} $).}
		\label{fig:partial_montecarlo_run_imu_position_monte_std_consistency_check}
	\end{subfigure}
	\caption{Monte Carlo runs for PU-MEKF}
	\label{fig:partial_montecarlo_run_imu_position}
\end{figure}

In Figure \ref{fig:rotation_error_single_run_comparison}, the MEKF estimates for the relative rotation between the camera and IMU are seen to be inconsistent as its estimates are outside the 3$\sigma$ bounds. This inconsistency stems from the incapacity of the MEKF to handle the relatively high initial uncertainties of this scenario. On the other hand, the PU-MEKF can handle the situation better, and not only do the estimates appear consistent, but the estimation errors are convergent to zero. Effectively, the PU-MEKF reduces initial overreactions of the filter and allows for more proper cross-correlations to be built while information is being gained, and overall helping to prevent divergence. In contrast, the MEKF early on in the simulation attempts large corrections when the uncertainty is high, and the cross-correlations are still being constructed. Although the PU-MEKF converges slowly, it can recover from the initial errors and better handle the nonlinearities than the MEKF. Moreover, once the filter recovers from the initial errors, it produces consistent estimates (estimated error covariance in the filter are well representing the actual errors, and the average error tends to zero as sample size increases), as seen from the 500 Monte Carlo runs in Figure \ref{fig:partial_montecarlo_run_imu_position}. 
Similarly, displayed in Figure \ref{fig:lever_arm_single_run_comparison} and \ref{fig:imu_global_position_comparison} results for the single run for the lever arm error and the global position estimates, respectively, are seen improved when the PU-MEKF is utilized.  

\begin{figure}[!h]
	\begin{subfigure}[b]{0.52\textwidth}
		\includegraphics[width=\textwidth,height=8cm]{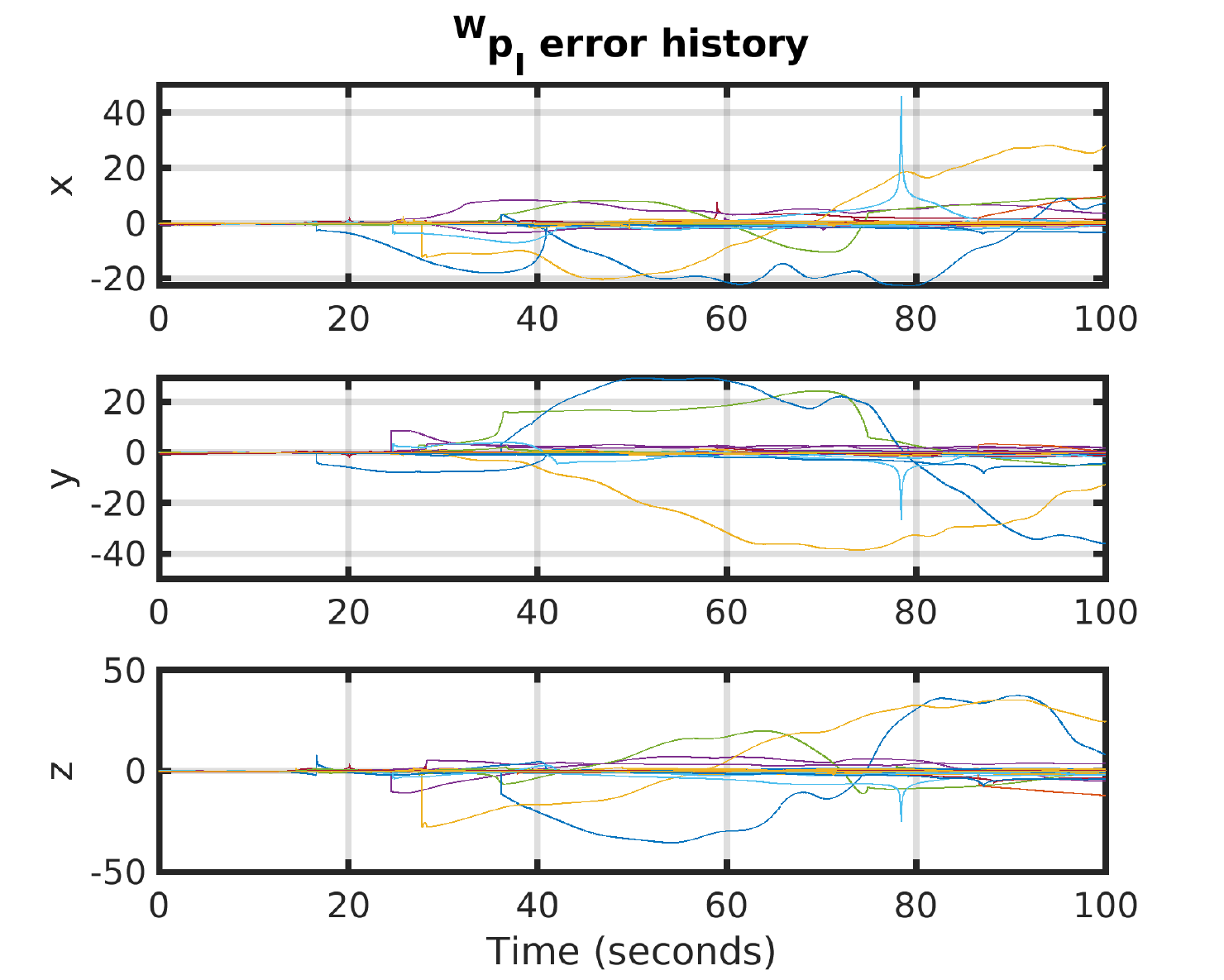}
		\caption{500 Monte Carlo runs for IMU global position \\ error (m).}    \label{fig:fig1}
	\end{subfigure}
	\hfill
	\begin{subfigure}[b]{0.48\textwidth}
		\includegraphics[width=\textwidth,height=8cm]{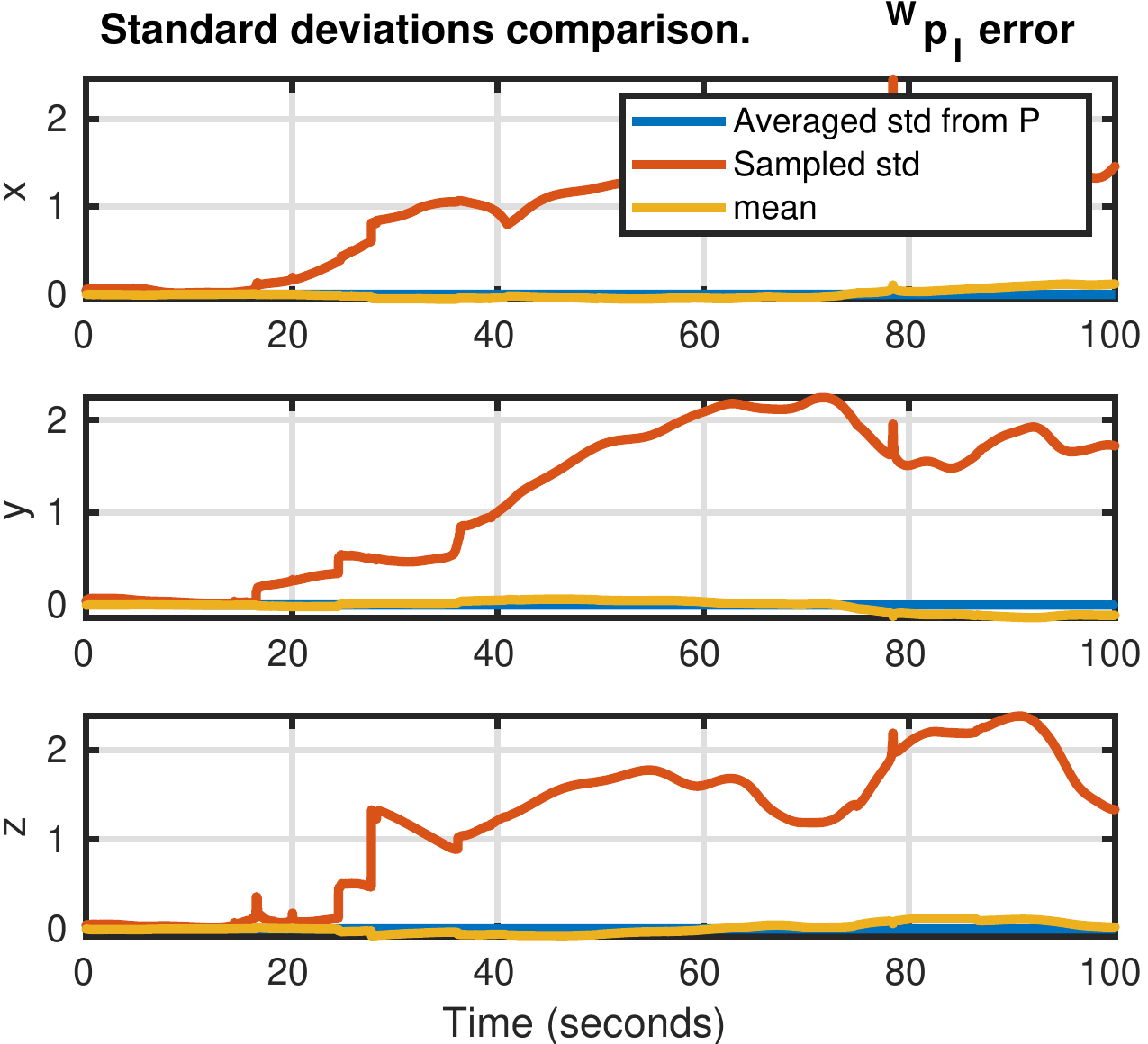}
		\caption{Estimated and actual standard deviation ($ \SI{}{\meter\squared} $) for 500 Monte Carlo runs, and filter mean errors (m).}  \label{fig:full_montecarlo_run_imu_position_monte_std_consistency_check}
	\end{subfigure}
	\caption{Monte Carlo runs for MEKF}
	\label{fig:full_montecarlo_run_imu_position}
\end{figure}

To further investigate the filters, a 500 runs Monte Carlo simulation utilizing the initial uncertainties from Table \ref{tab:initial_conditions_single_run} for the random draws, is performed.  Figures \ref{fig:full_montecarlo_run_imu_position} and \ref{fig:partial_montecarlo_run_imu_position} specifically show the position of the IMU with respect to the world frame ($\stdvec{\hat{\pchar}}[I][W]$) (for all 500 runs), which ultimately is what is sought to be improved by finding the IMU-camera calibration. These Monte Carlo results show that the MEKF, from Figure \ref{fig:full_montecarlo_run_imu_position}, it is unable to handle many of the initial conditions, causing navigation errors up to the order of meters or leading to divergence in many others. In contrast, the PU-MEKF shows a more robust behavior and more consistent estimates. In Figures \ref{fig:full_montecarlo_run_imu_position_monte_std_consistency_check} and \ref{fig:partial_montecarlo_run_imu_position_monte_std_consistency_check} the averaged standard deviation (computed using the filter covariance estimate) is compared with the sampled standard deviation (computed using the actual estimation errors) from all of the runs. It can be seen that the PU-MEKF estimated covariance, from Figure \ref{fig:partial_montecarlo_run_imu_position_monte_std_consistency_check}, is such that is very close to standard deviation of the true errors. In fact the PU-MEKF is slightly overconfident, but it is far better than the conventional MEKF results displayed in Figure \ref{fig:full_montecarlo_run_imu_position_monte_std_consistency_check}. The PU-MEKF error mean is also plotted in this figure.  It also can be seen that early in the simulation, as the filter attempts to estimate the state, the error mean is not precisely around zero (due to the large initial uncertainties, errors, and nonlinearities), however, as the filtering develops the mean of the error tends to zero. The fact that the PU-MEKF standard deviation essentially matches the actual standard deviation of the errors, and that the mean of the estimation errors tends to be zero over time, makes the PU-MEKF a consistent filter for this scenario.

The use of a PU-MEKF for the IMU-camera calibration problem does not remove the need for initializing the filter with proper values, nor suggests that better initialization methods should not be used. Instead, it gives far more room to have bad initial uncertainties while still providing converging results, relaxing the initialization methods and filter requirements. This capabilities of the PU-MEKF, and the fact that it is a sequential filter that runs online, makes it suitable for on-board-online calibration. Finally, these simulations demonstrated that the PU-MEKF could allow the same MEKF structure to broadens its applicability, and specifically for this case, enable its use for IMU-camera calibration. The next section presents the results of its hardware implementation.

\subsection{Hardware experiments} \label{sec:hardware}
\subsubsection{Setup}
The hardware implementation uses the PU-MEKF in UD factorized form. The experiments setup includes an Orbec Astra Pro RGB-D camera, and a VectorNav IMU VN-100 mounted on a rigid support. The arUco target used for calibration includes four markers, providing 16 known features. An Optitrack motion capture system is used to acquire ground reference data for the arUco markers and IMU, but it is not by any means used to validate the approach presented (Optitrack beacons are not aligned with the unknown IMU reference orientation). Instead, the re-projected pixel position error consistency, along with calibration parameter results, are used to validate the PU-MEKF approach. The IMU-camera calibration filter uses the partial-update weights of Equations (\ref{eq:cam_imu_weights_theta})-(\ref{eq:cam_imu_weights_alpha}).

The RGB-D camera images are captured at 20Hz, while the IMU sampling rate is 200 Hz. The calibration filter is run while the IMU-camera system was subject to general motion (as to excite all degrees of freedom) while arUco features are visible 99\% of the time. The profile of the position in the hardware experiments tries to mimic those of the simulations (Figure \ref{fig:camera_imu_motion_for_calibration}). Regarding magnitudes of motion velocities, at times, they are as large as the exposure time allows to obtain non-blurred images (maximum 1 m/s). The position amplitudes are as large as possible to allow the motion of the setup (by hand) with ease (approximately motions of up to 60 cm/s of amplitude). Rotational motion covers the range of up to about $\pm$ 45 deg for azimuth and elevation, and up to approximately $\pm$ 90 deg for rotations about the camera axis. Finally, the arUco calibration target is placed vertically and approximately aligned with the gravity vector with the use of the motion capture system (calibrated to make one of its axis reference frame align with the gravity vector).

\subsubsection{Filter Initialization}
To initialize the filter, specifically for IMU global position ($\stdvec{p}[I][W]$) and orientation $\quat[W][I]$, the following steps are taken. First, the perspective-n-point (PnP) algorithm is executed by using the known features positions and the detected arUco features positions. The PnP solution provides the position of the camera with respect to the world coordinatized in the camera frame, namely $[\stdvec{p}[W][C]]_C$, and the camera attitude with respect to the world frame, $\dcm[W][C]$. Then, as per Figure \ref{fig:cam_imu_feat_poses} the initial estimate for the IMU position vector is constructed as $\stdvec{p}[W][I]= -\dcm[W][I]\trans(\stdvec{p}[I][C]+\dcm[I][C]\trans[\stdvec{p}[W][C]]_C)$. Similarly, the initial global attitude of the IMU is obtained via $\dcm[W][I]=\dcm[I][C]\trans\dcm[W][C]$.

Rather than computing the PnP solution once, the initialization procedure averages 20 PnP solutions (Euler angles were used for attitude averaging) while the IMU-camera system remains still. The PnP computation is performed using the \code{solvePnP} function from OpenCV 3.3.1. Concerning the initial values for the IMU-camera lever arm, they are measured with a ruler for this setup (for the three axis). More complicated setups than the one utilized here may need to rely on 3D-CAD models, or even use additional motion-capture beacons to approximate the IMU-camera lever arm. The  IMU-camera relative attitude initialization was simplified by placing the camera and IMU, such that one plane of the IMU reference frame was approximately coincident with a plane of the camera reference frame. In this way, three simple rotations through three angles were eye-balled and rounded to a closed number. Specifically, the IMU to camera attitude initial estimate was constructed by composing three single rotations through Euler angles $[z:90 \ deg,y:0 \ deg,x:90 \ deg]$ (in that order with respect to the local axis). Note that for filter initialization,the quaternion parametrization is extracted from this constructed rotation matrix. 

Finally, the initial velocity was set to zero with very low uncertainty, while the biases for gyroscope and accelerometer were initialized with zero values, and their initial covariances are set based on the values included in the VectorNav-100 IMU's datasheet.

\subsubsection{Results}\label{sec:results}
In this section, the IMU-camera calibration results are presented. As the truth values are not available, the re-projection errors for the features are used as an indicator for calibration accuracy. For all the experiments performed, all IMU degrees of freedom are excited while the arUco features are on sight. The motion trajectories attempt to resemble those from the numerical simulations. The plots for the experiments using a full EKF are not included, but they were either divergent, similar to what is shown in Figure \ref{fig:lever_arm_single_run_comparison} or inconsistent as in the results shown in the simulations of chapter \ref{ch:square_root_pu_filter}.

\begin{figure}[h!]
\centering
\includegraphics[width=0.7\linewidth]{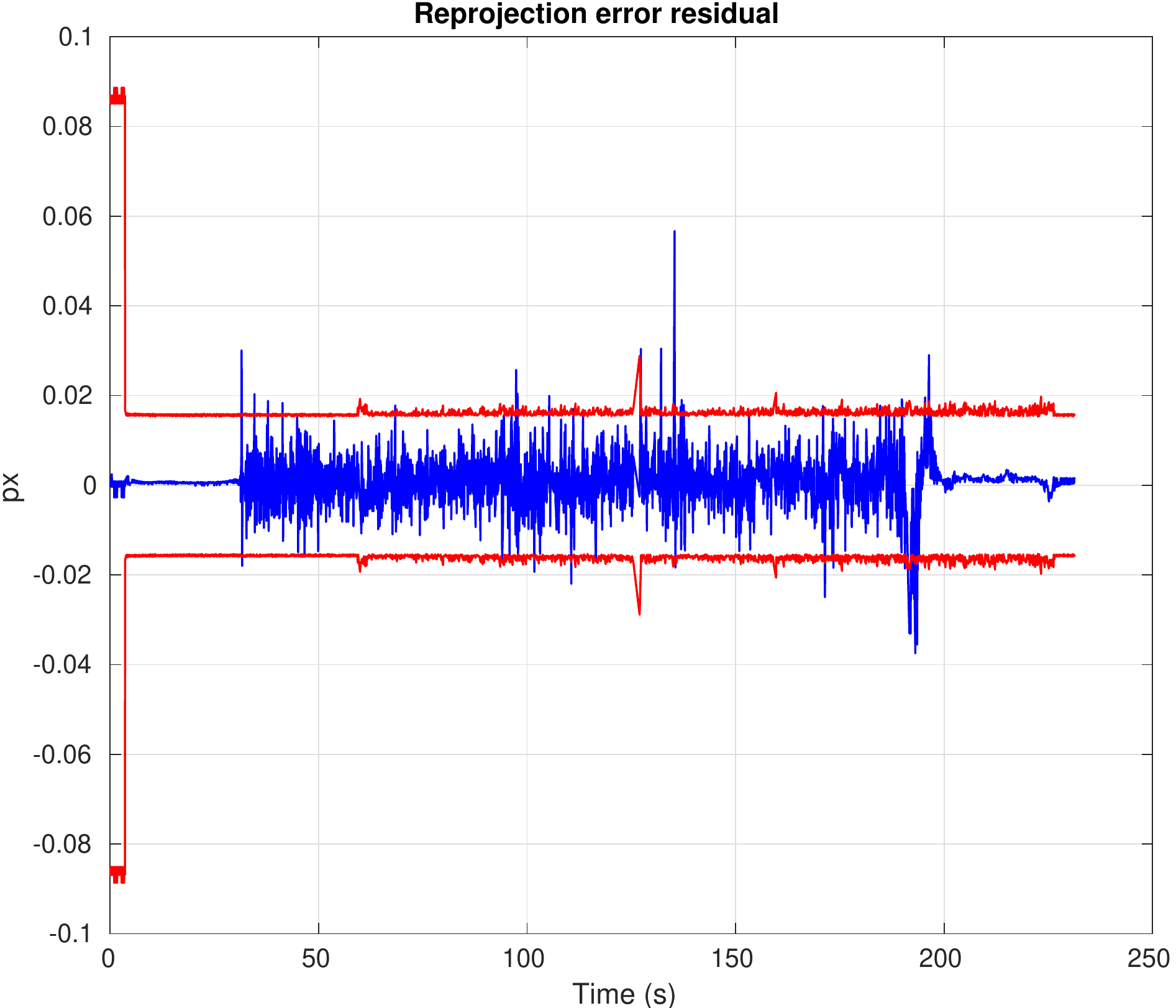}
\caption{Measurement residuals in pixels (px). The residuals indicate filter consistency. These residuals are employed to validate filter results since no true values for the calibration parameters are available.}
\label{fig:imu_camera_pixel_residuals}
\end{figure}

\begin{figure}[h!]
	\centering
	\includegraphics[width=0.8\columnwidth]{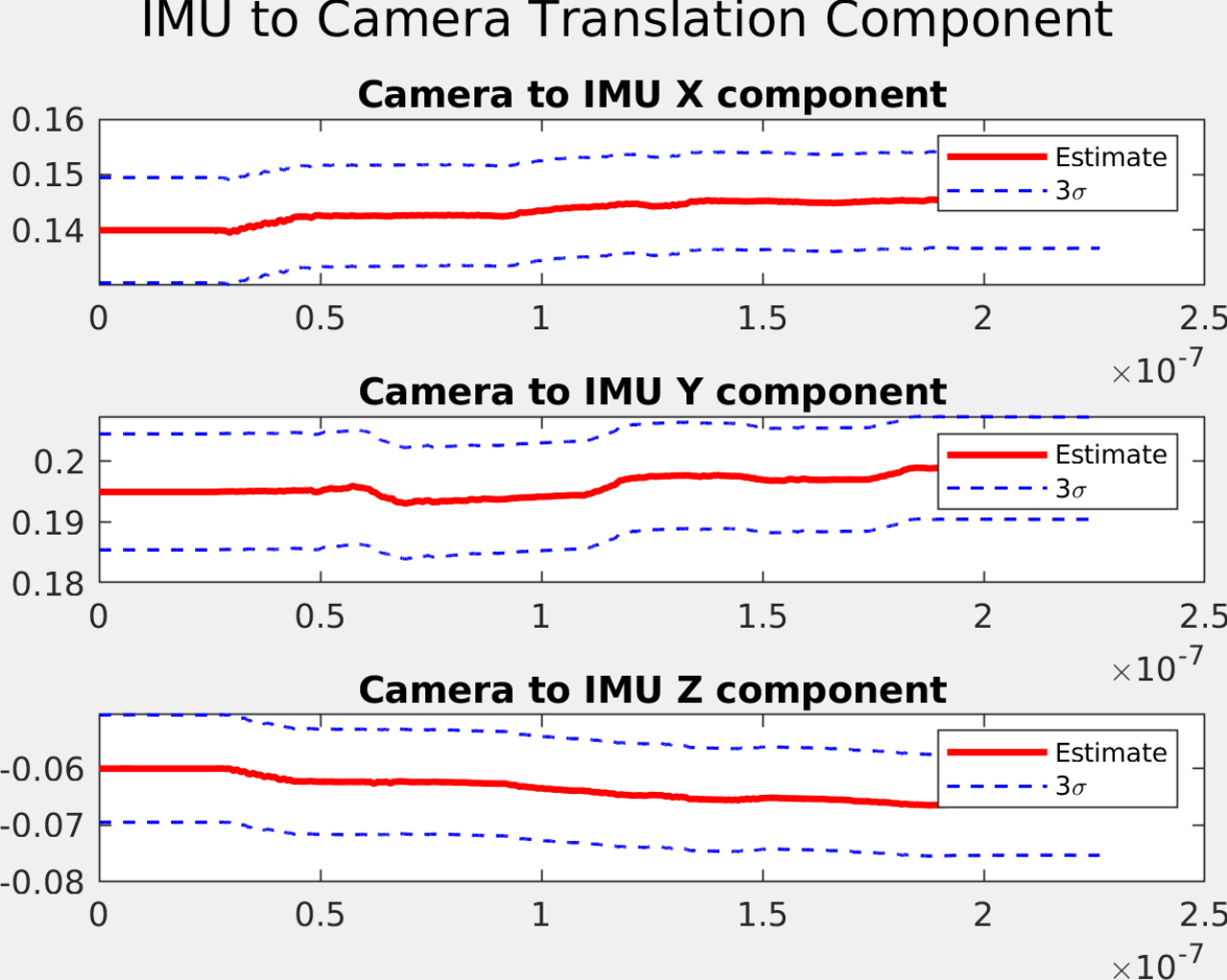}
	\caption{Lever arm hardware calibration result. The lever arm is considered to be the position vector of the camera with respect to the IMU reference frame. }
	\label{fig:imu_camera_hardware_result_lever_arm}
\end{figure}

\begin{figure}[h!]
	\centering
	\includegraphics[width=0.8\columnwidth]{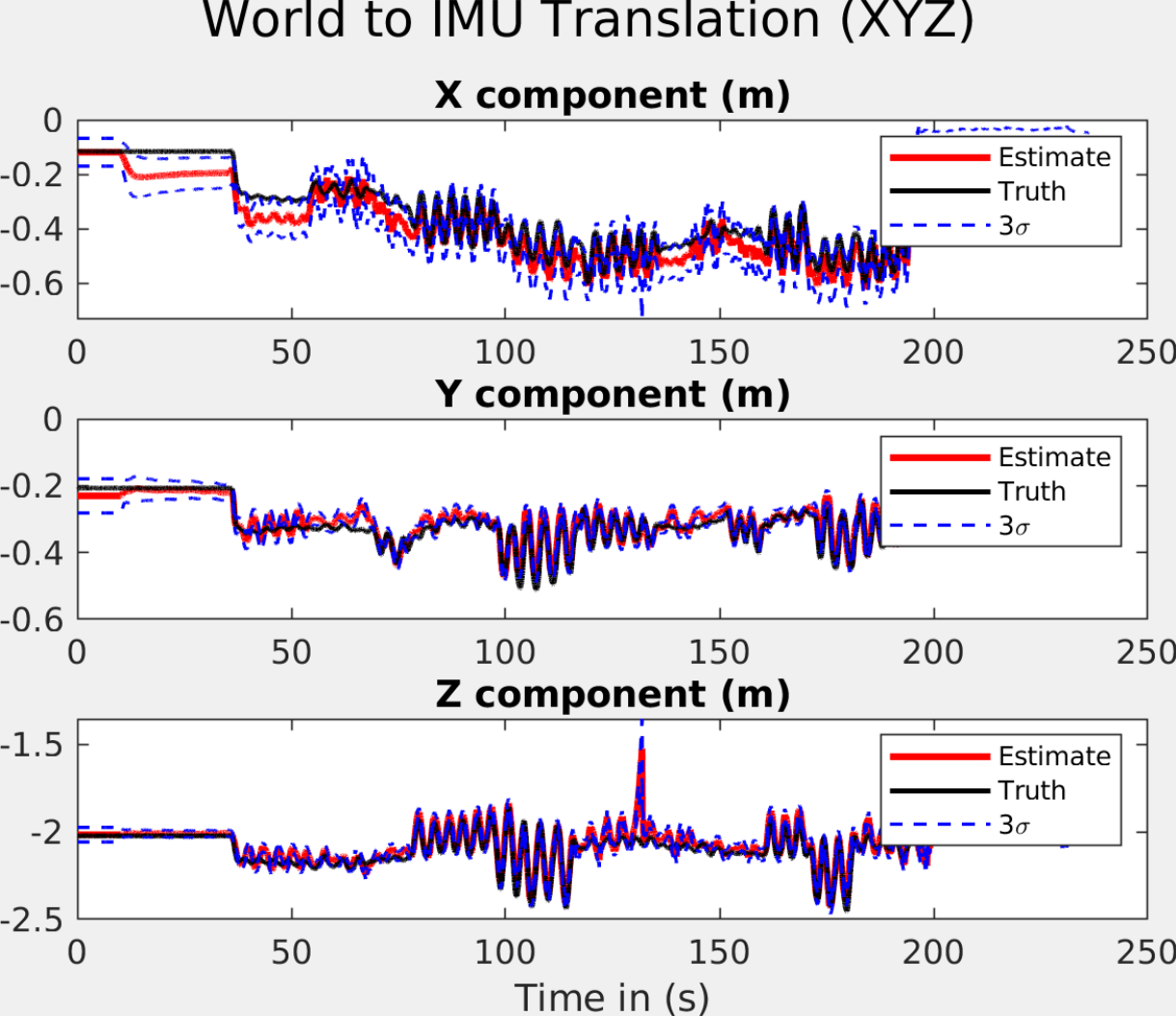}
	\caption{Position of the IMU with respect to the inertial frame hardware calibration result.}
	\label{fig:imu_camera_hardware_imu_position}
\end{figure}

\begin{figure}[h!]
	\centering
	\includegraphics[width=0.8\columnwidth]{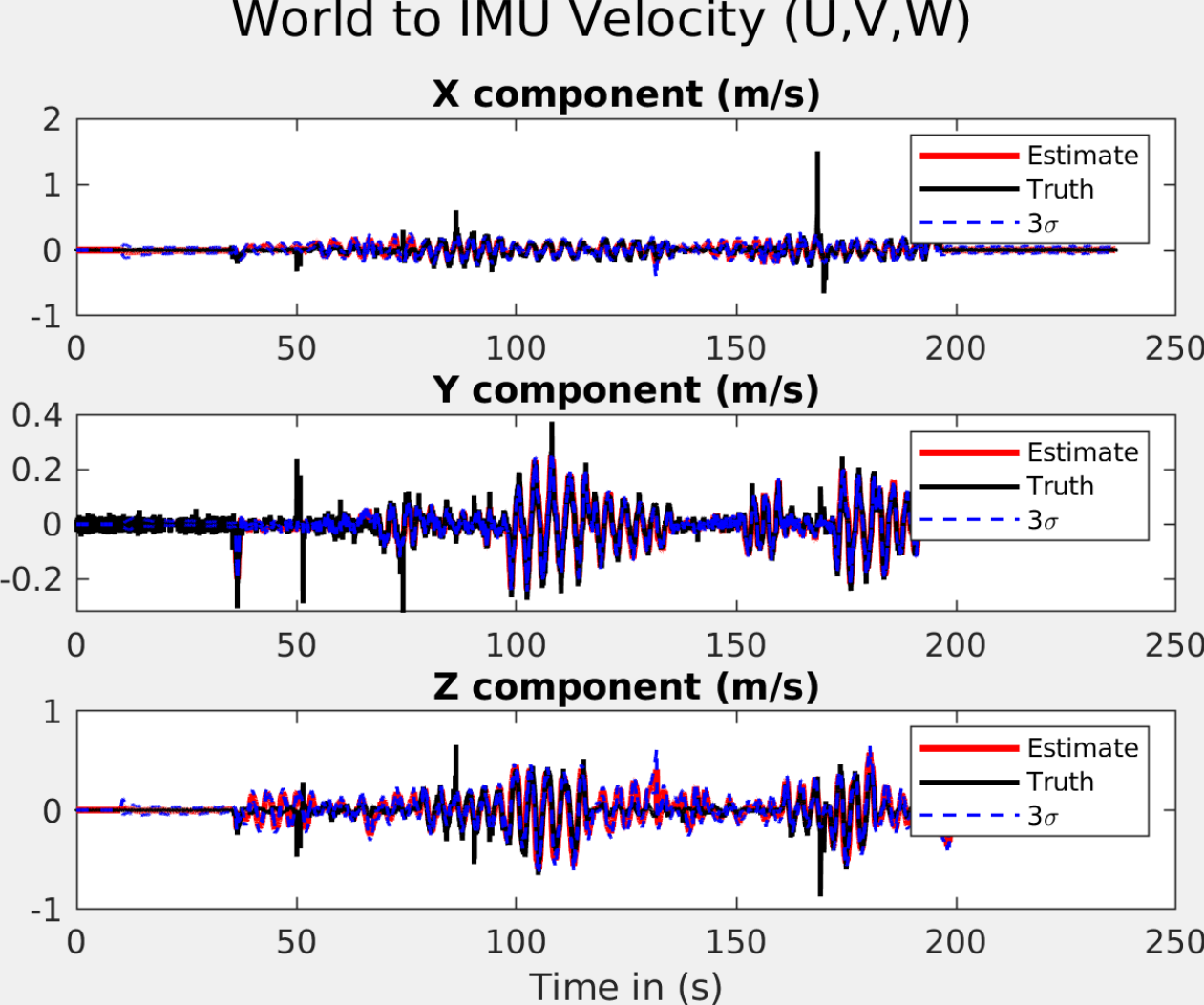}
	\caption{Velocity of the IMU with respect to the inertial frame hardware calibration result.}
	\label{fig:imu_camera_hardware_result_velocity}
\end{figure}

Calibration results for a typical hardware experiment are shown in Figures \ref{fig:imu_camera_hardware_result_lever_arm}-\ref{fig:imu_camera_hardware_result_velocity}. Figure \ref{fig:imu_camera_hardware_result_lever_arm} shows the IMU-camera lever-arm having a sub-centimeter accuracy and consistent estimate at the end of the experiment. Although the lever arm calibration is barely sub-centimeter accurate, it is seen that the filter is able to refine the initial estimate and despite the high initial uncertainties (recall that a MEKF is divergent under this scenario). The global IMU position results, which is what ultimately is sought to be improved by refining the IMU-camera parameters, are shown in Figure \ref{fig:imu_camera_hardware_imu_position}. From this figure, it is observed that at the beginning of the experiment, the global IMU position does not really "track" the Optitrack reference data; instead, it seems to diverge from it. This result is expected because the calibration parameters are initially \textit{unrefined}, in addition, the entire state vector is experiencing transients. By the middle of the run (approximately time $ t=60s $), the filter position estimates and ground reference appear to be more coincident in behavior with the ground reference, indicating that calibration parameters and biases are establishing in more appropriate values. As the experiment progresses towards its end, from Figure \ref{fig:imu_camera_hardware_result_lever_arm} can be observed that the lever arm evolution reaches a steady-state, while in Figure \ref{fig:imu_camera_hardware_imu_position} the IMU global position is seen to practically match the motion pattern of the ground reference. Again, although motion pattern matching is expected, this is not considered an indication of appropriate calibration (since motion capture markers and actual IMU frames alignment is unknown). Rather, the measurement residuals are used for consistency check and filter validation. The filter, in fact, shows consistency as from Figure \ref{fig:imu_camera_pixel_residuals}, where residuals shows to have zero mean, and remain within the 3-$\sigma$ bounds. Although true values for the calibration parameters are unavailable, the calibration parameters given by the filter were found realistic, and since the residuals were generally small and consistent, the PU-MEKF estimates were considered to be valid. The IMU velocity estimates are also showed agreement with ground truth data as seen in Figure \ref{fig:imu_camera_hardware_result_velocity}.



\subsection{Summary} \label{sec:conclusions}
This hardware application uses a PU-MEKF filter in UD factorization form. Through this application the PU-MEKF was shown to appropiately accommodate constant nuisance parameters in a Schmidt-like form while providing significant robustness against high uncertainty and nonlinearities. All, while retaining a familiar EKF framework. 

Numerical results showed that the PU-MEKF could achieve precise IMU-camera calibration while tracking core states. Further, Monte Carlo runs demonstrated that the proposed PU-MEKF is a consistent filter, and although the true parameters are unknown, filter consistency is corroborated by the measurement residuals being relatively small and consistent. Other than requiring exciting all IMU degrees of freedom, the calibration filter needs an arUco target for the feature extraction. Nevertheless, a pattern (either checkerboard or arUco or one of its variants) is often available as they are also used for intrinsic camera calibration. The filter simple initialization method and relatively low computational requirements, make the UD PU-MEKF suitable for on-board implementation, as demonstrated in this application. 


\section{A simplified Unmanned Aerial Vehicle state estimation framework}

\subsection{Introduction and Motivation} \label{introduction}
For many research projects that use aerial robots, significant effort is dedicated to building and sometimes adapting commercially available components to suit the use case. Furthermore, developing reliable flight software infrastructure diverts considerable amount of time from the main research topic. Attempting to overcome these bottlenecks, a few research groups have opted to use commercial-off-the-shelf (COTS) flight decks such as the PixHawk \cite{meier2012pixhawk} which is well suited for hobbyist-grade flight and way-point navigation in outdoor environments, but can be challenging to augment and customize due to their complex code base. Alternatively, commercial platforms such as the Parrot Bebop \cite{parrotbebop} and AR Drone \cite{Krajn2011} are also actively used, but proprietary hardware and software restrict their available functions and behavior. Some research groups have also created their own, very specific navigation solutions \cite{BYU1a}, \cite{gkasior2014estimation}, \cite{wang2012quadrotor}, but most of them tend to be highly specialized, and modifying them to meet the user needs can be time consuming and there is no guarantee of obtaining favorable results.

In order to mitigate such complications, the Autonomous Vehicle Laboratory (AVL) in the University of Florida's Research and Engineering Education Facility (REEF), and the Land Air and Space Robotics Laboratory (LASR) at Texas A\&M University, where researchers have also experienced technical difficulties, in a collaborative effort, have proposed a new starting point to establish flight capabilities: the \textit{REEF Estimator}. The REEF Estimator is an open-source and easy-to-use flight system that allows users to have a vehicle flying in a reliable manner without the need for GPS or motion capture (\textit{for the ease of understanding REEF Estimator refers to the whole system, but this includes the vehicle states estimator and its controller}). The REEF Estimator is not intended to demonstrate state-of-the-art flight capabilities, but rather to be a tractable, functional and easy-to-use implementation that offers new laboratories and students a solid \textit{launching} point for multirotor-based project development. Interested users can find the repository with the code base, flight simulator, hardware list and assembly instructions at \url{https://github.com/uf-reef-avl/reef_estimator_bundle}. 

The attractive features of the REEF estimator are the modularity and simplicity of the estimation approach. The multirotor estimator is broken into more accessible pieces: an attitude filter, a local-level-frame six-state velocity filter, and a three-state altitude filter, along with their associated controllers. Besides, any part can be used independently or replaced if needed. The REEF Estimator has been leveraged in several research problems providing a reliable and stable flight for data collection and visual odometry experiments, like the one reported in \cite{anderson2019real}. Also, the AVL has also used the three-state altitude estimator to introduce students to the EKF, C++ programming, the ROS environment, and even advanced control theory topics, to mention some. 

The altitude estimator and controller can be used independently of the lateral velocity estimator to remove throttle control from the hands of less experimented pilots. This setup alone has saved numerous flight vehicles from crashing and makes flying much more straightforward. If desired, the user can use the available REEF Estimator capabilities as-is to, for example, collect data in a more controlled flight setting (this supports several image processing efforts that require flights commensurate with stable autonomous vehicles), or exercising SLAM algorithms that generate maps and desired trajectories below allowable control rates. Laboratories at partnering universities are currently using this setup to support their research experiments.
``Out of the box," the REEF Estimator consists of a modular open-source code base written using the ROS framework \cite{quigley2009ros}, COTS components, and extensive documentation that makes it accessible to users of all levels of expertise. 

It is important to mention that the REEF estimator's simplicity and functionality rely entirely on the application of the partial-update filter.  The REEF estimator uses a UD partial-update filter to be specific. In contrast to the conventional EKF, the partial-update approach was able to maintain a consistent filter. It mainly prevents the mildly observable states IMU biases, the nuisance parameters for this application, from being over updated (in this case mainly due to model simplification) but still to estimate them. Consequently, allowing the filter core states to settle in more appropriate state and uncertainty values that produced a well-behaved estimator. The use of a conventional Schmidt or consider filter was not appropriate since it was unable to cope with the slowly-changing biases, especially for long flights.

The description of this hardware application is organized as follows. Section \ref{sec:Overview} gives an overview of the proposed solution. Sections \ref{sec:estimator_design} and \ref{sec:controller_design} discuss the details on the estimator and controller, respectively. The experiments used to validate the estimator and controller are discussed in Section \ref{sec:Result}. Finally, Section \ref{sec:Conclusion} includes a summary of the hardware implementation.

\subsection{High-Level System Overview} \label{sec:Overview}
The REEF Estimator is based on an existing attitude estimator available through the ROSFlight \cite{jackson2016rosflight} autopilot flight deck and a UD partial-update extended Kalman filter with a simplified model for local-level frame velocities and vehicle altitude estimation. Standard PID controllers are used to stabilizing the platform. The simplified state dynamics and controllers are chosen to keep the code readable and straightforward to implement, and while the proposed solution is not optimized in any way, the final product is intuitive to fly and easy to modify if needed.
Figure \ref{fig:FrameworkBlockDiagram} shows a block diagram of the framework presented in this section. 

\subsubsection{Autopilot}
The REEF estimator uses ROSFlight for attitude estimation. ROSFlight, which inherits the flexibility of ROS, is essentially a plug-and-play autopilot system. It runs on a Flip32 board that also serves as a physical interface between actuators and the on-board computer. Effectively, ROSFlight directly uses the data from the Flip32's on-board IMU to provide attitude estimates. The data coming from other sensors (e.g., sonar) connected to the Flip32 board are also available through the ROS interface (published by ROSFlight). In Figure \ref{fig:FrameworkBlockDiagram}, ROSFlight, and the Flip32 board are drawn inside the same box to denote them as the autopilot system. The REEF Estimator that runs on the on-board computer uses the sensor data and the ROSFlight attitude estimate to generate the velocity and altitude estimates of the vehicle.

\subsubsection{On-board computer}
\begin{figure*}[htp]
	\centering
	\includegraphics[width=6in]{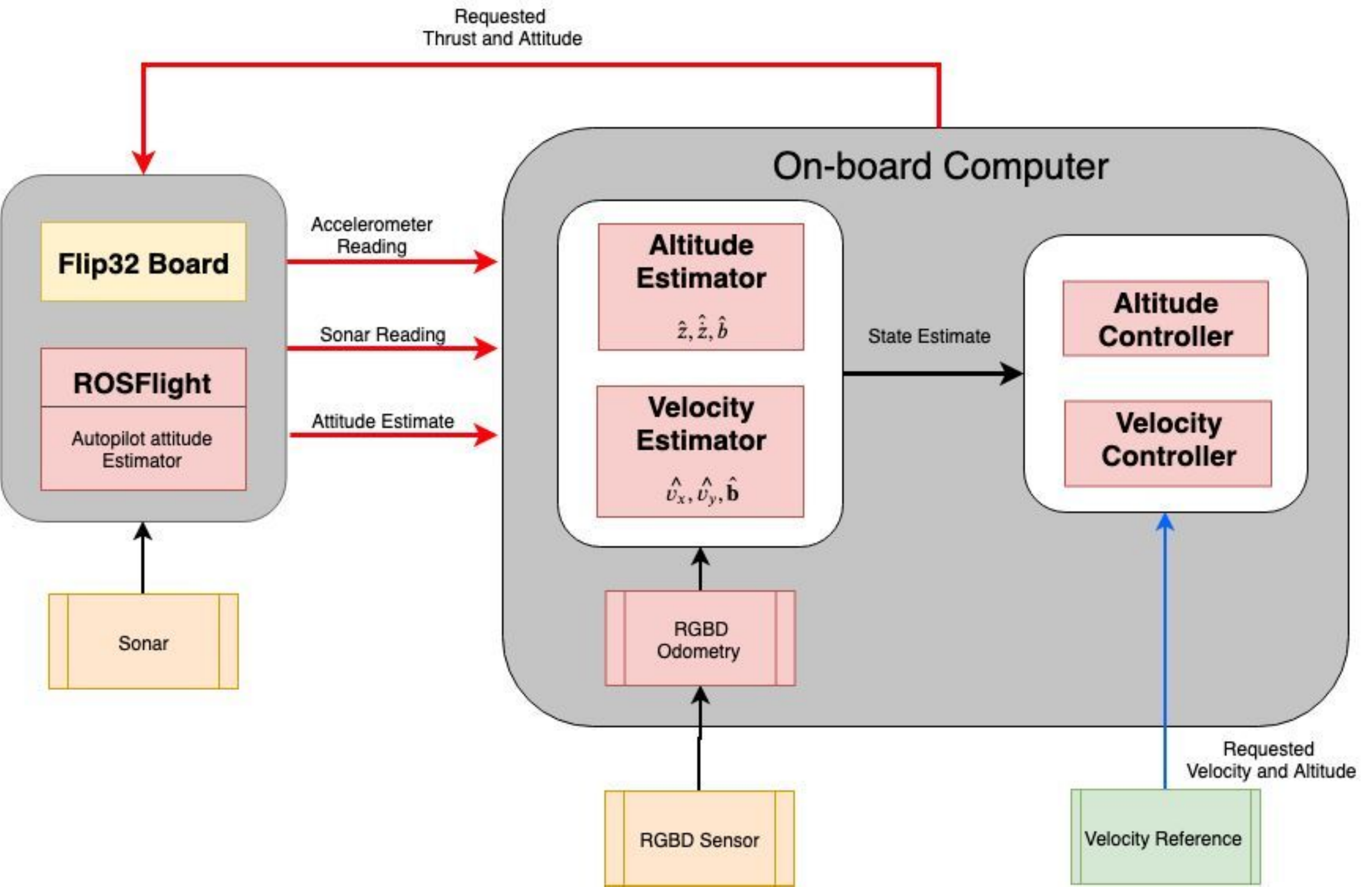}
	\caption{Block diagram of the REEF Estimator framework. Reprinted with permission from \cite{REEFEstimator}.}
	\label{fig:FrameworkBlockDiagram}
\end{figure*}

The estimators and controllers are designed and implemented to be executed on an on-board computer like Odroid or Intel NUC. The REEF Estimator (consisting of the estimators and controllers) is drawn in Figure \ref{fig:FrameworkBlockDiagram} as independent white boxes. Although the altitude and velocity estimator are contained in the same module, the user can use or replace them independently as required. As mentioned before, this can be useful if the human pilot desires to take control on the local-frame, leaving the altitude to be controlled by the computer. This mode is the most used at the AVL and LASR laboratories. This hardware implementation uses a commercial RGB-D sensor (to obtain velocity measurements indirectly) and a sonar (for altitude) to demonstrate the REFF estimator's functionality.

Inside the on-board computer box in Figure \ref{fig:FrameworkBlockDiagram}, the block called RGB-D Odometry receives RGB-D measurements (color images and depth information) and uses them to obtain velocity measurements. These measurements are then fed into the REEF Estimator. The details on how velocity estimates are generated from RGB-D data are given in Section \ref{subsub:Measurement update for the XY estimator}. Finally, a block called velocity reference is also shown in Figure \ref{fig:FrameworkBlockDiagram}. This block denotes that the on-board computer can also receive high-level velocity commands from an external source, which can be very useful if a third-party algorithm is generating the guidance information.

\subsection{Estimator Design}\label{sec:estimator_design}
\begin{figure}
	\centering
	\input{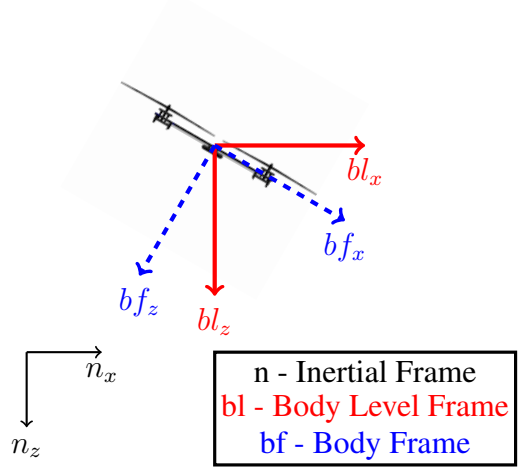}
	\caption{The body and body-level frames are co-located and share the same yaw angle. The XY plane in the body-level frame is parallel to the inertial frame XY plane. Reprinted with permission from \cite{REEFEstimator}.}
	\label{fig:body_level_frame}
\end{figure}

The REEF Estimator framework uses a UD partial-update EKF for XY local-level velocity estimator and a UD linear Kalman filter for the altitude Z. Local-level refers to the fact that the XY velocity estimation is performed in a body-fixed frame in which the XY plane is parallel to the XY plane of the inertial frame as pictured in Figure \ref{fig:body_level_frame}. The local-level frame is also referred to as the body-level frame in this implementation. From here on, the $x$ and $y$ velocity estimator is referred as the XY estimator, and the altitude estimator as the Z estimator. 

\subsubsection{Propagation of the Z estimator}
As illustrated in Figure \ref{fig:FrameworkBlockDiagram}, REEF Estimator relies on the attitude estimates coming from the sensors, ROSFlight autopilot, and the IMU readings coming from the Flip32. This data is used for the propagation step. The estimator performs Euler integration to propagate the state vector. The Z estimator state vector is comprised as follows: 
\begin{equation}\label{eq:z state}
\hat{\textbf{x}}_z=\begin{bmatrix}
\hat{z} & \hat{\dot{z}}  & \hat{b}_{a_z}
\end{bmatrix}^{T} \ .
\end{equation}
Where $\hat{z}$, $\hat{\dot{z}}$ and $\hat{b}_{a_z}$ is the altitude, vertical velocity and the accelerometer bias (in the direction of gravity) estimates, respectively. The Z estimator outputs are expressed in the gravity aligned inertial frame (positive direction being downwards). The altitude estimator process model is comprised as
\begin{equation}\label{eq:zdot dynamics}
\hat{\dot{\textbf{x}}}_z =
\begin{bmatrix}
\hat{\dot{{z}}}  \\
\tilde{a}_z+ \hat{b}_{a_z}-g\\
0\\
\end{bmatrix} \ ,
\end{equation}
where ,$\tilde{a}_z$, is the short notation for the third element in the vector $[\tilde{\textbf{a}}_z]_n$, that represents the accelerometer measurement in the $z$ direction expressed in the inertial frame ($n$ denoting the inertial frame). Specifically, $\tilde{\textbf{a}}_z$ is obtained by
\begin{equation}
[\tilde{\textbf{a}}_z]_n=\dcm[n][b]\trans[\tilde{\textbf{a}}_z]_b \ .
\end{equation}
Here, $\dcm[n][b]$ is the attitude matrix given by the ROSFlight autopilot. The $\dcm[n][b]$ is considered to be known without uncertainty, but notice that attitude bias states will be used to address autopilot attitude errors in the XY estimator (discussed in subsection \ref{subsub:Propagation of the XY estimator}). Finally, $g$ is the magnitude of the acceleration due to gravity. Covariance state propagation occurs according to the UD partial-update filter equations from Table \ref{table:UDPU_filter} from Chapter \ref{ch:ud_partial_update_filter}.


\subsubsection{Measurement update for the Z estimator}
The Z estimator implementation considers that direct altitude measurements are available, thus the measurement matrix is $\Hmat = \rowvec{1,0,0}$.
The measurement noise covariance, $\R$, is selected according to the sensor used. The measurement update is performed according to the UD partial-update filter equations from Table \ref{table:UDPU_filter} from Chapter \ref{ch:ud_partial_update_filter}.

\subsubsection{Propagation of the XY estimator}\label{subsub:Propagation of the XY estimator}
The XY velocity estimator consists of six states with the following state vector:
\begin{equation}\label{eq:vel_quad_state}
\hat{\textbf{x}}_{xy} = \begin{bmatrix}
\hat{v}_x & \hat{v}_y & \hat{b}_{q_x}& \hat{b}_{q_y} & \hat{b}_{a_x} & \hat{b}_{a_y}
\end{bmatrix}^T \ .
\end{equation}
Where $\hat{v}_x$ and $\hat{v}_y$ are the velocities in the $x$ and $ y $ directions in the body-level frame; $\hat{b}_{q_x}$ and $\hat{b}_{q_y}$ are the roll and pitch attitude bias; and $\hat{b}_{a_x}$ and $\hat{b}_{a_y}$ are the accelerometer biases in the $x$ and $y$ directions in the body-level frame. The reason for including the attitude bias states, $\hat{b}_{q_x}$ and $\hat{b}_{q_y}$, is to compensate for errors in the ROSFlight attitude estimate. The process model for the XY velocity estimator is the following:
\begin{equation}\label{eq:vdot dynamics}
\hat{\dot{\textbf{x}}}_{xy}=
\begin{bmatrix}
\dcm[b][bl]_{2\times3}(\atilde+\bhat_{a}+\dcm[n][b]\textbf{g})\\
\zerovec_{3\times1}\\
\zerovec_{3\times1}\\
\end{bmatrix} \ .
\end{equation}
Here, $\dcm[b][bl]$ is the attitude matrix that takes a vector represented in the body-fixed frame to its representation in the body-level frame. The term $\dcm[b][bl]_{2\times3}$ is the $2\times3$ top-left block of $\dcm[b][bl]$ matrix, hence $\dcm[b][bl]_{2\times3}(\atilde+\bhat_{a}+\dcm[n][b]\textbf{g})$ is a $2\times1$ acceleration vector in the body-level frame (acceleration in the $x$ and $y$ directions). The two zero vectors, $\zerovec_{3\times1}$, are the result of considering the biases as random walks with zero-mean Gaussian distribution. Equations (\ref{eq:zdot dynamics}) and (\ref{eq:vdot dynamics}) are directly used for state propagation, whereas the Jacobian matrix $ \Fk = \partder{\f_{k}}{\x}\Big\rvert_{\xprior_{k-1}}$ is computed at every time-step for propagating the covariance matrix. Notice that the rotation matrices from Equation (\ref{eq:vdot dynamics}) are given by ROSFlight, thus its notation omits the hat tilde. 

The simplification of the kinematics (with respect to a dynamical model) is apparent from \eqref{eq:vdot dynamics}. The fact that some modeling details like vehicle physics have not been considered in the process model may produce less accurate estimates than a complete model system. However, as pointed out before, the intention is not to create the perfect filter but rather a simple filter that is easy to understand and modify. The  shortcoming is acknowledged through the modularity of the code. If a particular use-case (like rapid motion) cannot be executed due to some of the modeling assumptions, the user can replace the model with a more suitable one, without disrupting the central architecture of the REEF Estimator. In fact, the modularity has been tested by exchanging the UD partial-update EKF based REEF Estimator with the square root form developed in Chapter \ref{ch:square_root_pu_filter}. The whole process involved a swap of files and changes to few lines of code.

\subsubsection{Measurement update for the XY estimator}\label{subsub:Measurement update for the XY estimator}
The XY filter relies on body-level velocity measurements. For the REEF Estimator code base the algorithm described in \cite{demo_RGBD} is used to obtain delta-pose estimates from RGB and depth images. The delta-pose is then numerically differentiated and transformed from the camera frame to the body-level frame. This transformation is straightforward since attitude estimates are readily available from ROSFlight. Since level-frame velocity measurements are generated, the $\Hmat$ matrix is considered as $\Hmat=[\Imat_{2\times2} ,\zerovec_{2\times6}]$. A fixed measurement error covariance was chosen experimentally.


\subsubsection{Partial-update filter modification}
Due to the use of Euler integration on attitude and accelerometer, apparent mismodelling of the system, nonlinearities and nuisance parameters present in the system, running the REEF Estimator with a conventional EKF was no functional. For some flight data the traditional EKF even divergent. Overall, the filter produced completely unrealistic covariance values for the velocity, and thus statistically inconsistent estimates. The first attempt at mitigating this behavior was to increase the process noise covariance (especially for the biases), however this led to poor performance when there were measurement outages, plus the estimates tended to be highly conservative. As an alternative solution, the partial-update Kalman filter was applied. For numerical robustness, the UD partial-update version was used.

Following the notation introduced in previous chapters, partial-update weights for the XY velocity estimator were selected to be close to 
\begin{align}
\bm{\beta_{XY}}&=\diag\rowvec{\beta_x,\beta_y,\beta_r,\beta_p,\beta_{b_{a_x}},\beta_{b_{a_y}}}\\\nonumber
&=\diag\rowvec{1,1, 0.01, 0.01, 0.01, 0.01} \ .
\end{align}
Using this partial-update configuration the vehicle to fly in a stable, reliable, and repeatable manner. In other words, a 100\% update for body-level frame $x$ and $y$ velocities was applied, whereas a 1\% update on the bias attitude (roll $\beta_r$ and pitch $\beta_p$) and accelerometer ($b_{a_x}$ and $b_{a_y}$) updates was used. Also with an appropriate flying behavior in mind, the altitude estimator Z was set to use percentage values of $\bm{\beta_Z}=diag\rowvec{\beta_z,\beta_{\dot{z}},\beta_{b_z}}=\diag\rowvec{1,1,0.5}$. That is, the updates percentages were 100 \% on $z$ position and $z$ velocity, while a 50$\%$ update  was applied to the bias state. 

\subsubsection{The XY and Z filters operation frequency}
Since the accelerometer measurements and attitude estimates are used to propagate the system Equation (\ref{eq:zdot dynamics}) and (\ref{eq:vdot dynamics}), the REEF Estimator runs at the rate at which the IMU delivers measurements. The measurement update on the other hand, is executed at measurement arrival. With the hardware setup presented here, the propagation step is executed at 500Hz while the updates are performed at 40 Hz for sonar and 20 Hz for the RGB-D  measurements. 


\subsection{Controller Design}\label{sec:controller_design}
\begin{figure}
	\centering
	\tikzstyle{decision} = [diamond, draw, fill=red!20, text badly centered]
\tikzstyle{block} = [rectangle split,rectangle split parts=2, draw, fill=blue!10, text badly centered, rounded corners]
\tikzstyle{line} = [draw,->,very thick]
\tikzstyle{cloud} = [draw, ellipse,fill=red!20,
    minimum height=2em]
    
\begin{tikzpicture}[auto]
\node [block] (gains) [text width=6cm,align=center , node distance= 2cm] {\textbf{Set Gains}\nodepart{second}Load tuned PID gain parameters for all controllers};

\node[block, below of=gains] (error) [text width=7cm,align=center , node distance= 3.5cm] {\textbf{Error Calculation}\nodepart{second} Calculate the errors:\\$\begin{aligned}
\dot{\textbf{z}}_{request} &= PID(\textbf{z}_{request} - \hat{\textbf{z}})\\
\ddot{\textbf{x}}_{request} &= PID(\dot{\textbf{x}}_{request} - \hat{\dot{\textbf{x}}})\\
\ddot{\textbf{y}}_{request} &= PID(\dot{\textbf{y}}_{request} - \hat{\dot{\textbf{y}}})\\
\ddot{\textbf{z}}_{request} &= PID(\dot{\textbf{z}}_{request} - \hat{\dot{\textbf{z}}})\\
PID(e) &= K_pe + K_d\dot{e} + K_i\sum e \ dt
\end{aligned}$};



\node [block, below of=error] (request) [text width=6cm,align=center , node distance= 4.5cm] {\textbf{Calculating Attitude and Thrust Requests} \nodepart{second} $\begin{aligned}
    thrust &= -\ddot{\textbf{z}}_{request}\\
    \phi &= \ddot{\textbf{y}}_{request}\\
    \theta &= -\ddot{\textbf{x}}_{request}\\
    \end{aligned}$ };
    
\node [cloud, below of=request] (pub) [text width=2cm,align=center, node distance=3.5cm] {ROSFlight};
    
\path [line] (gains) -- (error);
\path [line] (error) -- (request);
    \path [line] (request) -- node [text width=3cm] {Thrust \& Attitude request}(pub);
    \path [line] (pub) -- ++(-5.0cm,0) |- (error);
\end{tikzpicture}
	\caption{Flowchart for the XY and Z PID controllers. Reprinted with permission from \cite{REEFEstimator}.}
	\label{fig:controller_flowchart}
\end{figure}
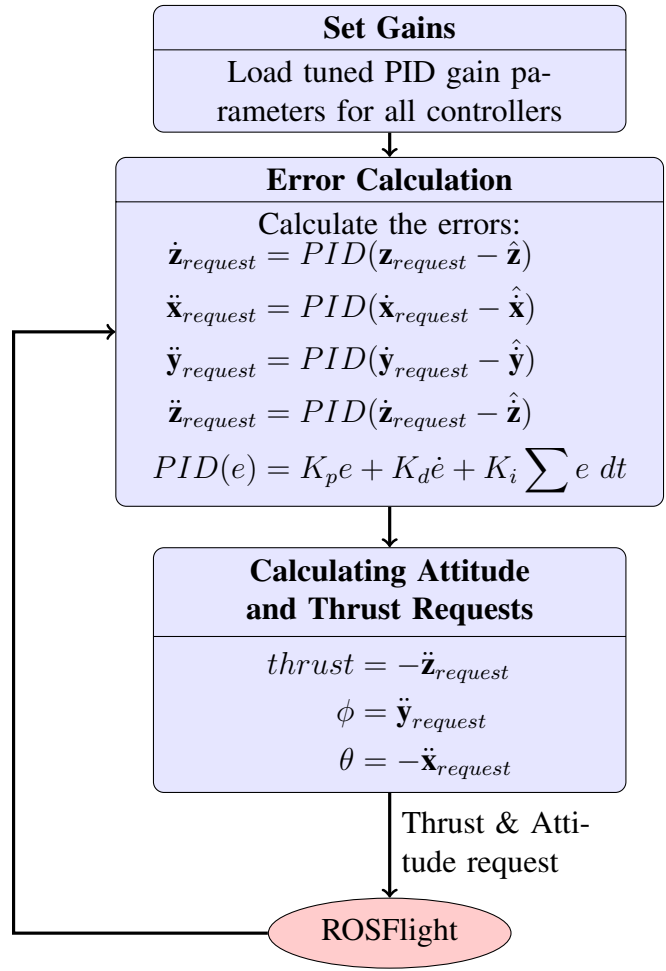

The REEF Estimator uses a cascaded Proportional-Integral-Derivative (PID) control implemented within the ROS environment. The PID controller is well established in the controls community and has been been widely used to control UAVs \cite{szafranski2011different}, \cite{salih2010flight}. Figure \ref{fig:controller_flowchart} shows how the cascading in the PID controller occurs (e.g. computations from $z_{request} $ to $\dot{z}_{request}$ to $\ddot{z}_{request}$). The PID controllers in the REEF Estimator receive altitude and body-level velocity requests along with the state estimates from the on-board computer. Then the altitude controller maps the altitude error (difference between the altitude request and current altitude) into a vertical velocity setpoint, which in turn is mapped to a thrust setpoint using the $ z $ velocity estimate. The $x$ and $y$ velocity requests are mapped to  attitude setpoints (pitch and roll, respectively) based on the error with respect to the velocity estimates. The yaw-rate requests are directly fed to the ROSFlight deck. The error calculations are performed according to the equations that appear in Figure \ref{fig:controller_flowchart}.
The mapping from error (difference between request and state estimate) into attitude and thrust setpoints is performed via the standard PID control law from the following equation: 
\begin{equation}\label{eq:PID}
u_k = K_pe_k + K_i\sum_{i=0}^{k} e_i\Delta t_i + K_d\frac{e_k-e_{k-1}}{\Delta t} \ ,
\end{equation}
where $e$ denotes the error value, and $u$ is the control signal to be applied (attitude or thrust). $K_p, K_i$, and $K_d$ denote the proportional, integral, and derivative gains, respectively, and $\Delta t$ is the time interval between estimates. In Figure \ref{fig:controller_flowchart}, the controller structure and flow of data through the system is shown. It is worth noting from the bottom box (where the attitude commands $\theta$ (pitch), $\phi$ (roll), and thrust request are calculated) that the thrust and pitch values are multiplied by a minus sign. This is just because the design of this platform follows the North-East-Down (NED) reference frame convention.



The controller framework within the REEF estimator was designed to be completely modular, with any software component, input, or output being replaceable or re-configurable with relative ease. For example, the velocity, altitude and yaw rate requests, can come from an RC or a gaming controller, or a high-level path planning algorithm. Additionally, the controller supports multiple modes of operation that intend to facilitate flight tasks. If users desire to use the system for data collection, using the altitude-hold mode would be recommended as they only have to focus on commanding motion on the XY plane. This alone is a big benefit when compared to flying with an RC transmitter commanding thrust and attitude simultaneously.


The current setup of cascaded PID control, while not robust to disturbances or capable of agile flight, is adequate for the intended purpose of expediting the process of establishing flying capabilities. Nevertheless, if the need for a different controller arises, the current code architecture enables its easy integration or substitution.

\subsection{Hardware implementation} \label{sec:Result}
\begin{figure}[htbp]
	\centering
	\includegraphics[width=3.5in]{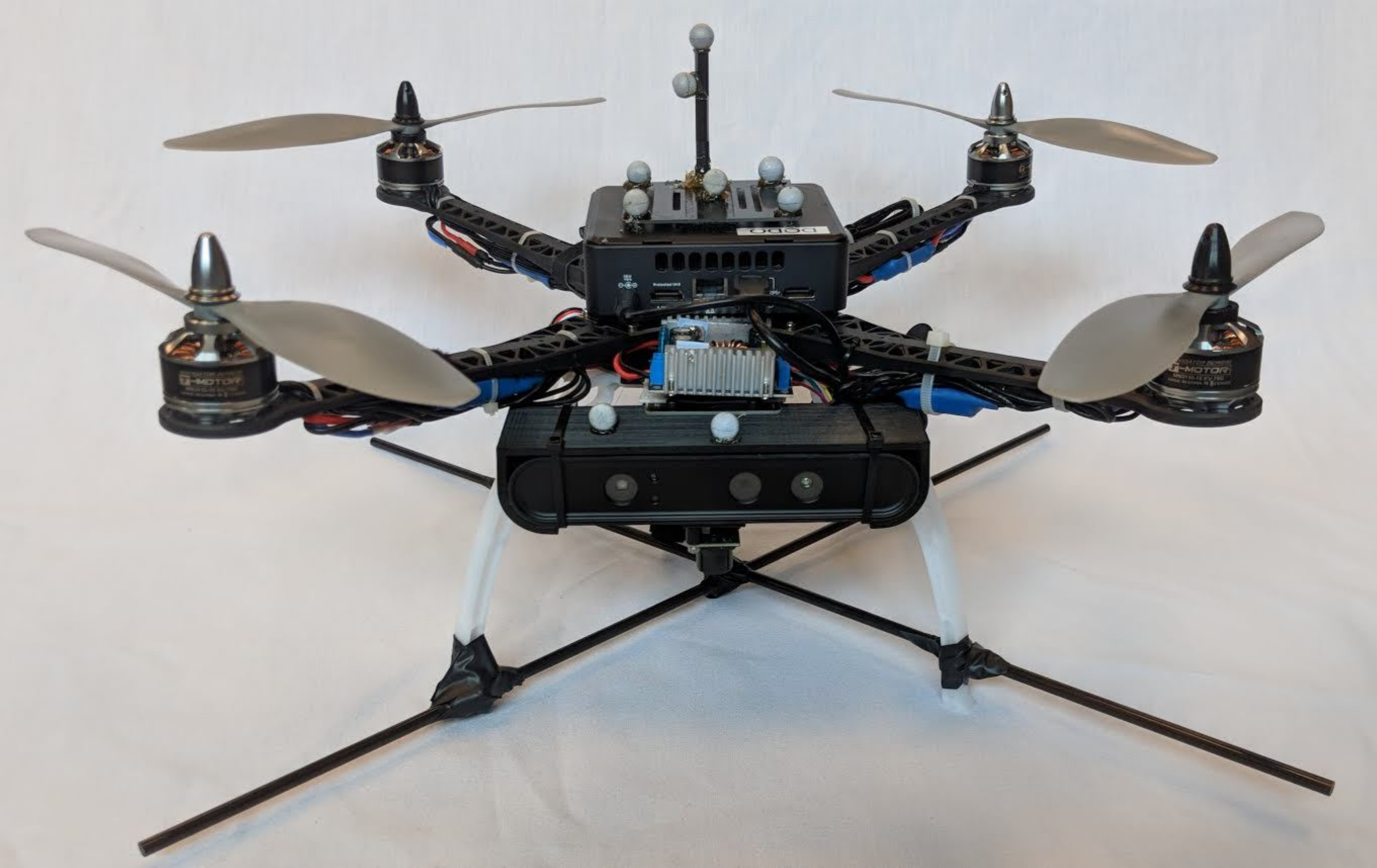}
	\caption{Experimental quadrotor platform. Reprinted with permission from \cite{REEFEstimator}.}
	\label{fig:quad hardware}
\end{figure}

The REEF Estimator was exercised on a hardware setup that uses a commercial S500 quadrotor frame, along with its corresponding brushless DC motors and motor controllers (ESCs). The RGB-D images used to estimate $x$ and $y$ velocity, come from an Orbec Astra Pro camera, while a MaxBotix MB1242 sonar provides altitude measurements. Since the purposes of the vehicle may vary from simple flight data collection, high load SLAM algorithms evaluation, to new control laws or estimators development, an Intel NUC computer with an i5 processor was incorporated. This on-board computer runs the Ubuntu 16.04 operating system and ROS Kinetic. Figure \ref{fig:quad hardware} shows a picture of the hardware setup utilized to generate the plots presented in this section. Motion capture markers were placed on the multirotor body in order to obtain ground truth data for validation only. However, motion capture data can easily replace velocity or altitude measurements or both if desired.

\subsubsection{Flight performance}
The results presented in this section are from a single flight in altitude-hold mode. The PID controller uses the REEF Estimator states as feedback while data from an OptiTrack motion capture system is recorded for ground truth reference. In this setup the PID controller receives velocity requests from a Logitech F710 gamepad. The appendix \ref{appendix:apendix_hardware} contains the values of all the parameters used by REEF Estimator when generating the plots shown in this section. 



Figure \ref{fig:xyz_estimate} shows a representative fragment of the XY velocity and Z altitude estimates from a flight. The true velocity (position numerically differentiated) from the motion capture system and the RGB-D velocity measurements are also included in the plot. From Figure \ref{fig:x_vel_estimate}, it can be observed that although the estimates are somewhat conservative, they track the ground truth very closely. The velocity filter in the $y$ direction, as depicted in Figure \ref{fig:y_vel_estimate} exhibits accurate tracking as well, and the estimated covariance appears lower than the $ x $ direction suggesting higher observability in the $ y $ (lateral) direction. The fact that the altitude is directly measured is reflected in the altitude filter estimates quality, as they track the ground truth very precisely, as observed in Figure \ref{fig:alt_estimate}. Overall, the UD partial-update filter is able to handle this scenario with varying nuisance parameters well, providing appropriate estimates for the core that allow stable and repeatable flight. Again, although a Schmidt filter was also implemented, it could handle the drifting biases (nuisance parameters) in general, producing divergent estimates in most of the experiments using the described hardware setup.

\begin{figure}[htbp]
	\begin{subfigure}{1\textwidth}\centering \includegraphics[width=0.8\textwidth]{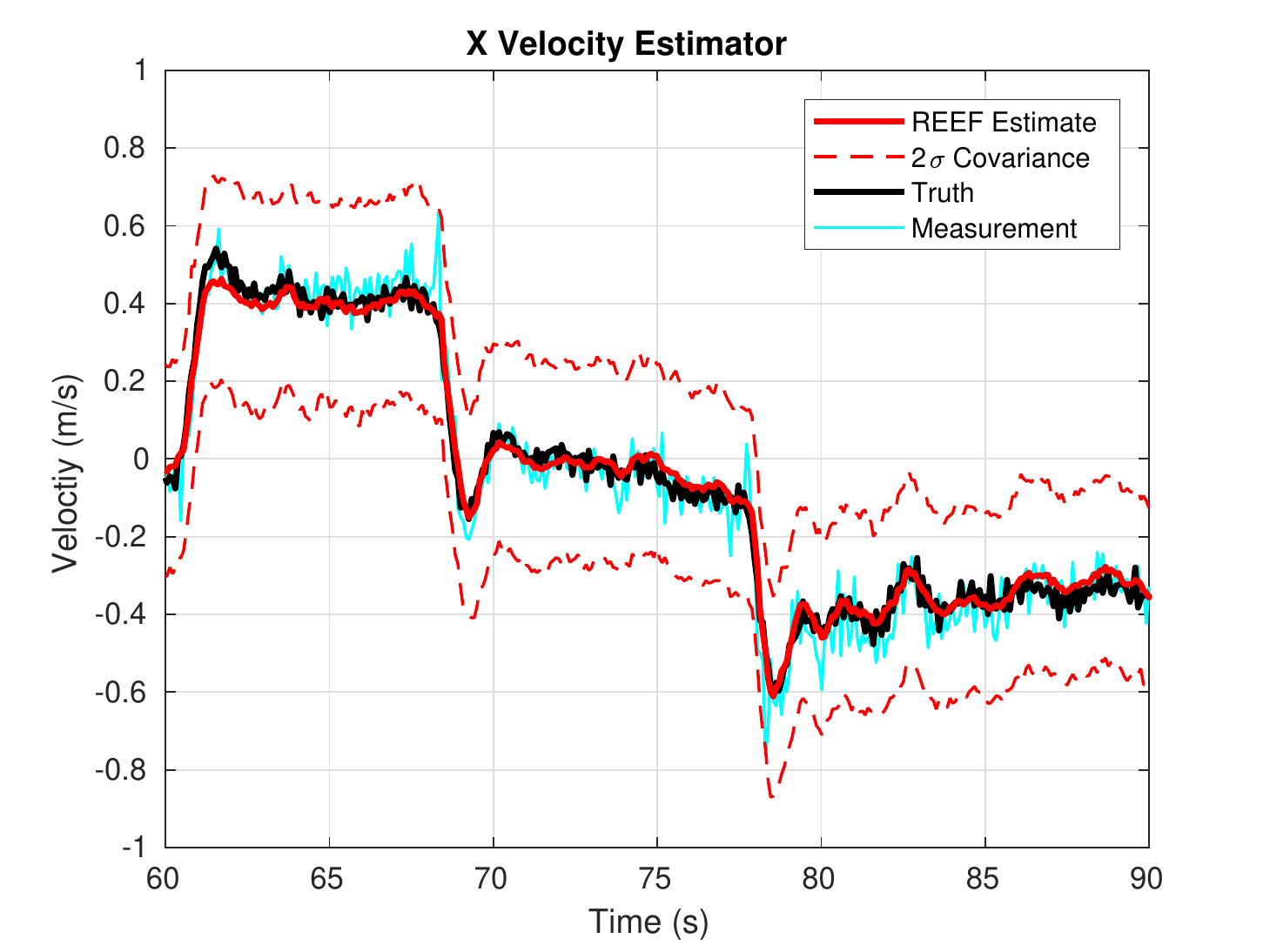}\caption{ } \label{fig:x_vel_estimate}
	\end{subfigure}
\end{figure}
\begin{figure}
	\ContinuedFloat
	\begin{subfigure}{1\textwidth}\centering \includegraphics[width=0.8\textwidth]{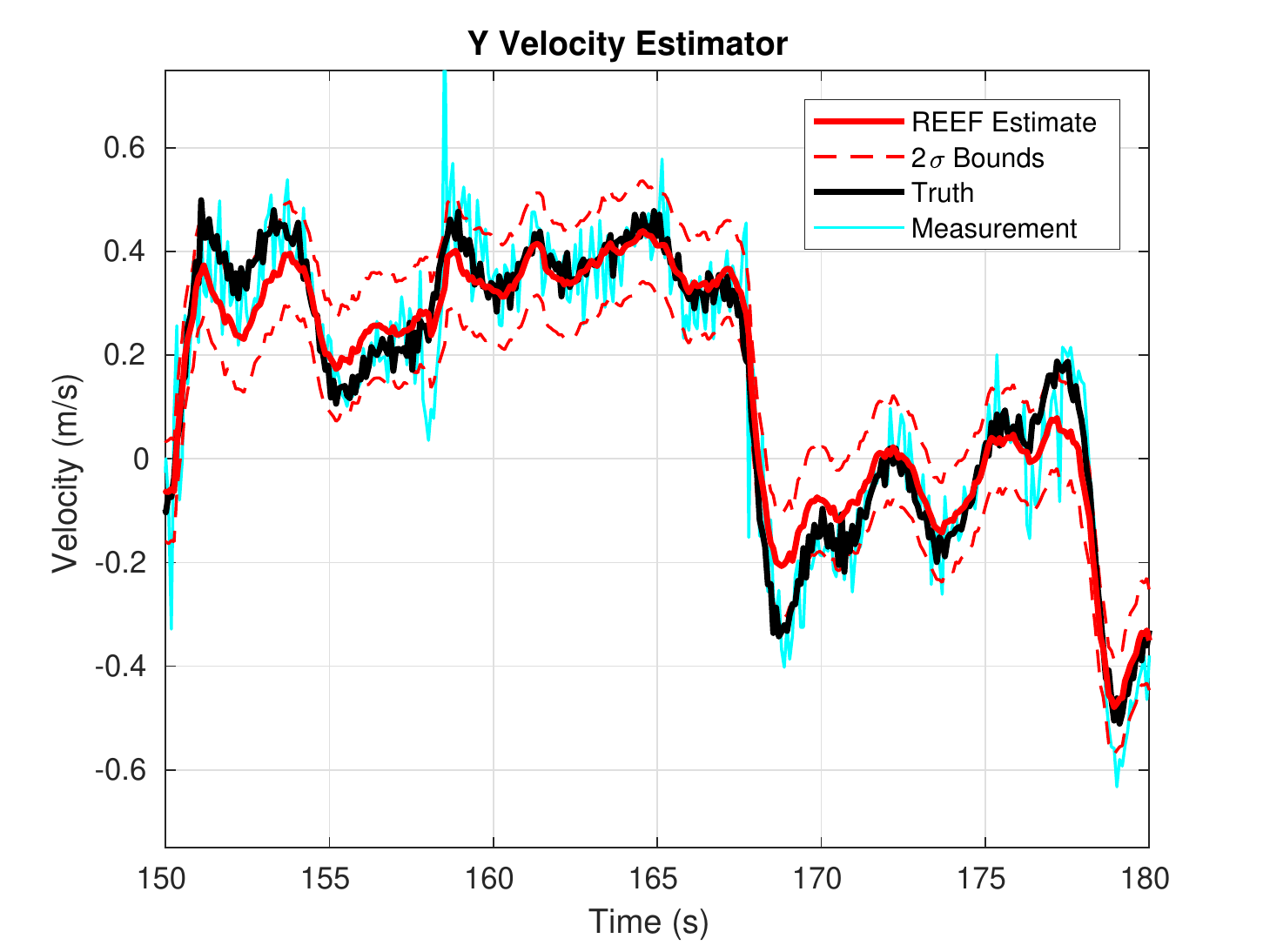}\caption{ } \label{fig:y_vel_estimate}
	\end{subfigure}
	\ContinuedFloat
	\begin{subfigure}{1\textwidth}\centering \includegraphics[width=0.8\textwidth]{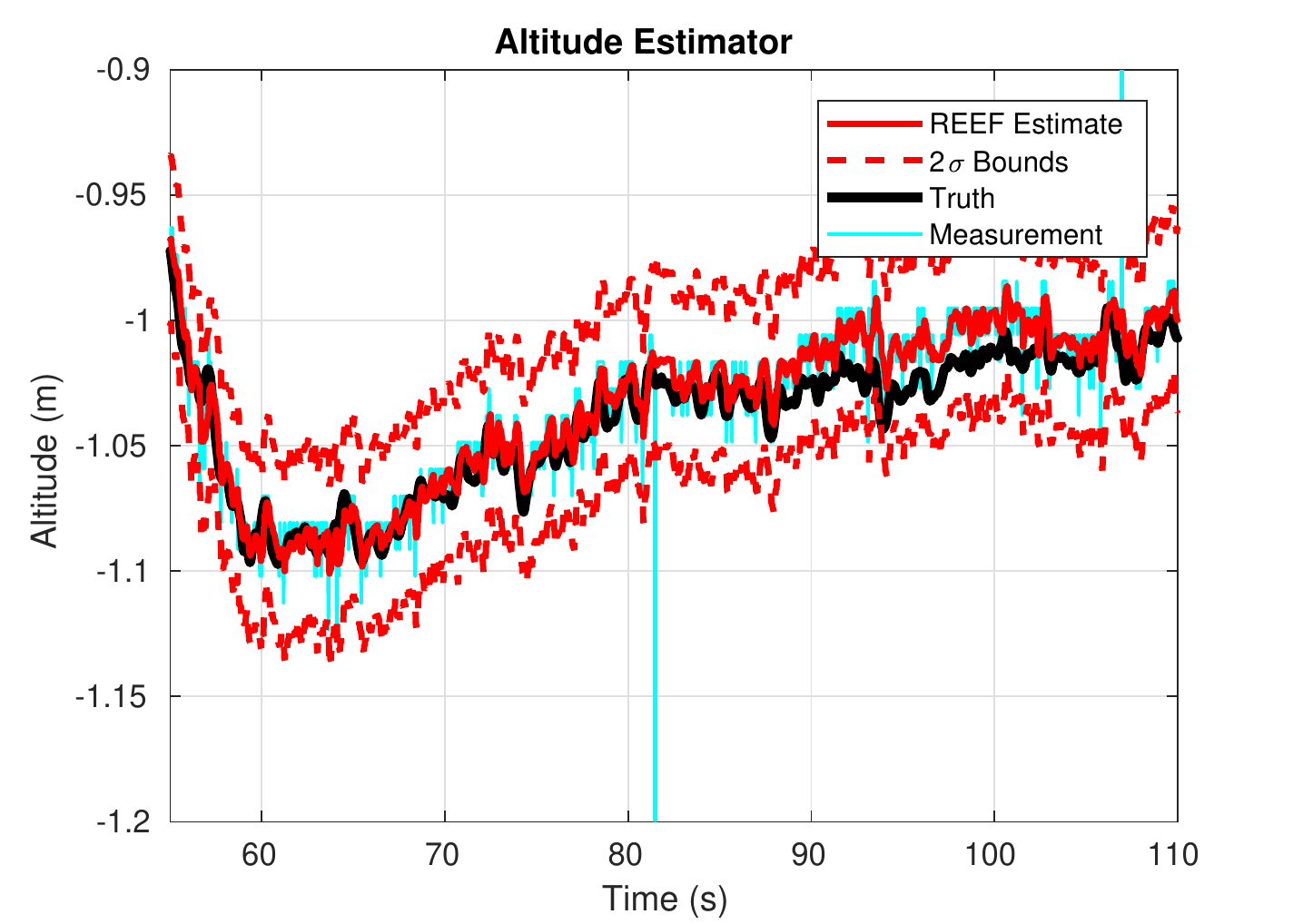}\caption{ } \label{fig:alt_estimate}
	\end{subfigure}
	\caption{XY velocity and altitude from the REEF Estimator. The estimates are compared with the ground truth from a motion capture system. Reprinted with permission from \cite{REEFEstimator}.}
	\label{fig:xyz_estimate}
\end{figure}

The tracking performance of the PID controller is shown in Figure \ref{fig:all_controllers}. Although the command tracking is far from being perfect for the $x$ and $y$ velocity controllers (as seen in Figures \ref{fig:x_vel_controller} and \ref{fig:y_vel_controller}, respectively), the intention is to provide a solid launching point for stable and repeatable flight. In Figure \ref{fig:z_controller}, the altitude controller performance is shown to be able to track the reference accurately. The first 20 seconds of Figure \ref{fig:z_controller} show the quadrotor to be on the ground waiting for the pilot to command the take-off. The multirotor starts tracking the reference of -1 meter once in the air (recall that NED frame is used). Notice that when the vehicle is on the ground, its minimum height is -0.25 meters (this is the mounting height for the altimeter).

The REEF Estimator is equipped with a $\chi^2$ rejection scheme based on the Mahalanobis distance \cite{de2000mahalanobis}. This scheme allows the filter to ignore erroneous and inconsistent altitude and velocity measurements (that may be due to the occasional generation of outliers).  A common outlier for a low-grade ultrasonic sensor (as the one used here) is observed in Figure \ref{fig:alt_estimate} at time $t\approx81s$. The rejection scheme, however, appropriately ignores the altitude measurement, as can be seen from the unaffected state estimate.

In the case of a velocity request of zero m/s, this controller has been seen to allow the vehicle to drift only about 20 cm over 60 seconds for the $x$ and $y$ axis, although this may be trim quality dependent. The vehicle's global position is not estimated; however, the current controller is written to receive position requests and use a motion capture system as feedback. This setting may be useful for specific experiments indoors as development and testing of guidance laws or SLAM algorithms.\vspace{-0.01cm}

\begin{figure}[htbp]
	\begin{subfigure}{1\textwidth}\centering \includegraphics[width=0.8\textwidth]{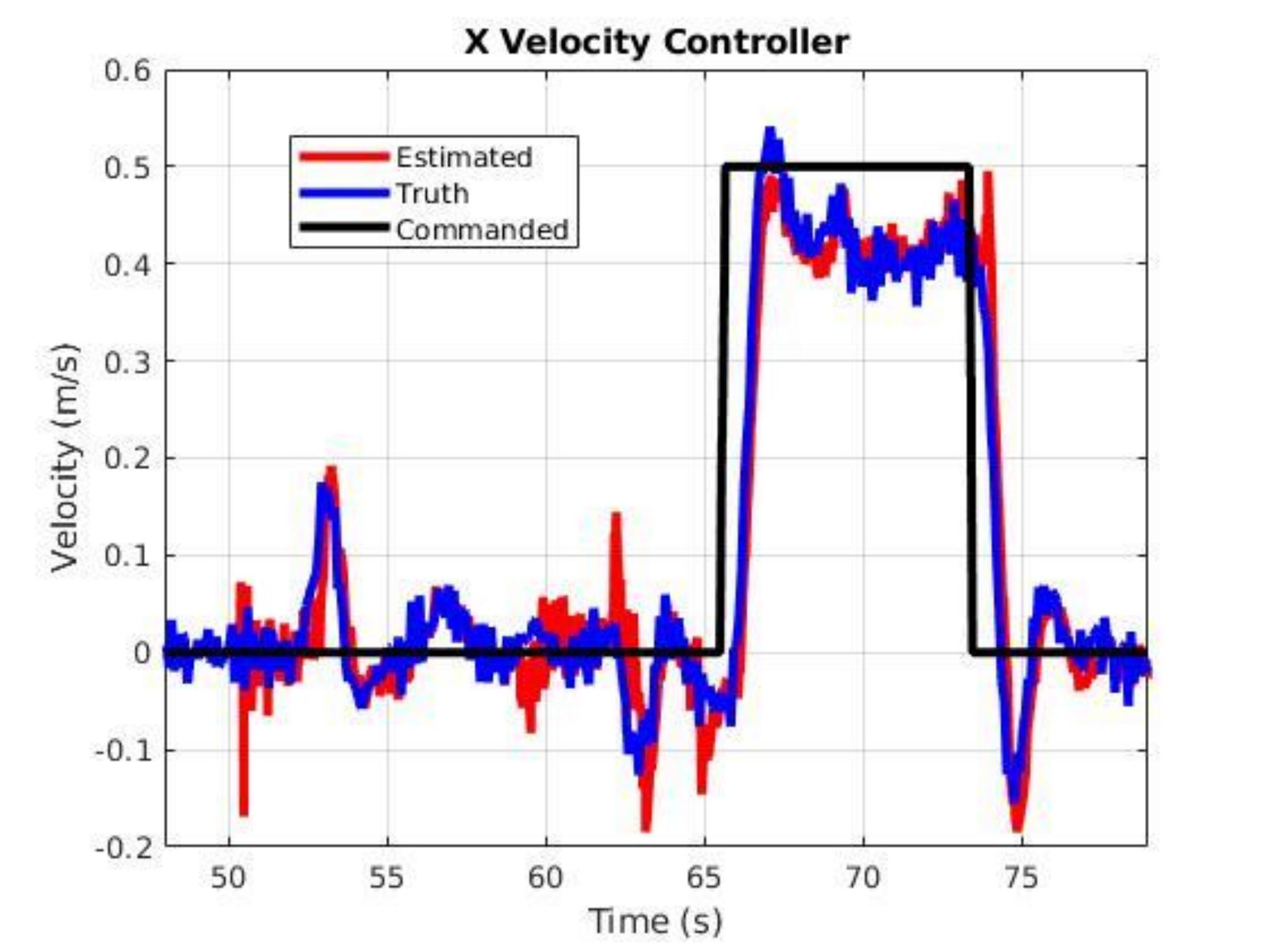}\caption{ } \label{fig:x_vel_controller}
	\end{subfigure}
	\begin{subfigure}{1\textwidth}\centering \includegraphics[width=0.8\textwidth]{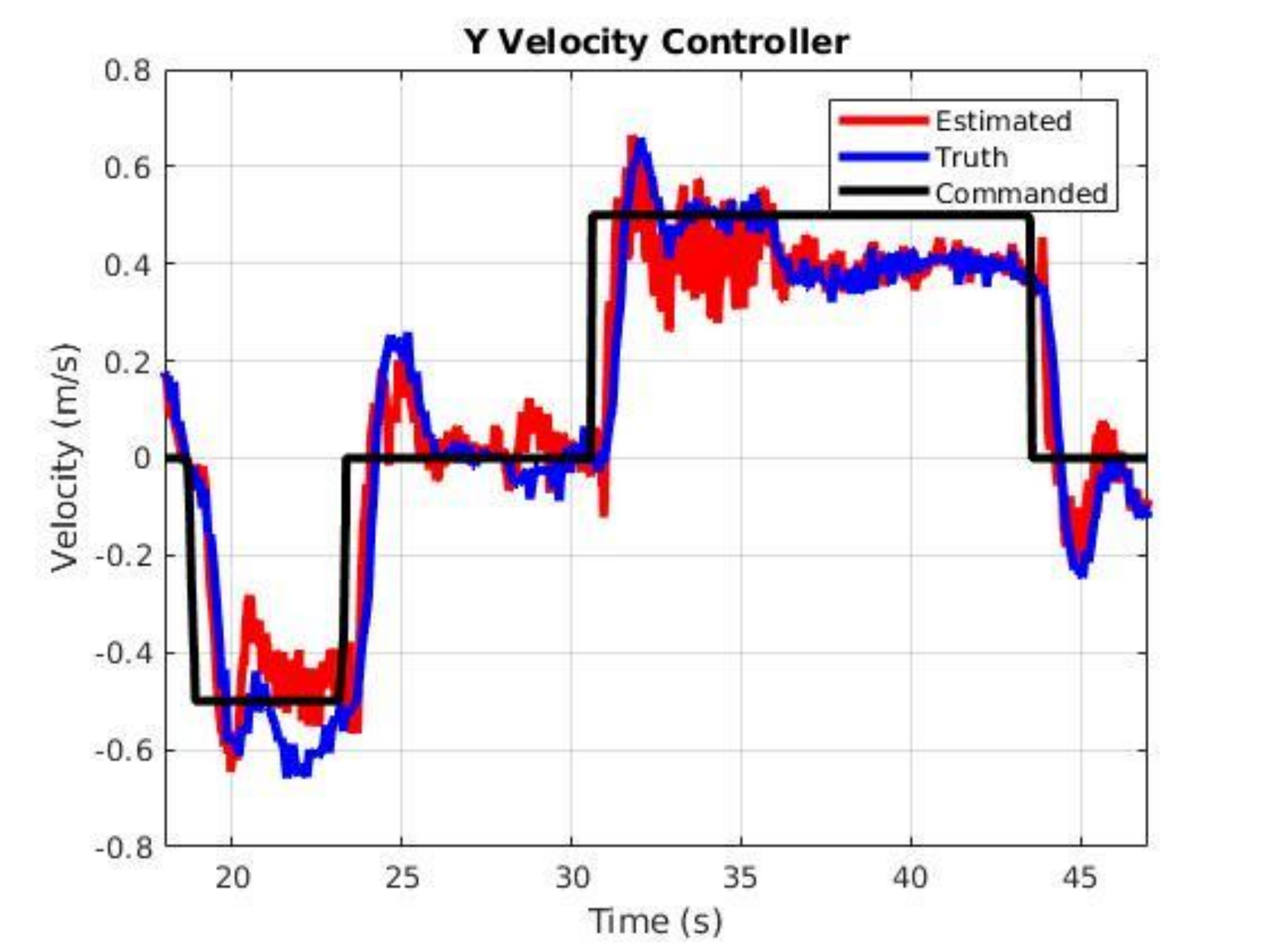}\caption{ } \label{fig:y_vel_controller}
	\end{subfigure}
\end{figure}
\begin{figure}
	\ContinuedFloat
	\begin{subfigure}{1\textwidth}\centering \includegraphics[width=0.8\textwidth]{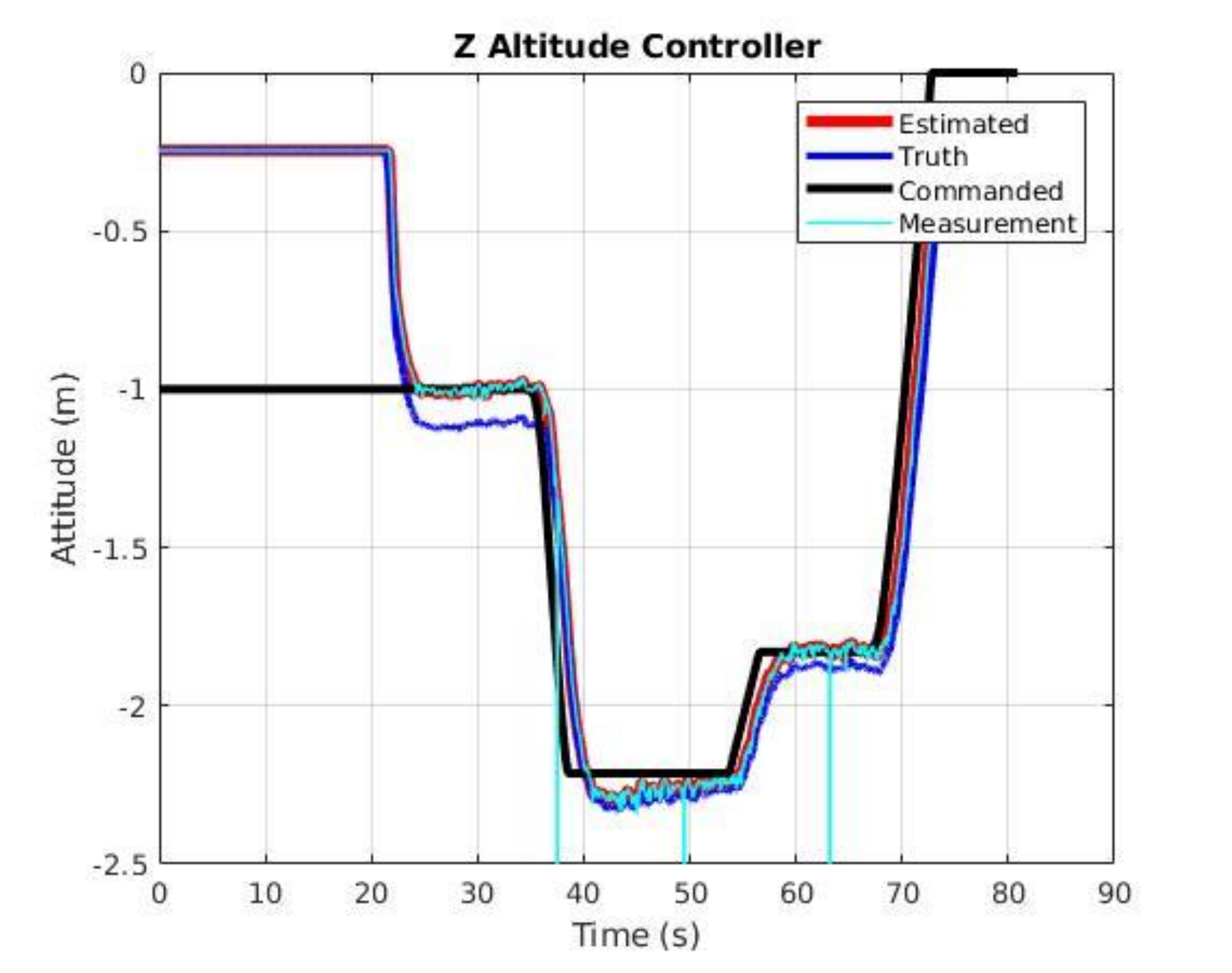}\caption{ } \label{fig:z_controller}
	\end{subfigure}
	\caption{Closed-loop performance of the multirotor. Reprinted with permission from \cite{REEFEstimator}.}
	\label{fig:all_controllers}
\end{figure}

\subsection{Summary}\label{sec:Conclusion}
The open-source REEF Estimator that leverages the partial-update capabilities to provide a simplified estimator for flying vehicles. The results shown used the UD partial-update filter, but the square root was also flown without complications. The main objective of this implementation is to expedite the establishment of flight capabilities that can support specialized research. The results for a typical flight showed the REEF Estimator's functionality and ability to cope with nonlinearities and varying nuisance parameters; inherited benefits from the partial-update approach. The REEF estimator package incorporates a PID controller that, in closed-loop, allows stable and repeatable flight. Compared to the IMU-camera calibration hardware from the previous section, a traditional Schmidt filter implementation is neither functional nor safe, especially for relatively long flights with low-grade IMU's, as is in this case. The reason is that varying biases (nuisance parameters) are not estimated, and if their drift is important, the risk of vehicle crash due to filter divergence increases.

The REEF Estimator is well suited for a large variety of end-users, starting for a beginner lacking experience in hardware, estimation, and control theory, to experts who have worked with similar platforms and want a system to support their work in a GPS or motion capture free environment. Since this solution does not depend on vehicle physics, it can be used on a different multirotor frame than the one shown here. In fact, this has been validated on hexacopters and even on ground robots (XY velocity estimator only). The full algorithms, flight simulator, and hardware list along with links to the associated code and further documentation are provided online at \url{https://github.com/uf-reef-avl/reef\_estimator\_bundle}.

%
\section{Angular rate estimation of a non-cooperative space body using RGB-D measurements}

\subsection{Introduction and motivation}

Recent work in \cite{woodbury2018} has studied the problem of determining if an observed orbital body is exerting internal control of its attitude.
This problem is of interest in space situational awareness domain, in which a bad actor might deceptively use a ``retired'' satellite for covert surveillance.
The task of classifying satellites based on measurements of their angular rate also has implications for orbital debris removal missions, in which mission planners may wish to prioritize debris targets whose motion closely resembles rigid body motion.
The work from \cite{woodbury2018} segmented the active control detection problem into two tasks: 1) estimation of motion and 2) classification of motion based on parameter estimation and statistical testing.
The initial work suggested a simple $\chi^2$ test for task \#2.
Nonlinear optimization was used to demonstrate task \#1, but is not suitable for real-time solutions.

This hardware application implements a vision-based technique to estimate the rigid-body angular rate of a target body.
This activity directly supports task \#1 in the active control detection problem.
Here as in \cite{woodbury2018}, any other torques that can potentially be present, such as those caused by gravity gradients or air drag, are assumed negligible over the experiment's time scale.
The only significant torques acting on the body should be assumed to come from internal control.

In this application, an RGB-D sensor is used as a stand-in for generic spacecraft measurements, which could come from a high-resolution ground-based radar or a stereo-vision system on a chasing spacecraft.
The measurements acquired by the RGB-D sensor are used to build estimates of the relative angular rate of the body being assessed. Since the primary objective of this implementation is to extract rotational kinematics information only, a dynamical model that includes mass properties is not necessary.

In order to get information about the rotational motion, visual features on the body are detected and then filtered by a selection criteria to eliminate low-quality features. The selected features are then tracked between sequential images and used by a UD partial-update EKF to generate estimates of the target's relative angular velocity. 

This hardware application involves nonlinear models and nuisance states as in the IMU-camera calibration and the REEF estimator, but in contrast, and very interestingly, this is a case where the nuisance states, namely the angular rates, are the states of interest; a case where the partial-update concept is highly useful. Due to the nature of this scenario, a conventional EKF was seen to be problematic and inconsistent. On the other hand, a Schmidt filter is not even an option because the angular rates would not be estimated at all.

The IMU-camera calibration previously presented may seem similar to the angular rates estimation problem. However, in the IMU-camera calibration application, although the nuisance parameters seem to be the sates of interest, they are not. In that example, the global navigation states are the states of interest, while the calibration parameters are a necessity to improve such navigation states. Put differently, the IMU-camera calibration parameters by themselves are not useful.
\subsection{Tracking and estimation system overview}\label{sec:system_overview}
Before going into the technical details, this section provides a brief description of the process utilized to obtain the relative angular rate estimates based on RGB-D measurements. Details on the actual implementation are given in the following subsection. 

The vision system uses a simple frame-to-frame estimation approach, in which changes between sequential images are processed to estimate the relative angular rate of the object of interest.
The basic operation of the vision system can be summarized as follows:

\begin{enumerate}
	\item Extraction of features: Salient features on the body are detected via a Shi-Tomasi corner detector.
	\item Selection of image features: Extracted features that fall outside a pre-selected 3D volume are eliminated. The goal is to avoid using features that are prone to rapidly move out of the image frame, and features that are far beyond the region of interest. Extracted features with no depth information are also eliminated.
	\item Tracking: The remaining complying features are tracked frame by frame with a Kanade-Lucas-Tomasi (KLT) tracker. 
	\item Mapping of image pixels to relative 3D position: The pixel coordinates $(u,v)$ given by the tracker, are mapped to the corresponding 3D positions vectors. The 3D position vectors are given on a algorithm-selected body-fixed frame, but coordinatized in the camera frame. Figure \ref{fig:reference_frames_used} shows the coordinates frames used in this implementation.
	\item Filtering: The 3D position vectors of the features are fed to the UD partial-update filter to perform relative angular velocity estimation.
	\item Filter re-initialization: Due to the body's rotation features initially detected are lost. The filter re-initializes features when available features are considered insufficient based on a threshold experimentally set.
\end{enumerate}

\subsection{Additional notation, modeling and feature treatment}
\subsubsection{Additional notation}
This hardware application follows the nomenclature used in previous chapters, along with the following additions. 
\begin{itemize}
	\item Rotation matrix that actively transforms a vector are expressed with $ \Rot $.
	\item When the context requires it, the matrix representation of a vector includes a subscript that indicates the reference frame in which the vector is resolved. In this application, the uppercase $ X , Y,$ and $ Z $ (not bold) letters are used to represent a 3D coordinate of a feature position. A vector representing the position of a feature, which is resolved in the $\cam$ frame, for instance, is written as follows:
	\begin{equation}
	\stdvecnb{\v}[C][F]=\colvec{X_F,Y_F,Z_F}_C \ .
	\end{equation}
	\item In this hardware application sub-indices $ i$ and $j$ are reserved for enumeration and matrices elements, whereas the sub-index $ k $ is used to denote time instance.
\end{itemize}

\subsubsection{Reference frames} \label{ssec:reference_frames_subsection}
\begin{figure}[tb!]
	\centering
	\includegraphics[width=0.9\linewidth]{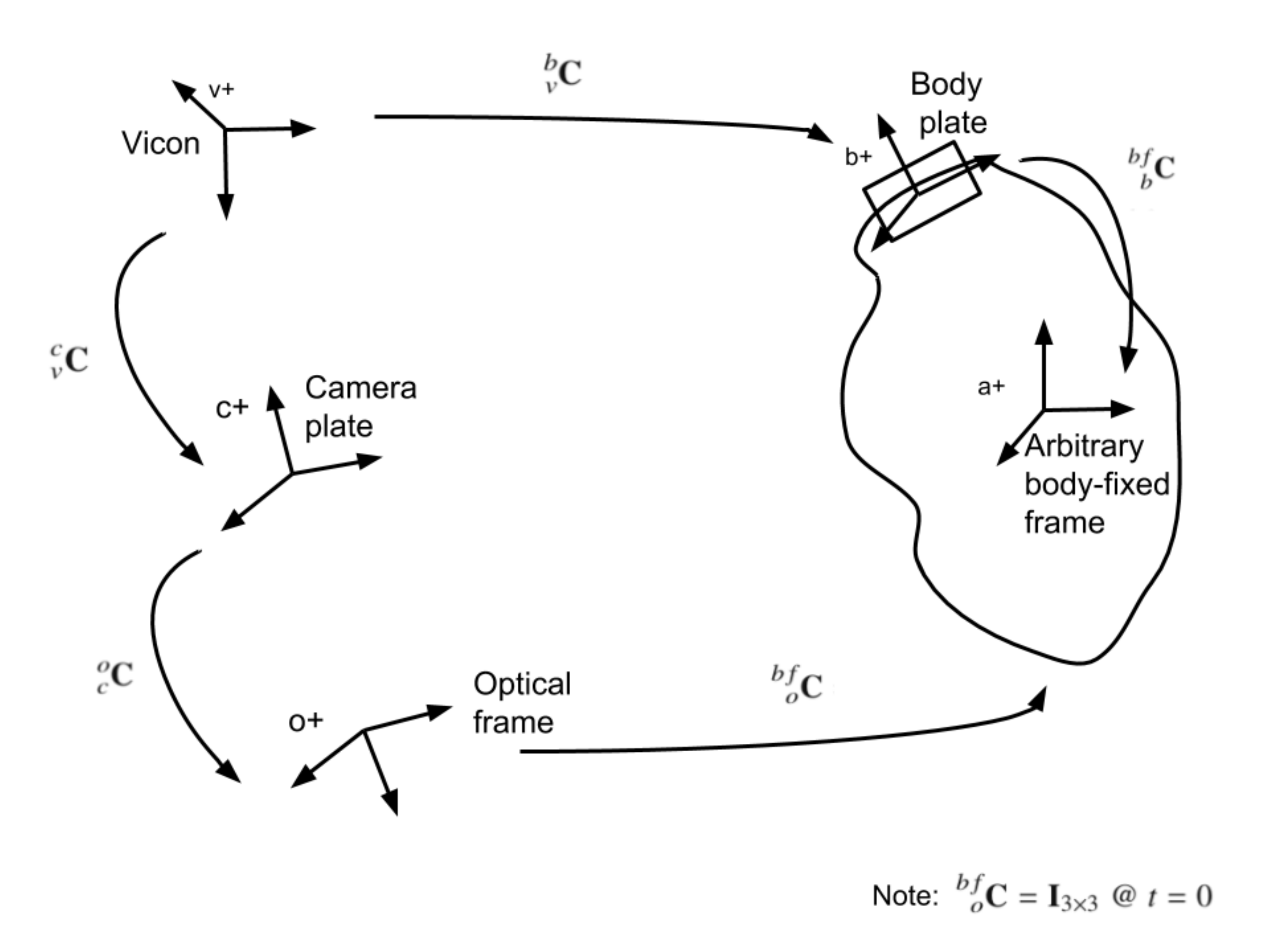}
	\caption{Coordinate reference frames utilized. Position of the features with respect to the body, attitude and angular velocities are all coordinatized in the camera frame. Reprinted with permission from \cite{Ramos2018}.}
	\label{fig:reference_frames_used}
\end{figure}

Two main reference frames are defined to perform the estimation of the body's rotational state: the camera optical frame $ \op $, which has its origin at the optical camera center, and an \textit{arbitrary} body-fixed reference frame $ \ar $.
The $ \ar $ frame, defined at the beginning of the experiment at time $ t = 0 s$, has its origin coincident with the position of the feature that is detected first. At this time, the $ \ar $ frame is set to be aligned with the optical frame. In practice, the "first" feature is the one that occupies the first position of the variable array containing all the detected feature positions. Figure \ref{fig:reference_frames_used} illustrates the relationships between the reference frames mentioned above.

In addition to the $ \op $ and $ \ar $ frames, for purposes of algorithm validation, a few extra reference frames are defined as, shown in Figure \ref{fig:reference_frames_used}. Reference frame $ \cam $ is the frame attached to the plate where the RGB-D sensor is mounted. Additionally, $ \bpframe $, is a frame attached to a plate placed on the body. This plate carries motion capture markers for obtaining a reference truth angular rate. Finally, the motion capture system frame, $\vicon $, is also shown in Figure \ref{fig:reference_frames_used}. The $\vicon $ frame represents the VICON motion capture system inertial frame in which ground reference values are obtained.

\subsubsection{Reprojection model}
\begin{figure}[tb!]
	\centering
	\includegraphics[width=0.9\linewidth]{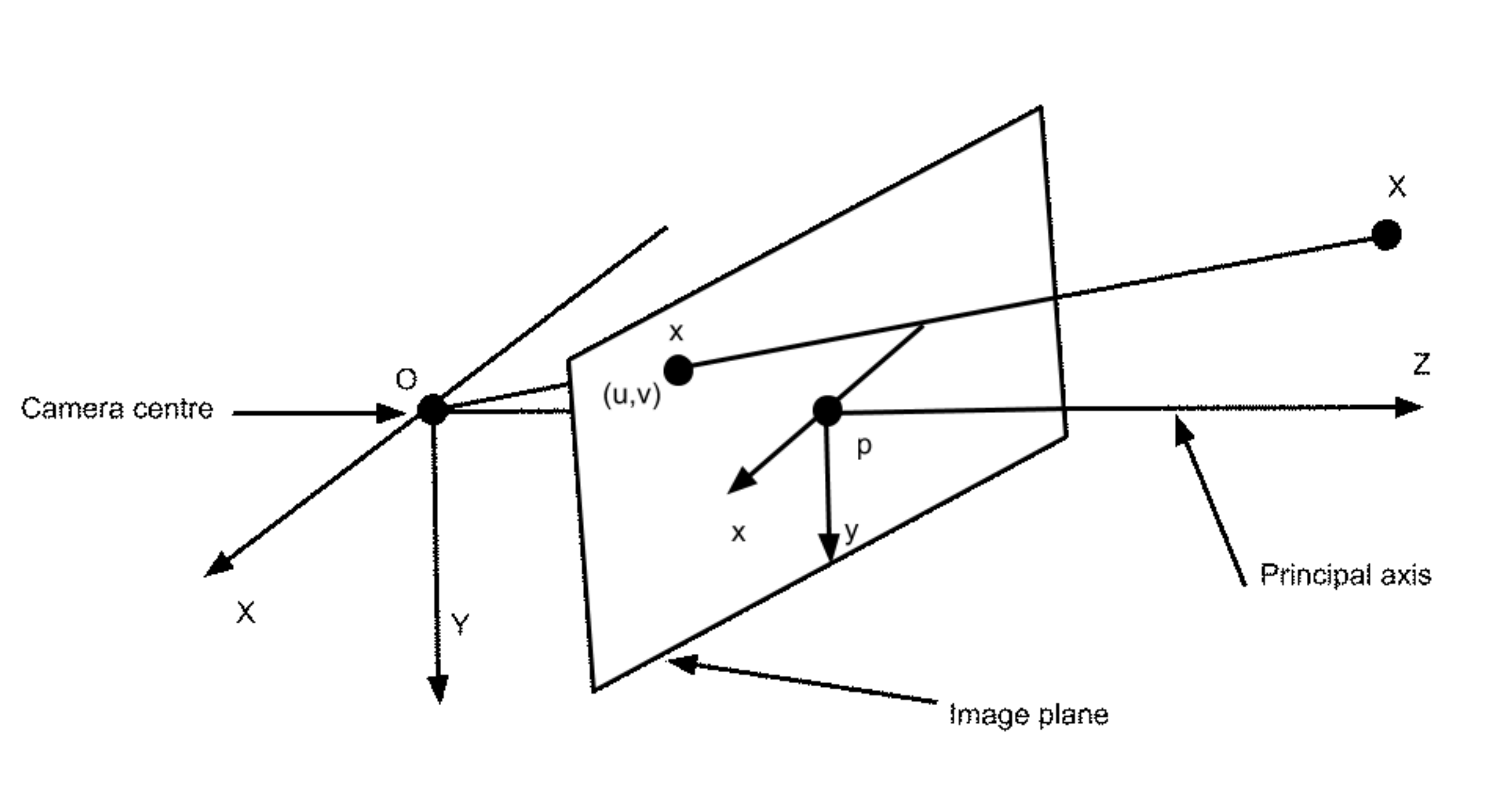}
	\caption{ The pinhole camera model projects a 3D coordinate $ \textbf{X} $ (resolved in the camera frame) into the image pixel coordinates ($ \um,\vm $). The camera center coordinates locate the point p on the image plane, which is the intersection of the image plane and the principal axis. The principal axis is along the Z axis of the camera, towards the scene being captured. Reprinted with permission from \cite{Ramos2018}.}
	\label{fig:camera_model}
\end{figure}

The positions of the body features with respect to the RGB-D sensor are obtained via the well-known pinhole camera model by leveraging the depth information. As previously described in the IMU-camera calibration hardware implementation, the pinhole camera model projects a 3D position $(X_i, Y_i, Z_i)$ that is resolved in the camera optical frame (collocated at the camera center), into the corresponding pixel position ($\um,\vm $) on the image. Since the RGB-D sensor \textit{generally} provides depth, $Z$ for a detected feature at ($\um,\vm $), its 3D position can be directly computed via the calibrated pinhole model

\begin{equation}\label{eq:reprojection_model1}
\yfim=\colvec{X_i,Y_i,Z_i}_O=\colvec{Z_i(\um - c_x)/f_x, Z_i(\vm - c_y)/f_y, Z_i}_O \ .
\end{equation}
Recall that $ f_x $ and $f_y$ are the focal length of the camera in the $ x$ and  $y$ axis, and that $c_x $ and $c_y $, are the coordinates of the camera center. Both focal lengths and camera center coordinates are quantities expressed in pixels, and their values for this application are obtained via standard OpenCV functions for camera calibration. Figure \ref{fig:camera_model} illustrates how a 3D coordinate for a feature, $ \textbf{X} $, is mapped into a pixel, $ \textbf{x} =[\um,\vm]^T $, on the image plane.

\subsubsection{Feature extraction}\label{ssec:feature_extraction}
The presented implementation, leverages the detected change in position of the image points from frame to frame to perform the estimation of the target (body) relative motion. In practice, this is often achieved by matching features in two consecutive frames and computing the rigid transformation (rotation and translation) that aligns the detected features. Whereas it is possible to use essentially all pixels in the frames to perform this computation, it is commonly more tractable to track visually distinctive features and use these only to compute the rigid transformation estimate. The latter approach is used in this hardware implementation as it is more suitable for real-time execution.

In this implementation the selection of features is done using a Shi-Tomasi \cite{shi1994good} corner detector, followed by a process that retains only \textit{good} features with the objective of increase overall feature traceability. These \textit{good} features are selected following the criteria presented in the following subsection.

Once the initially detected features have been refined, into a better sub-set, feature tracking is performed with a KLT Tracker \cite{lucas1981iterative}. The objective of the tracking the features is maintain the same good features from frame to frame to be able to estimate the underlying body's motion. The KLT tracker essentially estimates the displacement of features from frame to frame via an optimization process that minimizes an image intensity metric for the feature's neighborhood. This hardware implementation uses the KLT tracker function from the OpenCV 2.4.9.1 package \cite{opencv}.  

\subsubsection{Feature selection criteria}\label{ssec:feature_selection}
Feature extraction is restricted to a pre-defined searching volume. A polygon defines this volume on the image coordinates $ (u,v) $ and a pre-defined depth. For all the experiments included in this implementation, the maximum sensing depth is set to 3.5m, while the polygon on image coordinates is set to be a rectangle of enough size to cover the body of interest while it rotates. Although the selection of the region of interest can be automated using techniques as blob detection and image motion analysis, this implementation requires manual selection.

Within the polygon that covers the body's motion in analysis, a second region that encloses only features that are not close to body edges is also utilized. The objective of this second region is to help differentiate the long-term-to-short-term available features.
The short-term features are those that, due to the body's rotation and proximity to the body's edges, will move out of the image relatively soon after their potential detection occurs.
Long-term features are those that will remain visible for a longer period of time because they are relatively far from the body edges. This secondary region is selected manually as well.

Although color information could be used to extract good features from the complete detected set, the combination of the pre-selected volume and the built-in outlier rejection from the Kalman filter (described in Section \ref{sec:ekf}), showed to be enough for the purposes of this application. Direct use of color information is also less useful in space operations because available lighting tends to be of poor quality.

Features complying with the selection criteria outlined in this section are considered good quality features for body tracking and estimation purposes.

\subsubsection{Feature position vectors in the arbitrary body frame}\label{sec:Tracking of position vectors fixed to the target}
Once \textit{good} features are being tracked, the algorithm proceeds to compute the corresponding position vectors. The position vectors of the features, however, do not directly allow the estimation of the body's angular rate; instead, they are used to construct a set of vectors that resemble the body's rotational motion. These vectors defined in the body frame are referred to as the body vectors.

When the algorithm to estimate the angular rate begins, it fixes an arbitrary reference frame on the rotating body when the first measurement arrives. For this implementation, the feature that serves as the arbitrary origin on the body is the highest-quality feature produced by the OpenCV \code{GoodFeaturesToTrack} function, which implements the Shi-Tomasi corner detector. For this detector, the feature quality is related to the magnitude of the eigenvalues that capture the image motion \cite{shi1994good}.
At subsequent times (after the first measurement arrives), as the vision system provides the position of more features, they are re-expressed in the rotating body reference frame. This results in a set of body vectors that are tracked from frame to frame and fed as measurements to a partial-update EKF, which uses a motion model to obtain feature position estimates and angular rates of the body. This process is summarized in the next section.

The flow of the information, then, is as follows. The output of the feature detector and tracker is a set of feature positions $(u_i,v_i), \ i=1,\dots N$ on the image, and the corresponding depths $ Z_i $.
With this information in hand, Equation \ref{eq:reprojection_model1} is used to obtain the corresponding 3D coordinates $ (X_i,Y_i,Z_i) $ of each feature.
Then, the highest quality feature, named $ (X_o,Y_o,Z_o) $, is set as the origin ($ O $) of the reference frame that is attached to the body, and with subsequent features available, the body vectors $ \stdvecnb{\p}[O][F_i] $ are computed by vector subtraction as

\begin{equation}\label{eq:body_vectors}
\stdvecnb{\p}[O][F_i]  =\colvec{X_i,Y_i,Z_i}_O -  \colvec{X_o,Y_o,Z_o}_O \ .
\end{equation}
The body vectors, $ \stdvecnb{\p}[O][F_i] $, are the feature position vectors expressed from the body-fixed frame origin. These body vectors are then used as measurements for rotational motion estimation.

\begin{figure}[tb!]
	\centering
	\includegraphics[width=0.8\linewidth]{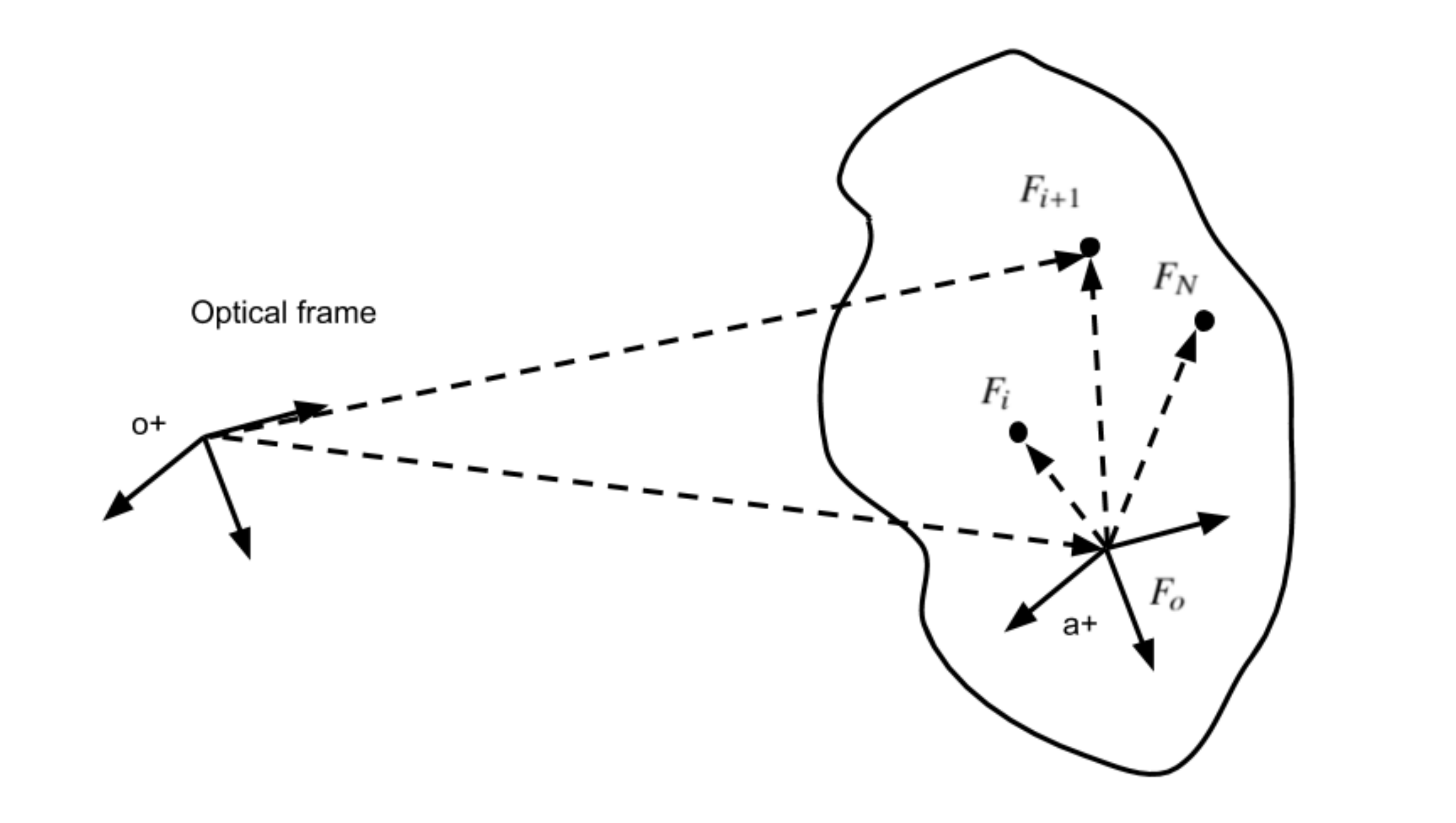}
	\caption{Coordinate reference frames utilized. Position of the features with respect to the body, attitude and angular velocities are all coordinatized in the camera frame. Reprinted with permission from \cite{Ramos2018}.}
	\label{fig:body_vectors}
\end{figure}

It is important to mention that although the $ i^{th} $ body vector, $ \stdvecnb{\p}[O][F_i] $, for the $ i^{th} $ feature, is expressed with respect to the arbitrary reference frame, $ \ar $, the $ \stdvecnb{\p}[O][F_i] $ vector is still resolved in the $\op$ frame.
Thus, the estimated relative angular rates and attitude will be naturally measured with respect to the $\op$ frame as well. Figure \ref{fig:body_vectors} shows the relationship between position of features $ F_i $ with respect to the optical frame $ \op $ and the established body-fixed reference frame $ \bframe $. The set of body vectors for a generic case are represented in Figure \ref{fig:body_vectors} as $\textbf{F}_1,\textbf{F}_2,\dots,\textbf{F}_N $ (omitting the notation $ \tilde{\textbf{p}} $ for clarity).
An additional observation is that the origin of the  $ \bframe $ frame is defined on the surface of the target body. Thus, it will exhibit translational motion with respect to the observer frame, however, for the experiments that were performed, the estimation of the angular rate is not ambiguous, since there is no relative translational velocity between the sensor and the body.

\subsection{Extended Kalman Filter}\label{sec:ekf}
An UD partial-update EKF that uses the body vectors that are obtained in Section \ref{sec:Tracking of position vectors fixed to the target} is implemented to estimate the relative angular rate of the target. This section presents the details.

\subsubsection{Process and measurement model}
The EKF state vector includes the body vectors, $ \stdvecnb{\p}[O][F_i] $ , of each tracked feature (obtained with equation \ref{eq:body_vectors}), and the relative angular velocity of the target body ,$ \stdvecnb{\angrate}[O][b] $ expressed in the RGB-D sensor frame $ \op $:

\begin{equation}\label{eq:state_vector}
\x=\rowvec{\stdvecnb{\p}[O][F][\rmT], \stdvecnb{\angrate}[O][b][\rmT]}\trans \ .
\end{equation}
Here, $ \stdvecnb{\p}[O][F][\rmT] $ encapsulates the N features' position vectors, as

\begin{equation}
\stdvecnb{\p}[O][F_i][\rmT] = \rowvec {X_i,Y_i,Z_i}\trans \ ,
\end{equation}

such that 

\begin{equation}
\stdvecnb{\p}[O][F][\rmT] = \rowvec{\stdvecnb{\p}[O][F_i][\rmT],\stdvecnb{\p}[O][F_{i+1}][\rmT],...,\stdvecnb{\p}[O][F_N][\rmT]} \ .
\end{equation}
More specifically, $ \stdvecnb{\p}[O][F_i][\rmT] $ is the (body) vector position of the $ i^{th}$ feature, measured from the arbitrary origin, $\ar $, on the body frame but coordinatized in the camera frame, $\op $. $ \stdvecnb{\angrate}[O][b][\rmT]$, the angular velocity of the body with respect to the camera frame, $ \op $, is composed as follows:

\begin{equation}
\stdvecnb{\angrate}[O][b][\rmT] = \rowvec{\angrate_x,\angrate_y,\angrate_z} \ .
\end{equation}
The sub-index for each component corresponds to the axis of rotation.
%
%

The discrete system process model is expressed as follows \cite{oumer20123d}:

\begin{equation}\label{eq:system_model}
\fxt_k=\colvec{X_1,Y_1,Z_1,X_{2},Y_{2},Z_{2},\vdots,X_{N},Y_{N},Z_{N},\angrate_x,\angrate_y,\angrate_z}
_k =
\begin{bmatrix}
X_1 - \wz\Delta t Y_1 + \wy\Delta t Z_1\\
\wz \Delta t X_1 + Y_1 - \wx \Delta t Z_1\\
-\wy \Delta t X_1 + \wx \Delta t Y_1 + Z_1\\
X_{2} - \wz\Delta t Y_{2} + \wy\Delta t Z_{2}\\
\wz \Delta t X_{2} + Y_{2} - \wx \Delta t Z_{2}\\
-\wy \Delta t X_{2} + \wx \Delta t Y_{2} + Z_{2}\\
\vdots\\
X_N - \wz\Delta t Y_N + \wy\Delta t Z_N\\
\wz \Delta t X_N + Y_N - \wx \Delta t Z_N\\
-\wy \Delta t X_N + \wx \Delta t Y_N + Z_N\\
\angrate_x\\
\angrate_y\\
\angrate_z\\
\end{bmatrix}_{k-1} +
 \ \w_{k-1} \ ,
\end{equation}
or
\begin{equation}
\fxt_k	= 
	\begin{bmatrix}
	(\eye{3} + \wskew\Delta t)\stdvecnb{\p}[O][F_1]  \\
	(\eye{3} + \wskew\Delta t)\stdvecnb{\p}[O][F_{2}]  \\
	\vdots\\
	(\eye{3} + \wskew\Delta t)\stdvecnb{\p}[O][F_{N}]  \\
\stdvecnb{\angrate}[O][b]
	\end{bmatrix}_{k-1}+\ \w_{k-1} \ .
\end{equation}
Where $ \Delta t $ is the propagation interval, $ \wskew $ is the skew symmetric matrix formed with elements $ \wx,\wy$ and $\wz $, and $ \eye{3} $ is a $ 3 \times 3 $ identity matrix.  $ X_i,Y_i$ and $Z_i $ are body vector components resolved in the $\op$ frame.

%

For this application, it is assumed that the RGB-D sensor provides direct measurements of the position of the features with respect to the current arbitrary origin $ X_o,Y_o,Z_o $. That is, that the camera reads the body vectors directly. Thus, the measurement model can be  written simply as

\begin{equation}\label{eq:measurement_model}
\yhat_{k} = \Hmat\xhat_{k}+\vk \ .
\end{equation}
Where $ \Hmat $ is defined as

\begin{equation}
\Hmat = 
\begin{bmatrix}
\eye{3N} & \zero[3N][6]\\
\end{bmatrix} \ ,
\end{equation}
here, $ \vk $ is a white noise process with covariance matrix $ \stdvec{R}[][F] = E[\stdvec{vv^T}]$ . For this work, it is assumed that the uncertainty levels for the measurements are known or can be computed from recorded data, and that they are the same for the three dimensions $ X,Y,Z $. However, it must be mentioned that as in \cite{willis2016benchmarking}, a more accurate noise model for a RGB-D type of sensor that considers the correlation between the three coordinates and uncertainty dependency on the depth value could be used.

To use the Extended Kalman Filter propagation equations for the error covariance, the state transition matrix $ \Fmat_{k-1} $ and the input noise matrix $ \Gmat_{k-1} $ are needed. By linearization of Equation (\ref{eq:system_model}), the state transition matrix is found to be in the following form:


\begin{equation}\label{eq:F_matrix}
\Fmat_{k-1}=
\begin{bmatrix}
(\eye{3} + \wskew\Delta t) & \zeromatthree &\zeromatthree& \dots & -\piskew\Delta t\\
\zeromatthree & (\eye{3} + \wskew\Delta t) &\zeromatthree &\dots&-\piplusskew\Delta t \\
\vdots&\vdots&\ddots&\dots&\vdots \\
\zeromatthree & \zeromatthree &\zeromatthree & (\eye{3} + \wskew\Delta t) & -\pnskew\Delta t  \\
\zeromatthree & \zeromatthree &\zeromatthree & \zeromatthree & \eye{3} \\
\end{bmatrix} \ .
\end{equation}
Where $ \Fmat $ is of dimensions $ (3N+3)\times(3N+3) $. Whereas $ \Gmat $, it is just a matrix with ones along the corresponding entries that map process noise into states. States with process noise are specified in the hardware implementation section. 
Finally, the EKF covariance is propagated according to the UD partial-update equations from Table \ref{table:UDPU_filter}.

\subsubsection{UD partial-update filter}
Due to the nature of this scenario, a conventional EKF was seen to be problematic, producing biased estimates for all state estimates and with overconfident covariance. Moreover, relatively large corrections were observed on the angular rates, even when they were known to be constant during experiments. This was also the case, even for correct initial conditions with low covariance.

As mentioned before, this hardware application involves nonlinear models and nuisance states as the previous applications, but in contrast, and very interestingly, this is a case where the nuisance states, namely the angular rates, are the states of interest; a case where the partial-update concept is highly useful. Specifically, this application demonstrates the successful use of the UD partial-update filter.

\subsection{Filter initialization}\label{ssec:filter intial}
In order to initialize the filter, the information provided by the feature tracker is leveraged to compute a rough estimate for the angular rates and their standard deviations. This subsection outlines the procedure. First, by using feature positions from one frame to the next, a rigid transformation that aligns two subsequent clouds of body vectors is obtained through Singular Value Decomposition (SVD)-based least-squares \cite{sorkine2017least}. Such transformation is computed to approximate the rotational change of the body in motion. Then, angle extraction from the computed rigid transformation is performed. This is, the change in the Euler angles that compose the rigid body transformation (locally), are obtained. Finally, the ratio of the change on the Euler angles to the time-step interval is used as a coarse approximation of the local angular velocity estimate. Based on the notation used in this hardware implementation, $ \stdvecnb{\p}[O][F][\rmT] = \rowvec{\stdvecnb{\p}[O][F_1][\rmT],\stdvecnb{\p}[O][F_{2}][\rmT],...,\stdvecnb{\p}[O][F_N][\rmT]} $ composes the first point cloud (set of body vectors in the previous frame), and $ \stdvecnb{\qvec}[O][F][\rmT] = \rowvec{\stdvecnb{\qvec}[O][F_1][\rmT],\stdvecnb{\qvec}[O][F_{2}][\rmT],...,\stdvecnb{\qvec}[O][F_N][\rmT]} $ the second one (body vectors in the current frame). The solution for the rotation is obtained by minimizing the cost function of Equation (\ref{eq:least_squares_cost_function}) subject to the constraint that the transformation $ \Rot $ is not a reflection. The implemented SVD-based solution of this problem, incorporates such constraint in the minimization process to search for rotations only \cite{sorkine2017least}. Due to the relatively high frequency of the RGB-D camera measurements and slow angular rate expected for the body, the rotation delivered by the least-squares approach is assumed to be well modeled as a differential rotation (from a frame to the next) $ \delta\Rot $ via :

\begin{equation}\label{eq:differential_rotation}
\delta\Rot = \eye{3} + 
\begin{bmatrix}
0 & -\wz \Delta t & \wy \Delta t \\
\wz\Delta t & 0 & -\wx\Delta t\\
-\wy\Delta t & \wx\Delta t & 0\\
\end{bmatrix} \ ,
\end{equation}
and
\begin{equation}\label{eq:least_squares_cost_function}
(\delta\Rot , \translation)= \arg\min_{\Rot\in SO(3),\translation\in\Rot^3}\sum_{i=1}^{n} \|(\Rot\p + \translation) - \qvec \|^2 \ .
\end{equation}
As mentioned before, $ \delta\Rot $ is chosen to be parameterized with Euler angles. Specifically, as an Euler 3-2-1 rotation through the angle set $ [(\psi,\theta,\phi)] $ (or yaw, pitch, and roll).

It is important to point out that since the pair of 3D point clouds, $ \p$ and $\qvec $, are resolved in the camera reference frame, the small change in attitude, $ \delta\Rot $, is given with respect to the camera frame as well. Again, once the small rotations are approximated, the time interval $ \Delta t $ utilized to estimate the angular rates.

\begin{figure}[H]
	\centering
	\includegraphics[width=\textwidth,trim=50pt 0pt 50pt 0pt, clip]{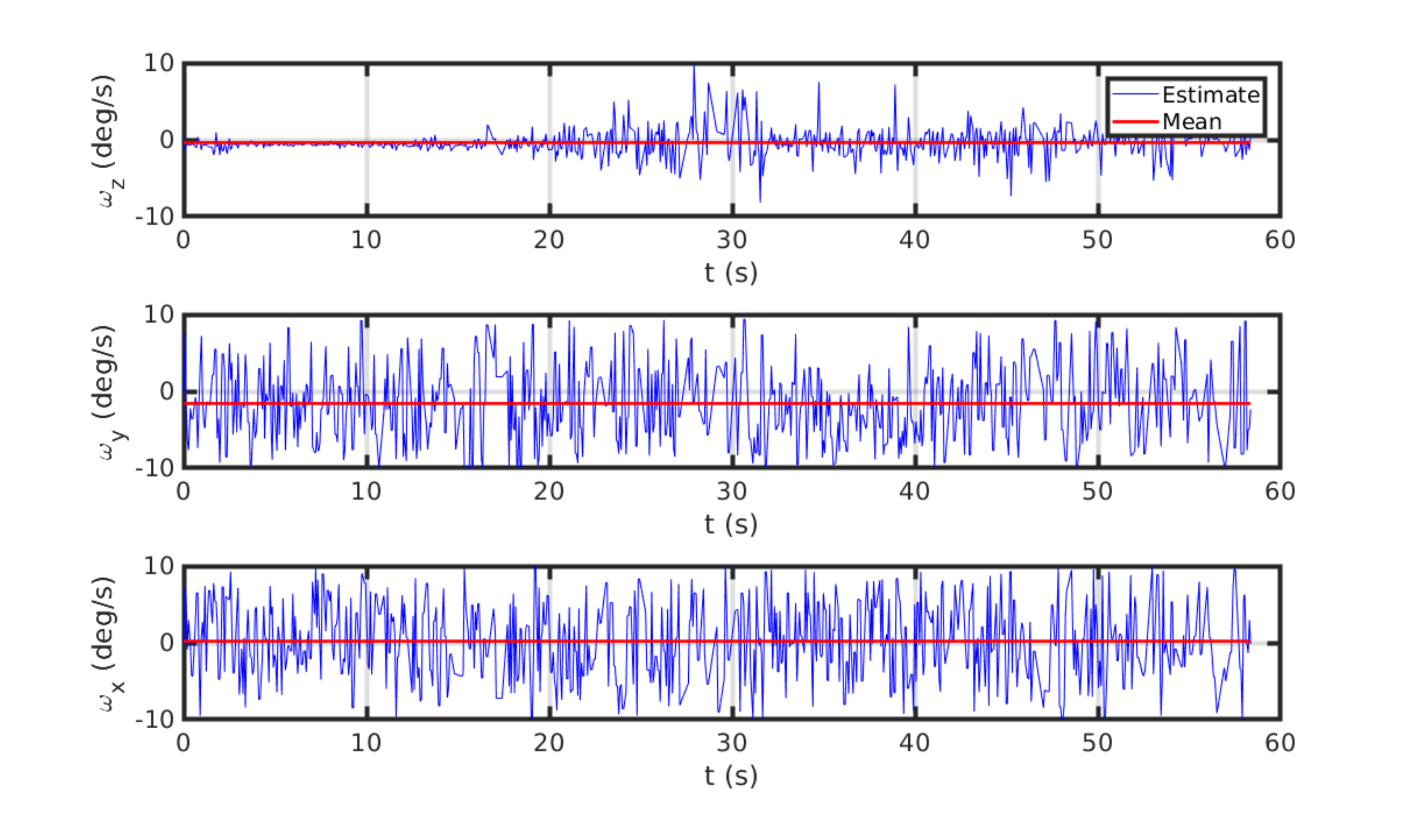}
	\caption{Angular rates obtained through numerical differentiation of differential rotation angles. The differential rotation angles are extracted from the estimated $ \delta\Rot $ from equation \ref{eq:differential_rotation}. In this figure a 60 seconds history of the coarse estimates for the angular velocity components is shown. The empirical average is shown in red. The truth value for this specific experiment was $ \omega_y = 2\;deg/s $ and $ \omega_x = \omega_z = 0 \;deg/s $. Reprinted with permission from \cite{Ramos2018}.}
	\label{fig:intial_omegas_plot}
\end{figure}

For this application, the coarse estimation of the angular rates is performed continuously during three seconds (as features become available), and then an empirical mean and standard deviation are computed from the collected data. These empirical values mean directly initialize the EKF. In Figure \ref{fig:intial_omegas_plot}, a typical history of the coarse angular rates estimates produced by this procedure is shown. Finally, the position of the body vectors is initialized using their computed values directly, and their initial covariance is set experimentally commensurate to the measurement noise values.

Since the tracked features eventually move out of the field of view due to the body rotation, the filter needs to be re-initialized when too few features remain in the frame. The filter is re-initialized according to the procedure that is discussed next.

\subsection{EKF re-initialization}\label{sssec:reinitialization}
Due to the body rotation or changes in illumination features' visibility can be lost. Thus, if one desires to maintain accurate estimates of the states, a re-initialization procedure is needed. In this implementation, the re-initialization of the partial-update EKF occurs when the number of tracked features falls below a threshold.

The re-initialization of the filter is performed according to the following procedure: 
\begin{itemize}
	\item The current values of the angular velocity are maintained along with their current covariances (cross-correlations are discarded).
	\item The corner detector is executed, and new features are extracted by following the selection procedure described in Sections \ref{ssec:feature_extraction} and \ref{ssec:feature_selection}. All features previously utilized are discarded.
	\item Given the position of the new features in image coordinates, predictions of their corresponding body vectors can be obtained via Equation \ref{eq:body_vectors} and a new $ \ar $ frame is set.
	Equal initial covariances are assigned to all new feature positions.
	\item The filter state with the previously estimated angular rate and the new features is then propagated and updated at the next measurement instance.
\end{itemize}

Since the EKF provides uncertainty information on the tracked features, it can help to the elimination of deceiving-quality features. If a feature is detected and tracked, and it disappears from view, its covariance will start to grow. This implementation, uses an empirical threshold on the covariance of the features such that \textit{high} covariance features are dropped. In this manner, only relatively low-covariance features participate in the filtering. Additionally, drifting features were filtered out by a $ \chi^2 $ rejection scheme.

\begin{figure}[p]
	\centering
	\includegraphics[width=0.7\linewidth]{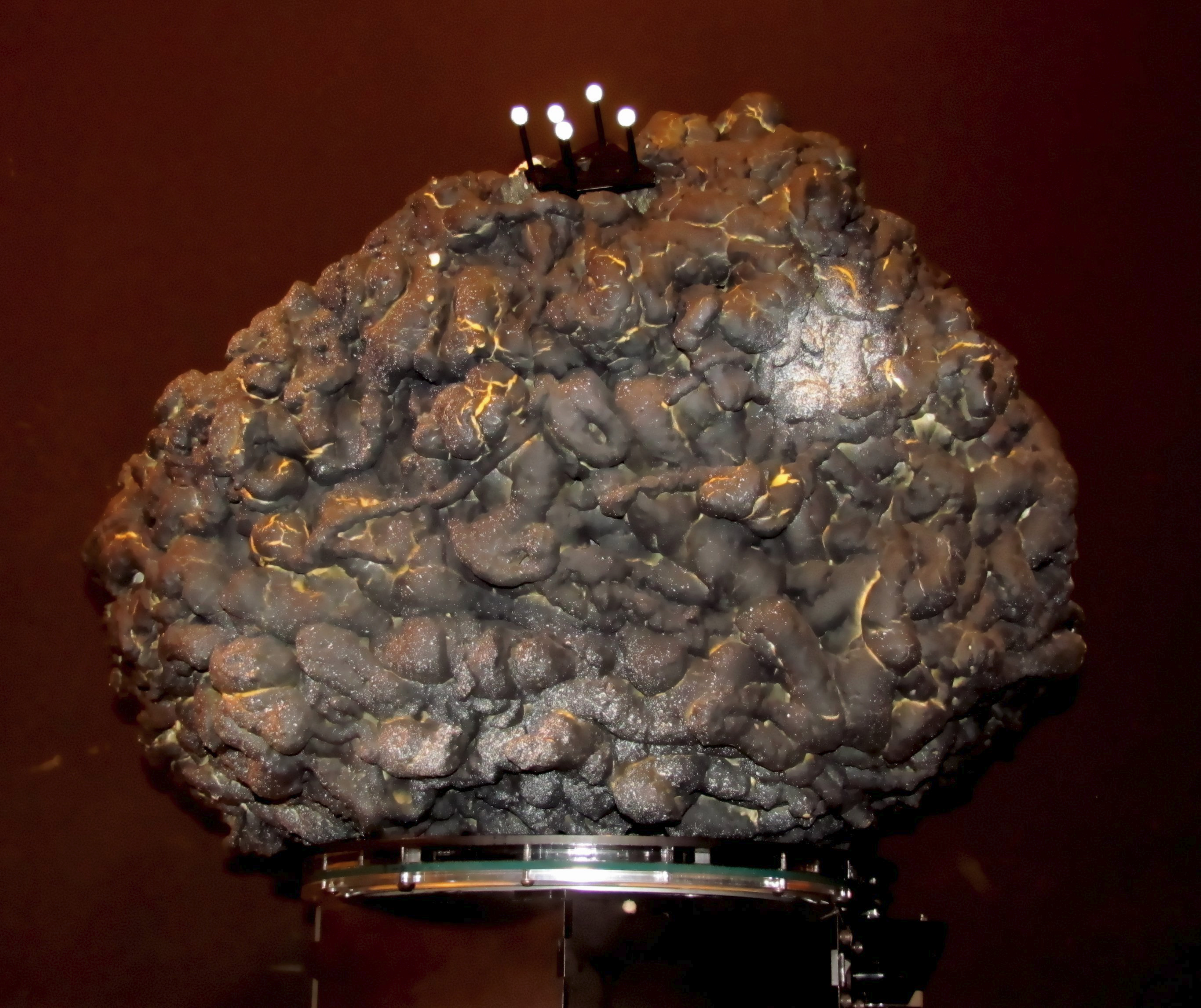}
	\caption{The mock asteroid placed on the turn table. The VICON dots can be observed placed at the top on a mounting plate. Reprinted with permission from \cite{Ramos2018}.}
	\label{fig:asteroid}
\end{figure}

\begin{figure}[p]
	\centering
	\includegraphics[width=0.7\linewidth]{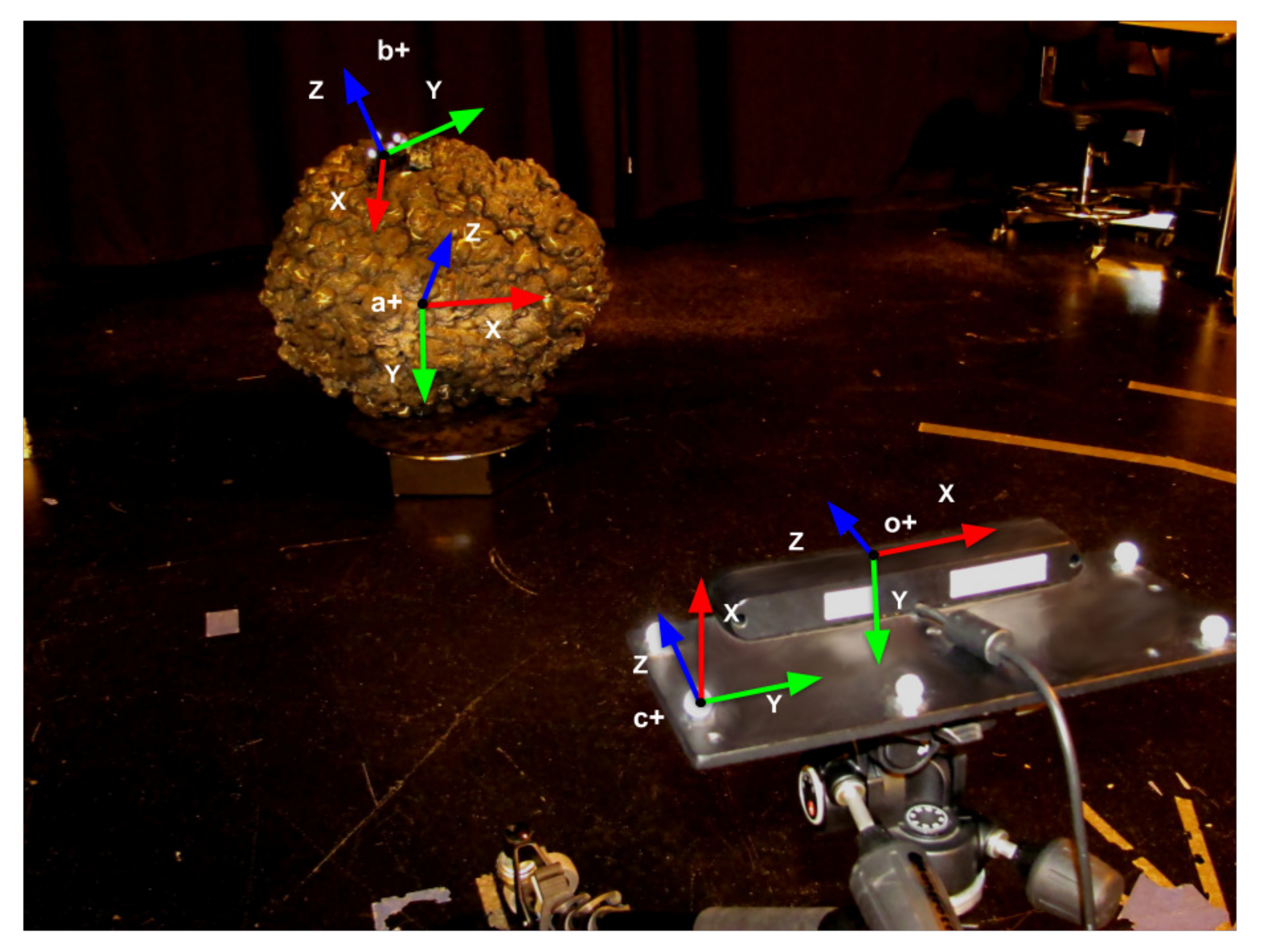}
	\caption{Xtion Live Pro camera and the mock asteroid while recording data. The arbitrary reference frame $ \ar $ (on the body), the body plate $ \bpframe $, the camera optical $ \op $ and the camera plate $ \cam $ reference frames, have been drawn. Reprinted with permission from \cite{Ramos2018}.}
	\label{fig:xtion_and_asteroid}
\end{figure}

\subsection{Hardware experiments}\label{sec:hardware experiments}
\subsubsection{Experimental setup} \label{ssec:hardware setup}
Details on the hardware experiments and validation of the vision system are now presented. A calibrated XtionPRO Live RGB-D sensor collects images of a non-cooperative body. The sensor is assumed to be on a satellite trying to detect active attitude control in a neighboring body, and it starts by estimating the body's angular rate (this application). An asteroid mockup, shown in Figure \ref{fig:asteroid}, is used as the target body for experiments. In order to impart rotational motion to the asteroid, a customized rate-configurable turn table is used. The turn table can be seen in Figure \ref{fig:asteroid} under the asteroid.
The vision system estimates for the angular rates are validated by comparing its outputs against a laboratory VICON motion capture system. VICON provides attitude measurements for the asteroid naturally resolved in the VICON $ \vicon $ frame. In order to evaluate angular velocity estimates, numerically differentiated attitude truth from VICON is used as the reference for truth angular rate. 
The VICON system tracks the body's motion by means of the plate attached to it carrying retro-reflective beacons.
Figure \ref{fig:asteroid} displays the plate with the markers placed at the top of the mockup.

The VICON system also tracks the truth pose of the RGB-D sensor. Both asteroid and RBG-D sensor are shown in Figure \ref{fig:xtion_and_asteroid}.
The reference frame on the camera plate is identified as $ \cam $. The calibration (rigid transformation) camera optical frame to the camera plate frame $ \dcm[O][C] $, as well as the RGB and IR ("depth") camera registration, are found beforehand. 
All reference frames used in the experiments are diagrammed in Figure \ref{fig:reference_frames_used}.

\subsubsection{Experiments} \label{ssec:tracking system test}
A series of experiments were performed using different rates for the turntable. For the results shown in this section, the turn table rate was set to 2 deg/s, but all other experiments with different rates (in the range from 1 to 10 deg/s) were found to behave similarly. The resolution of the RGB-D sensor is 640 x 480 pixels, and the frame-rate is 30 Hz. The RGB-D sensor is fixed on a tripod at 2 meters far from the center of the turn table. The video is processed as described in Section \ref{sec:system_overview}, and if needed, reinitialization of the system occurs automatically according to Section \ref{sssec:reinitialization}. The maximum sensing depth was limited to 3.5 meters, and the area for searching features was manually selected to cover the image area that the asteroid occupies when it rotates. The code was implemented in C++ with OpenCV 2.4.9.1 libraries under the Robot Operating System (ROS kinetic) framework \cite{quigley2009ros}.

The process and measurement noise values used to generate the results presented in the next subsections are

\begin{equation}\label{eq:process_noise_features}
\stdvec{Q}[][F] = \zero[3N][3N] \ ,
\end{equation}

\begin{equation}\label{eq:process_noise_omegas}
\stdvec{Q}[][\omega] = \zero[3][3] \ ,
\end{equation}

and
\begin{equation}\label{eq:measurement_noise_features}
\stdvec{R}[][F] = 0.005\eye{3N} \ .
\end{equation}

The propagation step $ \Delta t $ is simply taken as the inverse of the camera frame-rate, which in this case is 1/30 seconds. The selected $ \beta $ values for the partial-update are 
\begin{equation}
	\betamat = \diag\rowvec{1.0,1.0,1.0,\dots,0.05,0.05,0.05}\\ ,
\end{equation}
that is, full update for feature positions and 5\% of the nominal update for the angular rates.

\begin{figure}[htbp]
	\centering
	\begin{subfigure}[b]{0.475\textwidth}
		\centering\includegraphics[width=1\textwidth,]{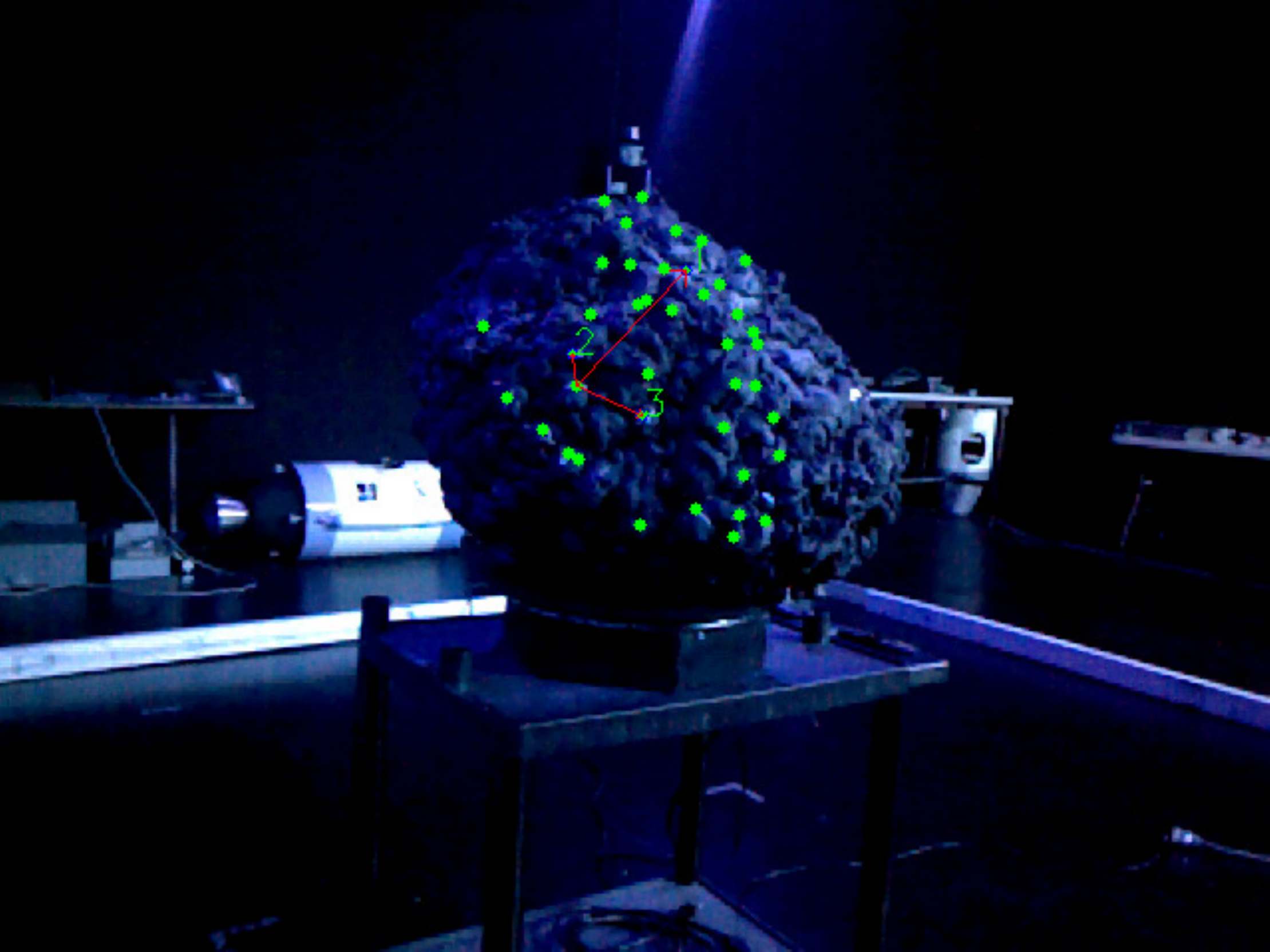}\caption{ }\label{fig:taba}
	\end{subfigure}
	\hfill
	\begin{subfigure}[b]{0.475\textwidth}
		\centering\includegraphics[width=1\textwidth]{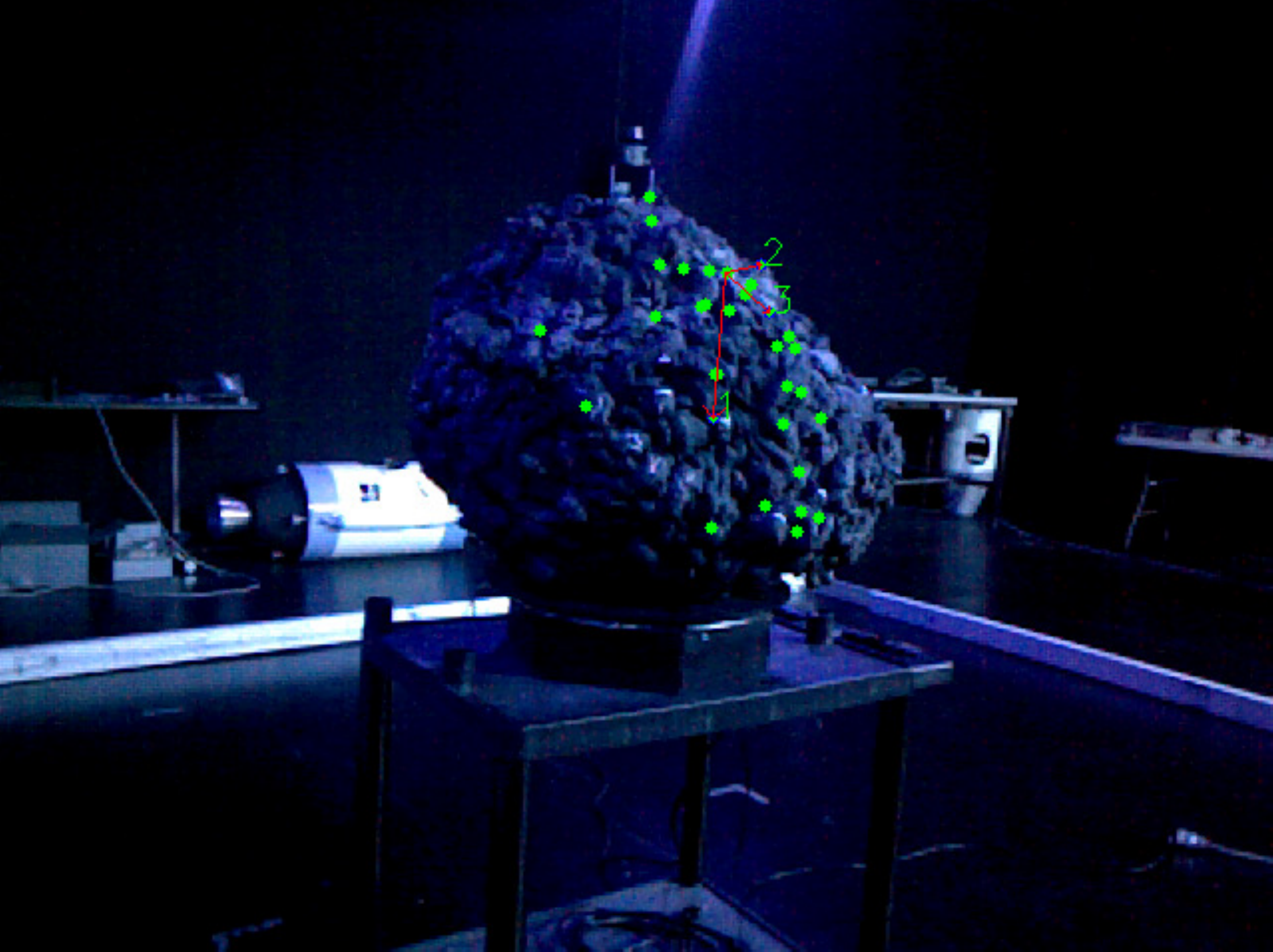}\caption{
		}\label{fig:tabb}
	\end{subfigure}
	\vskip\baselineskip 
	\begin{subfigure}[b]{0.475\textwidth}
		\centering\includegraphics[width=1\textwidth]{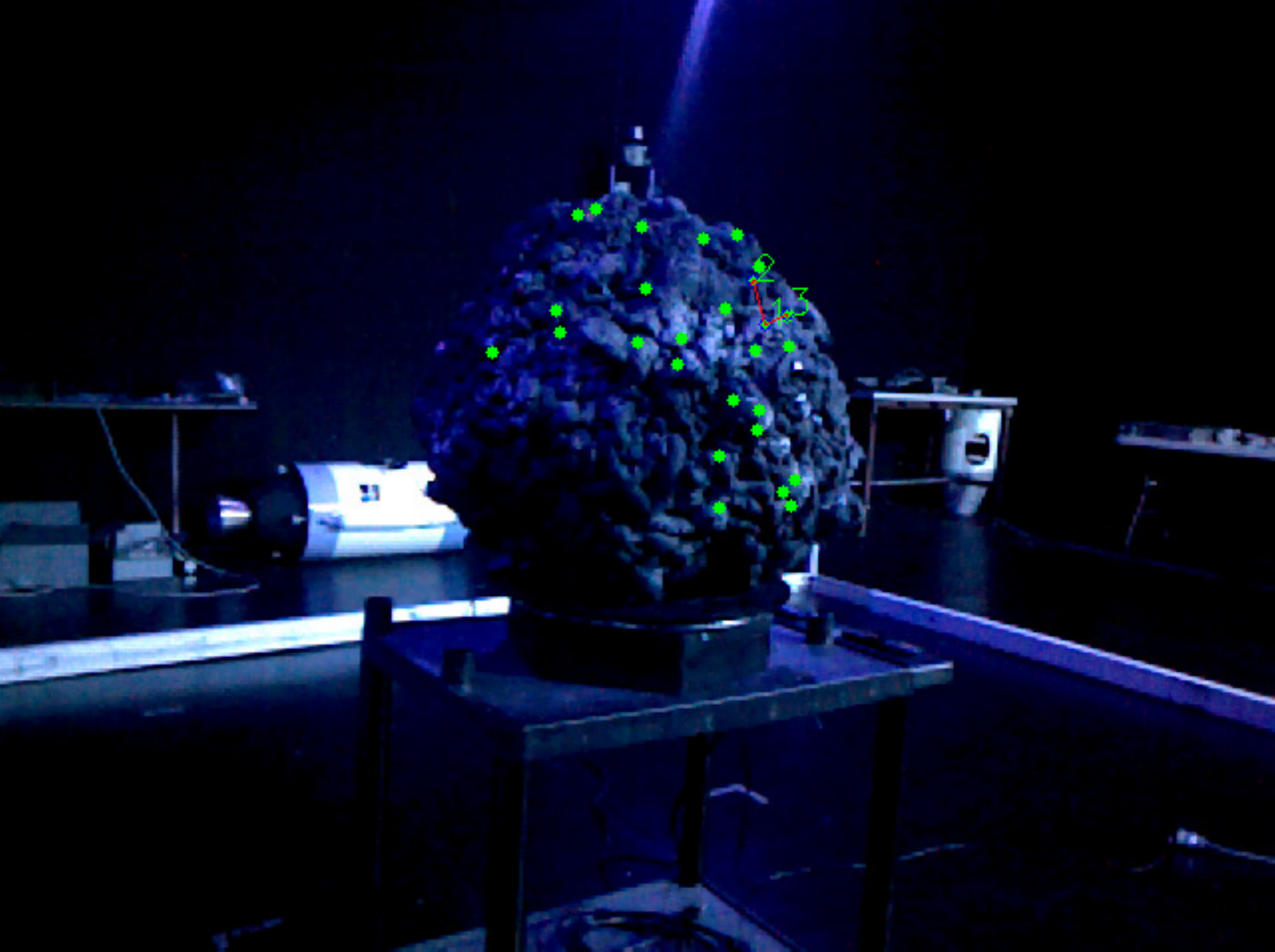}\caption{
		}\label{fig:tabc}
	\end{subfigure}
	\quad
	\begin{subfigure}[b]{0.475\textwidth}
		\centering\includegraphics[width=1\textwidth]{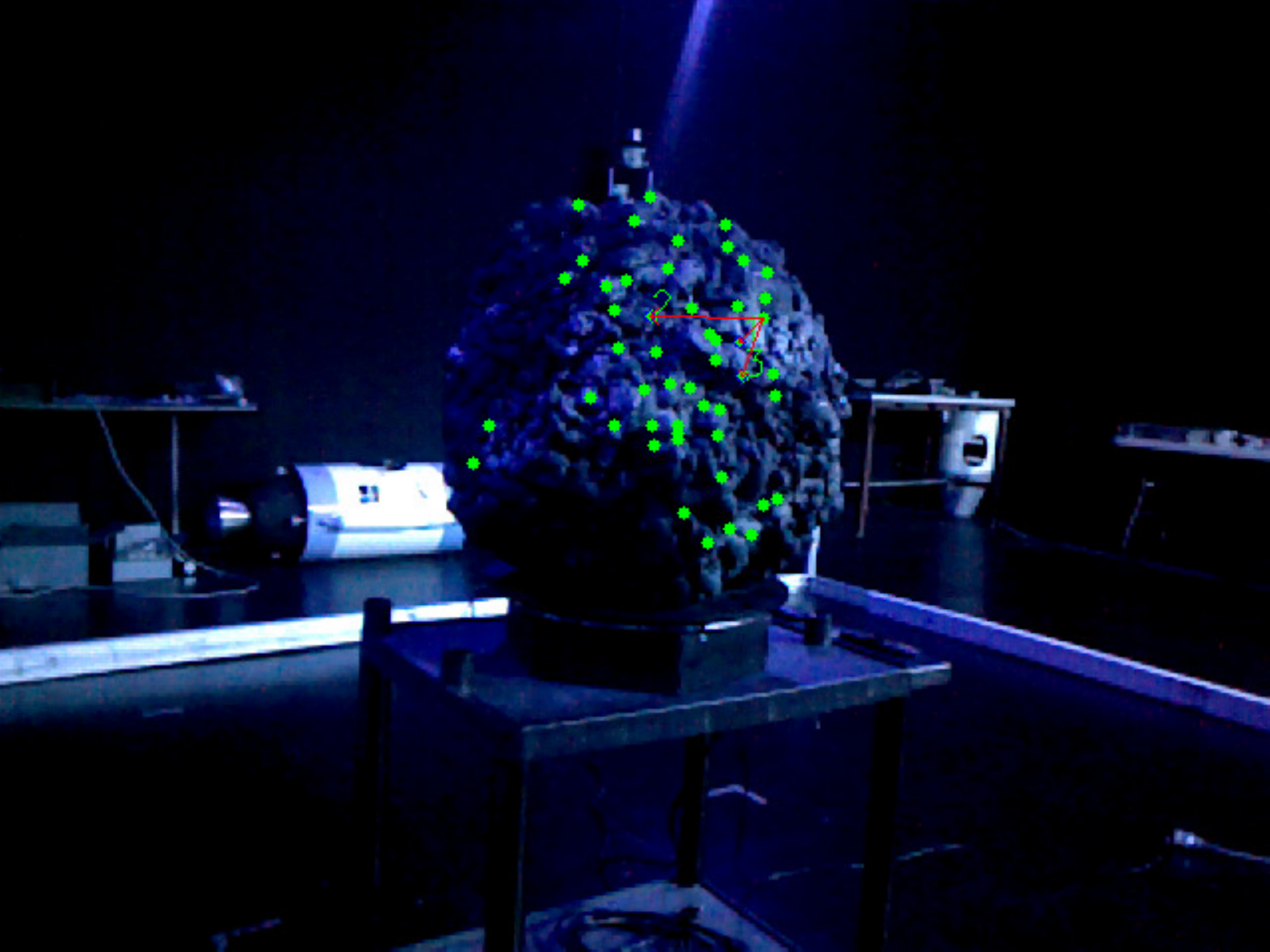}\caption{
		}\label{fig:taba2}
	\end{subfigure}
	\caption{Image sequence showing the detected features and three of the position vectors that are being tracked. Reprinted with permission from \cite{Ramos2018}.}
	\label{tab:klt_sequence}
\end{figure}

Figure \ref{tab:klt_sequence} shows a sequence of images (chronologically from a to d) from a single experiment. In each image, the good features to track are shown along with some body vectors in the current arbitrary reference frame, $ \ar $.
The detected features are marked as green dots, and the red vectors (arrows) are the estimated body vectors. The origin of the body vectors is at the arbitrary body-fixed reference frame, which corresponds to the best available feature as described in Section \ref{sec:Tracking of position vectors fixed to the target}. For clarity, vectors for the position of the estimated features are only shown for three of the features. The numbers that appear next to each body vector represent the feature quality, being the number one the highest-quality-feature after the origin.

In Figure \ref{tab:klt_sequence} it is possible to observe the vision system re-initialization being executed. Particularly, re-sampling of features occurs from (c) to (d), and a change of origin can be seen in the transition from (a) to (b). This confirms that the re-initialization detailed in Section \ref{sssec:reinitialization} behaves as expected in terms of sampling.

\subsubsection{Angular velocity estimates}
As mentioned in Section \ref{ssec:hardware setup}, to evaluate the angular velocity estimates, estimates are compared against numerically differentiated VICON attitude.

%

\begin{figure}[tb!]
	\centering
	\includegraphics[width=0.8\textwidth]{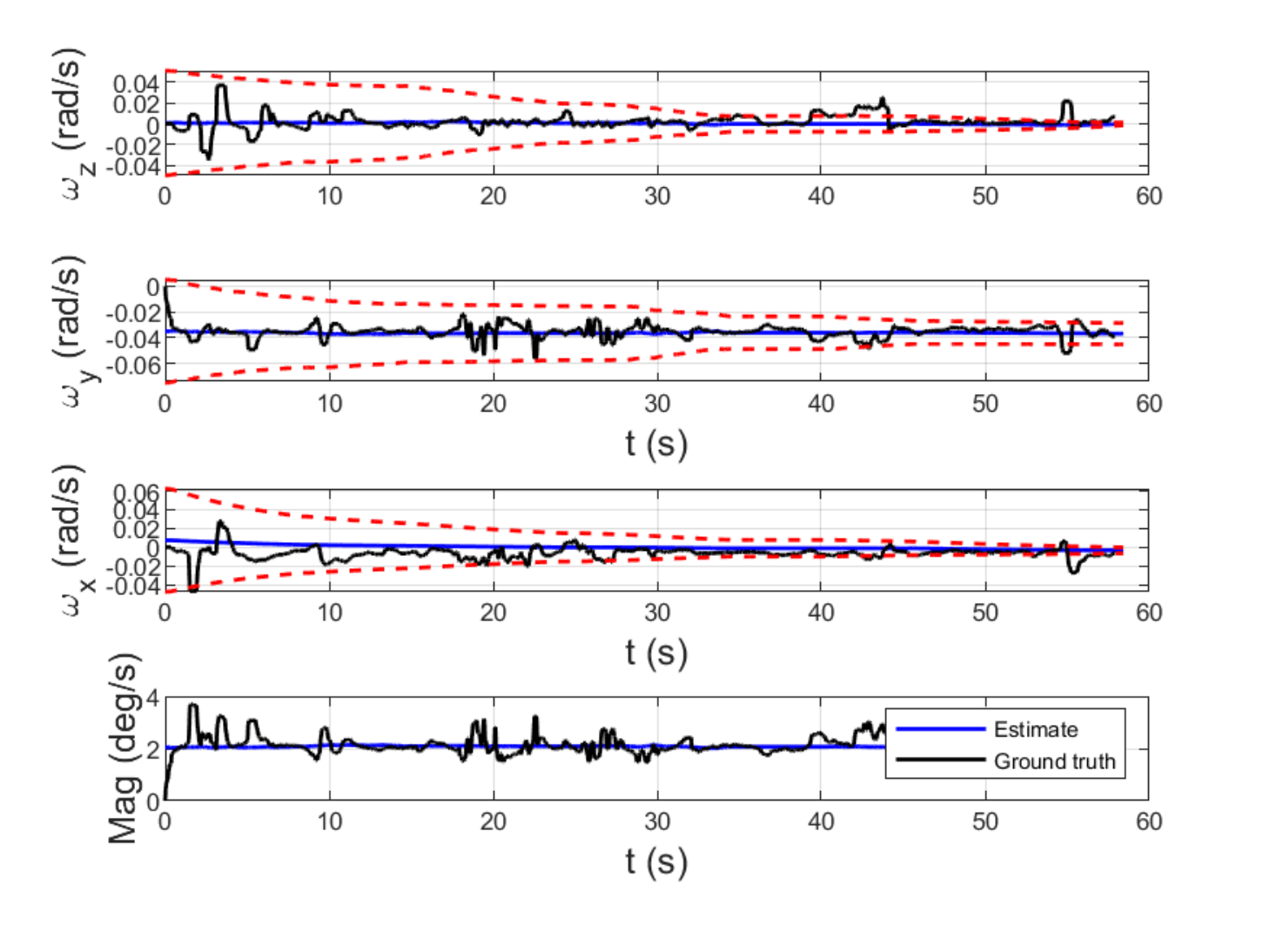}
	\caption{Estimated angular velocities for the target body when initial conditions for the angular rates are set to zero. The angular rates are resolved in the optical reference frame $ \op $. The bottom plot shows the comparison between angular velocity vector magnitudes. Reprinted with permission from \cite{Ramos2018}.}
	\label{fig:angular_rates_estimates_correct}
\end{figure}

\begin{figure}[tb!]
	\centering
	\includegraphics[width=0.8\textwidth]{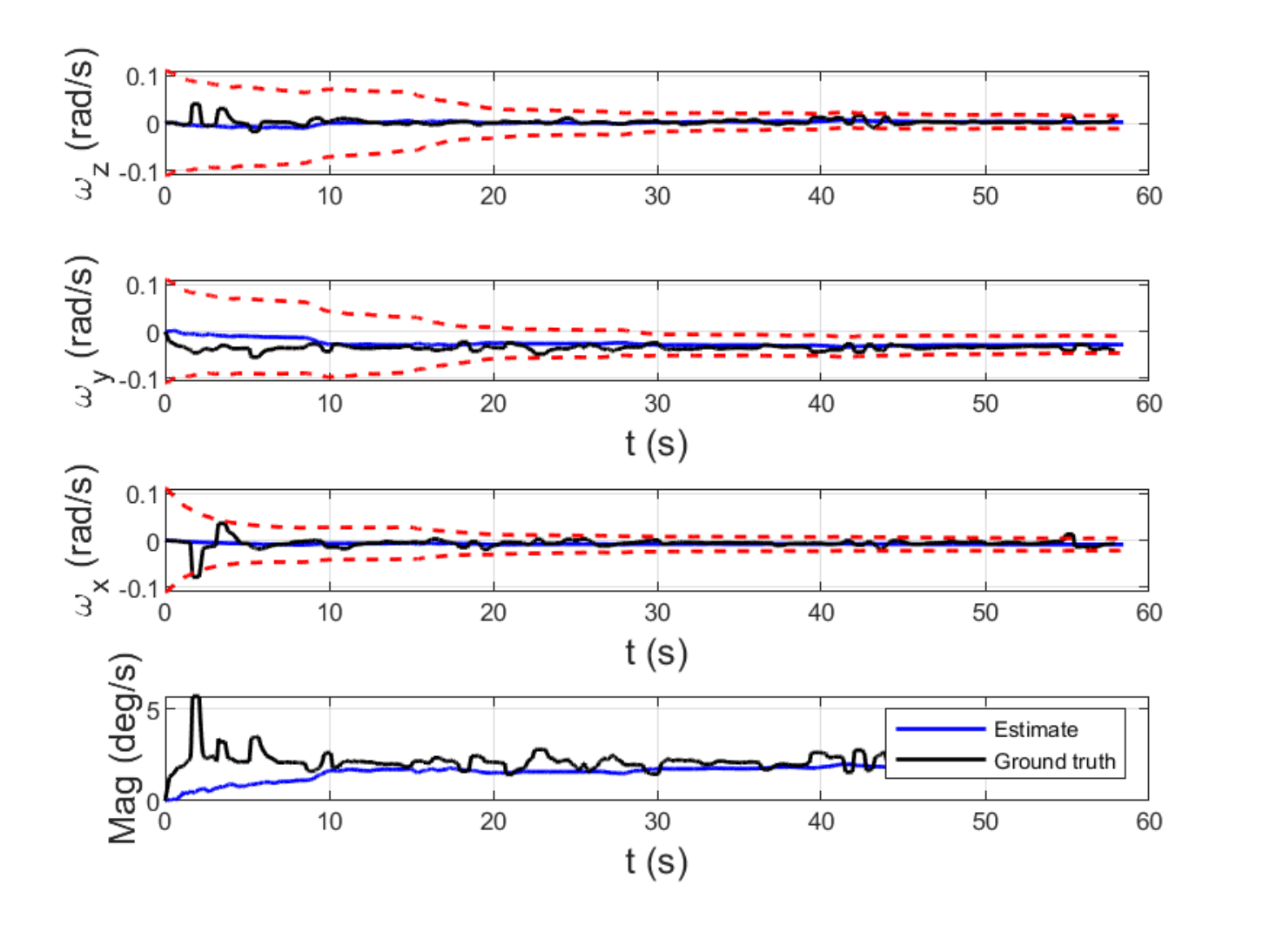}
	\caption{Estimated angular velocities for the target body when initial conditions for the angular rates are set to zero. The angular rates are resolved in the optical reference frame $ \op $. The bottom plot shows the comparison between angular velocity vector magnitudes. Reprinted with permission from \cite{Ramos2018}.}
	\label{fig:angular rates estimates zeros}
\end{figure}

Several experiments were conducted using the initial conditions generated by the initialization procedure outlined before. Overall, the filter was found to be well-behaved, and in contrast with the experiments that used the conventional EKF, the filter produced consistent and unbiased estimates. Figure \ref{fig:angular_rates_estimates_correct} shows the angular rates estimate time histories for a typical experiment. The estimates for a typical experiment when the angular rates are initialized as zeros are also included, and are depicted in Figure \ref{fig:angular rates estimates zeros} to show the UD partial-update EKF functionality.

\subsection{Summary}\label{sec:conclusions}
The hardware implementation of a vision system utilized for angular rate estimation of uncooperative bodies was described in this section. The UD partial-update approach was found to produce statistically consistent results where an EKF could not. Angular rate estimation results showed that although the UD partial-update EKF takes more time to converge, convergence is achieved, and consistent estimates are produced. In contrast with the first (with constant nuisance states) and second (varying nuisance states) partial-update filter application, this is a case where the nuisance states, namely the angular rates, are the states of interest, yet, the consistency of the results obtained via the UD partial-update filter demonstrated the power of the approach that still retains the EKF structure.

The results from all three hardware applications confirm that a partial-update EKF is a more flexible filter that accommodates real-world technical challenges in an effective and simple way. Overall, extending the EKF and Schmidt filter application capabilities, regardless of the nuisance states being constant, varying, and even if they are the states of interest.

%
%
%
%

\chapter{CONCLUSIONS \label{cha:Summary}}
In this dissertation, approaches to extend the practical applicability of the Extended Kalman filter (EKF) are presented. The proposed methods are based on a recent development called the partial-update filter, that already broadens the capabilities of the Schmidt-Kalman filter. The contributions of this work increase the partial-update filter robustness against numerical issues and high uncertainties, allowing the use of an extended Kalman filter framework where a more advanced filter may be needed, or where consistent estimates would not be obtained otherwise. Since the proposed techniques address fundamental problems within the Kalman filter, they apply widely just as the Schmidt modification, Joseph-update form, and factorized formulations for covariance propagation. Although this study focuses on the EKF, the findings can have a bearing on other filter forms where the measurement update appears linearly.

Based on the results of the partial-update factorized formulations presented in this dissertation, the UD form is recommended because it is more efficient and attractive for actual implementation. Although the factorized formulations require some extra work in terms of computer code, such code is not application-dependent. Thus, once extra code routines are available, the factorized filter implementations are not more complicated than a conventional filter formulation, and practically any user familiar with the Kalman filter can implement a partial-update factorized form. The findings show that the square root and UD partial-update formulations, increase the required computations of the conventional factorized forms due to the calculations related to partial-update terms, and this needs to be considered on the hardware selection process. However, the extra computational effort for relatively small systems is generally not significant to prevent the factorized partial-update formulations from their implementation on a system already using a conventional Kalman filter. 

This dissertation establishes two quantitative frameworks for online partial-update weight selection: The nonlinearity-aware and the covariance-aware method.  Both methods use a similar policy for update percentage selection, and although they have their virtues and limitations, the results of this research proved them to be functional and to extend the static partial-update approach's capabilities to handle high nonlinearities and uncertainties.  However, the implementation of the covariance-aware method for partial-update is recommended over the nonlinearity-aware method, especially for hardware applications where the system has a large number of parameters. Overall, this study strengthens the idea that it is beneficial to vary the amount of update applied to each element in the state vector. Furthermore, the insights gained from this study may be of assistance in developing more sophisticated methods for partial-update weight selection that can consider elements like the quality of cross-covariance terms and measurement sparsity. Since these dynamic methods are entirely based on quantities inherent to the Kalman filter, like the covariance matrix and process and measurement Jacobians, the dynamic partial-update concept can be adapted into any other Kalman filter variants. 

The deployment of the partial-update filter in real systems proved the filter functionality and higher consistency compared to the conventional Kalman filter. For most of the hardware applications where the partial-update filter has been used, a static weight was found via a trial and error tedious process. If the user intends to use the partial-update approach in any of its variants, it is recommended to implement a dynamic method to avoid weights tuning. Specifically, the UD covariance-aware dynamic method is recommended. Future work considers to investigate the methods for dynamic weight selection in hardware applications.

Although the partial-update filter has its limits, the findings of this research add to our understanding of the Schmidt filter and show the developments to augment the class of systems that the EKF structure can handle without over-specializing the filter formulation.

\let\oldbibitem\bibitem
\renewcommand{\bibitem}{\setlength{\itemsep}{0pt}\oldbibitem}
\bibliographystyle{ieeetr}

\phantomsection
\addcontentsline{toc}{chapter}{REFERENCES}

\renewcommand{\bibname}{{\normalsize\rm REFERENCES}}

\bibliography{Dissertation.bib}

%
%
%
%

\begin{appendices}
\titleformat{\chapter}{\centering\normalsize}{APPENDIX \thechapter}{0em}{\vskip .5\baselineskip\centering}
\renewcommand{\appendixname}{APPENDIX}

%
%
%
%


\phantomsection

\chapter{\uppercase {Hardware parameters }}\label{appendix:apendix_hardware}
The REEF estimator parameters used to generate the results shown in Section \ref{sec:Result} are included here. 

\textbf{NOTE:} The initial value of the Z position in the Z estimator is set at -0.25m since the sonar altimeter is mounted at a height of 0.25m from the ground. The sonar used in this application is only capable of measuring heights greater than 0.25m.
\begin{flalign*}
&\textbf{P}_{{xy}_o} = \diag \begin{bmatrix} 0.01& 0.01& 0.03& 0.02& 0.14& 0.14\end{bmatrix} \\
&\textbf{Q}_{xy} = \diag \begin{bmatrix}0 &0& 0.03& 0.03 &0.1& 0.1\end{bmatrix}  \\
&\textbf{x}_{{xy}_o}=\begin{bmatrix}0& 0& 0& 0& 0& 0\end{bmatrix}\\
&\bm{\beta}_{xy}= \begin{bmatrix}1&1& 0.01& 0.01& 0.01& 0.01\end{bmatrix}\\
&\textbf{P}_{{z}_o} = \diag\begin{bmatrix}0.025 & 1.0 & 0.09 \end{bmatrix} \\
&\textbf{Q}_{z} =  \diag\begin{bmatrix}0.03 & 0.001\end{bmatrix} \\ 
&\textbf{x}_{{z}_o}=\begin{bmatrix}-0.25& 0& 0\end{bmatrix}\\
&\bm{\beta}_{z} = \begin{bmatrix}1.0& 1.0& 0.5\end{bmatrix}\\
\end{flalign*}

%
%
%
%


\pagebreak{}

\end{appendices}

\end{document}